**FACULTY OF MATHEMATICS AND PHYSICS**
**Charles University**

MASTER THESIS

Jonáš Dujava

# Strongly Coupled Quantum Field Theory in Anti–de Sitter Spacetime

Institute of Particle and Nuclear Physics

Supervisor of the Thesis Mgr. Petr Vaško, Ph.D.
Study Programme Theoretical Physics

*I would like to express my deep gratitude to my advisor, Petr Vaško, for his guidance, patience, and invaluable advice during our collaboration.*

*Throughout my studies at Charles University I had the pleasure to interact with many great lecturers. In particular, I would like to thank Pavel Krtouš, whose style and dedication to clarity left an unmistakable imprint on my work.*

*These years would not have been the same without my friends — Matej, Marek, Tomáš (and their amazing girlfriends), Kubo, Jožko, Tomáš, Emil, Mišo, Adam, Tony, Jano, Robo, among others — whose companionship, endless conversations, and shared laughter filled them with joy and made them truly unforgettable.*

*I am especially grateful to my parents and family for their unwavering support, love, and encouragement throughout my academic journey and my whole life.*

*Miška, thank you for standing by my side and for bringing light and encouragement even in the most challenging moments.*

*Jonáš Dujava*

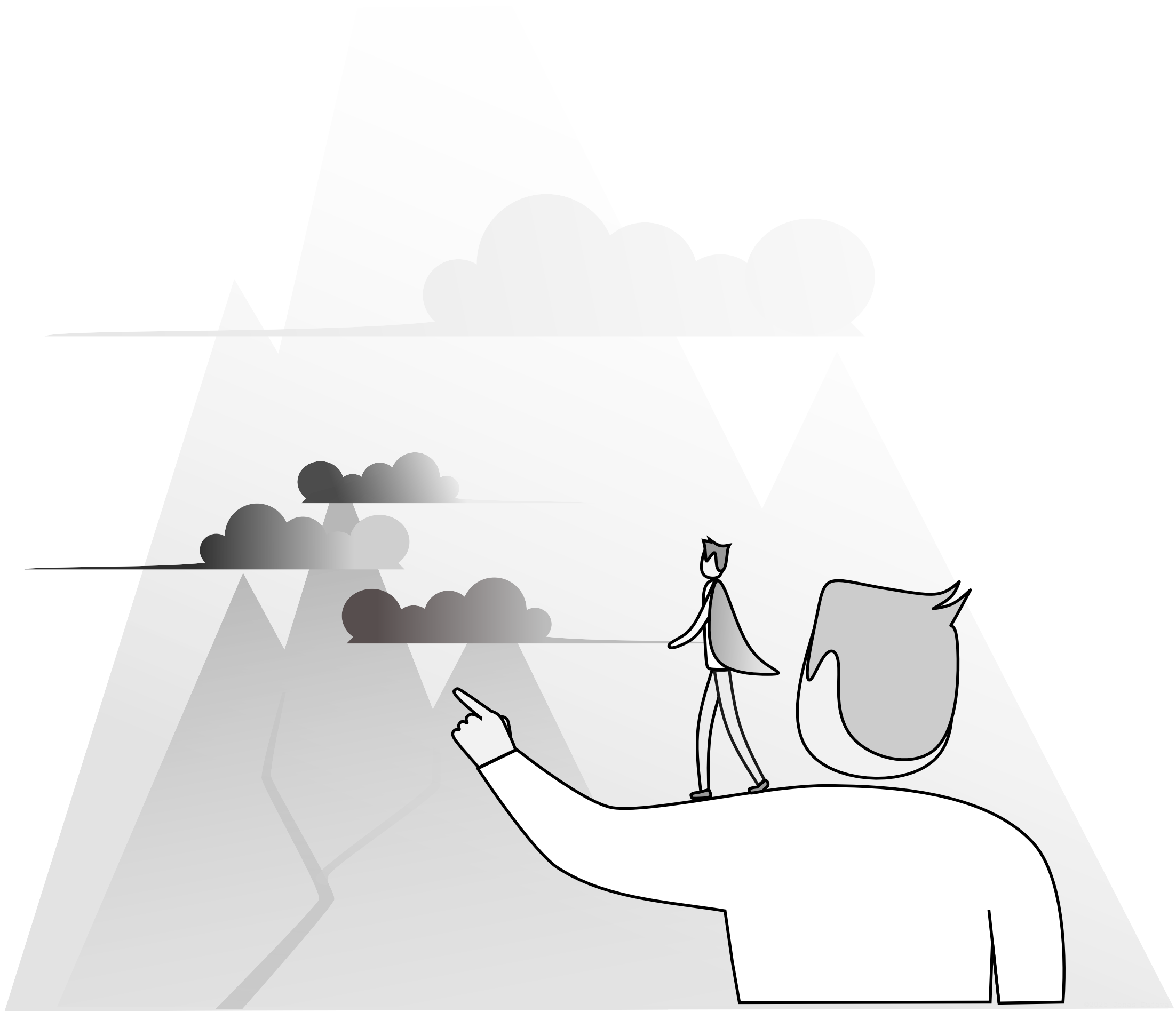

## Information

## Abstract

We introduce the framework of Quantum Field Theories in general backgrounds through the lens of the path integral, in the formulation known as the Functorial QFT. With the aim of studying properties of strongly coupled QFTs, we present key concepts and techniques such as the effective action, renormalization group flow, and the $1/N$ expansion.

At the fixed points of the RG flow we typically find Conformal Field Theories, which are symmetric under local rescaling of the metric. Among other things, CFTs play a central role in the context of QFT in Anti–de Sitter spacetime, through what is known as the AdS/CFT correspondence.

Finally, the developed formalism is applied to analyze the O($N$) model in AdS at finite coupling. In particular, we focus on the non-singlet sector, which requires an understanding of crossed-channel diagram contributions to the $s$-channel conformal block decomposition.

# Contents

# Introduction

Quantum Field Theory (QFT) stands as one of the cornerstones of modern theoretical physics. Its reach extends across a broad spectrum of disciplines, including particle physics, statistical mechanics, condensed matter physics, and even gravitational physics.

While its principles and methods are well-established and widely used, QFT still lacks a fully rigorous and comprehensive mathematical foundation. This becomes particularly evident when one attempts to formulate QFT in general curved spacetimes — an essential step for connecting it with cosmology and eventually *Quantum Gravity*.

Another major challenge lies in understanding *strongly coupled* regimes of QFTs, where standard perturbative techniques fail. In these cases, exact results are rare, and progress relies heavily on developing and studying special models. Such simplified yet nontrivial examples provide vital intuition and often serve as laboratories for testing and refining general ideas.

Throughout all of this, *symmetry* plays a central role. It not only guides the formulation of physical theories, but also provides strong constraints that make otherwise intractable problems manageable. Some symmetries can be emergent, as is the case for second order phase transitions, where the system becomes scale invariant. Moreover, they usually enjoy a further symmetry enhancement — they become conformally invariant, which means we can make even position-dependent scale transformations. Such critical point are described by Conformal Field Theories (CFTs), which due to their enlarged symmetry group admit powerful, non-perturbative techniques.

One of the most interesting features a physical/mathematical model or theory can have is the presence of a *duality* — see Figure 1 →p.vii. Different properties of the theory are often more naturally seen from different perspectives, which can offer us new intuitions, insights, or computational tools. One of the most famous examples is the AdS/CFT correspondence, which attempts to relate a Quantum Gravity in asymptotically AdS spacetime to a CFT on its boundary.

The final part of this thesis focuses on the study of the $\mathsf{O}(N)$ model in anti–de Sitter (AdS) spacetime. As it admits a large $N$ expansion, allowing us to compute at finite coupling, the $\mathsf{O}(N)$ model provides a fertile ground for exploring fundamental questions in QFT. Considering it in AdS spacetime, we can also leverage the AdS/CFT correspondence — a QFT in a fixed AdS background admits asymptotic observables, in particular boundary correlation functions, which satisfy essentially all CFT axioms. This leads to a rich interplay between standard QFT and CFT techniques, and allows us to study finite spectrum deformations of interacting CFTs.

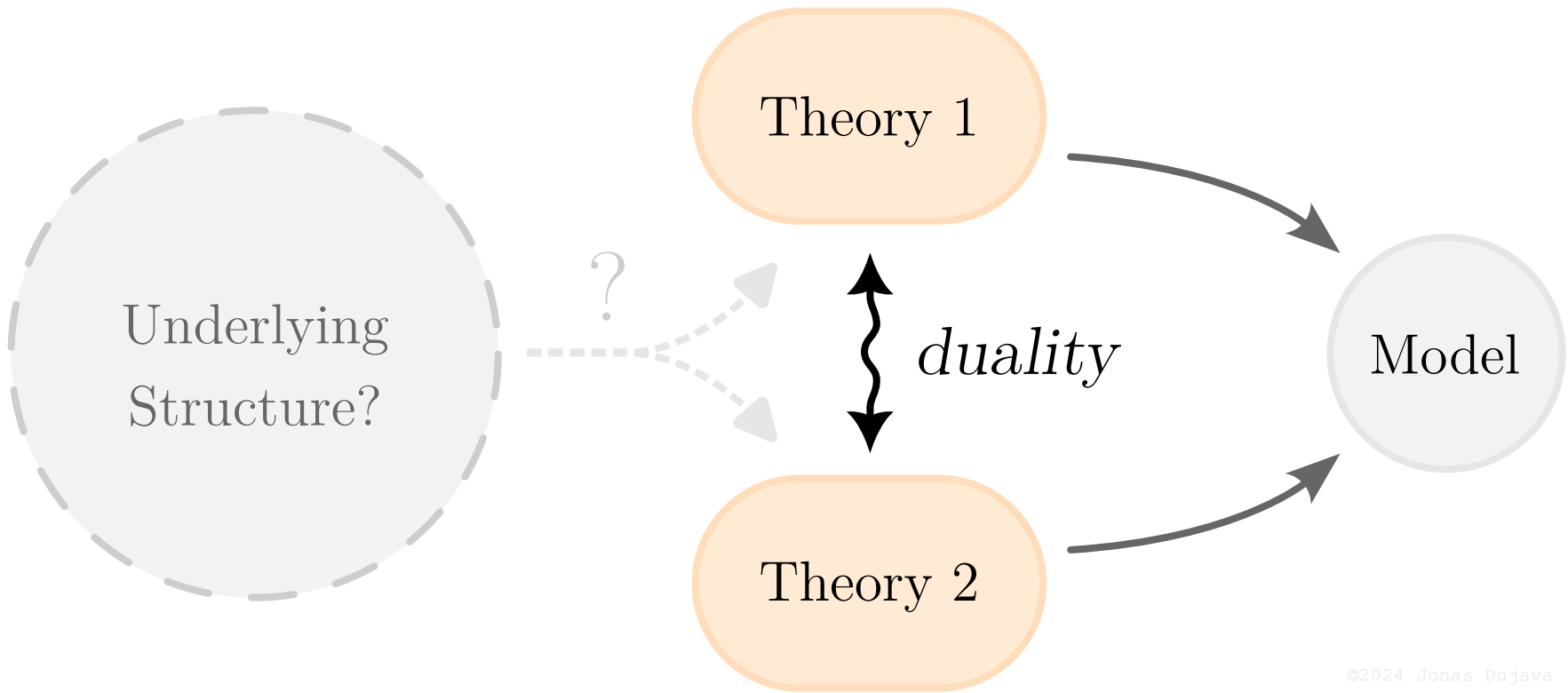

**Figure 1 /** A model (or its part) can sometimes be described by two distinct theories, with different starting principles and associated methods/techniques. Such duality hints at a possible underlying structure.

**Remark 2** (On Exposition)**.** The subject of QFTs can be hard to approach, not least due to the plethora of concepts and techniques one needs to know intimately before conceptual clarity can eventually be achieved. One of the objectives of this work is to provide — as much as possible — a self-contained account of various important topics that lie slightly beyond the standard curriculum of QFT.

While not all of the discussed aspects are used directly in the "application part" of this work — namely Chapter 5→p.100 — they serve to highlight and appreciate the interconnections between various notions and methods, which might otherwise seem unmotivated or *ad hoc*. ⌟

**Remark 3** (On Rigor)**.** The author's intention was to produce a clearly structured text, making it easy to locate and refer to almost any introduced concept. To this end, a "definition–theorem–proof–remark" structure was adopted, even though the work should not be considered fully mathematically rigorous.

As a byproduct, the structure makes it rather transparent which parts are more established and which remain more heuristic — whether due to gaps in the author's understanding, the absence of a rigorous mathematical underpinning, or simply because the necessary background would be too extensive to include.

This is considered a feature, since it avoids (at least partially) the pitfall of giving a false impression of understanding. A related and interesting read about the interaction between theoretical physics and mathematics can be found in [2]. ⌟

**Remark 4** (On Numbering)**.** To simplify the navigation, all types of document constructs — for example Definitions, Remarks, Figures, Theorems, and so on — are numbered within the same counter (inside the given chapter). To give you an idea, Figure 3.18→p.57 is shortly after Example 3.15→p.56. ⌟

# Quick Summary

In **Chapter 1** we briefly introduce the QFT framework, mainly through the lens of path integral. Its formalization is dubbed as the Functorial QFT (FQFT).

In **Chapter 2** we look at methods important to understand QFT at large — the notion of effective actions, and the renormalization group (RG) flow.

In **Chapter 3** we dive deep into the realm of Conformal Field Theories (CFTs), which usually appear at the fixed points of the RG flow. Apart from the scaling symmetry, they enjoy further symmetry enhancement to the conformal group.

In **Chapter 4** we discuss QFT on Anti–de Sitter spacetime (AdS), and the connection to CFTs through the AdS/CFT correspondence.

In **Chapter 5** we apply previously developed formalism to study the $\mathsf{O}(N)$ model in AdS at finite coupling, in particular its non-singlet spectrum.

We summarize the overarching narrative in **Conclusion**.

In **Appendix N** we collect some of the conventions and notation used throughout the work, mainly of various differential geometry objects.

# 1 Quantum Field Theory Formalism

We start by introducing some general aspects of QFT formalism, and in particular its formulation in general backgrounds. The most straightforward motivation is the fact that our Universe is not flat, so understanding effects of the spacetime curvature is required to study physics at cosmological scales, or in the presence of strong gravitational fields, for example near black holes.

Treating the spacetime as a classical static background may be viewed as an intermediate step towards *Quantum Gravity*. A proper understanding of all implications in such semi-classical approximation may contain valuable hints for the full theory.

Understanding QFT in general backgrounds can also bring other benefits, both of conceptual and computational nature. It allows us to consider models on manifolds with boundary or non-trivial topology, and enlightens the fundamental role of spacetime symmetries. Having such general formulation we can make statements about (and relate) theories in different backgrounds. In particular, we can get a better feeling for the connection between Lorentzian and Euclidean QFTs, the former describing relativistic particle or condensed matter systems, and the latter describing statistical systems (at large scales).

We will take the (heuristic) path integral approach as our starting point, and briefly introduce it in Section 1.1→p.5. Then we formalize the properties we usually require from the path integral into an axiomatic framework in Section 1.2→p.13. Symmetries in context of QFTs are treated in Section 1.3→p.22. As discussed in Section 1.4→p.29, they turn out to be central for building the usual "operator formalism" obtained by appropriately foliating the spacetime.

## 1.1 Path Integral Approach to QFT

A classical field theory describes a classical theory with infinite degrees of freedom. It is usually defined by a local action functional, which uniquely determines the evolution of the fields given the initial conditions.

Going to the quantum realm, that is considering QFT, the possibilities are much more diverse. Intuitively, the Feynman's "sum over histories" motto [3] says that we should integrate over all possible (spacetime) configurations of the field, with an appropriate measure depending on the action functional.

Performing such "path integral" exactly in a nontrivial interacting theory is unfeasible, so we often fall back to the perturbation theory. Various features are sometimes clearer by doing calculations in the Euclidean signature, and then analytically continuing the results back to the original Lorentzian case.

This section is inspired by [4] [5] [6] [7] [8], where you can find more details.

**Remark 1.1** (Attitude Towards the Path Integral). We view path integral as a powerful heuristic tool, which can be made mathematically rigorous only in some cases — in $d = 0$ it is just an ordinary (multidimensional) integral, in higher dimensions we obtain infinite-dimensional integration. We will not delve deeper concerning (possibly) rigorous definitions of path integral, but rather focus on the physical intuition behind the path integral. ⌟

## Classical Field Theory

Classical field theory is usually formulated by dynamical fields living on the spacetime manifold, whose dynamics is prescribed by the action functional. We will review the usual definitions to prepare for following sections, mainly to fix notations and consolidate understanding of the initial setup. For a more mathematically thorough account, see for example [6].

**Definition 1.2** (Spacetime Manifold, Background Fields). The field theory is formulated on a smooth manifold $\mathcal{M}$, which we call the *spacetime*. We allow a possibility of $\mathcal{M}$ having a *boundary*, or being endowed with an additional structure such as *orientation*, *metric*, or *spinor structure*.

Such extra structure is often understood as being part of the non-dynamical *background fields*, which also includes other background *sources*.

**Remark 1.3.** We usually assume orientability, and a metric usually of Lorentzian or Euclidean signature. Sometimes we explicitly write such structure as $(\mathcal{M}, \boldsymbol{g})$ for a pseudo-Riemannian manifold, or we implicitly include it in $\mathcal{M}$. ⌟

**Example 1.4.** For simplicity, most of the time we work with spacetimes enjoying maximal symmetry, examples being (with dimension $D \equiv \dim \mathcal{M}$):

- *Lorentzian signature*: Minkowski flat space $\mathsf{Mink}_D \equiv \mathbb{R}^{1,D-1}$, de Sitter $\mathsf{dS}_D$, Anti–de Sitter $\mathsf{AdS}_D$,
- *Euclidean signature*: Euclidean flat space $\mathbb{R}^D$, sphere $\mathbb{S}^D$, hyperboloid $\mathbb{H}^D$.

For Classical Mechanics (CM) or Quantum Mechanics (QM) we have $D = 1$, that is $\mathsf{Mink}_1 \simeq \mathbb{R}^1$ — there is just the (affine) *time*. ⌟

**Definition 1.5** (Dynamical Fields, Field Configurations). The space of *dynamical fields* (field configurations) on a manifold $\mathcal{M}$ is denoted by $\mathcal{Fields}_{\mathcal{M}}$. They are usually (smooth) sections of a given *field* fiber bundle $E \to \mathcal{M}$. We will collectively denote them by $\boldsymbol{\phi} \in \mathcal{Fields}_{\mathcal{M}}$.

**Example 1.6** ($\sigma$-Model, Real Scalar Field). A so-called $\sigma$-model field is given by a map $\phi\colon \mathcal{M} \to X$ for some auxiliary *target* manifold $X$. For a real scalar field we simply have $E = \mathcal{M} \times X$ with $X = \mathbb{R}$, whose sections are indeed functions

$C^\infty(\mathcal{M})$. An $N$-tuple of real scalar fields is obtained by taking $X = \mathbb{R}^N$. ⌟

**Remark 1.7** (Type of Field)**.** Typically, a field theory contains several fields $\boldsymbol{\phi} \equiv \{\phi_i\}$ with the corresponding field bundle $E = \bigtimes E_i$ being a fiber product.

Roughly speaking each $E_i$ decomposes into an intrinsic part times and extrinsic part. The extrinsic part is associated to the target values of the field, and the intrinsic part is associated to the principal frame bundle of $\mathcal{M}$ via a representation of GL (or Spin). This representation determines the *type* of the field, and in particular the transformation of its components under the change of frame/coordinates. For example, a trivial representation gives us a scalar field, a vector/spinor representation gives us a vector/spinor field, and so on. ⌟

**Example 1.8** (Gauge Theories)**.** In gauge theories, the basic field we consider is a connection on a principal $G$-bundle $P \to \mathcal{M}$ for some specified Lie group $G$. The matter fields are then additionally specified by a choice of representation under the gauge group $G$ — they are sections of some *associated bundle*. ⌟

**Definition 1.9** (Action, Lagrangian Density)**.** An *action* $\mathcal{S}_\mathcal{M}$ is a real functional of field configurations, that is $\mathcal{S}_\mathcal{M} \colon \mathit{Fields}_\mathcal{M} \to \mathbb{R}$. The *locality* of the theory is expressed by the form of the action

$$\mathcal{S}_\mathcal{M}[\boldsymbol{\phi}] = \int_\mathcal{M} \mathfrak{L}[\boldsymbol{\phi}] = \int_\mathcal{M} \mathfrak{L}(\phi, \nabla\phi, \ldots)$$

as a spacetime integral of a local functional $\mathfrak{L}$ called *Lagrangian density*, which at a given point $m \in \mathcal{M}$ depends only on a finite-order jet of $\boldsymbol{\phi}$ at $m$ (usually 1-jet) — just the values and derivatives of $\boldsymbol{\phi}$ at $m$. The dependence on background fields (such as the metric) is kept implicit.

**Remark 1.10** (Lagrangian)**.** We often factor out the metric density as $\mathfrak{L} \equiv \mathfrak{g}^{1/2} \mathcal{L}$, and work with the "Lagrangian" $\mathcal{L}$. ⌟

**Remark 1.11.** Strictly speaking, when $\mathcal{M}$ is not compact, $\mathcal{S}_\mathcal{M}$ will usually diverge. We should then work on some domain $\Omega \subset \mathcal{M}$. We will ignore this subtlety and often just write $\mathcal{S}$ instead of $\mathcal{S}_\mathcal{M}$, keeping the domain/manifold implicit. ⌟

**Remark 1.12** (Covariant Lagrangian Description)**.** Not all field theories admit a Lagrangian description, but we will assume it for the time being. Moreover, we assume a possibility of *covariant* formulation in a general background, that is for a diffeomorphism $f \colon \mathcal{M} \to \mathcal{M}'$ we have

$$\mathcal{S}_{(\mathcal{M},\boldsymbol{g})}[\boldsymbol{\phi}] = \mathcal{S}_{(\mathcal{M}',f_*\boldsymbol{g})}[f_*\boldsymbol{\phi}]\,,$$

where $\boldsymbol{g}$ labels all background fields, and $f_* \equiv (f^{-1})^*$ is the pushforward by $f$. ⌟

**Example 1.13** (Action of Free Scalar Field)**.** A *free* real scalar field $\phi$ has a quadratic

action (assuming Lorentzian signature)

$$\mathcal{S}_{\mathcal{M}}[\phi] = \int_{\mathcal{M}} \mathfrak{g}^{1/2} \Big( -\tfrac{1}{2} \boldsymbol{\nabla}_{\boldsymbol{a}} \phi \boldsymbol{\nabla}^{\boldsymbol{a}} \phi - \tfrac{1}{2} m^2 \phi^2 - \tfrac{1}{2} \xi \mathcal{R} \phi^2 \Big) ,$$

where the $m \geq 0$ is the "mass", and $\xi$ controls the coupling to the spacetime curvature through the Ricci scalar $\mathcal{R}$. Everything is written in terms of "geometric" objects, so the covariance Remark 1.12→p.3 is manifest. Since $\phi$ is a scalar, we could also write $\boldsymbol{\nabla} \phi = \boldsymbol{\partial} \phi \equiv \mathbf{d} \phi$, as it does not depend on the choice of metric connection. ⌟

**Remark 1.14** (Condensed DeWitt Notation)**.** To simplify the expressions, we will sometimes use the condensed "DeWitt notation", For example, the action of a free scalar field in Example 1.13→p.3 generalized to multiple flavors $\boldsymbol{\phi} = \{\phi^{\bullet}\}$ can be written as (after integrating by parts once and neglecting the boundary term)

$$\mathcal{S}_{\mathcal{M}}[\boldsymbol{\phi}] = -\tfrac{1}{2} \boldsymbol{\phi} \bullet \mathcal{F} \bullet \boldsymbol{\phi} , \quad \text{with } \mathcal{F} \equiv (-\Box + m^2 + \xi \mathcal{R}) \mathbf{1} ,$$

where $\Box \phi \equiv \boldsymbol{\nabla}^2 \phi \equiv \boldsymbol{g}^{\boldsymbol{ab}} \boldsymbol{\nabla}_{\boldsymbol{a}} \boldsymbol{\nabla}_{\boldsymbol{b}} \phi = \mathfrak{g}^{-1/2} \boldsymbol{\partial}_{\boldsymbol{a}} \Big( \mathfrak{g}^{1/2} \boldsymbol{g}^{\boldsymbol{ab}} \boldsymbol{\partial}_{\boldsymbol{b}} \phi \Big)$ is the d'Alembertian.

The dot "$\bullet$" denotes the contraction of both the "spacetime index" — representing integration over the manifold (when needed we include $\mathfrak{g}^{1/2}$) — and also possible flavor indices and/or field components. The differential operator $\mathcal{F}$ is understood to act on functions through the convolution with the corresponding kernel

$$\mathcal{F}(x,y) \equiv \Big[ (-\Box + m^2 + \xi \mathcal{R}) \mathbf{1} \Big] (x,y) \equiv (-\Box_x + m^2 + \xi \mathcal{R}) \delta(x,y) ,$$

where the $\delta$-distribution is appropriately normalized such that the identity operator $\mathbf{1}$ really acts on functions as an identity. ⌟

**Remark 1.15** (Conformal Coupling)**.** If we have a massless scalar field $m^2 = 0$, the choice of $\xi = \frac{1}{4} \frac{D-2}{D-1}$ (where $\dim \mathcal{M} \equiv D$) is called the *conformal coupling* — we say that we have a theory of a *conformally coupled scalar field*.

This refers to the fact that such theory is invariant with respect to Weyl transformations (see Definition 3.3→p.53) of the form

$$\boldsymbol{g} \mapsto \Omega^2 \boldsymbol{g} , \quad \phi \mapsto \Omega^{-\frac{D-2}{2}} \phi .$$

Invariance of $\mathcal{S}_{\mathcal{M}}$ under such transformation can be verified by utilizing an appropriate formula for the transformation of $\mathcal{R}$, or by computing the energy–momentum tensor Definition 1.74→p.25

$$\boldsymbol{T}_{\boldsymbol{ab}} = \boldsymbol{\nabla}_{\boldsymbol{a}} \phi \boldsymbol{\nabla}_{\boldsymbol{b}} \phi - \tfrac{1}{2} \boldsymbol{g}_{\boldsymbol{ab}} (\boldsymbol{\nabla}_{\bullet} \phi)^2 + \xi (\boldsymbol{G}_{\boldsymbol{ab}} + \boldsymbol{g}_{\boldsymbol{ab}} \Box - \boldsymbol{\nabla}_{\boldsymbol{a}} \boldsymbol{\nabla}_{\boldsymbol{b}}) \phi^2 ,$$

which can be checked to satisfy $\boldsymbol{T}^{\boldsymbol{a}}_{\boldsymbol{a}} \overset{\text{EL}}{\sim} 0$ for the chosen value of $\xi$. ⌟

**Definition 1.16** (Dynamics, Euler–Lagrange Equations Of Motion, On-Shell). Classical fields are governed by the *Euler–Lagrange* equations of motion given by the variational principle, which requires the stationarity of the action — $\delta\mathcal{S}_{\mathcal{M}} \overset{!}{=} 0$ — with some appropriate boundary conditions on $\partial\mathcal{M}$.

When we require fields to satisfy the Euler–Lagrange equations, we say that we are *on-shell*. We write $A \overset{\mathrm{EL}}{\sim} 0$ for a quantity $A$ which vanishes on-shell.

**Remark 1.17** (Variational Bicomplex). The formalism is naturally expressed using the *variational bicomplex* $\Omega^{\bullet,\bullet}_{\mathrm{loc}}(\mathcal{M} \times \mathit{Fields}_{\mathcal{M}})$. More details can be found in [6]. ⌟

**Example 1.18** (Euler–Lagrange EOM of Free Scalar Field). Using the symmetry of $\mathcal{F}$ in Remark 1.14 →p.4 (ignoring boundary terms), we have

$$\delta\mathcal{S}_{\mathcal{M}} = -\delta\phi \bullet \mathcal{F} \bullet \phi \overset{!}{=} 0 \quad \xRightarrow{\delta\phi \text{ arbitrary}} \quad \mathcal{F} \bullet \phi \equiv \left(-\Box + m^2 + \xi\mathcal{R}\right)\phi = 0\,.$$

⌟

## Path Integral Measure

In path integral approach to quantization, we imagine all possible (off-shell) evolutions of fields (with given boundary conditions), and sum/integrate over them with a complex functional "measure" written formally as

$$\int_{\mathit{Fields}} \mathcal{D}\phi \; \mathrm{e}^{\frac{i}{\hbar}\mathcal{S}[\phi]} \,\text{⬚}\,.$$

The classical field theory is then recovered in the "classical limit" $\hbar \to 0$, since by a stationary-phase/saddle-point argument, the path-integral contributions concentrate around the classical solution $\phi_{\mathrm{cl}}$ of the Euler–Lagrange equations Definition 1.16 →p.5, which are precisely the stationary condition of $\mathcal{S}$.

Before introducing some fundamentals of path integral in QFT, let us make some remark about the "path-integral measure" itself.

**Remark 1.19** (Path Integral in QM). In Quantum Mechanics (QM), the path integral can be made rather rigorous by using the theory of Gaussian measures on *Abstract Wiener spaces*. The idea is to work in Euclidean signature — to be discussed in Wick Rotation →p.12 — and discretize the time interval. We then define *cylindrical* functions depending only on finite number $N$ of points/times, and measure on them (just a finite-dimensional Lebesgue measure). Finally, we take the continuum limit $N \to \infty$, which however is less straightforward than one might initially think.

The reason is that neither $\mathcal{D}\phi$, nor $\mathrm{e}^{-\frac{1}{\hbar}\mathcal{S}[\phi]}$, exist in the continuum limit. Concerning the measure, see next Remark 1.20 →p.6. Concerning the action, we can clearly see that continuum limit of cylindrical function need not be smooth/differentiable, and thus the action is not well-defined.

It actually turns out that differentiable functions form a measure-zero set, and

↳ most important are nowhere differentiable continuous functions. This is closely tied to the non-commutativity of QM, since otherwise commutators would be trivial. So in the end we have some measure on *continuous* functions, even though we naively started with differentiable $\mathcal{F}ields$. ⌟

**Remark 1.20** (No Lebesgue Measure in Infinite-Dimensional Spaces)**.** There is no "Lebesgue measure" in infinite-dimensional Banach spaces, by which we mean it has the following properties — every point has a neighborhood of positive measure, every non-empty open set has positive finite measure, and it is translationally invariant. This is because we can fit infinitely many disjoint balls of some fixed radius around any point — each having the same positive measure by translation invariance — which is in contradiction with finite measure of the point neighborhood. ⌟

**Example 1.21** (Gaussian Measures)**.** One can however define *Gaussian measures* on infinite-dimensional spaces, which avoids the problem in Remark 1.20→p.6 by throwing away the translation invariance requirement.

These correspond to free theories, and we can formally work with them as if we could write $\mathcal{D}\mu(\phi) \simeq \mathcal{D}\phi\, \mathrm{e}^{-\frac{1}{2}\phi\bullet\mathcal{F}\bullet\phi}$, where we take $\mathcal{D}\phi$ to be translationally invariant and satisfy the usual substitution rules, just written in the functional language. ⌟

**Remark 1.22** (Interacting Measures)**.** In the case of interacting theories, the precise mathematical sense of the path integral is less clear. Two immediate approaches are:

- *Lattice Approach* — We discretize the spacetime manifold by replacing it with a lattice. Fields are then defined by values on the nodes of the lattice, and the action by some discretized version. If the manifold is compact, the discretized version has a finite number of nodes, and path integral becomes a finite-dimensional integral (but we manifestly violate Lorentz symmetry). Everything is then well-defined, and the only problem is whether appropriate continuum limit exists (in which we hope to recover Lorentz symmetry).
- *Perturbative Approach* — For interacting theory "close" to a free theory, we can view the measure as a perturbative deformation of the Gaussian measure. We will briefly discuss such approach below in Perturbation Theory→p.10. Let us quickly note that it requires some regularization and renormalization scheme, which is usually chosen to respect as many symmetries of the theory as possible, but generally not all classical symmetries can be preserved. ⌟

## Correlators

Now we will present basics of the path integral formalism, with examples being provided by the free scalar theory.

**Definition 1.23** (Field, Local Operator, Observable). A (complex topological) vector space $\mathcal{O}_x\mathcal{M}$ of *observables* at $x \in \mathcal{M}$ is the space of smooth functions $f\colon \mathcal{F}ields_{\mathcal{M}} \to \mathbb{C}$ such that $f[\boldsymbol{\phi}]$ depends only on the restriction of $\boldsymbol{\phi} \in \mathcal{F}ields_{\mathcal{M}}$ to an arbitrarily small neighborhood of $x$.

Spaces $\mathcal{O}_x\mathcal{M} \simeq V$ are isomorphic to some standard fiber of local/point observables $V$, and fit together to form a vector bundle $\mathcal{O}\mathcal{M}$ over $\mathcal{M}$, which is the space of *local operators/fields* of the theory.

**Example 1.24** (Fundamental, Elementary, Composite Operator). A field/operator $\mathcal{O}$ directly integrated over in the path integral is called *fundamental* or *elementary*, example being $\mathcal{O} = \phi$ for a scalar field $\phi$. Local functions of (multiple) fundamental fields are called *composite*, example being $\mathcal{O} \sim \phi^p$ with $p \geq 2$, where $p = 2$ corresponds to the mass term. ⌟

**Remark 1.25** (Nonlocal, Extended Operator). There are also important *nonlocal* operators, which are not localized at a single point, but for example on some surface — symmetry charges as spatial integrals of conserved currents — or a loop — Wilson loop as a holonomy of the gauge field around a closed loop. ⌟

In path integral approach, correlators are the fundamental objects in QFT.

**Definition 1.26** (Correlation Function, Correlator). The *correlation function* or simply *correlator* of an operator $\mathcal{O}\colon \mathcal{F}ields \to \mathbb{C}$ is defined as the functional average over the space of field configurations

$$\langle \mathcal{O} \rangle \equiv \frac{1}{\mathcal{Z}} \int\limits_{\mathcal{F}ields} \mathcal{D}\boldsymbol{\phi}\, \mathrm{e}^{\frac{i}{\hbar}\mathcal{S}[\boldsymbol{\phi}]}\, \mathcal{O}[\boldsymbol{\phi}]\,,$$

where $\mathcal{Z} \equiv \langle 1 \rangle$ is the normalization. We often suppress writing $\mathcal{F}ields$.

**Remark 1.27.** A lot of information about a QFT can be extracted from such correlators, for example flat-space scattering amplitudes are obtained via LSZ reduction formulas by looking at their poles in momentum representation.

Alternatively, in Euclidean QFTs describing continuum limit of statistical systems, one can read out correlation lengths or other measurable coefficients from their (asymptotic) behavior. ⌟

All correlators are compactly encoded in the following object of central importance.

**Definition 1.28** (Partition Function, Background Sources). The *partition function* depending on background *sources* $\boldsymbol{J}$ is defined as

$$\mathcal{Z}[\boldsymbol{J}] \equiv \left\langle \mathrm{e}^{\frac{i}{\hbar}\boldsymbol{J}\bullet\boldsymbol{\phi}} \right\rangle \equiv \int \mathcal{D}\boldsymbol{\phi}\, \mathrm{e}^{\frac{i}{\hbar}(\mathcal{S}[\boldsymbol{\phi}] + \boldsymbol{J}\bullet\boldsymbol{\phi})}\ .$$

Sometimes we implicitly include the source term $\boldsymbol{J} \bullet \boldsymbol{\phi}$ in the action, and we can then write $\mathcal{Z}[\boldsymbol{J}] = \langle 1 \rangle_{\boldsymbol{J}}$.

**Remark 1.29** (Normalization of Partition Function)**.** When working in a fixed background, we usually normalize the partition function such that $\mathcal{Z}[\mathbf{0}] = 1$. ⌟

**Example 1.30** (Partition Function for Free Scalar Field)**.** Completing the square in the exponent and assuming normalization as described above, we have

$$\begin{aligned}\mathcal{Z}[J] = \int \mathcal{D}\phi \, \mathrm{e}^{\frac{i}{\hbar}\left(-\frac{1}{2}\phi\bullet\mathcal{F}\bullet\phi + J\bullet\phi\right)} &= \int \mathcal{D}\phi \, \mathrm{e}^{\frac{i}{\hbar}\left(-\frac{1}{2}(\phi-\mathcal{F}^{-1}\bullet J)\bullet\mathcal{F}\bullet(\phi-\mathcal{F}^{-1}\bullet J)+\frac{1}{2}J\bullet\mathcal{F}^{-1}\bullet J\right)} \\ &= \left(\int \mathcal{D}\phi' \, \mathrm{e}^{\frac{i}{\hbar}\left(-\frac{1}{2}\phi'\bullet\mathcal{F}\bullet\phi'\right)}\right) \mathrm{e}^{\frac{i}{\hbar}\frac{1}{2}J\bullet\mathcal{F}^{-1}\bullet J} \equiv \mathrm{e}^{(\frac{i}{\hbar})^2\frac{1}{2}J\bullet G\bullet J}\end{aligned}$$

We used the translation invariance of the measure $\mathcal{D}\phi$ to shift the integration variable $\phi \mapsto \phi' \equiv \phi - \mathcal{F}^{-1} \bullet J$, after which the $J$-independent functional determinant obtained by Gaussian/Fresnel integration cancels with the normalization of the measure.

As we can see, such Gaussian integration is equivalent to evaluating the exponent at the solution of classical Euler–Lagrange EOM

$$-\mathcal{F} \bullet \phi + J = 0 \quad \Longleftrightarrow \quad \phi_J = \mathcal{F}^{-1} \bullet J \equiv \tfrac{i}{\hbar} G \bullet J\,.$$

⌟

**Remark 1.31** (Green's Functions)**.** The *Green's function* satisfying the equation

$$-\mathcal{F} \bullet G \overset{!}{=} i\hbar \mathbf{1}$$

is not uniquely defined in the Lorentzian signature, and one needs to provide appropriate "$i\varepsilon$ prescription". We will discuss this later in Wick Rotation →p.12. ⌟

The partition function $\mathcal{Z}[\boldsymbol{J}]$ is a generating functional for all correlators of local operators, in the following sense.

**Proposition 1.32** (Partition Function as Generating Functional)**.** Correlators of local functionals, which are monomials of fundamental fields $\boldsymbol{\phi}$, are given by appropriate functional derivatives of the partition function with respect to the sources $\boldsymbol{J}$. In particular, we have

$$\left\langle \boldsymbol{\phi}(x_1) \ldots \boldsymbol{\phi}(x_n) \right\rangle = \left(\frac{\hbar}{i}\right)^n \frac{1}{\mathcal{Z}[\boldsymbol{J}]} \frac{\delta}{\delta \boldsymbol{J}(x_1)} \cdots \frac{\delta}{\delta \boldsymbol{J}(x_n)} \mathcal{Z}[\boldsymbol{J}] \bigg|_{\boldsymbol{J}=0}.$$

**Proof.** Directly follows from $\frac{\delta}{\delta \boldsymbol{J}(x)}(\boldsymbol{J} \bullet \boldsymbol{\phi}) = \boldsymbol{\delta}_x \bullet \boldsymbol{\phi} = \boldsymbol{\phi}(x)$ and

$$\frac{\hbar}{i} \frac{\delta}{\delta \boldsymbol{J}(x)} \mathcal{Z}[\boldsymbol{J}] \equiv \frac{\hbar}{i} \frac{\delta}{\delta \boldsymbol{J}(x)} \left\langle \mathrm{e}^{\frac{i}{\hbar}\boldsymbol{J}\bullet\boldsymbol{\phi}} \right\rangle = \left\langle \boldsymbol{\phi}(x) \, \mathrm{e}^{\frac{i}{\hbar}\boldsymbol{J}\bullet\boldsymbol{\phi}} \right\rangle.$$

□

**Example 1.33** (Wick's theorem)**.** Using Proposition 1.32 →p.8 in the case of free scalar field Example 1.30 →p.8, we obtain

$$\left\langle \phi(x_1)\dots\phi(x_n)\right\rangle = \begin{cases} 0 & \text{for } n \text{ odd}\,, \\ G(x_1,x_2) & \text{for } n=2\,, \\ \displaystyle\sum_{\substack{\text{pairings} \\ \{(i_+,i_-)\}}} \prod_{(i_+,i_-)} G(x_{i_+},x_{i_-}) & \text{for } n \text{ even}\,, \end{cases}$$

since only when all $\frac{\delta}{\delta J(x_\bullet)}\frac{\delta}{\delta J(x_\bullet)}$ act in pairs on $\frac{1}{2} J \bullet G \bullet J$ we can obtain a nonzero contribution after finally evaluating at $J = 0$. ⌟

## Boundary States

Up to now we always imagined performing the path integral over all field configurations in the whole spacetime, and did not consider any boundaries. In their presence, we need to specify the boundary conditions on the fields, which usually means fixing values of the fields on the boundary, see Figure 1.34 →p.9 for visualization.

Such setting is well known from QM, where path integral with constrained positions at the ends gives us the kernel of the evolution operator in the position basis.

Similarly in QFT, constraining the field values on the boundary of the region, the path integral gives us the kernel of the evolution operator in the *field bases* of the Hilbert spaces of states associated to the boundaries. We will not delve into details of this here, since everything will be properly formalized in Section 1.2 →p.13.

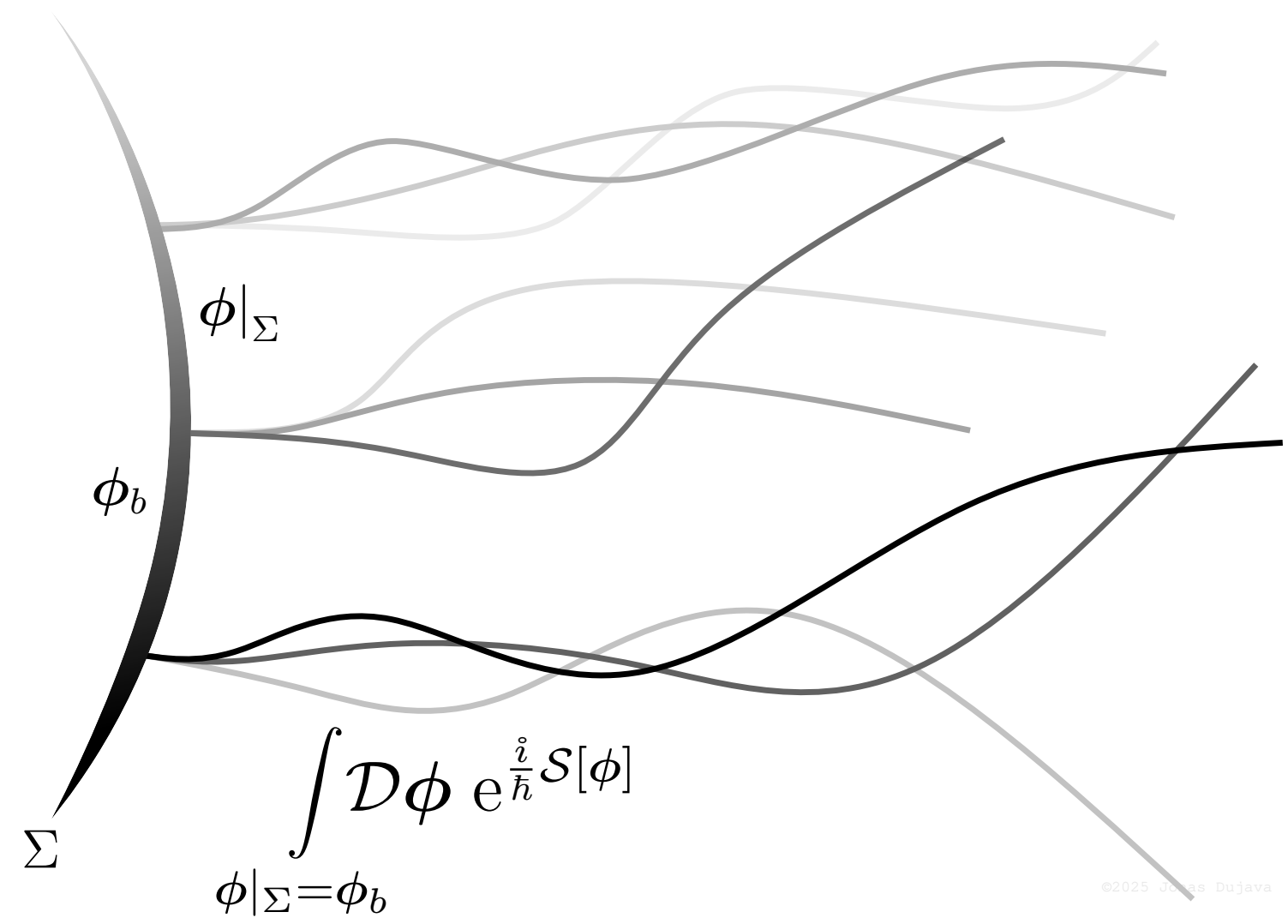

**Figure 1.34 /** Visualization of path integral with fixed boundary values $\phi_b$ on the boundary $\Sigma$. Only the field fluctuations are integrated over, the underlying structure of manifold is fixed.

## Perturbation Theory

Exact evaluation of the path integral is usually not feasible. Among few exceptions a special role is played by *free* theories, which have quadratic actions, leading to a Gaussian integration. In general, including any non-trivial interactions makes the explicit evaluation intractable. The most common way to proceed, applicable for *weak couplings*, is to employ *perturbation theory* around a free theory.

This amounts to splitting the action into *free* (quadratic) and *interaction* part, and then expanding the interaction part into series in the coupling.

**Example 1.35** (Interacting Scalar Field, Feynman Diagrams). For simplicity, consider an action symbolically of the form (integrations are done over $x \in \mathcal{M}$ with appropriate metric density)

$$\begin{aligned}\mathcal{S}[\phi] &= \mathcal{S}_{\text{free}}[\phi] + \mathcal{S}_{\text{int}}[\phi] \\ &\equiv -\tfrac{1}{2}\phi \bullet \mathcal{F} \bullet \phi - \lambda \int_x V(x)\,,\end{aligned}$$

where $\lambda$ is the coupling assumed to be small, and $V$ is some simple polynomial in $\phi$, for example $V(x) \sim \phi^4(x)$. Then the partition function or generating functional can be formally written as

$$\begin{aligned}\mathcal{Z}[J] &\simeq \int \mathcal{D}\phi\ \mathrm{e}^{\frac{i}{\hbar}\left(-\frac{1}{2}\phi\bullet\mathcal{F}\bullet\phi - \lambda\int_x V(x) + J\bullet\phi\right)} \\ &= \left\langle \mathrm{e}^{-\frac{i\lambda}{\hbar}\int_x V(x)}\,\mathrm{e}^{\frac{i}{\hbar}J\bullet\phi}\right\rangle_{\text{free}} = \sum_{n=0}^{\infty}\frac{1}{n!}\left(\tfrac{-i\lambda}{\hbar}\right)^n \int_{x_1,\ldots,x_n}\left\langle V(x_1)\cdots V(x_n)\,\mathrm{e}^{\frac{i}{\hbar}J\bullet\phi}\right\rangle_{\text{free}}.\end{aligned}$$

To compute arbitrary correlator, we take functional derivatives with respect to the source $J$ as in Proposition 1.32 →p.8, which brings down additional factors $\phi$ into the free correlator. These are then evaluated using Example 1.33 →p.9, just by pairing up all the $\phi$s with the free propagators $G$.

There is a well known diagrammatical technique — called *Feynman diagrams* — for organizing such an expansion. While more detail can be found in any QFT textbook, let us briefly summarize the main points.

To calculate a correlator $\langle\phi(y_1)\cdots\phi(y_n)\rangle$, we draw *external* (degree one) vertices for each $\phi(y_i)$, and then construct all diagrams by additionally inserting any number of *internal* vertices (with degree corresponding to the power of the polynomial $V$) and connecting them with simple edges/lines.

All such diagrams $\Gamma$ are summed up with the corresponding weights $w[\Gamma]$ — each insertion of internal vertex carries a factor of $\frac{-i\lambda}{\hbar}$, each edge between vertices at $x_+$ and $x_-$ corresponds to the free propagator $G(x_+, x_-)$, and all internal vertices are integrated over the spacetime manifold. One just needs to take care about combinatorial factors, which results in

$$\left\langle\phi(y_1)\cdots\phi(y_n)\right\rangle \simeq \frac{1}{\mathcal{Z}}\int_{\text{pert}}\mathcal{D}\phi\ \mathrm{e}^{\frac{i}{\hbar}\mathcal{S}[\phi]}\,\phi(y_1)\cdots\phi(y_n) \equiv \sum_{\Gamma}\frac{1}{|\mathrm{Aut}(\Gamma)|}\,w[\Gamma]\,,$$

where $|\mathrm{Aut}(\Gamma)|$ is the *symmetry factor* of the diagram $\Gamma$. Here $\Gamma$ is any diagram as described above, but we exclude *vacuum bubbles* — all interaction vertices are just connected to each other, and not to any external vertex — since they factor out and are cancelled by the normalization $\mathcal{Z}$. ⌟

**Remark 1.36** (Asymptotic Series)**.** Such perturbation expansions are typically not convergent for any finite value of the coupling constant $\lambda$, so should not be viewed as Taylor series. However, the formal manipulations are still justified in the sense of *asymptotic series*, which means that truncating it at any finite order gives increasingly good approximations when taking the limit $\lambda \to 0$. ⌟

**Remark 1.37** (Regularization, Counterterms, Renormalization)**.** In fact, the previous discussion swept under the rug the issue of *divergences*, which come from the coincidence limit of points during the integration over the internal vertices.

We should somehow regularize the integrals, and add appropriate *counterterms* in the interaction part such that in the end (after removing the regularization) we obtain finite results. This goes under the name of *renormalization*, and we will discuss it from slightly different perspective in Section 2.2→p.41. ⌟

To calculate the partition function $\mathcal{Z}[\boldsymbol{J}]$ in the perturbation theory, we draw all possible diagrams discussed above, but now external vertices come from expanding the exponential $\mathrm{e}^{\frac{i}{\hbar}\boldsymbol{J}\bullet\boldsymbol{\phi}}$, so they have an extra factor of $\frac{i}{\hbar}\boldsymbol{J}$ associated with them (and we integrate over them). We can simplify the diagrammatic expansion only to the connected diagrams — the motto is that *exponential map* generates all graphs/contributions from just connected graphs/contributions.

**Proposition 1.38** (Exponential and Connected Graphs)**.** Perturbative expansion of the partition function $\mathcal{Z}[\boldsymbol{J}]$ can be expressed as

$$\mathcal{Z}[\boldsymbol{J}] \simeq \sum_{\Gamma} \frac{1}{|\mathrm{Aut}(\Gamma)|} w[\Gamma;\boldsymbol{J}] = \exp\left(\sum_{\gamma \text{ connected}} \frac{1}{|\mathrm{Aut}(\gamma)|} w[\gamma;\boldsymbol{J}]\right) \cong \mathrm{e}^{\frac{i}{\hbar}\mathcal{W}[\boldsymbol{J}]} ,$$

where $\mathcal{W}[\boldsymbol{J}]$ is the generating functional for the *connected correlators*.

**Proof.** We can decompose any graph $\Gamma = \gamma_1^{\sqcup r_1} \sqcup \cdots \sqcup \gamma_k^{\sqcup r_k}$ into its connected components, $\gamma_\bullet$ being pairwise non-isomorphic connected graphs. Then we have for the symmetry factor $|\mathrm{Aut}(\Gamma)| = \prod_i r_i! |\mathrm{Aut}(\gamma_i)|^{r_i}$. Using the multiplicativity of the weights $w[\Gamma_1 \sqcup \Gamma_2] = w[\Gamma_1] w[\Gamma_2]$, we can write

$$\begin{aligned}\exp\left(\sum_{\gamma \text{ connected}} \frac{1}{|\mathrm{Aut}(\gamma)|} w[\gamma]\right) &= \prod_{\gamma \text{ connected}} \sum_{r=0}^{\infty} \frac{1}{r!|\mathrm{Aut}(\gamma)|^r} w[\gamma]^r \\ &= \sum_{\boldsymbol{r}=\boldsymbol{0}}^{\infty} \frac{1}{\prod_i r_i! |\mathrm{Aut}(\gamma_i)|^{r_i}} \prod_i w[\gamma_i]^{r_i} = \sum_{\Gamma} \frac{1}{|\mathrm{Aut}(\Gamma)|} w[\Gamma] .\end{aligned}$$

□

## Wick Rotation

As already noted in Remark 1.31 →p.8, the inverse of the differential operator $\mathcal{F}$ necessary for computing the functional Gaussian/Fresnel integrals is not unique in the Lorentzian signature.

**Remark 1.39** (Fresnel and Gauss Integrals). What the appropriate choice should be can be understood by considering one-dimensional Fresnel integral of the form

$$\int_{\mathbb{R}} \mathrm{d}x\, \mathrm{e}^{\mathring{\imath}x^2} = \sqrt{\pi}\, \mathrm{e}^{\frac{\mathring{\imath}\pi}{4}} \ .$$

The integrand is highly oscillatory, and the integral is not absolutely convergent. Still, it can be understood as a boundary value of the absolutely convergent Gauss integral (taking $\alpha \in \mathbb{C}$ with $\operatorname{Re}\alpha > 0$)

$$\int_{\mathbb{R}} \mathrm{d}x\, \mathrm{e}^{-\alpha x^2} = \sqrt{\frac{\pi}{\alpha}} \, ,$$

in particular taking the limit $\alpha \to -\mathring{\imath}$ gives us the above result. ⌟

We thus infer that the path integral in Lorentzian signature should be understood as a boundary value of an appropriate analytical continuation from the Euclidean signature, where the propagator is unique. In this way we recover the usual *Feynman propagator* corresponding to correlators understood as time-ordered vacuum expectation values. We will see this explicitly in Section 1.2 →p.13 and Section 1.4 →p.29.

**Example 1.40** (Wick Rotation in Flat Spacetime, Euclidean QFT). In flat spacetime, the so-called *Wick rotation* to the Euclidean signature is obtained by setting the time coordinate to be imaginary

$$t \equiv -\mathring{\imath}\tau \, , \quad \tau \in \mathbb{R} \, .$$

where $\tau$ is the *Euclidean time*. This can be seen as taking $t = \mathrm{e}^{-\mathring{\imath}\delta}\,\tau$ for $\delta \in [0, \frac{\pi}{2}]$ and $\tau \in \mathbb{R}$, where $\delta = 0$ corresponds to the Lorentzian signature, and $\delta = \frac{\pi}{2}$ to the Euclidean signature. For the action we have

$${}^{L}\mathcal{S} = \int \mathrm{d}t\, \mathrm{d}^{d-1}\mathbf{x}\; {}^{L}\mathcal{L} = -\mathring{\imath} \int \mathrm{d}\tau\, \mathrm{d}^{d-1}\mathbf{x}\; {}^{L}\mathcal{L}\Big|_{\substack{t=-\mathring{\imath}\tau \\ \partial_t = \mathring{\imath}\partial_\tau}} \equiv \mathring{\imath} \int \mathrm{d}\tau\, \mathrm{d}^{d-1}\mathbf{x}\; {}^{E}\mathcal{L} \equiv \mathring{\imath}\; {}^{E}\mathcal{S} \, ,$$

so under the path integral we obtain

$$\mathcal{D}\boldsymbol{\phi}\; \mathrm{e}^{\frac{\mathring{\imath}}{\hbar}{}^{L}\mathcal{S}} \longmapsto \mathcal{D}\boldsymbol{\phi}\; \mathrm{e}^{-\frac{1}{\hbar}{}^{E}\mathcal{S}} \ .$$

For example, the Euclidean action of a free scalar field is

$${}^{E}\mathcal{S}[\phi] = \int \mathrm{d}\tau\, \mathrm{d}^{d-1}\mathbf{x}\, \left(\tfrac{1}{2}(\boldsymbol{\partial}\phi)^2 + \tfrac{1}{2}m^2\phi^2\right) ,$$

where contractions are done using ${}^{E}\boldsymbol{\eta} \equiv \boldsymbol{\delta}$, so ${}^{E}\mathcal{S}$ is positive-definite functional, and actually it is the *energy* of the field configuration. We can thus conclude that a flat-space Lorentzian QFT is the Wick rotation of a Euclidean QFT describing a (classical) statistical system at "inverse temperature" $\beta = \frac{1}{\hbar}$. ⌟

We can slightly generalize this for arbitrary static spacetimes.

**Example 1.41** (Wick Rotation in Static Spacetime). Take a static Lorentzian spacetime of the form ${}^L\mathcal{M} = \mathbb{R} \times \Sigma$ with metric

$$ {}^L\boldsymbol{g} = -\,\mathbf{d}t^2 + \boldsymbol{g}_\Sigma \,. $$

We can consider its complexification $\mathcal{M}_\mathbb{C} = \mathbb{C} \times \Sigma$ with the metric as above, but now we take $t \in \mathbb{C}$. The real slice $t \in \mathbb{R} \subset \mathbb{C}$ of $\mathcal{M}_\mathbb{C}$ is the original Lorentzian ${}^L\mathcal{M}$, and taking the imaginary slice $t \in \imath\mathbb{R} \subset \mathbb{C} \Leftrightarrow \tau \equiv \imath t \in \mathbb{R}$ we obtain ${}^E\mathcal{M}$ which is topologically the same as ${}^L\mathcal{M}$, but with the Euclidean metric

$$ {}^E\boldsymbol{g} = \mathbf{d}\tau^2 + \boldsymbol{g}_\Sigma \,. $$

⌟

**Remark 1.42.** In the above example we complexified the manifold and identified its slices as our Lorentzian and Euclidean manifolds. One drawback of such approach is that such embedding into a complexified manifold is not generally available — however, when possible, it is a crucial tool. Alternative approach is to leave the spacetime fixed, and complexify the metric. See [9] for an attempt in the FQFT setting. ⌟

## 1.2 Functorial Quantum Field Theory

We will now formalize some of the ideas concerning the path integral we have presented and heuristically motivated in the previous sections. In the context of Functorial Quantum Field Theory (FQFT), we will take these properties as axioms of our QFT.

This will elucidate the structure behind QFTs, and provide us a more general framework to work with. Functorial viewpoint is especially useful for formulating different types of QFTs on manifolds with various structures, and for conceptual understanding of various "operator formalisms"/"operator quantizations" to be discussed in Section 1.4→ p.29.

Exposition of this chapter is mainly based on [4] [10] [11] [12] [9] [13].

**Remark 1.43.** FQFT has beginnings in the attempts of Greame Segal and Michael Atiyah to axiomatize CFTs [14] and TQFTs [15], and is most natural in that setting. Following should not be considered as a rigorous definition, but rather as a (sub)set of guiding principles and properties we consider when we talk about QFTs and work with ideas coming from the path integral viewpoint. ⌟

**Remark 1.44** (Other Axiomatic Frameworks). Apart from the FQFT approach, there are mainly two other types of axiomatic frameworks:

- Correlator-focused approaches (Wightman axioms, Osterwalder–Schrader) — various (related) axiomatic systems specifying properties of correlation functions of local operators acting on a Hilbert space,

- Algebraic Quantum Field Theory — emphasizes algebras of observables associated to local spacetime regions, with Hilbert spaces of states being a derived notion.

For more details, see the reviews [16] [17].

## Motivation

One key notion we would like to formalize is the *locality* of the path integral.

**Idea 1.45** (Cutting and Gluing)**.** Assuming a local Lagrangian theory, we can cut the manifold into pieces, compute the path integral individually on each piece (holding fixed boundary conditions), and then glue/sew them back together by finally integrating over the boundary configurations. In formula, this reads for $\mathcal{M} = \mathcal{M}_1 \cup_\Sigma \mathcal{M}_2$ as ($\phi_{1,2}$ are restrictions of $\phi$ to $\mathcal{M}_{1,2}$, respectively)

$$\int\limits_{\mathcal{F}ields_{\mathcal{M}}} \mathcal{D}\phi \, \mathrm{e}^{\frac{i}{\hbar}\mathcal{S}_{\mathcal{M}}[\phi]} = \int\limits_{\mathcal{F}ields_{\Sigma}} \mathcal{D}\phi_b \int\limits_{\mathcal{F}ields_{\mathcal{M}_1}, \phi_1|_\Sigma = \phi_b} \mathcal{D}\phi_1 \, \mathrm{e}^{\frac{i}{\hbar}\mathcal{S}_{\mathcal{M}_1}[\phi_1]} \int\limits_{\mathcal{F}ields_{\mathcal{M}_2}, \phi_2|_\Sigma = \phi_b} \mathcal{D}\phi_2 \, \mathrm{e}^{\frac{i}{\hbar}\mathcal{S}_{\mathcal{M}_2}[\phi_2]} ,$$

where we used additivity of the action $\mathcal{S}_{\mathcal{M}}[\phi] = \mathcal{S}_{\mathcal{M}_1}[\phi_1] + \mathcal{S}_{\mathcal{M}_2}[\phi_2]$.

In the well understood $(1+0)$–dimensional case — just QM — the *locality* of the time evolution is represented by the *functoriality* of going from time intervals to the corresponding evolution operators

$$[t_1, t_3] = [t_2, t_3] \cup [t_1, t_2] \quad \overset{\mathsf{QM}}{\rightsquigarrow} \quad U(t_3, t_1) = U(t_3, t_2) \circ U(t_2, t_1) \, .$$

Note that convention for writing intervals with $t_3 > t_2 > t_1$ is opposite to how we write and compose evolution operators, so we will sometimes use $[t_2|t_1] \equiv [t_1, t_2]$

We will generalize this idea to more general setting in higher dimensions.

## Cobordism Category

Oftentimes we imagine cutting/foliating the spacetime into pieces, and then study the dynamics in-between the cuts/leaves, where we appropriately specify some boundary conditions. Prototypical example is the evolution between chosen time-slices in a given coordinate system. To allow for more complicated situations, let us slightly formalize this idea.

**Definition 1.46** (Cobordism)**.** A *cobordism* $\Sigma_{\text{in}} \xrightarrow{\mathcal{M}} \Sigma_{\text{out}}$ is an oriented compact $(d+1)$–dimensional manifold $\mathcal{M}$ with boundary, together with smooth maps $\iota_{\text{in}} : \Sigma_{\text{in}} \hookrightarrow \mathcal{M} \hookleftarrow \Sigma_{\text{out}} : \iota_{\text{out}}$ with image in $\partial\mathcal{M}$ such that

$$(-\iota_{\text{in}}) \sqcup \iota_{\text{out}} : (-\Sigma_{\text{in}}) \sqcup \Sigma_{\text{out}} \xrightarrow{\sim} \partial\mathcal{M}$$

is an orientation-preserving diffeomorphism. The $-\Sigma_{\text{in}}$ denotes $\Sigma_{\text{in}}$ with the opposite orientation, and the above means we have the decomposition of the boundary $\partial\mathcal{M} = (-\Sigma_{\text{in}}) \sqcup \Sigma_{\text{out}}$ in disjoint $\binom{\text{in}}{\text{out}}$–components.

**Remark 1.47** (Direction)**.** Depending on the context, it may be more natural to write $\Sigma_{\text{out}} \xleftarrow{\mathcal{M}} \Sigma_{\text{in}}$ instead of $\Sigma_{\text{in}} \xrightarrow{\mathcal{M}} \Sigma_{\text{out}}$. ⌟

**Remark 1.48** (Additional Geometric Structure)**.** We will often consider cobordisms with additional geometric structure, such as a *metric* or *spin* structure. Then both the cobordism and boundaries should be equipped with the same geometric structure — more precisely, a boundary $\Sigma_\bullet$ should be a germ $\Sigma_\bullet \times (-\varepsilon, \varepsilon)$ of a $(d+1)$-dimensional manifold with the given geometric structure — and the inclusion maps $\iota_\bullet$ should be structure-preserving maps. ⌟

**Remark 1.49** (Gluing/Sewing of Cobordisms, Cobordism Category)**.** When out-boundary of one cobordism matches the in-boundary of another cobordism, we can glue/sew them together to obtain a new cobordism.

In equations, for cobordisms $\Sigma_2 \xleftarrow{\mathcal{M}_2} \Sigma_g$ and $\Sigma_g \xleftarrow{\mathcal{M}_1} \Sigma_1$, their *gluing along* $\Sigma_g$ is given by the cobordism $\Sigma_2 \xleftarrow{\mathcal{M} \equiv \mathcal{M}_2 \cup_{\Sigma_g} \mathcal{M}_1} \Sigma_1 \equiv \Sigma_2 \xleftarrow{\mathcal{M}_2} \Sigma_g \xleftarrow{\mathcal{M}_1} \Sigma_1$. This is illustrated in Figure 1.50→p.15.

We have thus the *geometric cobordism category* $\mathcal{Cob}$, whose objects are closed oriented $d$–manifolds and morphisms are $(d+1)$–dimensional cobordisms, where everything can be enriched with additional "geometric" structure. For nontrivial geometric structure, such as metric, $\mathcal{Cob}$ is actually a non-unital category, since there is no identity morphism. Instead, it has "almost identity" morphisms — the infinitesimally short cylinders. ⌟

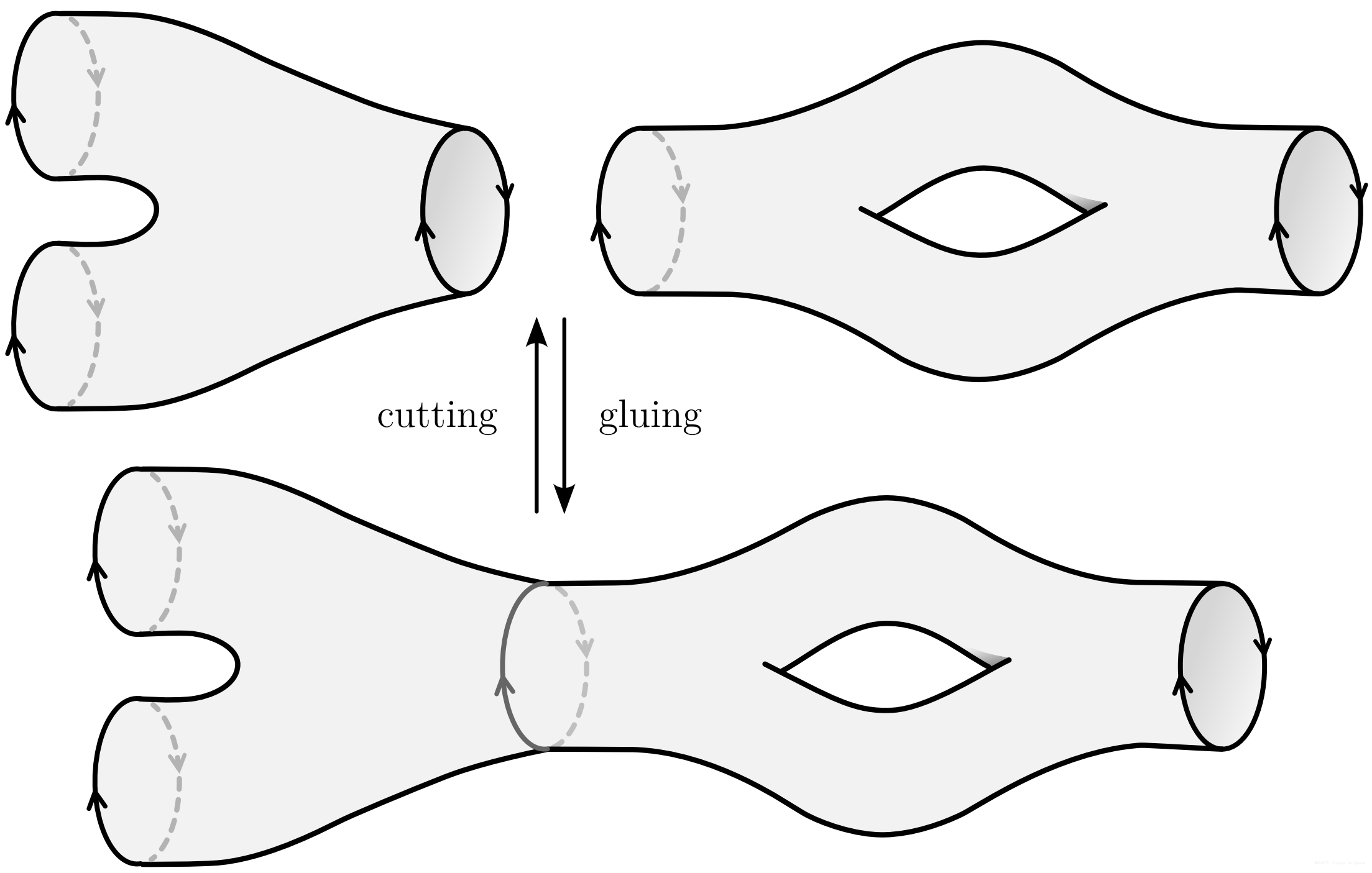

**Figure 1.50 /** Gluing and cutting of cobordisms.

## QFT as a Functor

**Definition 1.51** (Functorial QFT à la Atiyah–Segal). A $(d+1)$–dimensional $\mathsf{QFT}_{d+1} = (\mathcal{H}_{\circ}, \mathcal{Z}_{\circ})$ is defined by the following assignments:

- $\mathcal{H}_{\circ}$: **Boundary $\mapsto$ Space of Quantum States** — to each closed $d$-dimensional (codimension 1) manifold $\Sigma$ we assign a complex vector space $\mathcal{H}_\Sigma$ — the *space of quantum states*,
- $\mathcal{Z}_{\circ}$: **Cobordism $\mapsto$ Evolution Operator** — a $(d+1)$–dimensional cobordism $\Sigma_{\text{in}} \xrightarrow{\mathcal{M}} \Sigma_{\text{out}}$ is assigned a $\mathbb{C}$-linear map $\mathcal{Z}_\mathcal{M} : \mathcal{H}_{\Sigma_{\text{in}}} \to \mathcal{H}_{\Sigma_{\text{out}}}$ called an *evolution operator* (or sometimes partition function).

These furthermore satisfy following axioms:

(a) **Multiplicativity ($\sqcup \mapsto \otimes$):** Disjoint unions are mapped to tensor products, that is

$$\mathcal{H}_{\Sigma \sqcup \Sigma'} = \mathcal{H}_\Sigma \otimes \mathcal{H}_{\Sigma'} \,, \quad \mathcal{Z}_{\mathcal{M} \sqcup \mathcal{M}'} = \mathcal{Z}_\mathcal{M} \otimes \mathcal{Z}_{\mathcal{M}'} \,,$$

where for consistency an empty $d$-dimensional manifold $\varnothing$ is assigned a one-dimensional space of states $\mathcal{H}_\varnothing = \mathbb{C}$.

(b) **Gluing/Sewing ($\cup \mapsto \circ$):** Gluing of two cobordisms $\Sigma_2 \xleftarrow{\mathcal{M}_2} \Sigma_g$ and $\Sigma_g \xleftarrow{\mathcal{M}_1} \Sigma_1$ along $\Sigma_g$ is mapped to a composition of the evolution operators $\mathcal{Z}_{\mathcal{M}_2 \cup \mathcal{M}_1} = \mathcal{Z}_{\mathcal{M}_2} \circ \mathcal{Z}_{\mathcal{M}_1} : \mathcal{H}_{\Sigma_2} \leftarrow \mathcal{H}_{\Sigma_1}$.

(c) **Normalization:** We assume a "very short" (notion depending on the type of geometric structure on cobordisms) cylinder $\Sigma \xrightarrow{\Sigma \times [0,\varepsilon]} \Sigma$ over a closed $d$-manifold $\Sigma$ tends in the limit $\varepsilon \to 0$ to the identity map $\lim_{\varepsilon \to 0} \mathcal{Z}_{\Sigma \times [0,\varepsilon]} = \mathbf{1}_{\mathcal{H}_\Sigma} : \mathcal{H}_\Sigma \to \mathcal{H}_\Sigma$.

(d) **Naturality/Equivariance under Diffeomorphisms:** For a diffeomorphism of closed $d$–manifolds $\varphi : \Sigma \to \Sigma'$ we have an isomorphism $\rho_\varphi : \mathcal{H}_\Sigma \to \mathcal{H}_{\Sigma'}$ between the corresponding state-spaces. It is $\mathbb{C}$-linear if $\varphi$ is orientation-preserving, and $\mathbb{C}$-antilinear if $\varphi$ is orientation-reversing. Moreover, it is an action, that is $\rho_{\varphi' \circ \varphi} = \rho_{\varphi'} \circ \rho_\varphi$.

Given a diffeomorphism between cobordisms $\phi : \mathcal{M} \to \mathcal{M}'$, one has a commutative diagram (we denote $\phi|_{\text{in/out}}$ the restrictions of $\phi$ to $\Sigma_{\text{in/out}}$)

$$\begin{array}{ccc}
\mathcal{H}_{\Sigma_{\text{in}}} & \xrightarrow{\;\mathcal{Z}_{\mathcal{M},\boldsymbol{g}}\;} & \mathcal{H}_{\Sigma_{\text{out}}} \\
{\scriptstyle \rho_{\phi|_{\text{in}}}}\big\downarrow & & \big\downarrow{\scriptstyle \rho_{\phi|_{\text{out}}}} \\
\mathcal{H}_{\Sigma'_{\text{in}}} & \xrightarrow[\;\mathcal{Z}_{\mathcal{M}',\boldsymbol{g}' \equiv \phi_* \boldsymbol{g}}\;]{} & \mathcal{H}_{\Sigma'_{\text{out}}}
\end{array} ,$$

where we explicitly indicated how the background structure $\boldsymbol{g}$ on $\mathcal{M}$ is push-forwarded to $\boldsymbol{g}' \equiv \phi_* \boldsymbol{g}$ on $\mathcal{M}'$.

**Remark 1.52** (QFT as a Functor). The cobordism category $\mathcal{Cob}$ was already introduced in Remark 1.49→p.15, and the second category is the *vector space category* $\mathcal{Vect}_{\mathbb{C}}$, whose objects are complex vector spaces and morphisms are linear maps. They are even symmetric monoidal categories, monoidal product being disjoint union for $\mathcal{Cob}$ (with empty manifold being the unit) and tensor product for $\mathcal{Vect}_{\mathbb{C}}$ (with $\mathbb{C}$ being the unit).

Then a QFT can be understood as a functor of symmetric monoidal categories

$$\mathcal{Cob} \xrightarrow{\mathsf{QFT} \equiv (\mathcal{H}_{\square}, \mathcal{Z}_{\square})} \mathcal{Vect}_{\mathbb{C}}$$

natural under diffeomorphisms. Such a description neatly encompasses axioms listed in Definition 1.51→p.16. For additional category-theoretical details, refer to [18] [19]. ⌟

**Remark 1.53** (Closed Manifold, Partition Function). Since closed $(d+1)$–dimensional manifold $\mathcal{M}$ has no boundary, the evolution operator for cobordism $\varnothing \xrightarrow{\mathcal{M}} \varnothing$ is simply a map $\mathcal{Z}_{\mathcal{M}} : \mathbb{C} \to \mathbb{C}$. This is just a multiplication by a complex number, which we call the *partition function* (and use the same symbol $\mathcal{Z}_{\mathcal{M}}$). Oftentimes we omit $\mathcal{M}$ when it is clear from the context. ⌟

**Remark 1.54** (Topological QFT). For a topological QFT (TQFT) there is no geometric background structure on the cobordisms. Cylinders are thus mapped to identity operators, since any cylinder is "very short". Making a torus calculates the trace of the identity operator, which is the dimension of the Hilbert space. ⌟

**Remark 1.55** (Reversed Structure and Dual Space, $- \mapsto *$). Considering a "very short" cylinder $\Sigma \xrightarrow{\Sigma \times [0,\varepsilon]} \Sigma$ viewed as a cobordism $\Sigma \sqcup (-\Sigma) \xrightarrow{\Sigma \times [0,\varepsilon]} \varnothing$ yields in the $\varepsilon \to 0$ limit a bilinear pairing (for convenience we put $\mathcal{H}_{-\Sigma}$ first)

$$(\square, \square)_{\Sigma} : \mathcal{H}_{-\Sigma} \otimes \mathcal{H}_{\Sigma} \to \mathbb{C}\,,$$

which can be seen to be nondegenerate, and thus provides the identification $\mathcal{H}_{-\Sigma} \cong \mathcal{H}_{\Sigma}^*$ by $\mathcal{H}_{-\Sigma} \ni s \mapsto (s, \square) \in \mathcal{H}_{\Sigma}^*$, where $\mathcal{H}_{\Sigma}^*$ is the linear dual space to $\mathcal{H}_{\Sigma}$ (linear maps $\mathcal{H}_{\Sigma} \to \mathbb{C}$).

Nondegeneracy can be seen by taking a second short cylinder $\varnothing \xrightarrow{\Sigma \times [0,\varepsilon]} (-\Sigma) \sqcup \Sigma$, and gluing its out-boundary $(-\Sigma)$ to the in-boundary $(-\Sigma)$ of the first cylinder, obtaining thus a cylinder $\Sigma \xrightarrow{\Sigma \times [0,2\varepsilon]} \Sigma$ giving the identity in the limit $\varepsilon \to 0$ by the normalization axiom. The naturality axiom is tied to the assumption our functorial description describes already the *physical states* without any degeneracy. ⌟

**Remark 1.56** (Unitarity). Using the *tautological* orientation-reversing mapping $r : \Sigma \to -\Sigma$ mapping each point to itself and just reversing the geometrical data, we can define a nondegenerate sesquilinear form

$$\langle \square, \square \rangle_{\Sigma} : \mathcal{H}_{\Sigma} \otimes \mathcal{H}_{\Sigma} \xrightarrow{\rho(r) \otimes \mathbf{1}} \mathcal{H}_{-\Sigma} \otimes \mathcal{H}_{\Sigma} \xrightarrow{(\square, \square)_{\Sigma}} \mathbb{C}$$

Unitarity (in the Lorentzian signature) is then an additional (optional) collection

of assumptions on a QFT:

**(1)** For each $\Sigma$, the pair $(\mathcal{H}_\Sigma, \langle ⬚, ⬚ \rangle_\Sigma)$ forms a Hilbert space, possibly after an appropriate completion of $\mathcal{H}_\Sigma$. In particular, the corresponding sesquilinear form is positive definite.

**(2)** For any cylinder $\Sigma \xrightarrow{\Sigma\times[0,t]} \Sigma$ the evolution operator $\mathcal{Z}_{\Sigma\times[0,t]}\colon \mathcal{H}_\Sigma \to \mathcal{H}_\Sigma$ is a *unitary* operator, that is $\left\langle \mathcal{Z}_{\Sigma\times[0,t]}⬚, \mathcal{Z}_{\Sigma\times[0,t]}⬚ \right\rangle_\Sigma = \langle ⬚, ⬚\rangle_\Sigma$, or equivalently $\mathcal{Z}^\dagger_{\Sigma\times[0,t]} = \mathcal{Z}^{-1}_{\Sigma\times[0,t]}$.

**(3)** The representation $\rho$ of diffeomorphisms on the Hilbert spaces of states is *unitary*. More precisely, for any diffeomorphism $\phi\colon \Sigma \to \Sigma'$ the corresponding isomorphism $\rho_\phi\colon (\mathcal{H}_\Sigma, \langle ⬚, ⬚\rangle_\Sigma) \to (\mathcal{H}_{\Sigma'}, \langle ⬚, ⬚\rangle_{\Sigma'})$ is unitary, that is $\langle \rho_\phi ⬚, \rho_\phi ⬚\rangle_{\Sigma'} = \langle ⬚, ⬚\rangle_\Sigma$.

The Euclidean counterpart of the *unitarity* is the *reflection positivity* to be discussed at the end of Section 1.4 →p.29. For more FQFT details see [9].

As a simple example, lets look at Quantum Mechanics.

**Example 1.57** (Quantum Mechanics as FQFT)**.** Quantum Mechanics can be understood as a $(1+0)$–dimensional Riemannian Atiyah–Segal QFT. We will clearly see how the ordinary Schrödinger picture emerges.

Since $d = 0$, the objects from the cobordism/spacetime category $\mathcal{C}ob$ are oriented 0-dimensional manifolds, that is collections of points $\{\mathsf{pt}^\pm\}$ with orientation $\pm$. If we denote the space of states for $\mathsf{pt}^+$ by $\mathcal{H} \equiv \mathcal{H}_{\mathsf{pt}^+}$, we automatically have $\mathcal{H}_{\mathsf{pt}^-} = \mathcal{H}_{-\mathsf{pt}^+} = \mathcal{H}^*$ using Remark 1.55 →p.17.

Morphisms (1-cobordisms) are then collections of oriented intervals and circles equipped with a Riemannian metric. By naturality (see Definition 1.51 →p.16), the partition function depends only on the metric structure modulo diffeomorphisms, that is only on the lengths of the connected components.

We denote the evolution operator for an interval of length $t$ — viewed as a cobordism $\mathsf{pt}^+ \xrightarrow{[0,t]} \mathsf{pt}^+$ — by $\mathcal{Z}_t \equiv \mathcal{Z}_{[0,t]} \equiv U_t\colon \mathcal{H} \to \mathcal{H}$. Functoriality/gluing implies (semi)group law for evolution operators, that is

$$\mathcal{Z}_{t_2+t_1} \equiv \mathcal{Z}_{[0,t_2+t_1]} = \mathcal{Z}_{[t_1,t_1+t_2]\cup[0,t_1]} \stackrel{!}{=} \mathcal{Z}_{[t_1,t_1+t_2]} \circ \mathcal{Z}_{[0,t_1]} \equiv \mathcal{Z}_{t_2} \circ \mathcal{Z}_{t_1}.$$

The normalization axiom implies that the evolution operator in the limit of zero length interval is the identity operator, that is $\lim_{\varepsilon\to 0} \mathcal{Z}_\varepsilon = \mathbf{1}_\mathcal{H}$. Expanding around zero length (assuming analyticity), we obtain

$$\mathcal{Z}_\varepsilon \sim \mathbf{1}_\mathcal{H} - \frac{i}{\hbar} H\varepsilon + O\!\left(\varepsilon^2\right),$$

where $H \in \operatorname{End}(\mathcal{H})$ is the *Hamiltonian* operator — time translation generator. In Example 1.65 →p.20, we will discuss QM operators in more detail.

Either by forming a *Schrödinger* differential equation for $\mathcal{Z}_t \equiv U_t$ in the form

$$\frac{\partial}{\partial t} U_t \equiv \frac{\partial}{\partial \varepsilon} U_{\varepsilon+t}\Big|_{\varepsilon=0} \stackrel{!}{=} \frac{\partial}{\partial \varepsilon} U_\varepsilon \circ U_t\Big|_{\varepsilon=0} = -\frac{i}{\hbar} H U_t \quad \Longrightarrow \quad i\hbar \frac{\partial}{\partial t} U_t = H U_t \,,$$

or by using semigroup law iteratively as $\mathcal{Z}_t = \left(\mathcal{Z}_{\frac{t}{N}}\right)^N$ and taking the limit $N \to \infty$, we obtain the solution

$$\mathcal{Z}_t \equiv U_t = \mathrm{e}^{-\frac{i}{\hbar} H t} \,.$$

If we take our QM to be unitary, see Remark 1.56 →p.17, we require $\mathcal{H} \equiv \mathcal{H}_{\mathsf{pt}+}$ to be a Hilbert space with the natural inner product $\langle \cdot, \cdot \rangle \equiv \langle \cdot, \cdot \rangle_{\mathsf{pt}+}$, and the Hamiltonian $H$ to be Hermitian/self-adjoint, such that $\mathcal{Z}_t$ is unitary.

Considering a circle $\mathbb{S}^1_t$ of circumference $t$ — interval with the ends glued together — corresponds to making a trace in the end

$$\mathcal{Z}_{\mathbb{S}^1_t} = \mathrm{Tr}_{\mathcal{H}}\, \mathcal{Z}_t = \mathrm{Tr}_{\mathcal{H}}\, \mathrm{e}^{-\frac{i}{\hbar} H t} \,.$$

⌟

**Remark 1.58.** To describe Hamiltonians depending on time, we need to include extra structure to know where we are located with respect to this “global” time, on which the Hamiltonian parametrically depends. ⌟

**Remark 1.59** (Finite Temperature)**.** In the Euclidean signature we set $t = -i\tau$, so for a Euclidean circle $\mathbb{S}^1_\beta$ of circumference $\tau \equiv \beta > 0$ we have $(\hbar \equiv 1)$

$$\mathcal{Z}_{{}^E\mathbb{S}^1_\beta} = \mathrm{Tr}_{\mathcal{H}}\, {}^E\mathcal{Z}_\beta = \mathrm{Tr}_{\mathcal{H}}\, \mathrm{e}^{-\beta H} \,.$$

which is the partition function for QM system at inverse temperature $\frac{1}{T} \equiv \beta$. Considering a QFT at finite temperature thus geometrically corresponds to curling up the Euclidean time direction into a circle of circumference $\beta$. ⌟

## Quantum Observables

We will now show how to understand quantum operators/observables as insertions of states on the boundaries of infinitesimally small cuts of spacetime. First, we state the rather general definition, and then mention some important examples.

**Idea 1.60** (Operator as State Insertion on Cut)**.** Let $\Sigma_{\mathrm{in}} \xrightarrow{\mathcal{M}} \Sigma_{\mathrm{out}}$ be a cobordism, and let $\Gamma \subset \mathcal{M}$ be some CW subcomplex — collection of points, lines, surfaces, and so on in $\mathcal{M}$ — disjoint from $\partial\mathcal{M}$.

Consider a family of $\varepsilon$-thickenings $U_\varepsilon(\Gamma)$ of $\Gamma$ with $\varepsilon \in (0, \varepsilon_0)$, for example by equipping $\mathcal{M}$ with a metric, and taking all points with distance $\leq \varepsilon$ from $\Gamma$. Then $\mathcal{M} \backslash U_\varepsilon(\Gamma)$ is a cobordism, where we additionally made cuts around $\Gamma$ with some small padding parametrized by $\varepsilon$, so it has a boundary

$$\partial\Big(\mathcal{M} \backslash U_\varepsilon(\Gamma)\Big) = (-\partial U_\varepsilon(\Gamma)) \sqcup (-\Sigma_{\mathrm{in}}) \sqcup \Sigma_{\mathrm{out}} \,.$$

We can insert states on the boundaries of these cuts, which are viewed as extra pieces of the initial boundary, and in the limit $\varepsilon \to 0$ they will correspond to some local operators.

**Definition 1.61** (Insertion, Quantum Observable, Correlator, VEV). A *quantum observable* supported on $\Gamma$ is a family of elements $\mathcal{O}_{\Gamma,\varepsilon} \in \mathcal{H}_{\partial U_\varepsilon(\Gamma)}$ — states on the boundary of the $\varepsilon$-tube around $\Gamma$ — such that the *insertion/correlator* defined by the limit

$$\left\langle \mathcal{O}_\Gamma \right\rangle_{\mathcal{M}} \equiv \lim_{\varepsilon \to 0} \mathcal{Z}_{\mathcal{M} \setminus U_\varepsilon(\Gamma)} \left[ \mathcal{O}_{\Gamma,\varepsilon} \right] : \mathcal{H}_{\Sigma_{\text{in}}} \to \mathcal{H}_{\Sigma_{\text{out}}}$$

is well-defined. Here $\mathcal{Z}_{\mathcal{M} \setminus U_\varepsilon(\Gamma)}$ is a map $\mathcal{H}_{\partial U_\varepsilon(\Gamma)} \otimes \mathcal{H}_{\Sigma_{\text{in}}} \to \mathcal{H}_{\Sigma_{\text{out}}}$, and we substitute the states defining the operator into the first factor. An important case is when $\mathcal{M}$ is closed ($\Sigma_{\text{in}} = \Sigma_{\text{out}} = \varnothing = \partial\mathcal{M}$), in which case the correlator is just a complex number called *Vacuum Expectation Value* (VEV).

**Remark 1.62.** Usually we require for $\varepsilon_0 > \varepsilon' > \varepsilon > 0$ that

$$\mathcal{O}_{\Gamma,\varepsilon'} \stackrel{!}{=} \mathcal{Z}_{U_{\varepsilon'}(\Gamma) \setminus U_\varepsilon(\Gamma)} \left[ \mathcal{O}_{\Gamma,\varepsilon} \right],$$

so the states on a (bigger) $\varepsilon'$-thickenings are obtained by the "free" evolution (without additional insertions) from a (smaller) $\varepsilon$-thickening. In such a case, the expression under the limit in Definition 1.61→p.20 actually does not depend on $\varepsilon \in (0, \varepsilon_0)$ due to the gluing axiom in Definition 1.51→p.16.

**Remark 1.63** (Point Observables, Local Operators). Now we will compare the FQFT definition with Definition 1.23→p.7 and Definition 1.26→p.7. Consider a space with $n$ punctures $\{x_i\}$ understood as "infinitesimally" small balls $\{B_i\}$ centered on punctures with boundaries $\{S_i\}$. The corresponding local operators are $\mathcal{O}_{x_i}\mathcal{M} \cong \mathcal{H}_{S_i}$. If there is no other boundary (or it is infinitely far away), we can consider insertions of $\{\mathcal{O}_i\}$ as a vacuum $n$-point correlation function

$$\left\langle \mathcal{O}_1(x_1) \cdots \mathcal{O}_n(x_n) \right\rangle.$$

**Example 1.64** (Identity Operator). The *identity* operator/field $\mathbf{1} \equiv 1$ corresponds to a point observable satisfying

$$\langle \mathbf{1}(x)\, \mathcal{O}_\Gamma \rangle \equiv \langle \mathcal{O}_\Gamma \rangle$$

for any operator $\mathcal{O}_\Gamma$. It essentially means that putting the operator $\mathbf{1}$ on a puncture just fills the puncture as if it was never there. Alternatively, it corresponds to the vacuum vector $|\text{vac}\rangle \equiv \mathcal{Z}_B \in \mathcal{H}_S$ for an infinitesimally small ball $B$ with a boundary $S$ — it is the state obtained by path integrating over the ball $B$ inside $S$ without any additional operator insertions.

**Example 1.65** (Point Observables in QM — Operators). For an elementary but enlightening example, we will look at point observables in the context of Quantum Mechanics, see Example 1.57→p.18. To understand an observable supported on a single point $\Gamma \equiv t$ inside a cobordism (time interval) $\mathcal{M} = [t_{\text{in}}, t_{\text{out}}] \equiv [t_{\text{out}} | t_{\text{in}}]$, we consider the corresponding $\varepsilon$-thickenings

$$U_\varepsilon(\Gamma) \equiv U_\varepsilon(t) = (t+\varepsilon | t-\varepsilon) \equiv (t-\varepsilon, t+\varepsilon)$$

with boundaries

$$\partial U_\varepsilon(t) = \mathsf{pt}^+_{t+\varepsilon} \sqcup \mathsf{pt}^-_{t-\varepsilon} \,.$$

A quantum observable is then — in a limit $\varepsilon \to 0$, but we can take it to be constant even for nonzero $\varepsilon$ — an element

$$\mathcal{O} \in \mathcal{H}_{\partial U_\varepsilon(t)} = \mathcal{H}_{\mathsf{pt}^+ \sqcup \mathsf{pt}^-} = \mathcal{H}_{\mathsf{pt}^+} \otimes \mathcal{H}_{\mathsf{pt}^-} \equiv \mathcal{H} \otimes \mathcal{H}^* \cong \operatorname{End}(\mathcal{H}) \,,$$

so it is a linear operator acting on the space of states $\mathcal{H}$.

If we consider several point observables supported on $\Gamma = \{t_n, \dots, t_1\}$ with $t_{\text{out}} > t_n > \cdots > t_1 > t_{\text{in}}$, it is the same as choosing a collection of operators $\{\mathcal{O}_n, \dots, \mathcal{O}_1\} \subset \operatorname{End}(\mathcal{H})$. The "correlator" (or insertion into evolution operator) is then obtained by inserting these operators at the corresponding points (times), leading to

$$\begin{aligned}\Big\langle \mathcal{O}_n(t_n) \cdots \mathcal{O}_1(t_1) \Big\rangle_{\mathcal{M}} &= \mathcal{Z}_{[t_{\text{out}} | t_n]} \circ \mathcal{O}_n \circ \cdots \circ \mathcal{Z}_{[t_2 | t_1]} \circ \mathcal{O}_1 \circ \mathcal{Z}_{[t_1 | t_{\text{in}}]} \\ &= \mathrm{e}^{-\frac{i}{\hbar} H (t_{\text{out}} - t_n)} \circ \mathcal{O}_n \circ \cdots \circ \mathrm{e}^{-\frac{i}{\hbar} H (t_2 - t_1)} \circ \mathcal{O}_1 \circ \mathrm{e}^{-\frac{i}{\hbar} H (t_1 - t_{\text{in}})} \,.\end{aligned}$$

The order of operators on the left-hand side is not important, since they are by definition inserted at the appropriate times, automatically time-ordered. ⌟

**Example 1.66** (Topological Surface Operators)**.** Very important class of nonlocal operators are the so-called *conserved charges*. Usually, they are integrals of conserved currents (coming from Noether's theorem) over some codimension-1 submanifolds (spatial slices). Due to the conservation of the integrated current, these operators do not depend on the particular choice of the integration surface (as long as we do not cross other operator insertions), and are therefore called *topological surface operators*. We will discuss them in Section 1.3→p.28. ⌟

As we have seen, the Atiyah–Segal formulation of QFTs naturally generalizes the Schrödinger picture. In the following remark we briefly touch on the various "quantum pictures" in QM.

**Remark 1.67** (Various Quantum Pictures)**.** Special things happen when the spacetime has (continuous) symmetry (generating a family $\mathcal{F}$ of (space)slices foliating the spacetime) relating vector spaces $\mathcal{H}_\Sigma$ at different (space)slices $\Sigma \in \mathcal{F}$. Such isomorphic vector spaces can be then identified into one quantum state space $\mathcal{H}$, and similarly all operators can be viewed as acting on this one space. We will further discuss the notion of *Operator Formalism* in Section 1.4→p.29.

Such simplification precisely happens in QM, where the family of "(space)slices" is just collection of points along an interval $\mathcal{F} = \{\mathsf{pt}^+_t\}_{t \in I \subset \mathbb{R}}$, and corresponding Hilbert spaces (of quantum states) at different times are isomorphic. Put another way, they form a family $\mathcal{H}_{\mathcal{F}} \equiv \{\mathcal{H}_t\}_{t\in I\subset\mathbb{R}}$ linked by isomorphisms $\mathcal{H}_{t'} \xleftarrow{U(t',t)} \mathcal{H}_t$ generated by the time translation evolution operators $U(t',t)$. To describe the dynamics in such a setting, we can use various *Quantum Pictures*:

- *Schrödinger Picture*. Quantum states $|\phi(t)\rangle_{\mathrm{S}} \in \mathcal{H}$ evolve in time, operators $\mathcal{O}_{\mathrm{S}} \in \mathrm{End}(\mathcal{H})$ are fixed (their "time of insertion" is determined by the time parameter of the state) — closest to the FQFT picture, we just represent evolution on one fixed Hilbert space (because we can). Time evolution is given by the Schrödinger equation

  $$\mathrm{i}\hbar\frac{\mathrm{d}}{\mathrm{d}t}|\phi(t)\rangle_{\mathrm{S}} = H|\phi(t)\rangle_{\mathrm{S}}\,,$$

  where $H$ is the Hamiltonian operator.

- *Heisenberg Picture*. Quantum states $|\phi\rangle_{\mathrm{H}} \equiv |\phi(t_0)\rangle_{\mathrm{S}} \in \mathcal{H}$ are fixed and describe initial conditions at a reference time $t_0$, while operators $\mathcal{O}_{\mathrm{H}}(t) \in \mathrm{End}(\mathcal{H})$ evolve in time (time evolution pullback of some fixed "Schrödinger" operator acting on $\mathcal{H}_t$, so that it acts on reference Hilbert space $\mathcal{H}_{t_0}$)

  $$\begin{array}{ccc} \mathcal{H}_t & \xleftarrow{U(t,t_0)} & \mathcal{H}_{t_0} \\ \downarrow{\scriptstyle \mathcal{O}_{\mathrm{S}}} & & \downarrow{\scriptstyle \mathcal{O}_{\mathrm{H}}(t)} \\ \mathcal{H}_t & \xrightarrow[U(t,t_0)^{-1}]{} & \mathcal{H}_{t_0} \end{array} \qquad \begin{array}{rcccc} \mathcal{O}_{\mathrm{H}}(t) & \equiv & U(t,t_0)^{-1} & \mathcal{O}_{\mathrm{S}} & U(t,t_0) \\ \mathcal{O}_{\mathrm{H}}(t_0) & \equiv & \mathbf{1} & \mathcal{O}_{\mathrm{S}} & \mathbf{1} \end{array}\,.$$

  Infinitesimally, we obtain the evolution equation for Heisenberg operators

  $$\mathrm{i}\hbar\frac{\mathrm{d}}{\mathrm{d}t}\mathcal{O}_{\mathrm{H}}(t) = [\mathcal{O}_{\mathrm{H}}(t), H]\,.$$

  Comparing with Example 1.65→p.20, we see that matrix elements (after inserting states at the boundary of correlator/evolution function) calculated in either picture lead to the same result.

- *Interaction/Dirac Picture*. Sometimes it is useful to split the evolution operator into free (which we can explicitly solve) and interaction part. We evolve the operators with the "uninteresting" free part, and states with the interaction part. If the interaction is weak, we can conveniently do perturbation theory.

## 1.3 Symmetries in Quantum Field Theories

When we change the system (transform the fields), but the system nevertheless looks the same, we say that the given transformation is a *symmetry*. We will show how these often correspond to symmetries of the background structure.

Symmetries of a theory constrain the form of correlation functions. Infinitesimal form of these constraints are encoded in the so-called Ward–Takahashi identities.

Generators of symmetries are given by conserved currents, which are represented as topological (surface) operators in the theory. Algebra of generators encodes, how we should "commute" surfaces of these operators.

**Remark 1.68** (Types of Symmetries)**.** There are various types of symmetries — discrete or continuous, spacetime or internal, global or local, and so on. We will mainly focus on continuous spacetime symmetries, or their slight generalizations.

Internal symmetries correspond to the target values of the fields, that is to the extrinsic part of field bundle, as discussed in Remark 1.7→p.3. Global internal symmetries are conceptually pretty simple, especially when they are linear, see Remark 2.9→p.40.

Local internal symmetries are topic of gauge theories, and should be viewed more as a redundancy of description, similarly to diffeomorphism invariance. ⌟

## Finite Symmetry Transformations

Invariance of the functional measure+action under some transformation implies relations between correlation functions.

**Remark 1.69** (Correlation Relations from General Measure+Action Invariance)**.** Suppose a transformation of fundamental fields $\boldsymbol{\phi} \mapsto \boldsymbol{\phi}'$ together with background fields/structure $\boldsymbol{g} \mapsto \boldsymbol{g}'$ (for example metric) leaves the measure+action invariant, that is

$$[\mathcal{D}\boldsymbol{\phi}]_{\boldsymbol{g}}\, \mathrm{e}^{\frac{i}{\hbar}\mathcal{S}_{\boldsymbol{g}}[\boldsymbol{\phi}]} = [\mathcal{D}\boldsymbol{\phi}']_{\boldsymbol{g}'}\, \mathrm{e}^{\frac{i}{\hbar}\mathcal{S}_{\boldsymbol{g}'}[\boldsymbol{\phi}']} \,.$$

independent of other terms under the path integral. Then we have a relation for correlators ($X$ is an arbitrary operator, $X \mapsto X'$ is induced by $\boldsymbol{\phi} \mapsto \boldsymbol{\phi}'$)

$$\begin{aligned}\left\langle X' \right\rangle_{\boldsymbol{g}} &\equiv \int [\mathcal{D}\boldsymbol{\phi}]_{\boldsymbol{g}}\, \mathrm{e}^{\frac{i}{\hbar}\mathcal{S}_{\boldsymbol{g}}[\boldsymbol{\phi}]}\, X' \\ &= \int [\mathcal{D}\boldsymbol{\phi}']_{\boldsymbol{g}'}\, \mathrm{e}^{\frac{i}{\hbar}\mathcal{S}_{\boldsymbol{g}'}[\boldsymbol{\phi}']}\, X' \\ &= \int [\mathcal{D}\boldsymbol{\phi}]_{\boldsymbol{g}'}\, \mathrm{e}^{\frac{i}{\hbar}\mathcal{S}_{\boldsymbol{g}'}[\boldsymbol{\phi}]}\, X \equiv \left\langle X \right\rangle_{\boldsymbol{g}'} \,,\end{aligned}$$

where we first used the assumed invariance, and then we just relabeled the integration variables. We did not need to care of the normalizations $\mathcal{Z}_{\boldsymbol{g}}$ or $\mathcal{Z}_{\boldsymbol{g}'}$, since the invariance implies they are the same. ⌟

**Remark 1.70** (Diffeomorphism Relations, Spacetime Symmetries)**.** Assuming the theory is diffeomorphism-invariant, we can consider transformations induced by a diffeomorphism $f\colon \mathcal{M} \to \mathcal{M}$, that is $\mathcal{O} \mapsto f^*\mathcal{O}$ for any operator, and $\boldsymbol{g} \mapsto f_*\boldsymbol{g}$. The previous remark then gives us the following relation for any diffeomorphism $f$ in the form (to have both transformations on the same side, we pushforward $\boldsymbol{g}$)

$$\left\langle \mathcal{O}_1(x_1) \cdots \mathcal{O}_n(x_n) \right\rangle_{\boldsymbol{g}} = \left\langle [f^*\mathcal{O}_1](x_1) \cdots [f^*\mathcal{O}_n](x_n) \right\rangle_{f_*\boldsymbol{g}} \,.$$

This does not mean that we have an infinite number of symmetries, because a general $f$ relates correlators in different backgrounds — it should be understood

↳ as the usual redundancy in the choice of coordinates.

But when the chosen background $\boldsymbol{g}$ has some spacetime *symmetry*, that is there exists a (nontrivial) diffeomorphism $f$ preserving $\boldsymbol{g}$ as $f_*\boldsymbol{g} \overset{!}{=} \boldsymbol{g}$, we obtain

$$\Big\langle \mathcal{O}_1(x_1)\cdots\mathcal{O}_n(x_n)\Big\rangle_{\boldsymbol{g}} = \Big\langle [f^*\mathcal{O}_1](x_1)\cdots[f^*\mathcal{O}_n](x_n)\Big\rangle_{f_*\boldsymbol{g}} \overset{!}{=} \Big\langle [f^*\mathcal{O}_1](x_1)\cdots[f^*\mathcal{O}_n](x_n)\Big\rangle_{\boldsymbol{g}},$$

which is a nontrivial relation between correlators in the same background. ⌟

**Example 1.71** (Symmetries from Flat Space Isometries)**.** As a simple example we consider flat space $\mathcal{M} = \mathbb{R}^d$ with the usual flat metric, which we will suppress in the notation. The symmetries of the metric structure are called *isometries*, and in $\mathbb{R}^d$ we have the usual translations and rotations. For a translation $f\colon x \mapsto x + a$ we have the relation

$$\Big\langle \mathcal{O}_1(x_1)\cdots\mathcal{O}_n(x_n)\Big\rangle = \Big\langle \mathcal{O}_1(x_1+a)\cdots\mathcal{O}_n(x_n+a)\Big\rangle,$$

and similarly for rotations (where we also act by an appropriate representation of the rotation, recall Remark 1.7 →p.3). Translation symmetry implies correlators can depend only on the relative positions of the operators. Rotation symmetry is particularly simple for scalar operators, where we can conclude they depend only on the relative distances. ⌟

**Remark 1.72** (Symmetries of FQFT)**.** From the FQFT point of view, this is just the naturality axiom (see **(d)** of Definition 1.51 →p.16) restricted to the subgroup $\mathsf{Sym}(\mathcal{M}, \boldsymbol{g}) \subset \mathsf{Diff}(\mathcal{M})$ of diffeomorphisms $f\colon \mathcal{M} \to \mathcal{M}$ preserving the background structure $\boldsymbol{g}$, that is $f_*\boldsymbol{g} \overset{!}{=} \boldsymbol{g}$. For a cobordism $\Sigma_{\text{out}} \overset{\mathcal{M}}{\longleftarrow} \Sigma_{\text{in}}$, such a symmetry $f$ implies

$$\mathcal{Z}_{\mathcal{M},\boldsymbol{g}} = \rho_{f|_{\text{out}}} \circ \mathcal{Z}_{\mathcal{M},\boldsymbol{g}} \circ \rho^{-1}_{f|_{\text{in}}}.$$

One can try to think through how this works out when we include operator insertions along the lines of Definition 1.61 →p.20. ⌟

Now consider a slightly different situation. Suppose the theory is invariant under simultaneous active transformation of the operators and the background structure (now we allow different local operators $\mathcal{O}_\bullet$ to transform in different ways, not necessarily induced by the transformation of the fundamental fields)

$$\mathcal{O}_\bullet \mapsto \widetilde{\mathcal{O}}_\bullet, \quad \boldsymbol{g} \mapsto \tilde{\boldsymbol{g}},$$

that is, we have a relation

$$\Big\langle \mathcal{O}_1(x_1)\cdots\mathcal{O}_n(x_n)\Big\rangle_{\boldsymbol{g}} = \Big\langle \widetilde{\mathcal{O}}_1(x_1)\cdots\widetilde{\mathcal{O}}_n(x_n)\Big\rangle_{\tilde{\boldsymbol{g}}}.$$

When does such a relation imply a symmetry in a given background $\boldsymbol{g}$? The answer is that we must find a diffeomorphism $f$, with which we can pushforward $\tilde{\boldsymbol{g}}$ back to the original background $\boldsymbol{g} \stackrel{!}{=} f_*\tilde{\boldsymbol{g}}$. Then we obtain symmetry relation

$$\begin{aligned}\left\langle \mathcal{O}_1(x_1)\cdots\mathcal{O}_n(x_n)\right\rangle_{\boldsymbol{g}} &= \left\langle [f^*\widetilde{\mathcal{O}}_1](x_1)\cdots[f^*\widetilde{\mathcal{O}}_n](x_n)\right\rangle_{f_*\tilde{\boldsymbol{g}}} \\ &\stackrel{!}{=} \left\langle [f^*\widetilde{\mathcal{O}}_1](x_1)\cdots[f^*\widetilde{\mathcal{O}}_n](x_n)\right\rangle_{\boldsymbol{g}} .\end{aligned}$$

**Example 1.73** (Weyl Transformation)**.** Now we will consider a Weyl-invariant theory, so the correlators are invariant under an active Weyl transformation $\boldsymbol{g} \mapsto \tilde{\boldsymbol{g}} \equiv \Omega^2\boldsymbol{g}$ together with $\mathcal{O}_\bullet \mapsto \widetilde{\mathcal{O}}_\bullet \equiv \Omega^{\Delta_\bullet}\mathcal{O}_\bullet$ (we consider $\mathcal{O}_\bullet$ to be primary scalar operators with scaling dimension $\Delta_\bullet$, more details will be in Chapter 3 →p.51). For example, we have the following relation for 2-point functions

$$\left\langle \mathcal{O}_1(x)\mathcal{O}_2(y)\right\rangle_{\boldsymbol{g}} = \Omega(x)^{\Delta_1}\Omega(y)^{\Delta_2}\left\langle \mathcal{O}_1(x)\mathcal{O}_2(y)\right\rangle_{\Omega^2\boldsymbol{g}} .$$

When the same Weyl rescaled metric can be obtained also by diffeomorphism (which we then call *conformal transformation*, see Definition 3.6 →p.53), this induces relation between correlators on the same background.

Considering flat space $\mathbb{R}^d$ and scaling/dilatation $f : x \mapsto x' \equiv \lambda x$, simple calculation gives $f^*\boldsymbol{g} = \lambda^2\boldsymbol{g} \stackrel{!}{=} \Omega^2\boldsymbol{g} \equiv \tilde{\boldsymbol{g}}$, so we obtain

$$\left\langle \mathcal{O}_1(x)\mathcal{O}_2(y)\right\rangle = \lambda^{\Delta_1+\Delta_2}\left\langle \mathcal{O}_1(\lambda x)\mathcal{O}_2(\lambda y)\right\rangle ,$$

and thus the possibly consistent behavior of the 2-point function of a CFT is given by a power law (where we implicitly used the translational and rotational invariance)

$$\left\langle \mathcal{O}_1(x)\mathcal{O}_2(y)\right\rangle \propto \frac{1}{|x-y|^{\Delta_1+\Delta_2}} .$$

In Corollary 3.80 →p.77, by further considering "special conformal transformations", we will see that the correlator can be nontrivial only when $\Delta_1 = \Delta_2$. ⌟

## Energy–Momentum Tensor

In the following we will assume a classical theory with Lagrangian description on a manifold with a metric $\boldsymbol{g}$. Since spacetime symmetries are essentially symmetries of $\boldsymbol{g}$, they are intimately connected with the energy–momentum tensor.

**Definition 1.74** (Energy–Momentum Tensor)**.** The energy–momentum tensor $\boldsymbol{T}$ is defined as the functional derivative of the action with respect to the metric

$$\boldsymbol{T}^{ab} \equiv -2\,\frac{\delta\mathcal{S}_g}{\delta\boldsymbol{g}_{ab}} \quad\Longleftrightarrow\quad \delta_g\mathcal{S}_g \equiv -\frac{1}{2}\int_{\mathcal{M}} \mathfrak{g}^{1/2}\boldsymbol{T}^{ab}\delta\boldsymbol{g}_{ab} .$$

It is also often called the *stress tensor* or *stress–energy tensor*.

**Remark 1.75.** We will always work with such "Hilbert" energy–momentum tensor, as opposed to the "canonical" one, which is non-uniquely defined by the Noether's prescription. ⌟

**Proposition 1.76** (Properties of Hilbert Energy–Momentum Tensor)**.** We have

**(1)** $\boldsymbol{T}^{\bullet\bullet} = \boldsymbol{T}^{(\bullet\bullet)} \iff \boldsymbol{T}^{ab} = \boldsymbol{T}^{ba}$ — $\boldsymbol{T}^{\bullet\bullet}$ is a *symmetric* tensor.

**(2)** $\boldsymbol{\nabla}\cdot\boldsymbol{T} \overset{\text{EL}}{\sim} 0 \iff \boldsymbol{\nabla}_{\boldsymbol{a}}\boldsymbol{T}^{\boldsymbol{ab}} \overset{\text{EL}}{\sim} 0$ — $\boldsymbol{T}^{\bullet\bullet}$ is *conserved* on-shell.

**(3)** $\operatorname{Tr}\boldsymbol{T} \equiv \boldsymbol{T}^{\boldsymbol{a}}_{\boldsymbol{a}} \overset{\text{EL}}{\sim} 0$ — $\boldsymbol{T}^{\bullet\bullet}$ is *traceless* on-shell for Weyl-invariant theories.

**Proof.** The symmetry **(1)** follows from the definition and symmetry of $\boldsymbol{g}$.

The conservation property **(2)** follows from the fact that the action is invariant under diffeomorphisms. Indeed, using the covariance Remark 1.12→p.3 for automorphism generated by a vector field $\boldsymbol{\xi} \in \mathcal{TM}$ vanishing in a neighborhood of the boundary $\partial\mathcal{M}$, we have

$$0 \overset{!}{=} \delta_{\xi}\mathcal{S}_{\boldsymbol{g}}[\boldsymbol{\phi}] \equiv -\frac{1}{2}\int_{\mathcal{M}} \mathfrak{g}^{1/2}\boldsymbol{T}^{\boldsymbol{ab}}\delta_{\xi}\boldsymbol{g}_{\boldsymbol{ab}} + \int_{\mathcal{M}} \mathfrak{g}^{1/2}\,\frac{\delta\mathcal{S}_{\boldsymbol{g}}}{\delta\boldsymbol{\phi}}\,\delta_{\xi}\boldsymbol{\phi}$$
$$\overset{\text{EL}}{\sim} -\int_{\mathcal{M}} \mathfrak{g}^{1/2}\boldsymbol{T}^{\boldsymbol{ab}}\boldsymbol{\nabla}_{(\boldsymbol{a}}\boldsymbol{\xi}_{\boldsymbol{b})} + 0$$
$$\overset{\text{EL}}{\sim} \int_{\mathcal{M}} \mathfrak{g}^{1/2}(\boldsymbol{\nabla}_{\boldsymbol{a}}\boldsymbol{T}^{\boldsymbol{ab}})\boldsymbol{\xi}_{\boldsymbol{b}} + 0 \quad \xrightarrow{\boldsymbol{\xi}\text{ arbitrary}} \quad \boldsymbol{\nabla}\cdot\boldsymbol{T} \overset{\text{EL}}{\sim} 0\,.$$

We used definition of $\boldsymbol{T}$ and $\delta_{\xi}\boldsymbol{g}_{\boldsymbol{ab}} \equiv \mathcal{L}_{\xi}\boldsymbol{g}_{\boldsymbol{ab}} = \boldsymbol{\nabla}_{(\boldsymbol{a}}\boldsymbol{\xi}_{\boldsymbol{b})}$. In the last step we used symmetry of $\boldsymbol{T}$ and integrated by parts — the boundary term vanishes due to the assumption on $\boldsymbol{\xi}$. Since $\boldsymbol{\xi}$ is otherwise arbitrary, we obtain the conservation property **(2)**.

For Weyl transformation Example 1.73→p.25 we have the infinitesimal form $\delta_{\omega}\boldsymbol{g} = 2\omega\boldsymbol{g}$, so we calculate (similarly as previously for diffeomorphism)

$$0 \overset{!}{=} \delta_{\omega}\mathcal{S}_{\boldsymbol{g}}[\boldsymbol{\phi}] \equiv -\frac{1}{2}\int_{\mathcal{M}} \mathfrak{g}^{1/2}\boldsymbol{T}^{\boldsymbol{ab}}(2\omega\boldsymbol{g}_{\boldsymbol{ab}}) + \int_{\mathcal{M}} \frac{\delta\mathcal{S}_{\boldsymbol{g}}}{\delta\boldsymbol{\phi}}\,\delta_{\omega}\boldsymbol{\phi}$$
$$\overset{\text{EL}}{\sim} -\int_{\mathcal{M}} \mathfrak{g}^{1/2}\boldsymbol{T}^{\boldsymbol{a}}_{\boldsymbol{a}}\omega + 0 \quad \xrightarrow{\omega\text{ arbitrary}} \quad \operatorname{Tr}\boldsymbol{T} \overset{\text{EL}}{\sim} 0\,.$$

□

**Corollary 1.77** (Conserved Currents from Isometries)**.** Let $f\colon\mathcal{M}\to\mathcal{M}$ be an isometry generated by a (Killing) vector field $\boldsymbol{\xi} \in \mathcal{TM}$. Then $\boldsymbol{T}\cdot\boldsymbol{\xi}$ is a conserved current, that is $\boldsymbol{\nabla}\cdot(\boldsymbol{T}\cdot\boldsymbol{\xi}) \equiv \boldsymbol{\nabla}_{\boldsymbol{a}}(\boldsymbol{T}^{\boldsymbol{ab}}\boldsymbol{\xi}_{\boldsymbol{b}}) = 0$.

**Proof.** By direct calculation we have

$$\boldsymbol{\nabla}\cdot(\boldsymbol{T}\cdot\boldsymbol{\xi}) = (\boldsymbol{\nabla}_{\boldsymbol{a}}\boldsymbol{T}^{\boldsymbol{ab}})\boldsymbol{\xi}_{\boldsymbol{b}} + \boldsymbol{T}^{\boldsymbol{ab}}\boldsymbol{\nabla}_{\boldsymbol{a}}\boldsymbol{\xi}_{\boldsymbol{b}} = 0 + \boldsymbol{T}^{\boldsymbol{ab}}\boldsymbol{\nabla}_{(\boldsymbol{a}}\boldsymbol{\xi}_{\boldsymbol{b})} \overset{!}{=} 0\,,$$

where we used both the conservation and symmetry of $\boldsymbol{T}$, and finally the infinitesimal form of the isometry $\mathcal{L}_{\xi}\boldsymbol{g}_{\boldsymbol{ab}} = 2\boldsymbol{\nabla}_{(\boldsymbol{a}}\boldsymbol{\xi}_{\boldsymbol{b})} \overset{!}{=} 0$. □

**Remark 1.78** (Conserved Currents from Conformal Transformations)**.** Refer to Section 3.1→p.52 for the definition of conformal Killing vector fields (CKV), where we discuss in detail the *conformal structure* of manifolds. The infinitesimal condition for CKV is given by $\mathcal{L}_{\boldsymbol{\xi}}\boldsymbol{g} = \boldsymbol{\nabla}_{(\boldsymbol{a}}\boldsymbol{\xi}_{\boldsymbol{b})} \propto \boldsymbol{g}_{\boldsymbol{ab}}$, so assuming a traceless energy–momentum tensor, we again obtain a conserved current $\boldsymbol{T}\cdot\boldsymbol{\xi}$. Thus, a Weyl invariant theory has a conserved current for every CKV. ⌟

## Ward–Takahashi Identities

Now we will look at the infinitesimal forms of the continuous spacetime symmetries.

**Remark 1.79.** Even though it is not strictly necessary for our purposes, the following will be formulated in the "tetrad" formulation — see Appendix N→p.156. This would allow us to consider also spinors, since just the metric is not sufficient.

For example, the energy–momentum tensor is classically given by

$$\boldsymbol{T}^{\boldsymbol{a}}{}_{\boldsymbol{b}} \equiv \boldsymbol{T}^{\boldsymbol{a}}_{k}\boldsymbol{e}^{k}_{\boldsymbol{b}} \equiv -\frac{\delta\mathcal{S}}{\delta\boldsymbol{e}^{k}_{\boldsymbol{a}}}\,\boldsymbol{e}^{k}_{\boldsymbol{b}} = -2\,\frac{\delta\mathcal{S}}{\delta\boldsymbol{g}_{\boldsymbol{an}}}\,\boldsymbol{g}_{\boldsymbol{nb}}\,,$$

since $\boldsymbol{g}_{\boldsymbol{ab}} = \boldsymbol{\eta}_{kl}\boldsymbol{e}^{k}_{\boldsymbol{a}}\boldsymbol{e}^{l}_{\boldsymbol{b}}$. ⌟

In the quantum setting, we also include possible contribution from the variation of the (renormalized) path integral measure. For simplicity (to avoid factors of $\mathring{\imath}$) we will consider Euclidean signature and set $\hbar \equiv 1$.

**Definition 1.80** (Quantum Energy–Momentum Tensor)**.** We define the *quantum energy–momentum tensor* operator as the response of functional measure+action to an infinitesimal metric/tetrad variation

$$\delta_{\boldsymbol{e}}\Big([\mathcal{D}\boldsymbol{\phi}]_{\boldsymbol{e}}\,\mathrm{e}^{-\mathcal{S}_{\boldsymbol{e}}[\phi]}\Big) \equiv [\mathcal{D}\boldsymbol{\phi}]_{\boldsymbol{e}}\,\mathrm{e}^{-\mathcal{S}_{\boldsymbol{e}}[\phi]}\int_{\mathcal{M}}\mathfrak{e}\,\delta\boldsymbol{e}^{k}_{\boldsymbol{a}}\boldsymbol{T}^{\boldsymbol{a}}_{k} + O\Big((\delta e)^2\Big)\,,$$

by which we mean

$$\Big\langle\boldsymbol{T}^{\boldsymbol{a}}_{k}(x)X\Big\rangle_{\boldsymbol{e}} - \Big\langle\boldsymbol{T}^{\boldsymbol{a}}_{k}(x)\Big\rangle_{\boldsymbol{e}}\Big\langle X\Big\rangle_{\boldsymbol{e}} \equiv \frac{\delta}{\delta\boldsymbol{e}^{k}_{\boldsymbol{a}}(x)}\Big\langle X\Big\rangle_{\boldsymbol{e}}\,,$$

where the second term comes from the variation of normalization $\mathcal{Z}_{\boldsymbol{e}}$ in the definition of the correlator, because ($\mathcal{Z}_e \equiv \mathrm{e}^{-\mathcal{W}_e}$)

$$\frac{\delta\mathcal{Z}_{\boldsymbol{e}}}{\delta\boldsymbol{e}^{k}_{\boldsymbol{a}}(x)} \equiv \int[\mathcal{D}\boldsymbol{\phi}]_{\boldsymbol{e}}\,\mathrm{e}^{-\mathcal{S}_e[\phi]}\,\boldsymbol{T}^{\boldsymbol{a}}_{k} \equiv \mathcal{Z}_{\boldsymbol{e}}\Big\langle\boldsymbol{T}^{\boldsymbol{a}}_{k}(x)\Big\rangle_{\boldsymbol{e}} \iff \Big\langle\boldsymbol{T}^{\boldsymbol{a}}_{k}(x)\Big\rangle_{e} = -\frac{\delta\mathcal{W}_e}{\delta\boldsymbol{e}^{k}_{\boldsymbol{a}}(x)}\,.$$

**Calculation 1.81** (Ward–Takahashi Identities)**.** Recalling Remark 1.69→p.23 and Remark 1.70→p.23, for an infinitesimal diffeomorphism generated by a vector field $\boldsymbol{\xi}$ we have

$$\begin{aligned}\Big\langle\mathcal{L}_{\boldsymbol{\xi}}X\Big\rangle_{\boldsymbol{g}} = \frac{\mathrm{d}}{\mathrm{d}\varepsilon}\Big\langle X\Big\rangle_{f^*_\varepsilon\boldsymbol{g}}\Big|_{\varepsilon=0} &= \int_{\mathcal{M}}\mathfrak{g}^{1/2}\,\tfrac{1}{2}\mathcal{L}_{\boldsymbol{\xi}}\boldsymbol{g}_{\boldsymbol{ab}}\Big(\Big\langle\boldsymbol{T}^{\boldsymbol{ab}}X\Big\rangle_{\boldsymbol{g}} - \Big\langle\boldsymbol{T}^{\boldsymbol{ab}}\Big\rangle_{\boldsymbol{g}}\Big\langle X\Big\rangle_{\boldsymbol{g}}\Big)\\ &= -\int_{\mathcal{M}}\mathfrak{g}^{1/2}\,\boldsymbol{\xi}_{\boldsymbol{b}}\boldsymbol{\nabla}_{\boldsymbol{a}}\Big\langle\boldsymbol{T}^{\boldsymbol{ab}}X\Big\rangle_{\boldsymbol{g}} + 0\,,\end{aligned}$$

where we integrated by parts and used $\mathcal{Z}_{\boldsymbol{g}} = \mathcal{Z}_{\boldsymbol{g}+\mathcal{L}_{\boldsymbol{\xi}}\boldsymbol{g}} \Longrightarrow \boldsymbol{\nabla}\cdot\langle\boldsymbol{T}\rangle_{\boldsymbol{g}} = 0$.

↳ Thus, $\nabla\cdot T = 0$ up to contact terms (it is sometimes conventional to write $\nabla\cdot$ inside the correlator, although the calculation shows it should be understood to be outside). For example, in flat spacetime, and for scalar operators, we have

$$\partial_a\Big\langle T^{ab}(x)\mathcal{O}_1(x_1)\cdots\mathcal{O}_n(x_n)\Big\rangle = -\sum_{i=1}^{n}\delta(x,x_i)\partial^b_{(i)}\Big\langle\mathcal{O}_1(x_1)\cdots\mathcal{O}_n(x_n)\Big\rangle\,.$$

These are called the *Ward–Takahashi identities*. ⌟

## Symmetries as Topological Operators

As foreshadowed in Example 1.66→p.21, we can form symmetry generators (charges) as space-slice (surface) integrals of conserved currents, that is

$$Q_{\boldsymbol{\xi}}(\Sigma) \equiv -\int_\Sigma \boldsymbol{dS}_a \boldsymbol{T}^{ab}\boldsymbol{\xi}_b\,,$$

where now $\boldsymbol{\xi}$ is a Killing vector (or conformal Killing vector if $\boldsymbol{T}$ is traceless).

The Gauss law, Ward–Takahashi identities Calculation 1.81→p.27, and the conservation of currents Corollary 1.77→p.26 imply that the correlation function of $Q_{\boldsymbol{\xi}}(\Sigma)$ do not change as we deform the surface $\Sigma$, as long as we do not cross the support of other operators (see Figure 1.82→p.28).

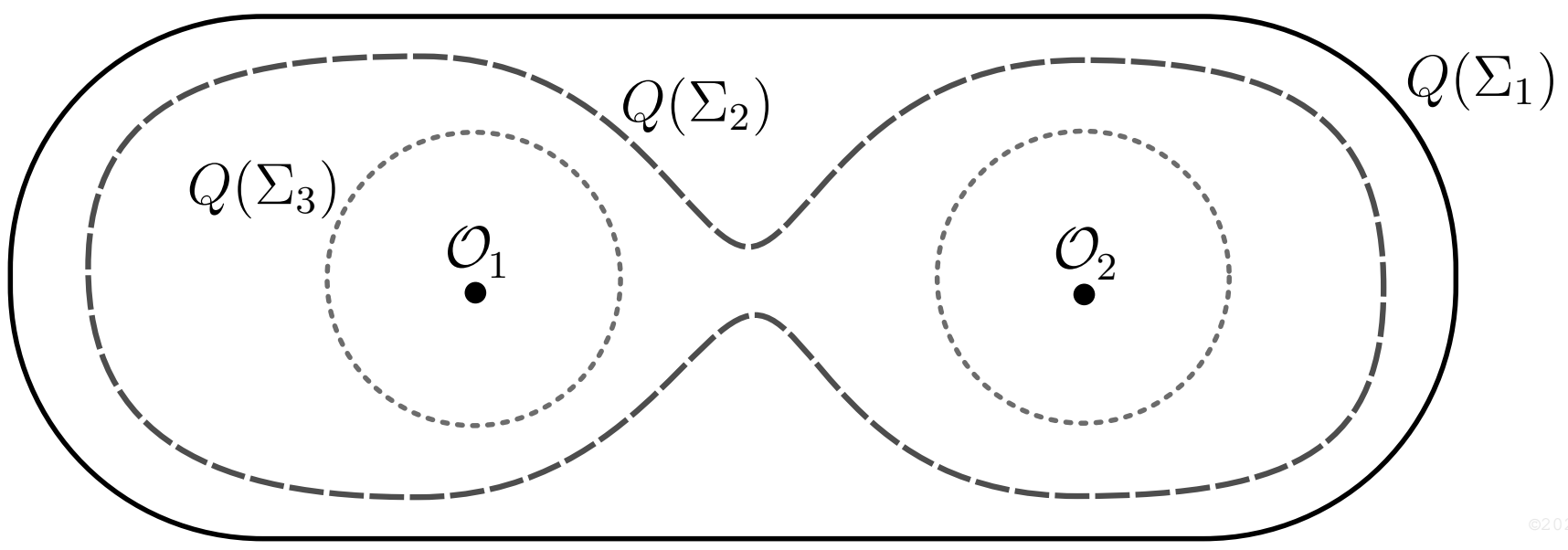

**Figure 1.82 /** Topological operator can be deformed to small surfaces around the operator insertions, so even though the operator is not local, the correlation functions depend locally on the points of insertions.

When we move the surface over the operator insertions, we obtain a contact term, which is the infinitesimal transformation of the operator under the symmetry generated by $Q_{\boldsymbol{\xi}}$. In particular, taking flat space and $\boldsymbol{\xi} \equiv \boldsymbol{\partial}_a$, we obtain momentum generators

$$\boldsymbol{P}_a(\Sigma) \equiv Q_{\boldsymbol{\partial}_a}(\Sigma) \equiv -\int_\Sigma dS_n \boldsymbol{T}^n{}_a\,,$$

which by Calculation 1.81→p.27 act as derivatives

$$\Big\langle \boldsymbol{P}_a(\Sigma)\mathcal{O}(x)\cdots\Big\rangle = \partial_a\Big\langle\mathcal{O}(x)\cdots\Big\rangle$$

where we took $\Sigma \equiv \partial B$ to be the boundary of a ball $B$ surrounding $x$ and no other operator insertions.

If we choose some particular foliation of spacetimes — as we will discuss below in Section 1.4 →p.29 — and define the symmetry charges as surface integrals over these slices, the above statement would be interpreted in an operator formalism as a commutation relation

$$[\boldsymbol{P}_a, \mathcal{O}(x)] = \partial_a \mathcal{O}(x)\,.$$

So even though the symmetry generators are not local operators, their commutators with local operators (or surrounding of local operators) are again local.

The algebra (commutation relations) of symmetry generators corresponds to how should one cross the surfaces of two topological operators. The important point of this discussion is, that this is a property independent of the particular foliation/operator formalism we choose.

## 1.4 Operator Formalism

Virtue of the functorial (path integral) formalism is that symmetries — both spacetime and internal ones — are usually manifest. Such a formulation is natural for conjuring up, comparing, and thinking about theories of various types, placed on backgrounds with diverse structures. Furthermore, calculations carried out via (formal) path-integral manipulations — oftentimes inspired by finite-dimensional analogs — can be especially convenient.

However, much less transparent are questions about the number of *physical* degrees of freedom, and the unitarity of the theory. It is necessary to address these issues on the level of Hilbert spaces and corresponding states/operators.

We will only briefly discuss such "operator formalisms", and by no means exhaustively treat the complicated subject of QFT in curved spacetimes.

### Foliation

Different foliations of spacetime in general give rise to different representations of the theory, since we obtain in general different Hilbert spaces of quantum states with different notions of "time evolution" and "time ordering" of operators. These can be sometimes related by symmetry transformations, but as we will briefly discuss below, in general they can be rather different.

By picking a specific foliation we always lose manifest covariance of the theory, but the spacetime symmetries naturally provide preferred foliations respecting given background structure.

**Example 1.83** (Static/Stationary Spacetimes). If the (Lorentzian) spacetime is static/stationary — we have a timelike Killing vector, which in static spaces is even surface orthogonal — then the preferred foliations (choice of time coordinate) respecting the time-translation symmetry share the same spatial geometry at each time slice. Corresponding Hilbert spaces are then naturally isomorphic as was discussed in Remark 1.67→p.21, and we can conveniently describe the system on one Hilbert space. ⌟

**Remark 1.84.** Without symmetries, we are quite lost, as computations become very complicated. This is why we usually work with spacetimes with maximum available symmetry — *maximally symmetric spacetimes*. One example is of course the Minkowski/Euclidean flat spacetime, and another one which we will cover in Chapter 4→p.91 is the *anti–de Sitter* spacetime. ⌟

## Vacuum and Particle Representation

In general, there is no preferred/distinguished ground state (vacuum vector) in Hilbert space $\mathcal{H}$, since there is no "natural" way to foliate spacetime — no preferred notion of time or energy (Hamiltonian). Usefulness of various descriptions depends on what questions we want to answer, which in turn depends on the observers asking those questions.

**Remark 1.85** (Particle Representations)**.** The notion of particles is tied to the choice of the vacuum state $|\text{vac}\rangle$, since particles are then excitations of this vacuum state.

In static/stationary spacetime (and for simplicity free theory), this is tied to the choice of positive-frequency solutions of Euler–Lagrange EOM — we pick out a subspace invariant under time-translation symmetry in space of complex solutions, to which we then assign annihilation/creation operators. This way we build up the Fock space of states over the vacuum state $|\text{vac}\rangle$.

In general, for different observers their natural particle representations are different, since foliations adapted to their worldlines are different, and their particle interpretations are different. In particular, a no-particle (vacuum) state in one representation can be viewed by another observer as a state with nonzero (or even infinite) number of particles. Such transformations between different representations are called *Bogoliubov transformations*.

So without enough symmetry, there is no natural way how to define what is a vacuum, or what is a particle. ⌟

**Remark 1.86** (Correlators as Vacuum Expectation Values)**.** Correlators in the sense of Definition 1.26→p.7 or Definition 1.61→p.20 are interpreted in the operator formalism as vacuum expectation values of time-ordered — perhaps better to say foliation ordered — products of operators, that is

$$\Big\langle \mathcal{O}_1(x_1)\cdots\mathcal{O}_n(x_n)\Big\rangle = \langle\text{vac}|\,\mathcal{T}\Big(\mathcal{O}_1(x_1)\cdots\mathcal{O}_n(x_n)\Big)\,|\text{vac}\rangle\,,$$

where $\mathcal{O}_\bullet$ are now represented on given Hilbert space $\mathcal{H}$ as elements of $\text{End}(\mathcal{H})$, and $\mathcal{T}$ orders them according to the chosen foliation. We have seen this in the context of QM in Example 1.65→p.20.

Let us stress again that different foliations give rise to different vacuum states $|\text{vac}\rangle$ and Hilbert space, different representations of operators on these Hilbert spaces, and different notions of time ordering. Nevertheless, in the end all of them calculate the same correlation functions. From this point of view, different operator formalisms are different interpretations of the same underlying physics, same underlying path integral.

**Remark 1.87** (Symmetry Charges). As mentioned briefly in Section 1.3→p.28, the charges corresponding to the symmetries of the theory are defined as integrals of the conserved currents over the slices of the foliation. Then the surrounding of some local operator $\mathcal{O}(x)$ by a charge $Q$ is represented as the commutator $[Q, \mathcal{O}(x)]$ — you can imagine how we can deform small surface around the operator insertion to a one surface above, and one below, with the orientation giving the relative minus sign. The finite exponentiated action (with some parameter $t$) is then given by

$$\mathrm{e}^{t[Q,\square]}\,\mathcal{O}(x) = \mathrm{e}^{tQ}\,\mathcal{O}(x)\,\mathrm{e}^{-tQ}\ .$$

**Example 1.88** (Minkowski Spacetime — Inertial Foliation). In flat Minkowski spacetime, we can foliate spacetime by hyperplanes of constant time, which corresponds to some class of inertial observers. This is adapted to the action of the Poincaré group (namely the subgroup of translations and spatial rotations). As usual, the vacuum state is invariant under the whole Poincaré group represented by unitary transformations. Foliation-ordering of operators is in this case given by the standard notion of time ordering (in the given time parameter tied to a chosen class of inertial observers).

Different choice of inertial observer corresponds to the foliations obtained by boosting the original foliation, see Figure 1.89→p.32 on the left. Since the vacuum is invariant, they share the same vacuum state, and thus the corresponding particle representation. We say that the particle interpretation is “unitarily equivalent” — descriptions of the same quantum state by different inertial observers are related by unitary transformation, with basis-independent characteristics (such as number of particles in a given state) staying the same.

**Example 1.90** (Minkowski Spacetime — Rindler Foliation). Suppose we wanted to describe the system from the point of view of uniformly accelerated observer. Observer then moves along an orbit of the boost action. A more natural foliation (adapted to the action of corresponding boost) is then the Rindler foliation (Rindler coordinates), see Figure 1.89→p.32 on the right.

Considering the causality structure of Minkowski, we see that there appear horizons (enclosing for example the right wedge containing the worldline of the observer), beyond which events cannot reach the observer.

To properly describe quantum field pervading the whole spacetime, we need to include into description also the left wedge — space of quantum states will be $\mathcal{H} \simeq \mathcal{H}_{\mathrm{L}} \otimes \mathcal{H}_{\mathrm{R}}$ — such that we can form Cauchy slice (with enough initial data for evolution of quantum field). At the end we can trace out through the left wedge $\mathcal{H}_{\mathrm{L}}$, and obtain density matrix operator carrying the whole information

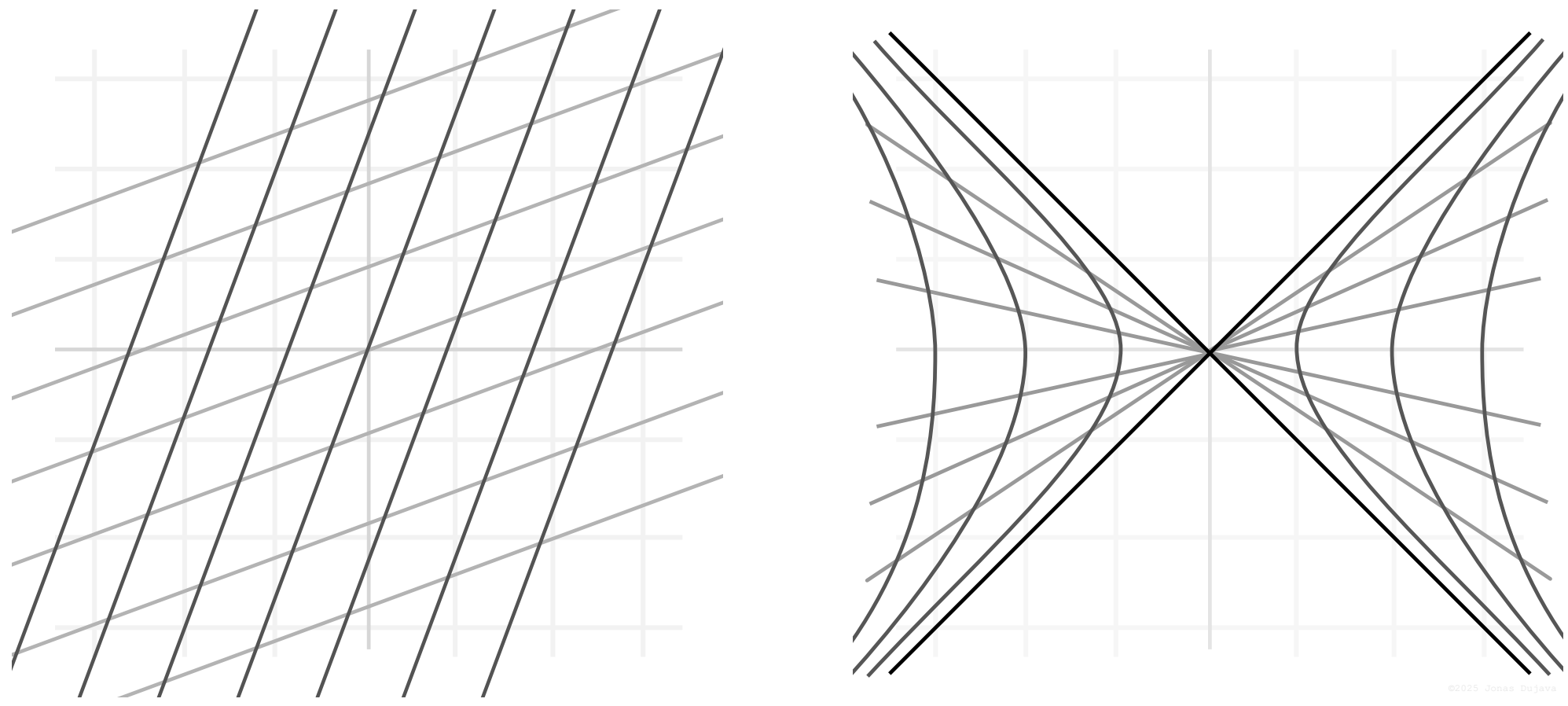

**Figure 1.89 /** Common foliations of Minkowski spacetime (or its part). The left figure shows the usual inertial foliation, while the right figure shows the Rindler foliation of the right/left wedges adapted to the trajectory of uniformly accelerated observer.

↳ which concerns the uniformly boosted observer in the right wedge.

From the point of view of the quantum field, the more natural choice is the inertial foliation discussed in Example 1.88→p.31. Most likely it rests in the corresponding *Minkowski* vacuum $|\mathrm{vac}\rangle_{\mathsf{Mink}}$, or some low excitation of it. How is such a state perceived by the accelerated observer? After partially tracing over the left wedge, we obtain a density matrix operator $\rho_\mathrm{R} \equiv \mathrm{Tr}_\mathrm{L}|\mathrm{vac}\rangle_{\mathsf{Mink}}\langle\mathrm{vac}|$, which actually turns out to be a thermal state — a uniformly accelerating observer interprets $|\mathrm{vac}\rangle_{\mathsf{Mink}}$ as a thermal bath — this is called the *Unruh effect*. ⌟

**Example 1.91** (Radial Quantization in CFTs)**.** For Conformal Field Theories (CFTs), which have in particular the scaling/dilatation symmetry, the common choice is the so-called *radial quantization*. Foliation by concentric spheres is adapted to the actions of rotations and dilatation. More on this in Section 3.4→p.78. ⌟

## Conjugation

The usual goal is to study unitary QFTs, but as we discussed in Section 1.1→p.12, when it is possible, it is often convenient to work in the Euclidean signature. In the following we will look at the Euclidean version of the Hermitian conjugation.

For the purposes of this and following chapter, we will explicitly indicate by a pre-superscript ${}^{L}/{}^{E}$ in what signature we consider the given object. As we will see shortly, the conjugation in the Euclidean signature slightly differs from the usual one in Lorentzian signature — a Euclidean version of Hermitian operator is typically not Hermitian in the usual sense. To avoid confusion, it is helpful to introduce extra nomenclature.

**Definition 1.92** (Real and Complex Operators)**.** Considering a standard inertial quantization in the Lorentzian signature, we say that an operator is *real* if it is Hermitian, that is ${}^{L}\mathcal{O}^\dagger = {}^{L}\mathcal{O}$. Otherwise, we say it is *complex* and denote its Lorentzian conjugation by ${}^{L}\mathcal{O}^{*} \equiv {}^{L}\mathcal{O}^{\dagger}$, and its continuation to Euclidean signature similarly by ${}^{E}\mathcal{O}^{*}$.

**Remark 1.93.** Alternatively, we can think of ${}^{L}\mathcal{O}^{*}$ as first performing the complex conjugation of the underlying classical field, and only then representing it as an operator on the Hilbert space, giving us precisely ${}^{L}\mathcal{O}^{\dagger}$. ⌟

**Remark 1.94.** In a unitary Lorentzian QFT, the symmetry generators are represented as conserved charges which are (anti)unitary operators — depending on whether we include $\mathring{\imath}$ in the finite exponentiated transformation — such that their exponentiated actions are unitary transformations of the Hilbert space. In particular, the energy–momentum tensor is real. ⌟

**Example 1.95** (Conjugation of Scalar Operators in Euclidean Signature)**.** Starting with a Lorentzian theory where time translations are generated by a (Hermitian) Hamiltonian $H$ as

$$ {}^{L}\mathcal{O}(t,\mathbf{x}) = \mathrm{e}^{\mathring{\imath}Ht}\,{}^{L}\mathcal{O}(0,\mathbf{x})\,\mathrm{e}^{-\mathring{\imath}Ht}\ , $$

an analytical continuation to the Euclidean time $\tau \equiv t_{\mathrm{E}} \equiv \mathring{\imath}t$ gives

$$ \begin{aligned} {}^{E}\mathcal{O}(\tau,\mathbf{x}) &\equiv {}^{L}\mathcal{O}(t=-\mathring{\imath}\tau,\mathbf{x}) = \mathrm{e}^{\tau H}\,{}^{L}\mathcal{O}(0,\mathbf{x})\,\mathrm{e}^{-\tau H}\,, \\ \Longrightarrow {}^{E}\mathcal{O}(\tau,\mathbf{x})^{\dagger} &= \mathrm{e}^{-\tau H}\,{}^{L}\mathcal{O}^{*}(0,\mathbf{x})\,\mathrm{e}^{\tau H} = {}^{E}\mathcal{O}^{*}(-\tau,\mathbf{x})\,, \end{aligned} $$

so the conjugation in Euclidean signature contains extra time reflection. ⌟

**Example 1.96** (Euclidean Version of Tensor Operators)**.** Before discussing the conjugation of tensor operators, let us first comment on an additional detail in their Wick-rotation. Since $\mathbf{d}t \equiv \mathbf{d}(-\mathring{\imath}\tau) = -\mathring{\imath}\,\mathbf{d}\tau\,,\ \frac{\boldsymbol{\partial}}{\boldsymbol{\partial}t} \equiv \frac{\boldsymbol{\partial}}{\boldsymbol{\partial}(-\mathring{\imath}\tau)} = \mathring{\imath}\frac{\boldsymbol{\partial}}{\boldsymbol{\partial}\tau}$, we typically add factors of $\mp\mathring{\imath}$ to the time components of tensors. In this way Lorentzian correlators covariant under $\mathsf{SO}(1,d-1)$ become (after analytical continuation) Euclidean correlators covariant under $\mathsf{SO}(d)$ rotations.

As an example, for a vector operator we have

$$ \begin{aligned} {}^{E}\mathcal{O}^{0}(\tau,\mathbf{x}) &\equiv -\mathring{\imath}\,{}^{L}\mathcal{O}^{0}(-\mathring{\imath}\tau,\mathbf{x})\,, \\ {}^{E}\mathcal{O}^{i}(\tau,\mathbf{x}) &\equiv \quad {}^{L}\mathcal{O}^{i}(-\mathring{\imath}\tau,\mathbf{x})\,. \end{aligned} $$

⌟

The following definition slightly generalizes examples provided above.

**Definition 1.97** (Euclidean Conjugation, Time Reflection)**.** A local Euclidean tensor operator with $l$ spacetime indices behaves under conjugation as

$$ {}^{E}\mathcal{O}^{a_1\ldots a_l}(\tau,\mathbf{x})^{\dagger} = \Theta^{a_1}_{b_1}\cdots\Theta^{a_l}_{b_l}\,{}^{E}\mathcal{O}^{b_1\ldots b_l *}(-\tau,\mathbf{x}) \equiv \Theta\Big[{}^{E}\mathcal{O}^{a_1\ldots a_l *}(\tau,\mathbf{x})\Big]\,, $$

where $\Theta^{a}_{b} \equiv \delta^{a}_{b} - 2\delta^{a}_{0}\delta^{0}_{b}$, so the *time-reflection* operation $\Theta$ in addition to performing $(\tau \mapsto -\tau)$ includes a factor of $(-1)^{N_\perp}$ with $N_\perp$ being the number of time components — indices perpendicular to the reflection plane $\tau = 0$.

**Remark 1.98.** We could freely take all indices to be the upper ones, since we are in the Euclidean signature. ⌟

**Remark 1.99.** Such conjugation is in particular consistent for operators whose indices come from derivatives, because

$$\begin{aligned}\partial_\tau\Big({}^{E}\mathcal{O}(\tau,\mathbf{x})^\dagger\Big) &= \partial_\tau\Big({}^{E}\mathcal{O}^*(-\tau,\mathbf{x})\Big) = -{}^{E}\partial_\tau\mathcal{O}^*(-\tau,\mathbf{x}) = \Big({}^{E}\partial_\tau\mathcal{O}(\tau,\mathbf{x})\Big)^\dagger,\\ \partial_{x^i}\Big({}^{E}\mathcal{O}(\tau,\mathbf{x})^\dagger\Big) &= \partial_{x^i}\Big({}^{E}\mathcal{O}^*(-\tau,\mathbf{x})\Big) = {}^{E}\partial_{x^i}\mathcal{O}^*(-\tau,\mathbf{x}) = \Big({}^{E}\partial_{x^i}\mathcal{O}(\tau,\mathbf{x})\Big)^\dagger.\end{aligned}$$

⌟

**Remark 1.100** (Conjugation of Conserved Charges)**.** One needs to be slightly careful when applying Euclidean conjugation to conserved charges, which are nonlocal operators. However, since they are integrals of local energy–momentum tensor, we can calculate (for simplicity we consider flat space $\mathbb{R}^d$)

$$\begin{aligned}{}^{E}Q^\dagger_{\boldsymbol{\xi}} &\equiv \Big(-\int_{\tau=\text{const.}}\mathrm{d}^{d-1}\mathbf{x}\;{}^{E}\boldsymbol{T}^0{}_b(\tau,\mathbf{x})\,\xi^b(\tau,\mathbf{x})\Big)^\dagger = -\int_{\tau=\text{const.}}\mathrm{d}^{d-1}\mathbf{x}\;\Big(-{}^{E}\boldsymbol{T}^0{}_a(-\tau,\mathbf{x})\Theta^a_b\Big)\,\xi^b(\tau,\mathbf{x})\\ &= -\int_{\tau=-\text{const.}}\mathrm{d}^{d-1}\mathbf{x}\;{}^{E}\boldsymbol{T}^0{}_a(\tau,\mathbf{x})\Big(-\Theta^a_b\xi^b(-\tau,\mathbf{x})\Big) \equiv {}^{E}Q_{-\mathrm{Ad}_\Theta\boldsymbol{\xi}} \equiv {}^{E}Q_{-\Theta\boldsymbol{\xi}\Theta}\,.\end{aligned}$$

where the steps are given as

**(1)** definition of Euclidean charge ${}^{E}Q_{\boldsymbol{\xi}}$,

**(2)** reality of $\boldsymbol{T}^{\bullet}_{\bullet}$ and $\xi^{\bullet}$, and Definition 1.97 →p.33 which gives a minus sign for the upper index "0", and $\Theta^{\bullet}_{\bullet}$ for the lower one,

**(3)** exchange ($\tau \mapsto -\tau$) and simple rearrangement,

**(4)** conserved charge being topological, it does not depend on the choice of $\tau$ "const.", and definition of $\mathrm{Ad}_\Theta\boldsymbol{\xi} \equiv \Theta_*\boldsymbol{\xi}$,

**(5)** understanding both $\boldsymbol{\xi}$ and $\Theta$ as operators on functions, and $\Theta^{-1} = \Theta$. ⌟

**Example 1.101** (Conjugation of Momentum Operators)**.** Either by using the previous Remark 1.100 →p.34 for isometry generators — $\mathrm{Ad}_\Theta\frac{\partial}{\partial\tau} = -\frac{\partial}{\partial\tau}$ and $\mathrm{Ad}_\Theta\frac{\partial}{\partial x^i} = \frac{\partial}{\partial x^i}$ — or by writing

$${}^{E}\boldsymbol{P}_a \equiv -\int_{\tau=\text{const.}}\mathrm{d}^{d-1}\mathbf{x}\;{}^{E}\boldsymbol{T}^0{}_a(\tau,\mathbf{x})\,,$$

and applying Definition 1.97 →p.33, it follows that ${}^{E}\boldsymbol{P}_0$ is Hermitian and ${}^{E}\boldsymbol{P}_i$ are anti-Hermitian. So by identifying

$${}^{E}\boldsymbol{P}_0 \equiv H \equiv {}^{L}\boldsymbol{P}_0\,, \qquad {}^{E}\boldsymbol{P}_i \equiv -\mathring{\imath}\,{}^{L}\boldsymbol{P}_i\,,$$

we recover the standard Lorentzian (Hermitian) momentum operators, together with agreement of translation formulas (for scalar operator)

$${}^{E}\mathcal{O}(x) = \mathrm{e}^{x\cdot{}^{E}\boldsymbol{P}}\,{}^{E}\mathcal{O}(0)\,\mathrm{e}^{-x\cdot{}^{E}\boldsymbol{P}} \iff {}^{E}\mathcal{O}(\tau,\mathbf{x}) = \mathrm{e}^{\tau H-\mathring{\imath}\mathbf{x}\cdot{}^{L}\boldsymbol{P}}\,{}^{L}\mathcal{O}(0,\mathbf{0})\,\mathrm{e}^{-\tau H+\mathring{\imath}\mathbf{x}\cdot{}^{L}\boldsymbol{P}}\,,$$

the first using directly the Euclidean generators of translations, and the second obtained by Wick rotation of the standard Lorentzian formula. ⌟

**Remark 1.102.** From the Lorentzian point of view the time-direction is clearly special, and this is reflected in the way Euclidean conjugation acts. If we take as a starting point the Euclidean theory, there is initially no preferred direction. It only appears when we choose a direction we call "time" — we conventionally label the associated index with "0" — and attempt to quantize it in a way such that it can be analytically continued to a unitary Lorentzian theory. ⌟

## Unitarity and Reflection Positivity

We have seen that different foliations lead to different Hilbert spaces, and to possibly different notions of conjugation. As shown in previous subsection, when we Wick-rotate a Lorentzian theory to the Euclidean signature, we obtain a specific type of conjugation which reflects the time direction.

Can we also turn this around? Given a Euclidean path integral, how do we know it calculates Wick-rotated correlators of a unitary Lorentzian theory?

The necessary condition for a unitary theory is that the norms $\langle\psi|\psi\rangle$ for every $\psi \in \mathcal{H}$ must be non-negative. It is reflected in the Euclidean theory by the so-called *reflection positivity* condition.

**Definition 1.103** (Reflection Positivity)**.** Let $P$ be a polynomial of Euclidean operators ${}^{E}\mathcal{O}$ (possibly smeared) in a half-space with $\tau < 0$. Then we say that the Euclidean theory is *reflection positive* if all correlation functions satisfy

$$\Big\langle \Theta[P^*]\, P \Big\rangle \geq 0\,.$$

**Remark 1.104** (Lorentzian Positivity)**.** The Lorentzian counterpart of positivity is simply $\langle P^* P\rangle \geq 0$, where $P$ is build out of ${}^{L}\mathcal{O}$. ⌟

**Proposition 1.105** (Unitarity Leads to Reflection Positivity)**.** A Euclidean theory obtained from a Wick-rotated unitary Lorentzian theory is reflection positive.

**Proof.** Consider a state given by a time-ordered product of Euclidean operators acting on the vacuum

$$|\psi\rangle \equiv {}^{E}\mathcal{O}_1(x_1)\cdots {}^{E}\mathcal{O}_n(x_n)|\text{vac}\rangle \equiv P|\text{vac}\rangle$$

with $0 > \tau_1 \geq \cdots \geq \tau_n$. Since Definition 1.97→p.33 gives us

$$\begin{aligned}\langle\psi| &\equiv \Big({}^{E}\mathcal{O}_1(x_1)\cdots {}^{E}\mathcal{O}_n(x_n)|\text{vac}\rangle\Big)^\dagger \\ &= \langle\text{vac}|\Theta\Big[{}^{E}\mathcal{O}_n^*(x_n)\Big]\cdots\Theta\Big[{}^{E}\mathcal{O}_1^*(x_1)\Big] \equiv \langle\text{vac}|\Theta[P^*]\,,\end{aligned}$$

we obtain the positivity reflection condition for the correlator as

$$0 \leq \langle\psi|\psi\rangle = \langle\text{vac}|\,\Theta[P^*]\,P\,|\text{vac}\rangle \equiv \Big\langle\Theta[P^*]\,P\Big\rangle \overset{!}{\geq} 0\,.$$

Note that inside VEV was a correctly time-ordered product of operators. □

**Remark 1.106** (Reconstruction Theorems)**.** For a unitary QFT in a flat space, the famous *reconstruction theorems* assert that the correlation functions — called *Wightman distributions* in the Lorentzian $\mathsf{Mink}_d \equiv \mathbb{R}^{1,d-1}$ and *Schwinger functions* in the Euclidean $\mathbb{R}^d$ — completely determine the quantum theory, including its Hilbert space.

The positivity requirements mentioned above are absolutely crucial for such reconstruction (in addition to other technical assumptions), otherwise one could not invert the Wick-rotation to obtain a local unitary QFT in Lorentzian signature, as there would be no probability interpretation for states in $\mathcal{H}$ [20] [21] [22]. ⌟

As unitarity and reflection positivity in the respective signatures are essentially equivalent, we will often refer to both as "unitarity", even though most of the time we will be working in the Euclidean signature. Some of their implications in the context of CFTs will be discussed in Section 3.4→p.78 and Section 3.5→p.86.

Finally, let us remark that — at least on a formal level — the Euclidean path integral is reflection-positive.

**Remark 1.107** (Reflection Positivity of Euclidean Path Integral)**.** Using the locality as in Idea 1.45→p.14, we can split the path integral into positive ($\tau > 0$) and negative ($\tau > 0$) Euclidean "time" parts. Assuming that the Euclidean path integral measure $\mathcal{D}\phi \, \mathrm{e}^{-\overset{E}{\mathcal{S}}[\phi]}$ is invariant under the combined action of the time-reflection and complex conjugation of all fields $\phi$, we can write

$$\Big\langle \Theta[P^*]\, P \Big\rangle = \int\limits_{\mathit{Fields}} \mathcal{D}\phi \; \mathrm{e}^{-\overset{E}{\mathcal{S}}[\phi]} \, P[\Theta\phi^*] P[\phi] = \int\limits_{\mathit{Fields}_{\tau=0}} \mathcal{D}\phi_0 \, \left| \int\limits_{\mathit{Fields}_{\tau<0},\, \phi_-|_{\tau=0}=\phi_0} \mathcal{D}\phi_- \; \mathrm{e}^{-\overset{E}{\mathcal{S}}_{\tau<0}[\phi_-]} \, P[\phi_-] \right|^2 \geq 0 \, .$$

Alternatively, since the action is real, we can separate and split the interaction part of the action as (again the locality is crucial)

$$\mathrm{e}^{-\overset{E}{\mathcal{S}}{}^{\mathrm{int}}[\phi]} = \mathrm{e}^{-\overset{E}{\mathcal{S}}{}^{\mathrm{int}}_{\tau>0}[\phi]} \, \mathrm{e}^{-\overset{E}{\mathcal{S}}{}^{\mathrm{int}}_{\tau<0}[\phi]} \equiv \Theta[F^*]\, F \, , \quad \text{with } F \equiv \mathrm{e}^{-\overset{E}{\mathcal{S}}{}^{\mathrm{int}}_{\tau<0}[\phi]} \, ,$$

giving us (assuming the reflection positivity of the free theory)

$$\Big\langle \Theta[P^*]\, P \Big\rangle = \Big\langle \Theta\big[(FP)^*\big]\, (FP) \Big\rangle_{\mathrm{free}} \geq 0 \, .$$

Of course, while this is formally true for local Euclidean theories, difficulties can (and do) arise when one tries to turn these expressions into well-defined quantities. ⌟

# 2 Beyond Basic Perturbation Theory

It is often the case that a "basic" perturbation theory — Feynman diagrams to some finite order — is not enough to describe the phenomena we are interested in. Alternatively, we want to know which properties are present in the full theory, and perturbation theory is sometimes able to give us some hints, but more often falls short of giving us an adequate picture.

Thus, it is highly desirable to have some non-perturbative methods — or ones which enhance the ordinary perturbation theory. In this chapter we will only briefly comment on some notions and concepts which are useful in this context, and for understanding QFT in general. For much more thorough accounts and deeper insights we refer the reader to the literature, for example [5] [23] [24] [25] [26] [27] [28].

## 2.1 Effective Action

In the classical theory, the action $\mathcal{S}$ directly governs the dynamics of the system, in particular the classical fields $\phi_{\mathrm{cl}}$ must extremize the action. Things complicate in the quantum theory, where we need to perform path integrals over all possible field configurations.

In a free theory, or in the classical limit $\hbar \to 0$, the path integral is dominated by the classical solutions, and we can directly use the classical action $\mathcal{S}$ to analyze the system. However, for an interacting QFT (outside the classical limit), we encounter non-classical contributions to $\mathcal{Z}$ from "quantum fluctuations", which can have huge impact on the physics.

Is there a way to package (at least partially) quantum effects into some kind of effective action, for which we can again use a classical intuition? Access to such an object would allow us to properly study various aspects of QFTs — the phase structure, symmetries, or renormalizability.

### Different Types of Effective Actions

Depending on the context, the term "effective action" can mean different things:

**(1)** Wilsonian low-energy effective action — We integrate out the "heavy" fields (or high-momentum field modes), leaving us with an effective action for the remaining degrees of freedom.

Such effective action contains contributions from the high-energy physics, and defines an *Effective Field Theory* (EFT) valid only for the description of low-energy phenomena. We will discuss this more in Section 2.2 → p.41.

**(2)** Quantum (1PI) effective action — We "integrate out the loops", and obtain effective action including the quantum corrections.

By examining its form, we can read out exact/full propagators and 1PI vertices, reducing the calculation to drawing only "classical" tree graphs. We focus on this type of effective action in the following.

**Remark 2.1.** As we can see, effective actions correspond to a certain partial evaluation of the path integral. For interacting theories, even such partial path-integration is usually not feasible to perform exactly, and we have to resort to some perturbation scheme. ⌟

## Quantum Effective Action

For discussing the phase structure — and possible *spontaneous symmetry breaking* — we ultimately want to know whether field $\boldsymbol{\phi}$ obtains nonzero VEV, so we want to study the expectation values of the fields. It turns out to be fruitful to take into consideration the presence of the sources $\boldsymbol{J}$.

**Definition 2.2** (Expectation Values). The *expectation values* $\boldsymbol{\Phi}$ of the fields $\boldsymbol{\phi}$ in the presence of sources $\boldsymbol{J}$ are given by

$$\boldsymbol{\Phi}_{\boldsymbol{J}}(x) \equiv \left\langle \boldsymbol{\phi}(x) \right\rangle_{\boldsymbol{J}} = \frac{\hbar}{i} \frac{1}{\mathcal{Z}[\boldsymbol{J}]} \frac{\delta}{\delta \boldsymbol{J}(x)} \mathcal{Z}[\boldsymbol{J}] \equiv \frac{\delta \mathcal{W}[\boldsymbol{J}]}{\delta \boldsymbol{J}(x)} .$$

Our goal is to obtain Quantum Effective Action $\boldsymbol{\Gamma}$, which would govern $\boldsymbol{\Phi}$ similarly to how $\mathcal{S}$ governs $\boldsymbol{\phi}_{\text{cl}}$. We thus promote $\boldsymbol{\Phi}_{\boldsymbol{J}}$ to a primary variable $\boldsymbol{\Phi}$ and demote $\boldsymbol{J}$ to a derived one $\boldsymbol{J}_{\boldsymbol{\Phi}}$, which is (implicitly) defined such that

$$\boldsymbol{\Phi}_{\boldsymbol{J}} = \frac{\delta \mathcal{W}}{\delta \boldsymbol{J}} \quad \leftrightsquigarrow \quad \boldsymbol{\Phi} = \frac{\delta \mathcal{W}}{\delta \boldsymbol{J}} \bigg|_{\boldsymbol{J}=\boldsymbol{J}_{\boldsymbol{\Phi}}}$$

holds.

Inspired by similar scenarios in

- Classical Mechanics — Lagrangian $L \underset{\text{velocities } \dot{\boldsymbol{q}} \longleftrightarrow \boldsymbol{p} \text{ momenta}}{\overset{\text{Legendre Transform}}{\leftrightsquigarrow}} H$ Hamiltonian ,
- Thermodynamics — Helmholtz Energy $F \underset{\text{volume } V \longleftrightarrow p \text{ pressure}}{\overset{\text{Legendre Transform}}{\leftrightsquigarrow}} G$ Gibbs Energy ,

we come to the following definition.

**Definition 2.3** (Quantum Effective Action). The *Quantum Effective Action* $\boldsymbol{\Gamma}$ is defined by the following relation

$$\text{Connected Generating Functional } \mathcal{W} \underset{\text{sources } \boldsymbol{J} \longleftrightarrow \boldsymbol{\Phi} \text{ expectation values}}{\overset{\text{Legendre Transform}}{\leftrightsquigarrow}} \boldsymbol{\Gamma} \text{ Quantum Effective Action} ,$$

that is

$$\boldsymbol{\Gamma}[\boldsymbol{\Phi}] = \Big( \mathcal{W}[\boldsymbol{J}] - \boldsymbol{J} \bullet \boldsymbol{\Phi} \Big) \bigg|_{\boldsymbol{J}=\boldsymbol{J}_{\boldsymbol{\Phi}} \,:\, \boldsymbol{\Phi}=\frac{\delta \mathcal{W}}{\delta \boldsymbol{J}}} ,$$

or conversely

$$\mathcal{W}[\boldsymbol{J}] = \Big( \boldsymbol{\Gamma}[\boldsymbol{\Phi}] + \boldsymbol{J} \bullet \boldsymbol{\Phi} \Big) \bigg|_{\boldsymbol{\Phi}=\boldsymbol{\Phi}_{\boldsymbol{J}} \,:\, \boldsymbol{J}=-\frac{\delta \boldsymbol{\Gamma}}{\delta \boldsymbol{\Phi}}} .$$

**Remark 2.4.** As is usual — both in the context of classical mechanics and statistical physics/thermodynamics — in the following we will oftentimes omit the dependence of the variables, since it will be clear from the context if we should express $\boldsymbol{J}$ as function of $\boldsymbol{\Phi}$, or vice versa. ⌟

**Remark 2.5.** As is usual for Legendre Transform, we have (using the chain rule)

$$\frac{\delta \boldsymbol{\Gamma}}{\delta \boldsymbol{\Phi}} = \left( \frac{\delta \mathcal{W}}{\delta \boldsymbol{J}} \bullet \frac{\delta \boldsymbol{J}}{\delta \boldsymbol{\Phi}} - \frac{\delta \boldsymbol{J}}{\delta \boldsymbol{\Phi}} \bullet \boldsymbol{\Phi} \right) - \boldsymbol{J} = 0 - \boldsymbol{J} \,,$$

so in the absence of sources $\boldsymbol{J} = 0$, we obtain equation for $\boldsymbol{\Phi}$ as a stationary point of $\boldsymbol{\Gamma}$. This is one sense in which $\boldsymbol{\Gamma}$ is seen as the quantum counterpart of $\mathcal{S}$. ⌟

## 1PI Effective Action

Can we take this analogy even further? It turns out interesting to take $\boldsymbol{\Gamma}$ as a full-fledged classical action on its own, and consider the corresponding quantum theory given by

$$\mathrm{e}^{\frac{i}{\hbar}\mathcal{W}_{\boldsymbol{\Gamma}}[\boldsymbol{J}]} \equiv \int \mathcal{D}\boldsymbol{\Phi} \ \mathrm{e}^{\frac{i}{\hbar}(\boldsymbol{\Gamma}[\boldsymbol{\Phi}] + \boldsymbol{J}\bullet\boldsymbol{\Phi})} \,,$$

where we keep the “Planck constant” $\hbar$ distinct from $\hbar$ to avoid mixing them, since $\hbar$ is already present in $\boldsymbol{\Gamma}$.

**Proposition 2.6** (Quantum Effective Action as 1PI Effective Action)**.** The quantum effective action $\boldsymbol{\Gamma}$ is also the One-Particle-Irreducible (1PI) effective action, meaning that $\mathcal{W}$ is (perturbatively) given by all connected *tree* graphs, with vertices and propagators given by $\boldsymbol{\Gamma}$. Schematically, we have

$$i\mathcal{W}[\boldsymbol{J}] \simeq \int_{\substack{\text{connected}\\\text{tree graphs}}} \mathcal{D}\boldsymbol{\Phi} \ \mathrm{e}^{i(\boldsymbol{\Gamma}[\boldsymbol{\Phi}] + \boldsymbol{J}\bullet\boldsymbol{\Phi})} \quad \Longleftrightarrow \quad \frac{i}{\hbar}\boldsymbol{\Gamma}[\boldsymbol{\Phi}] \simeq \int_{\substack{\text{connected}\\\text{1PI graphs}}} \mathcal{D}\phi \ \mathrm{e}^{\frac{i}{\hbar}\mathcal{S}[\phi+\boldsymbol{\Phi}]} \,.$$

**Idea of the Proof.** See any of [26] [27] [5] [10] for the detailed proof. The idea is that connected graphs contributing to $\mathcal{W}_{\boldsymbol{\Gamma}}[\boldsymbol{J}]$, see Proposition 1.38 →p.11, have the following dependence on $\hbar$ — propagator is inverse of the quadratic term in $\frac{1}{\hbar}\boldsymbol{\Gamma}$ and hence proportional to $\hbar$, while each interaction vertex carries a factor of $1/\hbar$ from the $\frac{i}{\hbar}$ in front of $\boldsymbol{\Gamma}$ — so a graph with $I$ internal lines and $V$ vertices is proportional to $\hbar^{I-V} = \hbar^{L-1}$, where $L = I - V + 1$ is the number of loops for a connected graph.

The leading contribution $\mathcal{W}^{(0)}_{\boldsymbol{\Gamma}} \propto \hbar^0$ for $\hbar \to 0$ is thus given by the *connected tree graphs*, but alternatively it is given by the stationary-phase/saddle-point approximation of the path integral, giving us

$$\mathcal{W}^{(0)}_{\boldsymbol{\Gamma}}[\boldsymbol{J}] = \Big(\boldsymbol{\Gamma}[\boldsymbol{\Phi}] + \boldsymbol{J}\bullet\boldsymbol{\Phi}\Big)\Big|_{\boldsymbol{\Phi}=\boldsymbol{\Phi}_{\boldsymbol{J}}\,:\,\boldsymbol{J}=-\frac{\delta\boldsymbol{\Gamma}}{\delta\boldsymbol{\Phi}}} \equiv \mathcal{W}[\boldsymbol{J}] \,,$$

where we recognized the Legendre transform relation between $\boldsymbol{\Gamma}$ and $\mathcal{W}$. Since any connected graph can be understood as a tree graph made with vertices representing ⤵

1PI graphs, the interpretation of $\boldsymbol{\Gamma}$ as the 1PI effective action follows. □

**Remark 2.7** (1PI Vertices, Exact Propagator). In the light of Proposition 2.6→p.39, we can see that $\boldsymbol{\Gamma}$ is the quantum corrected counterpart of $\mathcal{S}$, having for the interactions the renormalized 1PI vertices, and for the propagator the exact/full propagator. Indeed, using Remark 2.5→p.39 and Definition 2.2→p.38 we have

$$\frac{\delta^2 \boldsymbol{\Gamma}}{\delta \boldsymbol{\Phi} \delta \boldsymbol{\Phi}} = -\frac{\delta \boldsymbol{J}}{\delta \boldsymbol{\Phi}} = -\left(\frac{\delta \boldsymbol{\Phi}}{\delta \boldsymbol{J}}\right)^{-1} = -\left(\frac{\delta^2 \mathcal{W}}{\delta \boldsymbol{J} \delta \boldsymbol{J}}\right)^{-1}$$

so the quadratic part of $\boldsymbol{\Gamma}$ is the inverse of the full propagator

$$\big\langle \boldsymbol{\phi}\boldsymbol{\phi} \big\rangle_{\text{conn}} = \frac{\hbar}{i} \frac{\delta^2 \mathcal{W}}{\delta \boldsymbol{J} \delta \boldsymbol{J}}\bigg|_{\boldsymbol{J}=0} = \left(-\frac{i}{\hbar} \frac{\delta^2 \boldsymbol{\Gamma}}{\delta \boldsymbol{\Phi} \delta \boldsymbol{\Phi}}\right)^{-1}\bigg|_{\boldsymbol{\Phi}=0} .$$

⌟

**Remark 2.8** (Energy Interpretation, Effective Potential). There is also very important energy interpretation of $\boldsymbol{\Gamma}$. For constant $\boldsymbol{\Phi} = \boldsymbol{\Phi}_0$ we can write

$$\boldsymbol{\Gamma}[\boldsymbol{\Phi}_0] = -\mathcal{V}_{\mathcal{M}} V(\boldsymbol{\Phi}_0) \,,$$

where we factored out the volume $\mathcal{V}_{\mathcal{M}} \equiv \operatorname{vol}(\mathcal{M})$ of the spacetime, and $V(\boldsymbol{\Phi}_0)$ is the *effective potential* — for constant $\boldsymbol{\Phi} = \boldsymbol{\Phi}_0$ we are only left with potential terms in the effective Lagrangian.

It can be then shown that the value of $V(\boldsymbol{\Phi}_0)$ is the *minimum* of the expectation value of the *energy density* among all states satisfying $\langle \boldsymbol{\phi} \rangle_\Omega \equiv \langle \Omega | \boldsymbol{\phi} | \Omega \rangle = \boldsymbol{\Phi}_0$. The ground state (vacuum) is thus found by minimizing the effective potential $V$.

Say a minimum is at $\boldsymbol{\Phi}_g$. To avoid working with linear "tadpole" terms in the effective action $\boldsymbol{\Gamma}$ (or fields with nonzero VEV), the calculations naturally proceed in terms of the shifted fields $\widetilde{\boldsymbol{\phi}} \equiv \boldsymbol{\phi} - \boldsymbol{\Phi}_g$. If the minimum $\boldsymbol{\Phi}_g$ is not invariant under some symmetry of the original action $\mathcal{S}$, we say that the symmetry is *spontaneously broken*. ⌟

**Remark 2.9** (Slavnov–Taylor Identities). Consider a continuous symmetry transformation generated by ($\boldsymbol{F}$ depends locally on $\boldsymbol{\phi}$)

$$\boldsymbol{\phi} \longmapsto \boldsymbol{\phi}' \equiv \boldsymbol{\phi} + \varepsilon \boldsymbol{F}[\boldsymbol{\phi}] \,,$$

which leaves the (path-integral measure×action) invariant, that is

$$\mathcal{D}\boldsymbol{\phi}' \, \mathrm{e}^{\frac{i}{\hbar}\mathcal{S}[\boldsymbol{\phi}']} = \mathcal{D}\boldsymbol{\phi} \, \mathrm{e}^{\frac{i}{\hbar}\mathcal{S}[\boldsymbol{\phi}]} \,.$$

Then we have

$$\begin{aligned}
\mathcal{Z}[\boldsymbol{J}] &= \int \mathcal{D}\boldsymbol{\phi}' \, \mathrm{e}^{\frac{i}{\hbar}(\mathcal{S}[\boldsymbol{\phi}'] + \boldsymbol{J}\bullet\boldsymbol{\phi}')} = \int \mathcal{D}\boldsymbol{\phi} \, \mathrm{e}^{\frac{i}{\hbar}(\mathcal{S}[\boldsymbol{\phi}] + \boldsymbol{J}\bullet\boldsymbol{\phi} + \varepsilon \boldsymbol{J}\bullet\boldsymbol{F}[\boldsymbol{\phi}])} \\
&= \int \mathcal{D}\boldsymbol{\phi} \, \mathrm{e}^{\frac{i}{\hbar}(\mathcal{S}[\boldsymbol{\phi}] + \boldsymbol{J}\bullet\boldsymbol{\phi})} \Big(1 + \varepsilon \tfrac{i}{\hbar} \boldsymbol{J} \bullet \boldsymbol{F}[\boldsymbol{\phi}] + \ldots\Big) \\
&\equiv \mathcal{Z}[\boldsymbol{J}]\Big(1 + \varepsilon \tfrac{i}{\hbar} \boldsymbol{J} \bullet \big\langle \boldsymbol{F}[\boldsymbol{\phi}] \big\rangle_{\boldsymbol{J}} + \ldots\Big) \quad \Longrightarrow \quad \boldsymbol{J} \bullet \big\langle \boldsymbol{F}[\boldsymbol{\phi}] \big\rangle_{\boldsymbol{J}} = 0 \,.
\end{aligned}$$

↳ Setting $\boldsymbol{J} = \boldsymbol{J_\Phi}$, and using Remark 2.5→p.39, we can write

$$\frac{\delta \boldsymbol{\Gamma}}{\delta \boldsymbol{\Phi}} \bullet \left\langle \boldsymbol{F}[\phi] \right\rangle_{\boldsymbol{J}_\Phi} = 0\,.$$

In other words, the effective action $\boldsymbol{\Gamma}[\boldsymbol{\Phi}]$ is invariant under the transformation

$$\boldsymbol{\Phi} \longmapsto \boldsymbol{\Phi}' \equiv \boldsymbol{\Phi} + \varepsilon \left\langle \boldsymbol{F}[\phi] \right\rangle_{\boldsymbol{J}_\Phi}\,.$$

Such symmetries are known as *Slavnov–Taylor identities*.

In the important special case where $\boldsymbol{F}$ is linear, we have $\left\langle \boldsymbol{F}[\phi] \right\rangle_{\boldsymbol{J}_\Phi} = \boldsymbol{F}[\boldsymbol{\Phi}]$, and the classical symmetry is (at least formally) preserved also at the quantum level.

Quantum corrections in such a case do not generate terms in $\boldsymbol{\Gamma}$ which would break the symmetry, so we do not have to worry about renormalizing them. Since we can remove all ultraviolet divergences by adding appropriate counterterms *before* we shift the fields — see Remark 2.8→p.40 — the possible spontaneous symmetry breakdown does not affect the divergence structure of the theory.

For non-linear $\boldsymbol{F}$, the obtained symmetry of $\boldsymbol{\Gamma}$ is not generally the same as the original symmetry of $\mathcal{S}$. One important example is the BRST transformation encountered in gauge theories, which complicates the analysis of their renormalization. ⌟

## 2.2 Renormalization Flow

Since we do not have access to arbitrary energies to probe the high-energy "ultraviolet" (UV) physics, our experience mostly lies in the large-distance "infrared" (IR) domain. In the following we will look on the big picture ideas concerning effective description of such long-distance physics.

**Idea 2.10** (Renormalization Group Framework)**.** A common idea of *Renormalization Group* techniques is the following:

**(1)** have a system, and identify the physical aspects of interest,

**(2)** re-express the parameters in terms of another set — usually we choose simpler ones by some kind of coarse-graining, thus decimating certain degrees of freedom — while keeping the physical aspects of interest *unchanged*,

**(3)** study how the description of the system changes under the corresponding *Renormalization Flow*, and extract information about the desired physical phenomena.

⌟

### Effective Field Theory Picture

It is unreasonable to expect that any theory at our current disposal is accurately applicable at arbitrarily high energy/momenta, or equivalently arbitrarily small distances.

Unavoidably, every (realistic) model works with particular set of relevant degrees of freedom (DOF), while the rest can at most induce effective interactions between our chosen DOF. Thus, it is natural that our *effective theory* has finite domain of applicability, beyond which it is not able to encompass all interesting/relevant phenomena.

**Remark 2.11.** Why is it only in QFT that such detailed discussion is necessary? Why did not questions of similar nature appear already in Classical Mechanics, Relativity, or ordinary Quantum Mechanics? The answer lies in the unique combination of crucial aspects present in Relativity and Quantum Mechanics:

- Classical Mechanics (CM) — There is not much to discuss, since our system is described precisely by its location in the phase space, and the evolution is deterministic — not really space for crazy stuff.
- Relativistic Mechanics (RM) — *new aspect enters* — Mass is just a form of energy, allowing us to create and destroy particles. Still, energy being conserved, we have only so much options, and we are not bothered by any DOF lying (energetically) far beyond our reach.
- Quantum Mechanics (QM) — *new aspect enters* — Evolution is not deterministic in the strong sense of CM, since from path integral point of view all possible evolutions contribute in certain sense. Yes, the physical language and intuitions drastically change, nonetheless, choosing basically any particular QM system we obtain a self-consistent description.
- Quantum Field Theory (QFT) — Only when we combine RM and QM we face a fundamental difficulty. By QM our system goes through all available processes, which by RM can include producing particles of essentially arbitrary type (supposing the fields are not in totally decoupled sectors). We thus necessarily encounter cross-linking and interaction of various DOF, which would be otherwise (classically) unreachable. ⌟

After reading the last point, it could seem hopeless to obtain any reasonably predictive theory without "knowing everything". This is in stark contrast with our experience, where even idealized physical models prove immensely useful.

This is rooted in the fact that vast amount of high-energy physics effects (say at energy scales higher than $\Lambda$) essentially decouples. Certain aspects of this can be illuminated in the framework colloquially known as *renormalization theory*, to be discussed below.

**Remark 2.12** (Lattice Models, Condensed Matter Physics)**.** All of this is conceptually clear in lattice models or condensed matter physics, where the field theory description is effective from the start, and valid only on length scales $|x| \gg a$ much larger than the lattice spacing $a$. ⌟

## Wilsonian Effective Action

We will now look in more detail on how the renormalization idea Idea 2.10 →p.41 plays out in (flat-space) QFT. For better contact with statistical physics we work throughout in Euclidean signature with $\hbar \equiv 1$, and from start we impose a UV cut-off $\Lambda_0$.

Including appropriate sources $\boldsymbol{J}$ in the action $\mathcal{S}^{\Lambda_0}$, the partition function

$$\mathcal{Z} \equiv \int [\mathcal{D}\boldsymbol{\phi}]_{k<\Lambda_0} \, \mathrm{e}^{-\mathcal{S}^{\Lambda_0}[\boldsymbol{\phi}]}$$

carries all information about the theory for energies/momenta $k \equiv |k| < \Lambda_0$, or equivalently for distances $|x| > L_0 \equiv \Lambda_0^{-1}$. If we are interested in low-energy large-distance physics, we can try to simplify our description by integrating out the high-energy degrees of freedom.

**Definition 2.13** (Wilsonian Effective Action)**.** Let $\Lambda < \Lambda_0$ be a lower cut-off, and split the fields into long-wavelength/low-momentum modes $\boldsymbol{\phi}^-$ with $k < \Lambda$, and short-wavelength/high-momentum modes $\boldsymbol{\phi}^+$ with $k > \Lambda$. Since

$$\mathcal{Z} = \int [\mathcal{D}\boldsymbol{\phi}]_{k<\Lambda_0} \, \mathrm{e}^{-\mathcal{S}^{\Lambda_0}[\boldsymbol{\phi}]} = \int [\mathcal{D}\boldsymbol{\phi}^-]_{k<\Lambda} \int [\mathcal{D}\boldsymbol{\phi}^+]_{\Lambda<k<\Lambda_0} \, \mathrm{e}^{-\mathcal{S}^{\Lambda_0}[\boldsymbol{\phi}^- + \boldsymbol{\phi}^+]} \,,$$

the *Wilsonian effective action* defined by

$$\mathrm{e}^{-\mathcal{S}^{\Lambda}_{\mathrm{eff}}[\boldsymbol{\phi}^-]} \equiv \int [\mathcal{D}\boldsymbol{\phi}^+]_{\Lambda<k<\Lambda_0} \, \mathrm{e}^{-\mathcal{S}^{\Lambda_0}[\boldsymbol{\phi}^- + \boldsymbol{\phi}^+]}$$

describes the same physics of the low-momentum modes $\boldsymbol{\phi}^-$ as the original action $\mathcal{S}^{\Lambda_0}$.

**Remark 2.14** (Partial Trace Analogy, Real-Space Coarse-Graining)**.** In lattice models we naturally encounter real-space variant of the above. By grouping multiple lattice sites into blocks, we coarse-grain by partially tracing over the internal degrees of freedom inside the block. Only thing left is a block variable describing the overall/mean/dominant state of the block. ⌟

**Remark 2.15** (Locality of Effective Action)**.** Integrating out high-momentum modes $\boldsymbol{\phi}^+$ generally introduces non-local interactions for $\boldsymbol{\phi}^-$. However, these are suppressed by the high-momentum scale $\Lambda$ — coming from their propagator — so as long as we want effective low-energy description for $E \ll \Lambda$, we can write the effective Lagrangian as a series in "$E/\Lambda$" of local terms.

The same happens in lattice models. Starting with nearest-neighbor interactions, coarse-graining into blocks introduces next-to-nearest–neighbor interactions. ⌟

**Remark 2.16** (Wilson Renormalization Group)**.** We can iterate this procedure, and go to a still lower scale $\Lambda' < \Lambda$ — it yields the same effective action as going directly from $\Lambda_0$ to $\Lambda'$. For this reason it is known as the *renormalization group*, even though "renormalization semigroup" would be more appropriate. ⌟

## Flow in the Space of Couplings

Having fixed the underlying degrees of freedom, even if we start with a particular set of couplings, the effective theory (after integrating out heavy modes) will have different couplings. In general, it can generate all possible terms compatible with symmetries of the theory.

**Definition 2.17** (Space of Couplings, Most General Effective Action)**.** Fixing a set of relevant degrees of freedom $\boldsymbol{\phi}$, we consider the most general effective action (respecting the locality and symmetries) in the form

$$\mathcal{S}_{\boldsymbol{c}}^{\Lambda_0}[\boldsymbol{\phi}] = \int_{\mathcal{M}} \mathrm{d}^d x \left[\mathcal{L}_{\text{kin}}[\boldsymbol{\phi}] + \sum_i \frac{c_i}{\Lambda_0^{d_i - d}} \mathcal{O}_i[\boldsymbol{\phi}]\right].$$

As such, we have allowed all possible local (Lorentz-invariant) operators $\mathcal{O}_i$ of classical mass dimensions $d_i$. For convenience, we have included explicit factors of $\Lambda_0$ such that the couplings $\boldsymbol{c} = \{c_i\}$ are dimensionless.

With such generic initial action (for simplicity we consider a single real scalar $\phi$), the Wilsonian effective action for some cutoff $\Lambda$ will be of the form

$$\mathcal{S}_{\text{eff}}^{\Lambda}[\varphi] = \int_{\mathcal{M}} \mathrm{d}^d x \left[\frac{Z_\Lambda}{2} (\boldsymbol{\partial}\varphi)^2 + \sum_i \frac{c_i(\Lambda)}{\Lambda^{d_i - d}} Z_\Lambda^{n_i/2} \mathcal{O}_i[\varphi]\right],$$

where the *field renormalization factor* $Z_\Lambda$ accounts for the possible corrections to the kinetic term. We also introduced a factor $Z_\Lambda^{n_i/2}$ before $\mathcal{O}_i \propto \varphi^{n_i}$, such that in terms of the *renormalized field*

$$\phi \equiv Z_\Lambda^{1/2} \varphi$$

we have again a canonically normalized kinetic term, and the whole effective action of the form $\mathcal{S}_{\text{eff}}^{\Lambda}[\varphi] = \mathcal{S}_{\boldsymbol{c}(\Lambda)}^{\Lambda}[\phi]$ as defined in Definition 2.17 →p.44.

By Definition 2.13 →p.43, changing the cutoff scale together with the appropriate "running" of the couplings leaves the low-energy physics unchanged.

**Definition 2.18** (Renormalization Flow, Beta Functions)**.** The *renormalization flow* of the couplings

$$\boldsymbol{c}(\Lambda) \longmapsto \boldsymbol{c}(\Lambda') \equiv \mathcal{R}_{\Lambda' \leftarrow \Lambda}[\boldsymbol{c}(\Lambda)],$$

is infinitesimally generated by the so-called *$\beta$-functions*

$$\beta_i\big(\boldsymbol{c}(\Lambda)\big) \equiv \Lambda \frac{\partial c_i(\Lambda)}{\partial \Lambda} \equiv \frac{\partial}{\partial \ln \Lambda} c_i(\Lambda),$$

which can be viewed as a vector field $\boldsymbol{\beta}$ on the space of couplings.

**Remark 2.19.** For the dimensionless couplings $\{c_i\}$ it is convenient to write

$$\beta_i\big(\boldsymbol{c}(\Lambda)\big) = (d_i - d)c_i(\Lambda) + \beta_i^{\text{quant}}\big(\boldsymbol{c}(\Lambda)\big)\,,$$

where the first term compensates the explicit powers of $\Lambda$ in Definition 2.17→p.44, and the second term represents the quantum effect of integrating out the high-momentum modes.

One can see that we have $\boldsymbol{\beta}^{\text{quant}} = 0$ in a free (massless) theory with $\boldsymbol{c} = 0$, since the high-momentum modes are completely decoupled from the low-momentum ones. Thus, $\boldsymbol{\beta} = 0$, and the couplings are constant. We will discuss such fixed points more generally in a moment. ⌟

Suppose we want to calculate $n$-point correlators inserted at well separated points $x_1, \ldots, x_n \in \mathcal{M}$. Since high-energy modes do not play much of a role, we can use the effective action at some scale $\Lambda$ as

$$\Big\langle \varphi(x_1) \cdots \varphi(x_n) \Big\rangle_{\Lambda,\boldsymbol{c}(\Lambda)} \equiv \frac{1}{\mathcal{Z}} \int [\mathcal{D}\varphi]_{k<\Lambda}\, \mathrm{e}^{-\mathcal{S}^{\Lambda}_{\boldsymbol{c}(\Lambda)}[Z_\Lambda^{1/2}\varphi]}\, \varphi(x_1) \cdots \varphi(x_n)\,,$$

where we allowed the possibility that we did not yet canonically normalize the field. For the renormalized field $\phi \equiv Z_\Lambda^{1/2}\varphi$ we define *renormalized correlation* functions

$$\Gamma^{(n)}_{\Lambda,\boldsymbol{c}(\Lambda)}(x_1, \ldots, x_n) \equiv \Big\langle \phi(x_1) \cdots \phi(x_n) \Big\rangle_{\Lambda,\boldsymbol{c}(\Lambda)} \equiv Z_\Lambda^{n/2} \Big\langle \varphi(x_1) \cdots \varphi(x_n) \Big\rangle_{\Lambda,\boldsymbol{c}(\Lambda)}\,.$$

If the field insertions are really far apart — equivalently involve modes with energies $E \ll \Lambda$ — we would obtain the same (nonrenormalized) correlators if we first integrated out modes in the range $(s\Lambda, \Lambda]$ for some $s < 1$. For an arbitrary configuration of (distinct) points, we can choose $s$ sufficiently close to 1 such that effect of integrating out the high-momentum modes is still negligible.

**Proposition 2.20** (Callan–Symanzik Equation, Anomalous Dimension)**.** The renormalized correlation function obey the *Callan–Symanzik equation*

$$\left(\Lambda \frac{\partial}{\partial \Lambda} + \boldsymbol{\beta} \cdot \frac{\partial}{\partial \boldsymbol{c}} + n\gamma_\phi\right) \Gamma^{(n)}_{\Lambda,\boldsymbol{c}(\Lambda)}(x_1, \ldots, x_n) = 0\,,$$

where we defined the *anomalous dimension* of the field $\phi$ as

$$\gamma_\phi(\Lambda) \equiv -\frac{1}{2}\Lambda \frac{\partial \ln Z_\Lambda}{\partial \Lambda} \equiv \frac{\partial \ln Z_\Lambda^{-1/2}}{\partial \ln \Lambda}\,.$$

**Proof.** From the discussion above we have ($s$ infinitesimally close to one)

$$\Big\langle \varphi(x_1) \cdots \varphi(x_n) \Big\rangle_{\Lambda,\boldsymbol{c}(\Lambda)} = Z_\Lambda^{-n/2} \Gamma^{(n)}_{\Lambda,\boldsymbol{c}(\Lambda)}(x_1, \ldots, x_n) \overset{!}{=} Z_{s\Lambda}^{-n/2} \Gamma^{(n)}_{s\Lambda,\boldsymbol{c}(s\Lambda)}(x_1, \ldots, x_n)\,,$$

and the result is obtained by performing $\frac{\mathrm{d}}{\mathrm{d}s}\big|_{s=1}$ and using the chain rule together with the definition of $\boldsymbol{\beta}$ and $\gamma_\phi$. □

## Scale-Dependence of Theories

Up to now we discussed effective descriptions (with different UV cutoffs) of the same low-energy theory. As we will see shortly, this is connected to the behavior at different length scales.

**Remark 2.21** (RG Flow and Scaling). Suppose that after integrating out modes in the range $(s\Lambda, \Lambda]$ as above, we additionally perform a scaling transformation

$$\begin{aligned} x &\longmapsto x' \equiv sx\,, & \mathrm{d}^d x &\longmapsto \mathrm{d}^d x' = s^d\,\mathrm{d}^d x\,, \\ \Lambda &\longmapsto \Lambda' \equiv \tfrac{1}{s}\Lambda\,, & \partial &\longmapsto \partial' = \tfrac{1}{s}\partial\,, \\ \phi &\longmapsto \phi'(sx) \equiv s^{-d_\phi}\phi(x) \equiv s^{-\frac{d-2}{2}}\phi(x)\,, \end{aligned}$$

under which the action in Definition 2.17→p.44 is invariant. Therefore, we have

$$\begin{aligned} \Gamma^{(n)}_{\Lambda,\boldsymbol{c}(\Lambda)}\left(\frac{x_1}{s},\ldots,\frac{x_n}{s}\right) &= \left[\frac{Z_\Lambda}{Z_{s\Lambda}}\right]^{n/2} \Gamma^{(n)}_{s\Lambda,\boldsymbol{c}(s\Lambda)}\left(\frac{x_1}{s},\ldots,\frac{x_n}{s}\right) \\ &= \left[\frac{Z_\Lambda}{Z_{s\Lambda}}\right]^{n/2} s^{\frac{d-2}{2}n}\,\Gamma^{(n)}_{\Lambda,\boldsymbol{c}(s\Lambda)}(x_1,\ldots,x_n)\,, \end{aligned}$$

where in the second line we just performed the scaling transformation, which does not affect the couplings $\boldsymbol{c}(s\Lambda)$, as they are dimensionless.

In this way we ended up with the same UV cutoff $\Lambda$, and related correlation functions at different scales. When $s \to 0$, the left-hand side probes correlations at longer and longer distances, while the distances on the right-hand side are kept constant, but (apart from the prefactor) we let the coupling run to the IR.

This is what we would expect — the IR behavior is governed by the low-energy effective theory (and corresponding couplings which survive) obtained after we integrate out all of the high-energy modes. ⌟

**Remark 2.22** (Interpretation of Anomalous Dimension, Scaling Dimension). For each field, the correlator above in Remark 2.21→p.46 has a scaling factor on the right-hand side, which for $s = 1 + \delta s$ with $|\delta s| \ll 1$ is approximately

$$\left[\frac{Z_\Lambda}{Z_{s\Lambda}}\right]^{1/2} s^{\frac{d-2}{2}} \approx 1 + \left(\frac{d-2}{2} + \gamma_\phi\right)\delta s + \cdots\,.$$

The scaling is thus determined by the *scaling dimension* of the field $\phi$ defined as

$$\Delta_\phi \equiv \frac{d-2}{2} + \gamma_\phi = d_\phi + \gamma_\phi\,,$$

so in addition to the classical mass dimension $d_\phi = (d-2)/2$ we also need to include the anomalous dimension $\gamma_\phi$ defined in Proposition 2.20→p.45. It generally depends on the energy scale $\Lambda$ and the couplings $\boldsymbol{c}(\Lambda)$, and similarly to $\beta$-functions, it is zero in the free massless theory, see Remark 2.19→p.45. ⌟

## Fixed Points

Very special things happen, when the $\beta$-functions vanish at some point in the space of couplings.

**Definition 2.23** (Fixed Point). A point $\boldsymbol{c}_*$ in the space of couplings is called a *fixed point* if

$$\boldsymbol{\beta}(\boldsymbol{c}_*) = 0 \quad \Longrightarrow \boldsymbol{c}_*(\Lambda) = \text{const.}\,.$$

The corresponding theory is then invariant under the renormalization group flow, or equivalently under the scaling transformations.

**Remark 2.24** (Correlation Length, Mass Gap, Critical Point). The (connected) two-point correlators at large distances typically decay as

$$\Big\langle \phi(x)\phi(y) \Big\rangle_{\text{conn}} \sim \frac{1}{|x-y|^{\vartheta}}\, \mathrm{e}^{-\frac{|x-y|}{\xi}}\,,$$

where $\xi$ is the *correlation length* of the theory, and $\vartheta$ is some exponent.

When the theory has a "mass gap" — that is $m_{\text{gap}} \equiv E_1 - E_0 > 0$, where $E_0, E_1$ are the energies of the ground state and first excited state, respectively — the correlation length is finite and given by

$$\xi \equiv \frac{1}{m_{\text{gap}}}\,.$$

This is because the dominant contribution comes from the propagation of the lightest particle/state (apart from the vacuum).

The fixed points however can not have any characteristic scale (such as correlation length or mass), since they would necessarily change under the renormalization flow or scaling. Only possibilities are thus:

**(1)** $\xi = 0 \Longleftrightarrow m_{\text{gap}} = \infty$ — Correlation functions of local operators vanish, and we are left with a TQFT. This is what typically happens for generic statistical and condensed matter systems, since microscopic correlation lengths are typically $\xi \sim a$, which is negligible at large distances. It can be understood as flowing off to the infinity of the coupling space.

**(2)** $\xi = \infty \Longleftrightarrow m_{\text{gap}} = 0$ — For specially tuned couplings it can happen we flow into a *critical point*, where correlation functions do not decay exponentially at large distances, but rather as a power law. This is the case of the *second-order phase transitions*.

⌟

**Remark 2.25** (Two-Point Function at Criticality). Since $\boldsymbol{c}_* = \text{const.}$, also the anomalous dimension $\gamma_\phi(\boldsymbol{c}_*) \equiv \gamma_\phi^*$ is constant. The scaling in Remark 2.22 →p.46 can be then easily exponentiated as

$$\left[\frac{Z_\Lambda}{Z_{s\Lambda}}\right]^{1/2} s^{\frac{d-2}{2}} = s^{\Delta_\phi}\,.$$

↳ Therefore, for the 2-point function we have scaling symmetry Remark 2.21→p.46, which together with the translation/rotation invariance gives

$$\Big\langle \phi\Big(\frac{x}{s}\Big)\phi\Big(\frac{y}{s}\Big)\Big\rangle = s^{2\Delta_\phi}\Big\langle\phi(x)\phi(y)\Big\rangle \quad\Longrightarrow\quad \Big\langle\phi(x)\phi(y)\Big\rangle \sim \frac{1}{|x-y|^{2\Delta_\phi}}\,.$$

Alternatively, we could obtain the same result using the Callan–Symanzik equation Proposition 2.20→p.45.

**Example 2.26** (Free Theory, Generalized Free Field, Mean Field Theory)**.** As already mentioned in Remark 2.19→p.45 and Remark 2.22→p.46, the free massless theory is a fixed point of the renormalization group flow, and thus should have 2-point function

$$\Big\langle\phi(x)\phi(y)\Big\rangle \sim \frac{1}{|x-y|^{2\Delta_\phi^{\text{free}}}} \equiv \frac{1}{|x-y|^{d-2}}\,.$$

Indeed, this is a solution of the free equation of motion $\Box\phi = 0$. The higher-order correlators are then obtained by Wick's theorem Example 1.33→p.9.

In the context of QFT in AdS to be discussed in Chapter 4→p.91, we naturally encounter *Generalized Free Fields* (GFFs) — they share the same factorization property of correlators as the free theory, but have a generic (noninteger) scaling dimension $\Delta_\phi$.

We will use term *Mean Field Theory* (MFT) for theories built out of GFFs. Fields in such theories do not obey simple local equations of motion, so they do not admit a local Lagrangian description. In particular, MFTs do not have a conserved energy–momentum tensor.

**Remark 2.27** (Weyl and Conformal Symmetry at Critical Point)**.** The same correlation function was already obtained in Example 1.73→p.25 using a stronger Weyl symmetry, which contains arbitrary *local* scaling transformations. In fact, this is what often happens at critical points — if a theory is local, and also scale invariant by virtue of being critical, it generically enjoys further symmetry enhancement — it becomes *conformally invariant*. In the next Chapter 3→p.51, we will study such Conformal Field Theories (CFT) in more detail.

## Universality

Now we will look at theories near, but not directly at, a fixed point $\boldsymbol{c}_*$ of RG flow. Since by definition $\boldsymbol{\beta}(\boldsymbol{c}_*) = 0$, we can linearize the flow around the fixed point, and write

$$\Lambda\,\frac{\partial \boldsymbol{c}}{\partial\Lambda}\bigg|_{\boldsymbol{c}\equiv\boldsymbol{c}_*+\delta\boldsymbol{c}} \approx \boldsymbol{M}\cdot\delta\boldsymbol{c} + O(\delta\boldsymbol{c}^2)\,, \qquad M_{ij} \equiv \frac{\partial\beta_i}{\partial c_j}\bigg|_{\boldsymbol{c}_*}\,.$$

Assuming the matrix $\boldsymbol{M}$ can be diagonalized with eigenvectors $u_i$ and eigenvalues $\lambda_i \equiv \Delta_i - d$, we can write

$$\Lambda\,\frac{\partial u_i}{\partial\Lambda} = \lambda_i u_i + \dots \quad\Longrightarrow\quad u_i(\Lambda) = \bigg(\frac{\Lambda}{\Lambda_0}\bigg)^{\Delta_i - d} u_i(\Lambda_0) + \dots\,.$$

Recalling Remark 2.19$^{\to p.45}$, we would classically expect $\lambda_i \stackrel{?}{=} d_i - d$, but interactions usually induce some anomalous dimension $\gamma_i \equiv \Delta_i - d_i$. Similarly to Remark 2.22$^{\to p.46}$, $\Delta_i = d_i + \gamma_i$ is the *scaling dimension* of the operator $\mathcal{O}_i$ corresponding to the coupling $u_i$.

**Definition 2.28** (Relevant, Irrelevant, Marginal Couplings and Operators). Depending on the sign of $\lambda_i \equiv \Delta_i - d$, we distinguish three cases:

- $\lambda_i > 0 \iff \Delta_i > d$ — the coupling $u_i$ is *irrelevant*, and decays under the RG flow (lowering the scale $\Lambda$). Any such deformation from the fixed point is washed away at large distances $\Lambda \to 0$.
- $\lambda_i < 0 \iff \Delta_i < d$ — the coupling $u_i$ is *relevant*, and grows under the RG flow. Deforming the theory in such direction drives it away from the fixed point as we go into the IR.
- $\lambda_i = 0 \iff \Delta_i = d$ — the coupling $u_i$ is *marginal*, and the fate of such deformation depends on higher-order terms in the $\beta$-functions.

Correspondingly, the operators $\mathcal{O}_i$ associated to couplings $u_i$ are also called *irrelevant*, *relevant*, or *marginal*.

**Example 2.29** (Mass Term). For a free scalar field $\phi$, the mass term corresponds to the operator $\phi^2$ with the scaling dimension equal to the classical mass dimension

$$\Delta_{\phi^2} = d_{\phi^2} = 2\,\frac{d-2}{2} = d-2\,.$$

This can be seen from Remark 2.19$^{\to p.45}$, where for free theory we have $\beta^{\text{quant}}_{\phi^2} = 0$, and only the classical scaling due to explicit power of $\Lambda$ appears

$$\Lambda \frac{\partial m^2}{\partial \Lambda} = (d_{\phi^2} - d)m^2 = -2m^2\,.$$

Clearly, it is a relevant deformation, so as we lower the energy scale $\Lambda$, we flow away from the free *Gaussian critical point* with $m^2 = 0$. Eventually we reach trivial theory with $m_{\text{gap}} = \infty$ — see Remark 2.24$^{\to p.47}$ — since every massive mode gets integrated out, and we are left with the vacuum. ⌟

**Remark 2.30** (Finite Number of Relevant Operators). Since each derivative or new field adds to the dimension of the operator — anomalous dimensions do not overwhelm the classical contribution — there can be only finitely many (and typically just a few) relevant operators.

As a consequence, even though a generic (effective) QFT is parametrized by infinitely many couplings, the vast majority is irrelevant and gets quickly suppressed in the IR. This is the reason why Effective Field Theories are really *effective*, as only a handful of couplings govern the low-energy physics. The rest gives subleading corrections, and if needed, they can be systematically accounted for. ⌟

**Remark 2.31** (Universality, Critical Exponents)**.** Another facet of the same idea is the *universality* of the critical points. Many systems differing in their microscopic description — when appropriately tuned — flow to the same critical/fixed point.

This explains why critical behavior at second-order phase transitions (captured by the critical exponents) is shared by many systems. Depending on the fixed point, they can be categorized into *universality classes*. Theories in the same universality class are described by the same CFT and by the same few relevant operators, which is also reason for various relationships between the critical exponents. ⌟

**Example 2.32** (Ising Model in $d = 3$, Liquid–Vapor Transition, $\phi^4$ Theory)**.** In $d = 3$, the Ising model (uniaxial magnet) at critical temperature, critical point of vapor–liquid transition, and critical $\phi^4$ theory, all fall into the same universality class.

While the $\mathbb{Z}_2$ symmetry of the Ising model and $\phi^4$ theory is manifest microscopically, for the liquid–vapor it is only emergent at the critical point. On the other hand, water (or $\phi^4$ theory) has full rotational symmetry, which is not manifest for the lattice Ising model. ⌟

# 3 Conformal Field Theory

The importance of Conformal Field Theories (CFTs) in both physics and mathematics can be hardly overstated. Let us briefly mention some of the connections and motivations to study them (in no particular order):

- At critical points of statistical systems — second order transitions — the correlation length diverges, and theories become scale-invariant. By this we mean that fluctuations and their correlations are no longer exponentially suppressed at long distances, but follow power-laws characteristic to critical phenomena.

  Furthermore, it generically happens that *local* theories enjoy an additional symmetry enhancement — they become *conformally invariant*, and are thus described by Euclidean CFTs. Note that emergence of such symmetries at large distances is rather nontrivial for discrete models — imagine the Ising model on a square lattice — as they initially do not even posses symmetry under arbitrary rotations.

- Connected to the previous point, the fixed points of the RG flow are usually described by CFTs. Being such special landmarks in the space of QFTs, they are important for their study — oftentimes, we can understand a UV-complete QFT as an RG flow $\mathsf{CFT}_{\mathrm{UV}} \rightsquigarrow \mathsf{CFT}_{\mathrm{IR}}$ triggered by deforming the $\mathsf{CFT}_{\mathrm{UV}}$.

- Having a larger symmetry group (compared to standard isometries), we obtain stronger constrains on the theory. This makes non-perturbative "bootstrap" approaches based mainly on the symmetry and consistency of the underlying theories much more potent. Such program is in particular successful for $d = 2$, where the conformal algebra becomes infinite-dimensional, sometimes allowing to completely solve particular classes of theories [29].

- String theory falls under the purview of two-dimensional CFTs, since in the standard formulation it is "just" a specific $\sigma$-model on a two-dimensional $\mathsf{CFT}_{\mathrm{worldsheet}}$.

- There are also fruitful relations to TQFT, which can be exploited to study topological invariants [30].

- Last, but not least, there is the AdS/CFT correspondence, also known as holography or gauge/gravity duality. The original formulation by Maldacena [31] related a theory with dynamical gravity in asymptotically AdS spacetime to a non-gravitational CFT in one less dimension "living" on the boundary. Since then, it has become much more general.

  Even without dynamical gravity, a QFT in fixed AdS background admits asymptotic observables (boundary correlation functions) satisfying basically all of the CFT axioms.

**Remark 3.1** (Gapped QFT vs CFT). The language and methods used to study CFTs are very different from the ones used for QFTs with a mass gap (or QFTs with finitely many massless particles in the IR).

In the latter case, due to the mass gap, widely separated particles cease to interact. We can thus define asymptotic states behaving like free, non-interacting particles, and study their scattering amplitudes/$S$-matrix (with appropriate care, it is also possible to treat massless particles).

On the other hand, in CFTs, it is impossible to disentangle the continuum of massless excitations — interactions never "turn off" completely at large distances — so we cannot define asymptotic states and the $S$-matrix. Instead, we are mainly interested in the correlation functions of local operators, and utilize the stronger conformal symmetry to study their structure. ⌟

**Remark 3.2** (Coleman–Mandula Theorem). Considering flat space, the spacetime and internal symmetries of gapped interacting theories (nontrivial $S$-matrix) can only combine in a trivial way — the statement of *Coleman–Mandula theorem* is that the symmetry group is a direct product $G = G_{\text{Poincaré}} \times G_{\text{internal}}$.

Notable "exceptions" are supersymmetry — includes Grassman (anticommuting) charges — and conformal symmetry. Thus, in a certain sense, CFTs provide examples of most symmetric, yet interacting theories. ⌟

To get a good grasp on CFTs, in Section 3.1→p.52 we will first study the conformal structure of manifolds in general. Afterward, we focus on flat spaces specifically in Section 3.2→p.61. Initial implications of the conformal symmetry for the correlation functions of local operators in CFTs will be explored in Section 3.3→p.69. In Section 3.4→p.78 we will study in more detail the structure of CFTs by choosing a particular operator formalism — the Radial Quantization. We will see that CFTs have a special property — quantum states correspond to local operators. Its particular important consequence, the existence of Operator Product Expansion, will be discussed in Section 3.5→p.86. Finally, in Section 3.6→p.88 we will briefly comment on the structure of 4-point correlators.

Exposition of this chapter was strongly influenced by [32] [11] [4], and also [33] [34] [35] [36].

## 3.1 Conformal Structure Generalities

We start by studying conformal structure of manifolds in general. Conformal-structure preserving maps are called *conformal transformations*. As is usual, the analysis of such global transformations can be simplified by studying corresponding infinitesimal actions, which are generated by *conformal Killing vectors*. Conformal automorphisms of a manifold form a *conformal group*, and to not lose out on transformations which are singular at some points, we will perform the *conformal compactification*.

## Conformal Transformations

It is natural that a local field theory invariant under global scaling transformations will be also invariant under local scaling transformations. By this we mean that the induced mapping leaves the metric invariant up to a conformal factor, which can depend on the spacetime point.

**Definition 3.3** (Weyl Transformation, Conformal Structure)**.** A *Weyl transformation* is an active transformation of the metric on a pseudo-Riemannian manifold $(\mathcal{M}, \boldsymbol{g}) \mapsto (\mathcal{M}, \tilde{\boldsymbol{g}} \equiv \Omega^2 \boldsymbol{g})$ by scaling it with a (square of) smooth positive function $\Omega \in C^\infty_{>0}(\mathcal{M})$, called *conformal* or *Weyl factor*.

Two metrics on $\mathcal{M}$ related by a Weyl transformation are said to be *conformally equivalent*, and the equivalence class $\mathfrak{c} \equiv \boldsymbol{g}/_{C^\infty_{>0}(\mathcal{M})}$ of metrics under Weyl transformations is called a *conformal structure* on $\mathcal{M}$.

**Remark 3.4.** After a Weyl transformation with a conformal factor $\Omega$, the lengths of vectors at a point $x \in \mathcal{M}$ are multiplied by $\Omega(x)$. Thus, conformal structure enables us to locally measure angles and ratios of lengths, but not lengths themselves. ⌟

**Remark 3.5.** Every pseudo-Riemannian manifold has automatically a conformal structure. However, we can also consider manifolds which intrinsically have just the conformal structure, with no canonical/preferred metric.

Examples are boundaries of pseudo-Riemannian manifolds after a compactification à la Penrose [37]. The metric initially does not extend over the boundary where it diverges, and to make it regular we need to perform a Weyl transformation which itself vanishes at the boundary. Since there is no natural choice of such prescription, we are left with the whole equivalence class of metrics, giving us just the conformal structure. ⌟

**Definition 3.6** (Conformal Transformation, Conformal Map)**.** A smooth map $\phi\colon (\mathcal{M}, \mathfrak{c}) \to (\mathcal{M}', \mathfrak{c}')$ between manifolds equipped with conformal structures is called a *conformal transformation* or *conformal map* if it preserves the conformal structure — it is a *homomorphism* of conformal structures.

In the case of pseudo-Riemannian manifolds $(\mathcal{M}, \boldsymbol{g})$ and $(\mathcal{M}', \boldsymbol{g}')$, this means that the metrics (after pullback) are related by a Weyl transformation, that is

$$\phi^* \boldsymbol{g}' \overset{!}{=} \Omega^2 \boldsymbol{g} \equiv \mathrm{e}^{2\omega} \boldsymbol{g}\,,$$

where $\phi^*$ is pullback map corresponding to $\phi$, and $\Omega \equiv \mathrm{e}^{\omega}$ is the associated conformal factor. Specially, if $\Omega = 1 \Leftrightarrow \omega = 0$, the map is called an *isometry*, as it preserves the metric itself.

**Remark 3.7** (Properties of Conformal Maps, Conformal Equivalence)**.** The identity map is trivially conformal, and composition of conformal maps yields another conformal map. If a conformal map is a diffeomorphism, we call it a *conformal*

↳ *diffeomorphism*, and we easily see that its inverse is also conformal. It thus makes sense to say that two manifolds are *conformally equivalent* if there exists a conformal diffeomorphism between them.

In the case of pseudo-Riemannian manifolds, the associated conformal factors for composition of conformal maps and inverse of conformal diffeomorphism are straightforward to calculate from definition. Manifolds with conformally equivalent metrics — see Definition 3.3 →p.53 — are automatically conformally equivalent, where the conformal diffeomorphism is provided by the identity map. ⌟

**Remark 3.8.** Since null geodesics are mapped to null geodesics by conformal maps of a Lorentzian manifold, we can say — ignoring the possible time-reversals — that conformal transformations preserve the causal structure of the manifold. ⌟

**Example 3.9** (Dilatation in Flat Space)**.** Apart from the usual isometries of the flat space $\mathbb{R}^{p,q}$ (with usual metric) — translations and pseudorotations — other prominent conformal transformations are the *dilatations* (scaling transformations). These are given in the usual Cartesian coordinates by

$$\begin{aligned} D_\lambda \colon \mathbb{R}^{p,q} &\longrightarrow \mathbb{R}^{p,q} \\ x &\longmapsto x' \equiv D_\lambda(x) \equiv \lambda x \,, \end{aligned}$$

and can be easily seen to be conformal with $\Omega = \lambda$ since $\frac{\partial x'}{\partial x} = \lambda \mathbf{1}$. All of such simple conformal transformations with constant $\Omega$ are illustrated in Figure 3.10 →p.54. ⌟

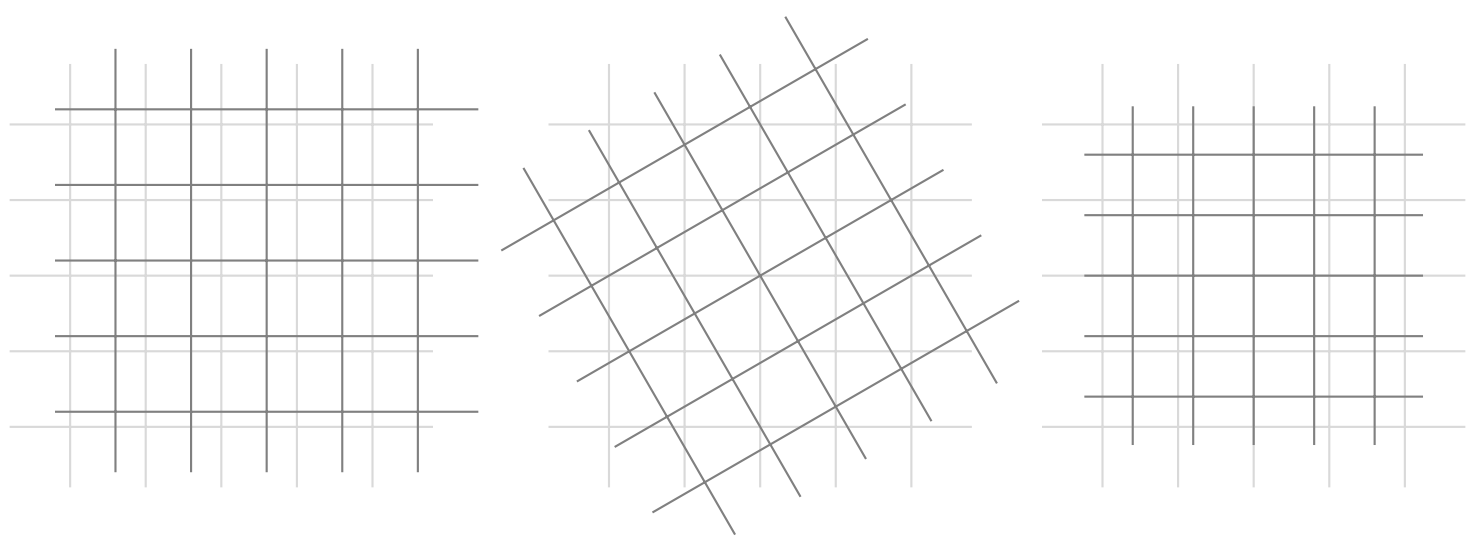

**Figure 3.10 /** Simple conformal transformations of a flat plane $\mathbb{R}^2$ — translation, rotation, and scaling/dilatation.

**Example 3.11** (Stereographic Projection)**.** Consider a unit $n$-sphere $\mathbb{S}^n$ with metric induced by the standard embedding in $\mathbb{R}^{n+1}$ (with usual metric). Denoting the *north pole* as $\mathsf{N} \equiv (1, 0, \ldots, 0)$ — expressed in Cartesian coordinates $(\mathsf{x}^0, \mathsf{x}^1, \ldots, \mathsf{x}^n)$ on $\mathbb{R}^{n+1}$ — we can define the *stereographic projection* $\phi$ as the map sending every point $\mathsf{x} \in \mathbb{S}^n \backslash \{\mathsf{N}\}$ to the intersection of the line through $\mathsf{N}$ and $\mathsf{x}$ with the equatorial plane $\{(0, x^1, \ldots, x^n)\} \simeq \mathbb{R}^n$. For a visualization, see Figure 3.12 →p.55.

In detail, using the embedding coordinates it is given by

$$\begin{aligned} \phi \colon \qquad \mathbb{S}^n \backslash \{\mathsf{N}\} &\longrightarrow \qquad \mathbb{R}^n \\ \mathsf{x} \equiv \left( \mathsf{x}^0, \mathsf{x}^1, \ldots, \mathsf{x}^n \;\middle|\; \sum_{i=0}^{n} (\mathsf{x}^i)^2 = 1 \right) &\longmapsto x \equiv \phi(\mathsf{x}) \equiv \tfrac{1}{1-\mathsf{x}^0} \left( \mathsf{x}^1, \ldots, \mathsf{x}^n \right), \end{aligned}$$

↴

where we considered affine combination $x_t \equiv (1-t)\mathsf{N} + t\mathsf{x}$ of $\mathsf{N}$ and $\mathsf{x}$ in the embedding space, and found the intersection $x_{t_*}$ with the equatorial plane at $(1-t_*) + t_*\mathsf{x}^0 = 0 \Leftrightarrow t_* = \frac{1}{1-\mathsf{x}^0}$, giving us the desired formula for $x$ as the last $n$ components of $x_{t_*}$. Note that we implicitly understand $\mathsf{x}^0$ as a function of $\mathsf{x}^i$.

The map $\phi$ is a diffeomorphism, since $\mathsf{x}^0$ can be calculated as a smooth function of $x$ (outside of $\mathsf{N}$) by

$$x^2 \equiv |x|^2 = \frac{1}{(1-\mathsf{x}^0)^2}\sum_{i=1}^{n}(\mathsf{x}^i)^2 = \frac{1-(\mathsf{x}^0)^2}{(1-\mathsf{x}^0)^2} = \frac{1+\mathsf{x}^0}{1-\mathsf{x}^0} \iff \mathsf{x}^0 = \frac{x^2-1}{x^2+1}\ ,$$

giving rest of the components as $\mathsf{x}^i = (1-\mathsf{x}^0)x^i = \frac{2x^i}{x^2+1}$. We can quickly check that the origin $(0,\dots,0) \in \mathbb{R}^n$ corresponds to the south pole $\mathsf{S} \equiv (-1, 0, \dots, 0) \in \mathbb{S}^n$, and the unit circle $\{x^2 = 1\} \subset \mathbb{R}^n$ is precisely the equator $\{\mathsf{x}^0 = 0\} \subset \mathbb{S}^n$.

A direct computation verifies that it is conformal, so it maps circles to circles, and circles through $\mathsf{N}$ to straight lines. ⌟

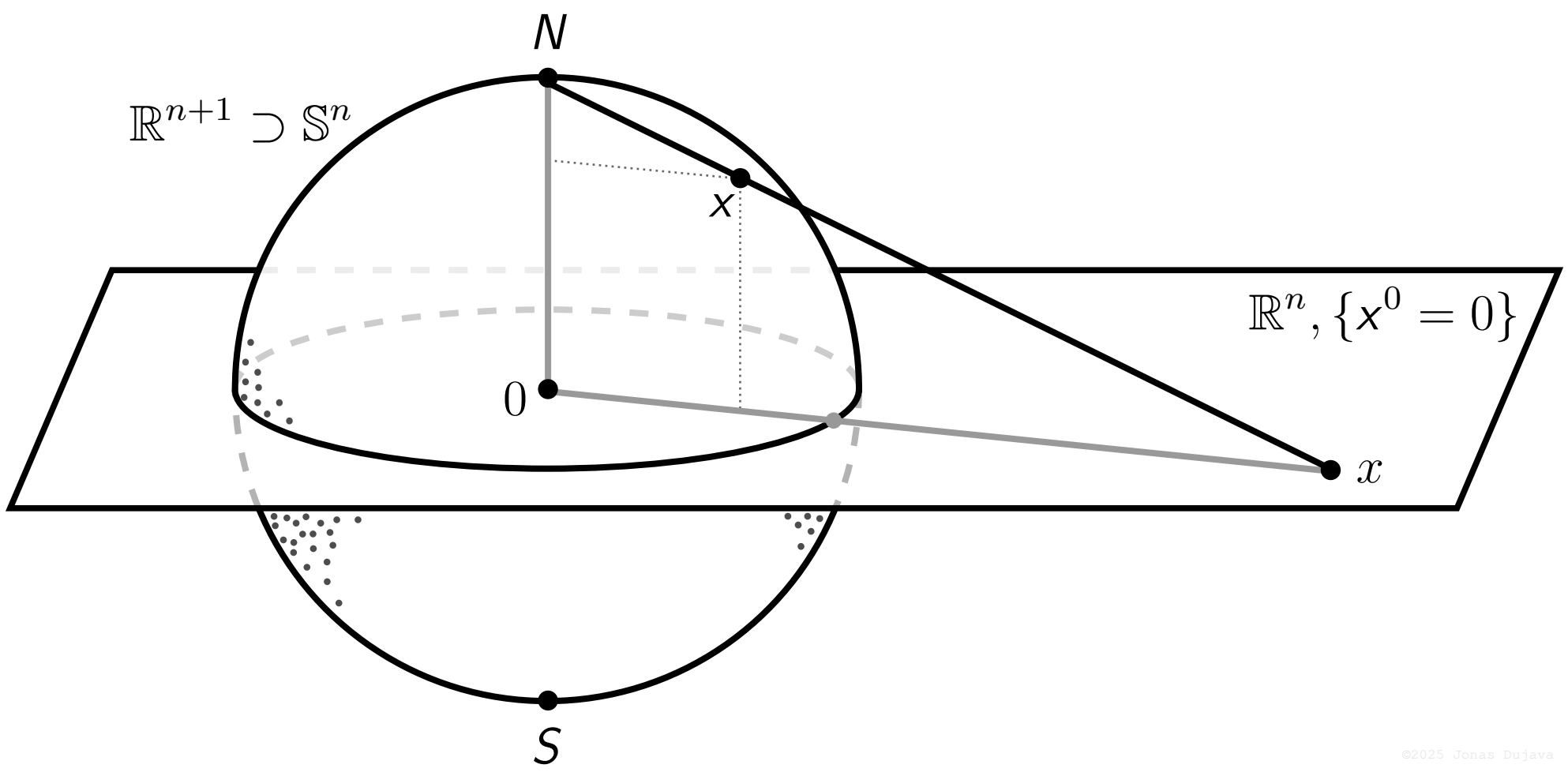

**Figure 3.12 /** Stereographic projection $\mathbb{S}^n \backslash \{\mathsf{N}\} \to \mathbb{R}^n$.

**Remark 3.13.** Conformal maps in $\mathbb{R}^d$ can be viewed through the lens of such stereographic projections. The idea is to make the inverse stereographic projection $\mathbb{R}^d \to \mathbb{S}^d \hookrightarrow \mathbb{R}^{d+1}$, then some isometry of $\mathbb{R}^{d+1}$ (rotation+translation), and finally the stereographic projection back to $\mathbb{R}^d$, for which we need to formulate stereographic projection for arbitrary positioning of the sphere in $\mathbb{R}^{d+1}$. For more details see Section 1.2 of [38]. ⌟

We will be mainly interested in conformal transformations for $d \geq 3$ spacetimes. Nevertheless, let us briefly mention the specialties of $d = 1$ and $d = 2$.

**Example 3.14** (Conformal Transformations in $d = 1$)**.** Since in $d = 1$ there is no angle dependence and metrics have only one component, any diffeomorphism is automatically conformal. For example a diffeomorphism $\phi\colon \mathbb{R} \to \mathbb{R}$ — endowing the real line with the standard metric $\boldsymbol{g} = \mathbf{d}x^2$ — is conformal with $\Omega = \frac{\mathrm{d}\phi}{\mathrm{d}x}$. ⌟

**Example 3.15** (Conformal Transformations in $d = 2$ as Holomorphic Maps)**.** In $d = 2$ we already obtain nontrivial restrictions on conformal maps. Identifying $\mathbb{R}^2$ with $\mathbb{C}$, it can be quickly verified that any holomorphic or antiholomorphic map $\phi\colon D \to \mathbb{C}$ locally invertible on open set $D \subset \mathbb{C}$ is conformal, as illustrated in Figure 3.16→p.56. For more details see for example [4] or [39]. ⌟

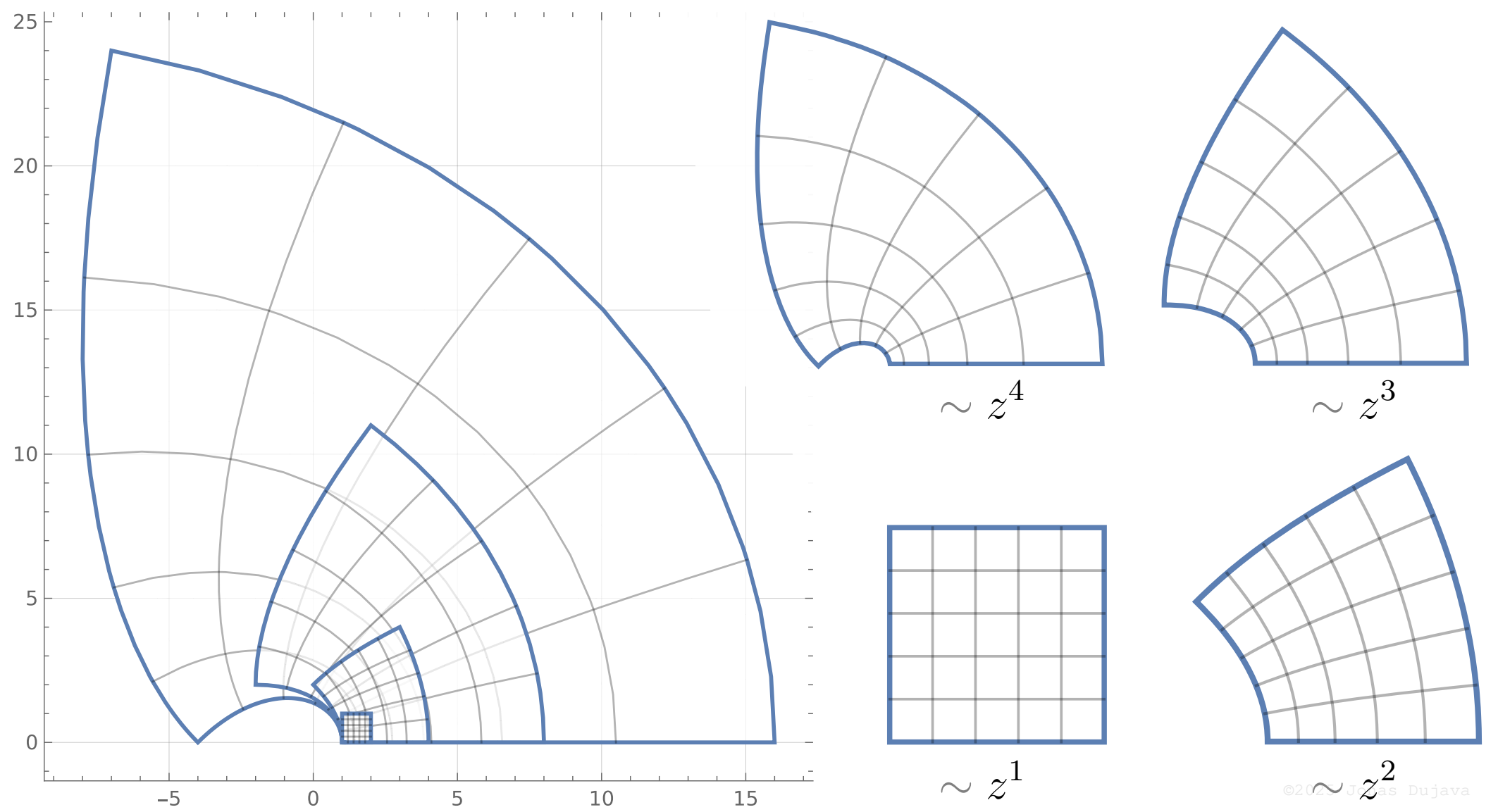

**Figure 3.16** / Conformal transformations of a square $D \equiv [1,2] \times [0,1] \subset \mathbb{R}^2 \simeq \mathbb{C}$ induced by holomorphic maps $z^n\colon \mathbb{C} \to \mathbb{C}$ for $n = \{1,2,3,4\}$. On the right we display the shape of $z^n(D)$ after some rescaling (such that bottom edge has the same length).

Together with the Riemann mapping theorem, this provides a powerful tool for studying for example classical electrostatics or ideal hydrodynamics in $d = 2$, as is showcased in the following.

**Example 3.17** (Parallel Plate Capacitor in $d = 2$)**.** Consider a (finite) parallel plate capacitor in $\mathbb{R}^2$. How can we find (at least approximately) the electric potential $\phi$ in the region near end of the plates, which satisfies Laplace equation $\triangle\phi = 0$ and boundary conditions $\phi|_{\mathsf{plate}^\pm} = \pm V$ on plates?

We can utilize that conformal transformations map solutions of Laplace equation to solutions of Laplace equation in $d = 2$, since $\triangle = \mathfrak{g}^{-1/2}\boldsymbol{\partial}_a(\mathfrak{g}^{1/2}\boldsymbol{g}^{ab}\boldsymbol{\partial}_b\,\cdot\,)$ and the expression $\mathfrak{g}^{1/2}\boldsymbol{g}^{ab}$ is invariant with respect to Weyl transformations of the metric in $d = 2$. This allows us to map an easily found solution with infinite plates to the one where plates are finite (from one side).

Indeed, the holomorphic map $\phi\colon \mathbb{C} \to \mathbb{C}$ given by $w \mapsto \phi(w) \equiv z \equiv 1 + w + \exp(w)$ transforms the strip $S \equiv \{|\mathrm{Im}(w)| < \pi\}$ to $\mathbb{C}\backslash\{x \pm i\pi \mid x \in \mathbb{R}_{\leq 0}\}$. One can check that boundaries $\partial^\pm S = \{u \pm i\pi \mid u \in \mathbb{R}\}$ of the strip are mapped to the $\mathsf{plate}^\pm \equiv \{x \pm i\pi \mid x \in \mathbb{R}_{\leq 0}\}$ by calculating $\mathrm{Im}(\phi(u \pm i\pi))$ and finding the maximal value of $\mathrm{Re}(\phi(u \pm i\pi))$. For example, the region just below $\partial^+ S$ and $\mathrm{Re}(w) < 0$ stays below the $\mathsf{plate}^+$, while $\mathrm{Re}(w) > 0$ part wraps back ending just above $\mathsf{plate}^+$.

It is therefore enough to solve Laplace equation on $S$ with boundary conditions ↴

↳ $\phi|_{\partial^\pm S} = \pm V$ — just a linear dependence on $\mathrm{Im}(w)$ — and then transform the solution with the conformal map $\phi$ to obtain the solution of the original problem. See Figure 3.18 →p.57 for the visualization of resulting field lines and equipotentials.

This works approximately even for $d = 3$ in the region near edge (not corner) assuming the distance between plates is much smaller than their extent in the other two directions. ⌟

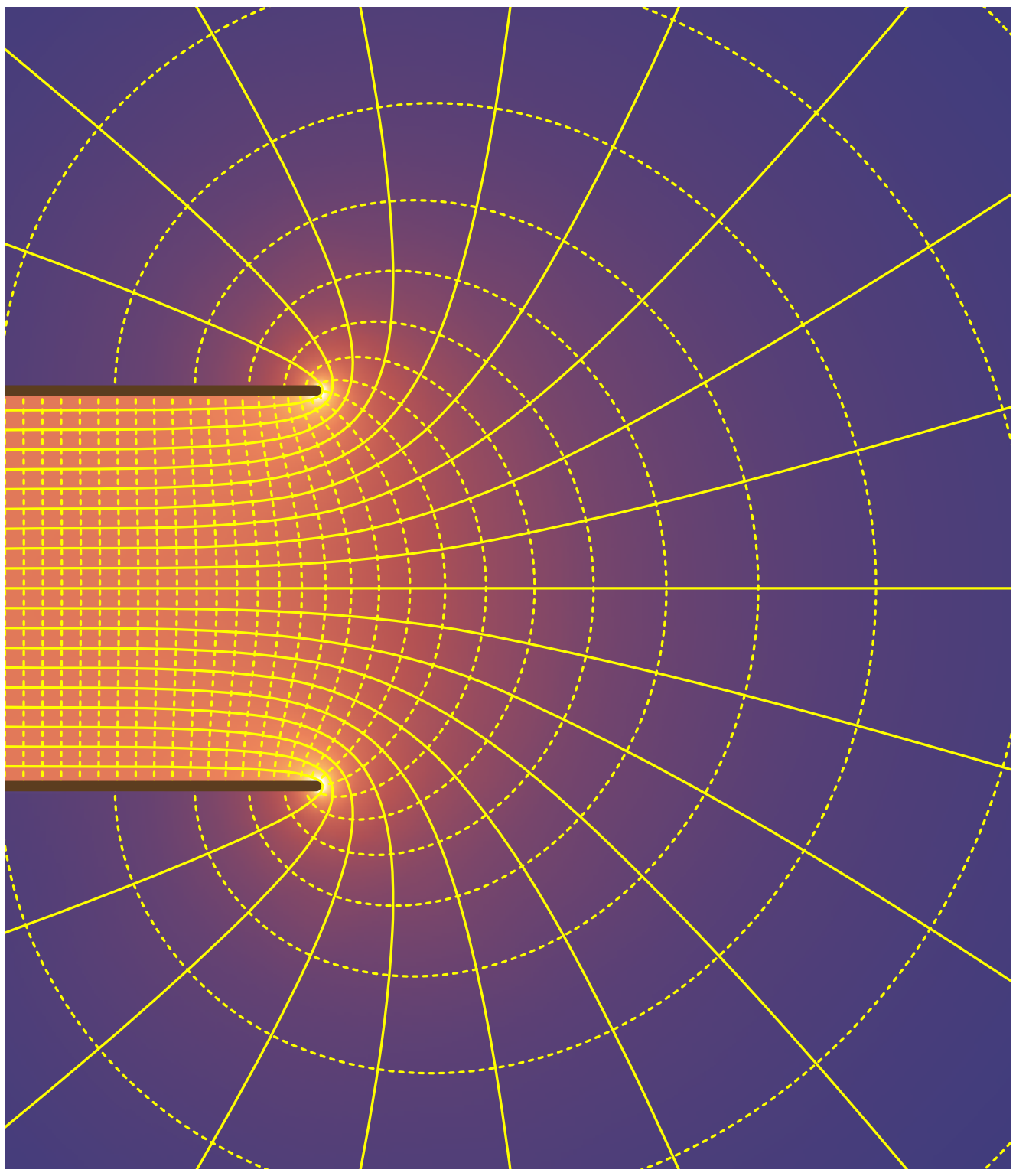

**Figure 3.18** / Two-dimensional electric field near one end of a parallel plate capacitor.

It is obtained by conformal transformation of field between two infinite plates.

**Legend:**

- equipotentials
- field lines
- capacitor plate

## Conformal Group

We usually work with just one underlying manifold as our spacetime, so most interesting conformal transformations in our case have the same manifold both for the domain and the target. As already mentioned in Remark 3.7 →p.53, the identity map is always conformal and composition/inverse of a conformal diffeomorphism stays conformal, so we clearly see that conformal automorphisms form a group.

**Definition 3.19** (Conformal Group, Isometry Group)**.** Let $\mathcal{M}$ be a manifold with a conformal structure. The group of all conformal automorphisms $\mathsf{Conf}(\mathcal{M})$ is called the *conformal group* of $\mathcal{M}$. If $\mathcal{M}$ is a pseudo-Riemannian manifold, all its *isometries* — conformal automorphisms with $\Omega = 1 \Leftrightarrow \omega = 0$ — form a subgroup $\mathsf{Iso}(\mathcal{M}) \subset \mathsf{Conf}(\mathcal{M})$ preserving the metric structure.

**Remark 3.20.** We will usually not be careful to distinguish between $\mathsf{Conf}(\mathcal{M})$ (which includes discrete transformations) and its connected component of the identity $\mathsf{Conf}_0(\mathcal{M})$. ⌟

**Remark 3.21** (Left Action of Conformal Group)**.** We have a natural left action $\triangleright\colon \mathsf{Conf}(\mathcal{M}) \times \mathcal{M} \to \mathcal{M}$, $(g, x) \mapsto g \triangleright x \equiv g(x)$, where we first think of $g$ as some abstract group element, and then as a conformal automorphism of $\mathcal{M}$. By action being *left* we mean the property $g \triangleright (h \triangleright x) = (gh) \triangleright x = g(h(x)) \equiv (g \circ h)(x)$. ⌟

**Remark 3.22** (Fundamental Vector Field)**.** A Lie algebra element $X \in \mathsf{Lie}(\mathsf{Conf}(\mathcal{M}))$ generates a 1-parametric subgroup of $\mathsf{Conf}(\mathcal{M})$ through the exponential map $\phi_\varepsilon \equiv \mathrm{e}^{\varepsilon X} \in \mathsf{Conf}(\mathcal{M})$, $\varepsilon \in \mathbb{R}$. The flow $\phi_{\circ}\colon \mathbb{R} \times \mathcal{M} \to \mathcal{M}$ is in turn generated by a vector field $\boldsymbol{\xi}_X \equiv \frac{\mathbf{D}}{\mathrm{d}\varepsilon}\phi_\varepsilon|_{\varepsilon=0} \in \mathcal{T}\mathcal{M}$, which is called the *fundamental vector field* corresponding to $\mathsf{X}$.

Generally, having a group $G$ with a left action on $\mathcal{M}$, it can be shown that the map $\mathsf{Lie}(G) \equiv \mathfrak{g} \to \mathcal{T}\mathcal{M}$, $X \mapsto \boldsymbol{\xi}_X$ assigning the fundamental vector field is a Lie algebra antihomomorphism, that is $[\![\boldsymbol{\xi}_X, \boldsymbol{\xi}_Y]\!] = \boldsymbol{\xi}_{-[X,Y]}$, where $[\![\bullet, \bullet]\!]$ is the Lie bracket of vector fields and $[\bullet, \bullet]$ is the Lie bracket of $\mathfrak{g}$. ⌟

## Conformal Killing Vector Fields

Since most interesting (for us) are continuous symmetries and corresponding conservation laws, we will now focus on conformal automorphisms continuously connected to the identity. As usual, it is convenient to first study the generators of infinitesimal conformal transformations.

**Definition 3.23** (Conformal Killing Vector Field)**.** A vector field $\boldsymbol{\xi} \in \mathcal{T}\mathcal{M}$ is called a *conformal Killing vector field* (CKV) if it satisfies (with $\phi_\varepsilon$ being the infinitesimal flow generated by $\boldsymbol{\xi}$, that is $\boldsymbol{\xi} \equiv \frac{\mathbf{D}}{\mathrm{d}\varepsilon}\phi_\varepsilon|_{\varepsilon=0}$)

$$\mathcal{L}_{\boldsymbol{\xi}} \boldsymbol{g}_{\boldsymbol{ab}} \equiv \frac{\mathrm{d}}{\mathrm{d}\varepsilon}\phi_\varepsilon^* \boldsymbol{g}\Big|_{\varepsilon=0} = \boldsymbol{\nabla}_{\boldsymbol{a}} \boldsymbol{\xi}_{\boldsymbol{b}} + \boldsymbol{\nabla}_{\boldsymbol{b}} \boldsymbol{\xi}_{\boldsymbol{a}} = 2\kappa\, \boldsymbol{g}_{\boldsymbol{ab}}\,,$$

for some function $\kappa \in C^\infty(\mathcal{M})$. The second equality is an identity for $\mathcal{L}_{\boldsymbol{\xi}}\boldsymbol{g}$ in terms of the Levi-Civita connection $\boldsymbol{\nabla}$ corresponding to $\boldsymbol{g}$. In the special case where $\kappa = 0$, we have a *Killing vector field* corresponding to an isometry.

**Remark 3.24.** Denoting $\Omega_\varepsilon \equiv \mathrm{e}^{\omega_\varepsilon}$ the conformal factor associated to $\phi_\varepsilon$, we have $\kappa = \frac{\mathrm{d}}{\mathrm{d}\varepsilon}\omega_\varepsilon|_{\varepsilon=0}$ (since $\Omega_0 = 1 \Leftrightarrow \omega_0 = 0$), which need not be positive. Also, by contracting indices of the CKV equation with the inverse metric $\boldsymbol{g}^{\boldsymbol{ab}}$, we obtain

$$\kappa = \frac{1}{d}\,\boldsymbol{\nabla}\boldsymbol{\cdot}\boldsymbol{\xi} \equiv \frac{1}{d}\,\boldsymbol{\nabla}_{\boldsymbol{n}}\boldsymbol{\xi}^{\boldsymbol{n}}\,, \quad (d = \mathrm{Tr}(\boldsymbol{\delta}) \equiv \boldsymbol{\delta}_{\boldsymbol{k}}^{\boldsymbol{k}} \equiv \boldsymbol{g}_{\boldsymbol{kl}}\boldsymbol{g}^{\boldsymbol{kl}})\,.$$

⌟

**Remark 3.25** (Lie Algebra of Conformal Killing Vectors)**.** The set of all conformal Killing vector fields forms a Lie algebra closed under linear combinations and Lie brackets. This follows from $\mathcal{L}_{\boldsymbol{\xi}+\boldsymbol{\zeta}} = \mathcal{L}_{\boldsymbol{\xi}} + \mathcal{L}_{\boldsymbol{\zeta}}$ and $\mathcal{L}_{[\![\boldsymbol{\xi},\boldsymbol{\zeta}]\!]} = [\mathcal{L}_{\boldsymbol{\xi}}, \mathcal{L}_{\boldsymbol{\zeta}}]$, where the first bracket $[\![\bullet, \bullet]\!]$ is Lie bracket of vector fields, and second one is just a commutator.

We denote set of all CKVs by $\mathfrak{conf}(\mathcal{M})$, and we have the following sequence

$$\mathfrak{iso}(\mathcal{M}) \subset \mathfrak{conf}(\mathcal{M}) \subset \mathfrak{T}\mathcal{M}\,,$$

where the $\mathfrak{iso}(\mathcal{M})$ is the Lie subalgebra of Killing vectors ($\kappa = 0$), and $\mathcal{TM}$ is the infinite-dimensional Lie algebra of smooth vector fields on $\mathcal{M}$. ⌟

**Remark 3.26** (Lie Algebra of Conformal Group versus CKVs). By Remark 3.22→p.58 we have a natural inclusion $\mathsf{Lie}(\mathsf{Conf}(\mathcal{M})) \hookrightarrow \mathfrak{conf}(\mathcal{M})$ of the Lie algebra of the conformal automorphism group into the Lie algebra of CKVs.

For $\mathcal{M}$ compact, it is actually an isomorphism, since the flow of a CKV can be globally integrated, yielding a 1-parametric subgroup of $\mathsf{Conf}(\mathcal{M})$. However, in general $\mathfrak{conf}(\mathcal{M})$ can be bigger than CKVs corresponding to global conformal transformations, since it can happen that the differential equation defining the flow can blow up for some value of the affine parameter. ⌟

**Remark 3.27.** Conformally equivalent metrics share their CKVs, since we have $\mathcal{L}_{\boldsymbol{\xi}} \boldsymbol{g}' \equiv \mathcal{L}_{\boldsymbol{\xi}}(\mathrm{e}^{2\omega}\,\boldsymbol{g}) = (2\boldsymbol{\xi}(\omega)\,\mathrm{e}^{2\omega} + 2\kappa\,\mathrm{e}^{2\omega})\boldsymbol{g} = 2(\boldsymbol{\xi}(\omega) + \kappa)\boldsymbol{g}' \equiv 2\kappa'\boldsymbol{g}'$. ⌟

**Remark 3.28.** Conformal Killing vectors have constant "nature" — they stay *timelike*, *spacelike*, or *null* — along their integral curves. We can see this by differentiating $f \equiv \boldsymbol{\xi}^2 \equiv \boldsymbol{g}(\boldsymbol{\xi}, \boldsymbol{\xi})$ along the associated integral curve $x^{\boldsymbol{\xi}}(t)$. We obtain $\frac{\mathrm{d}f}{\mathrm{d}t} = \mathcal{L}_{\boldsymbol{\xi}} f = 2\kappa f$ using $\mathcal{L}_{\boldsymbol{\xi}} \boldsymbol{\xi} = [\![\boldsymbol{\xi}, \boldsymbol{\xi}]\!] = 0$. By uniqueness of the solution of smooth first-order differential equations we infer that $f$ has a constant sign, since if it is zero at any point, it stays zero along the whole integral curve. ⌟

**Proposition 3.29** (Further Equations for Conformal Killing Vectors). Let $\boldsymbol{\xi}$ be a conformal Killing vector field with a conformal factor $\kappa$. Then it also satisfies the following equations

$$\begin{aligned} \boldsymbol{\nabla}_a \boldsymbol{\nabla}_b \boldsymbol{\xi}_c &= \boldsymbol{R}_{bc}{}^{n}{}_{a} \boldsymbol{\xi}_n + (\boldsymbol{\partial}_a \kappa)\boldsymbol{g}_{bc} + (\boldsymbol{\partial}_b \kappa)\boldsymbol{g}_{ca} - (\boldsymbol{\partial}_c \kappa)\boldsymbol{g}_{ab}\,, \\ \Box \boldsymbol{\xi}_c &= -\mathbf{Ric}_c{}^{n} \boldsymbol{\xi}_n + (2 - d)\boldsymbol{\partial}_c \kappa\,, \end{aligned}$$

where $\Box \equiv \boldsymbol{g}^{ab} \boldsymbol{\nabla}_a \boldsymbol{\nabla}_b$ and $\mathbf{Ric}_{ab} \equiv \boldsymbol{R}_{na}{}^{n}{}_{b}$.

**Proof.** Writing the covariant derivative of CKV equation in Definition 3.23→p.58 three times with permuted indices, and summing them in a particular way gives

$$\left.\begin{aligned} \boldsymbol{\nabla}_c(\boldsymbol{\nabla}_a \boldsymbol{\xi}_b + \boldsymbol{\nabla}_b \boldsymbol{\xi}_a) &= 2(\boldsymbol{\partial}_c \kappa)\boldsymbol{g}_{ab} \quad (-) \\ \boldsymbol{\nabla}_a(\boldsymbol{\nabla}_b \boldsymbol{\xi}_c + \boldsymbol{\nabla}_c \boldsymbol{\xi}_b) &= 2(\boldsymbol{\partial}_a \kappa)\boldsymbol{g}_{bc} \quad (+) \\ \boldsymbol{\nabla}_b(\boldsymbol{\nabla}_c \boldsymbol{\xi}_a + \boldsymbol{\nabla}_a \boldsymbol{\xi}_c) &= 2(\boldsymbol{\partial}_b \kappa)\boldsymbol{g}_{ca} \quad (+) \end{aligned}\right\} \Rightarrow \begin{aligned} &(\boldsymbol{\nabla}_a \boldsymbol{\nabla}_b + \boldsymbol{\nabla}_b \boldsymbol{\nabla}_a)\boldsymbol{\xi}_c - (\boldsymbol{R}_{ac}{}^{n}{}_{b} + \boldsymbol{R}_{bc}{}^{n}{}_{a})\boldsymbol{\xi}_n \\ &\quad = 2\Big[(\boldsymbol{\partial}_a \kappa)\boldsymbol{g}_{bc} + (\boldsymbol{\partial}_b \kappa)\boldsymbol{g}_{ca} - (\boldsymbol{\partial}_c \kappa)\boldsymbol{g}_{ab}\Big]\,, \end{aligned}$$

where we used the Ricci identity $(\boldsymbol{\nabla}_a \boldsymbol{\nabla}_b - \boldsymbol{\nabla}_b \boldsymbol{\nabla}_a)\boldsymbol{\xi}_c = -\boldsymbol{R}_{ab}{}^{n}{}_{c} \boldsymbol{\xi}_n$. Using it once more to commute indices $\boldsymbol{ba}$ in the term $\boldsymbol{\nabla}_b \boldsymbol{\nabla}_a \boldsymbol{\xi}_c$, and utilizing the Bianchi identity

$$\boldsymbol{R}_{[ab}{}^{n}{}_{c]} = 0 \iff \boldsymbol{R}_{ab}{}^{n}{}_{c} + \boldsymbol{R}_{ca}{}^{n}{}_{b} + \boldsymbol{R}_{bc}{}^{n}{}_{a} = 0 \iff \boldsymbol{R}_{ab}{}^{n}{}_{c} - \boldsymbol{R}_{ac}{}^{n}{}_{b} = -\boldsymbol{R}_{bc}{}^{n}{}_{a}$$

we obtain the desired formula after a simple manipulation. □

**Corollary 3.30** (Maximal Number of Conformal Killing Vector Fields)**.** Let $\mathcal{M}$ be a $d$-dimensional connected pseudo-Riemannian manifold with $d \geq 3$. Then the number of independent conformal Killing vector fields is bounded by

$$\dim \mathfrak{conf}(\mathcal{M}) \leq \frac{(d+1)(d+2)}{2} ,$$

and in the case of Killing vectors we have (holds for arbitrary $d$)

$$\dim \mathfrak{iso}(\mathcal{M}) \leq \frac{d(d+1)}{2} .$$

**Idea of the Proof.** We have a system of first-order linear partial differential equations for CKV data $\{\boldsymbol{\xi_a}, \boldsymbol{\omega_{ab}} \equiv \boldsymbol{\nabla}_{[a}\boldsymbol{\xi}_{b]}, \kappa \sim \boldsymbol{\nabla}\cdot\boldsymbol{\xi}, \boldsymbol{F_a} \equiv \boldsymbol{\nabla_a}\kappa \sim \boldsymbol{\nabla_a}\boldsymbol{\nabla}\cdot\boldsymbol{\xi}\}$ in Definition 3.23→p.58 and Proposition 3.29→p.59 — for $d \geq 3$ it is also possible to obtain equation for $\boldsymbol{\nabla_a F_b}$ in terms of other variables — so assuming connectedness it is enough to specify initial conditions at a single point. This gives us maximally $d + \frac{d(d-1)}{2} + 1 + d = \frac{(d+1)(d+2)}{2}$ parameters. In the case of ordinary Killing vectors (isometries) we have $\kappa \equiv 0$, so we are left with just $d + \frac{d(d-1)}{2} = \frac{d(d+1)}{2}$ parameters. For further details see [40]. □

**Remark 3.31** (Maximally Conformally-Symmetric Spacetimes)**.** In the following Section 3.2→p.61 we will explicitly see that flat spaces $\mathbb{R}^{p,q}$ saturate this bound. Moreover, by studying the integrability conditions of the equations for CKV data, it can be shown that any manifold with maximal number of independent CKVs is automatically conformally flat [40]. ⌟

**Remark 3.32.** The finite bound for CKVs in Corollary 3.30→p.60 fails for $d \leq 2$. The technical reason is that it is not possible to constrain $\boldsymbol{\nabla_a F_b} \equiv \boldsymbol{\nabla_a \nabla_b}\kappa$, thus we have more freedom allowing possibly infinite independent CKVs. We will see this explicitly in context of flat space during the proof of Proposition 3.37→p.62. ⌟

## Conformal Compactification

As already mentioned in Remark 3.26→p.59, not all CKVs can be integrated to obtain global conformal transformations. Since we want to retain as many symmetries as possible — and it is anyway mathematically elegant — we will always try to complete $\mathcal{M}$ such that every local conformal transformation can be extended to a global one.

Such completion would be achieved by embedding $\mathcal{M}$ into a compact manifold $\mathcal{M}'$ (by adding measure zero set of points) with the compatible conformal structure, since then every CKV can be globally integrated [41].

**Definition 3.33** (Conformal Compactification)**.** Let $(\mathcal{M}, \mathfrak{c})$ be a manifold equipped with a conformal structure. A compact manifold $(\mathcal{M}', \mathfrak{c}')$ with conformal structure is called a *conformal compactification* of $\mathcal{M}$ when:

**(1)** There is a conformal embedding $\iota\colon (\mathcal{M}, \mathfrak{c}) \hookrightarrow (\mathcal{M}', \mathfrak{c}')$ with dense image in $\mathcal{M}'$, that is $\overline{\iota(\mathcal{M})} = \mathcal{M}'$.

**(2)** All conformal vector fields on $(\mathcal{M}, \mathfrak{c})$ extend to ones on $(\mathcal{M}', \mathfrak{c}')$, and thus can be automatically integrated to global conformal automorphisms.

**Remark 3.34.** Conformal compactifications in this strong sense exist only for $d \geq 3$, since for example in $d = 2$ it is not possible to continue all local conformal transformations to global ones.

If the conformal compactification of $\mathcal{M}$ exists, it is unique (up to diffeomorphism), and we denote it with $\overline{\mathcal{M}}$. Oftentimes we will omit such distinction and implicitly mean $\overline{\mathcal{M}}$, for example when we write $\mathsf{Conf}(\mathcal{M})$. Furthermore, by Remark 3.26 →p.59 we can now identify $\mathsf{Lie}(\mathsf{Conf}(\mathcal{M}))$ with $\mathfrak{conf}(\mathcal{M})$, but remember that corresponding Lie brackets differ by extra minus sign. ⌟

**Remark 3.35.** Usually we compactify just by adding some points at "infinity", which is naturally done for a certain conformally equivalent manifold. It is then obvious that both conditions in Definition 3.33 →p.60 are satisfied. A prime example is a sphere $\mathbb{S}^n$ as a one-point conformal compactification of a plane $\mathbb{R}^n$ via the inverse stereographic projection, see Example 3.11 →p.54. There is only one extra point in $\mathbb{S}^n$, namely the north pole $N$ corresponding to "infinity" of $\overline{\mathbb{R}^n}$. ⌟

## 3.2 Flat-Space Conformal Structure

From now on we restrict our attention to flat spaces $\mathbb{R}^{p,q}$ with $d \equiv p + q \geq 3$. For simplicity, we will often consider the Euclidean $\mathbb{R}^d$. As already noted in Remark 3.31 →p.60, this is essentially the same as focusing on maximally conformally-symmetric spacetimes.

### Conformal Transformations in Flat Space

Essentially by Definition 3.6 →p.53, a conformal transformation of a flat space looks locally like translation+rotation+dilatation.

**Remark 3.36** (Jacobian Matrix of Flat-Space Conformal Transformation)**.** For any conformal transformation $g \in \mathsf{Conf}(\mathbb{R}^d)\colon x \mapsto x' \equiv g(x)$ we can write

$$\frac{\partial x'}{\partial x} \equiv \Omega_g(x) R_g(x), \quad R_g(x) \in \mathsf{SO}(d) \ \text{or} \ \mathsf{O}(d)\,.$$

Alternatively, the scaling factor $\Omega_g$ is related to the Jacobian by

$$\Omega_g(x) = \left|\frac{\partial x'}{\partial x}\right|^{\frac{1}{d}} \equiv \det{}^{\frac{1}{d}}\left(\frac{\partial g(x)}{\partial x}\right).$$

⌟

Due to the simple geometry of $\mathbb{R}^{p,q}$, we can write down the general form of conformal Killing vector fields $\mathfrak{conf}(\mathbb{R}^{p,q})$ by explicitly solving the CKV equation.

**Proposition 3.37** (General Form of CKV in Flat Space)**.** A general solution of the conformal Killing equation in flat space $\mathbb{R}^{p,q}$ of dimension $d \equiv p+q \geq 3$ can be written as

$$\boldsymbol{\xi} \equiv \xi^a \frac{\boldsymbol{\partial}}{\boldsymbol{\partial} x^a} \equiv \xi^a \boldsymbol{\partial}_a = t^a \boldsymbol{p}_a + \tfrac{1}{2}\omega^{ab}\boldsymbol{m}_{ab} + \sigma \boldsymbol{d} + s^a \boldsymbol{k}_a ,$$

with the generators explicitly given by

$$\begin{aligned} \boldsymbol{p}_a &\equiv \boldsymbol{\partial}_a , \quad \boldsymbol{m}_{ab} \equiv -x_a \boldsymbol{\partial}_b + x_b \boldsymbol{\partial}_a = \boldsymbol{m}_{[ab]} , \\ \boldsymbol{d} &\equiv x^n \boldsymbol{\partial}_n , \quad \boldsymbol{k}_a \equiv 2x_a x^n \boldsymbol{\partial}_n - x^2 \boldsymbol{\partial}_a , \end{aligned}$$

where $t^{\bullet}, \omega^{\bullet\bullet} \equiv \omega^{[\bullet\bullet]}, \sigma, s^{\bullet}$ are constants parametrizing the CKVs. In other words, the Lie algebra of conformal vector fields on $\mathbb{R}^{p,q}$ — on the level of vector spaces — decomposes into direct sum of subalgebras

$$\mathfrak{conf}(\mathbb{R}^{p,q}) = \underbrace{\{\text{translations}\}}_{\simeq \mathbb{R}^{p,q}} \oplus \underbrace{\{\text{pseudorotations}\}}_{\simeq \mathfrak{so}(p,q)} \oplus \underbrace{\{\text{dilatations}\}}_{\simeq \mathbb{R} \simeq \mathfrak{so}(1,1)} \oplus \underbrace{\{\text{SCTs}\}}_{\simeq \mathbb{R}^{p,q}} ,$$

where SCTs are the so-called *Special Conformal Transformations*.

**Proof.** In the flat space we have $\boldsymbol{g} \equiv \boldsymbol{\eta}$, $\boldsymbol{\nabla} \to \boldsymbol{\partial}$, and $\boldsymbol{R}_{\bullet\bullet}{}^{\bullet}{}_{\bullet} = 0$, which simplifies the analysis of the conformal Killing equation in Definition 3.23 → p.58. A chain of manipulations gives (throughout we will use Cartesian coordinates)

$$\begin{aligned}
[\text{CKV}]_{ab} \equiv \quad & \partial_a \xi_b + \partial_b \xi_a = 2\kappa\, \eta_{ab} \\
\partial^b [\text{CKV}]_{ab} \implies \quad & d\partial_a \kappa + \Box \xi_a = 2\partial_a \kappa \\
\partial^a \partial^b [\text{CKV}]_{ab} \implies \quad & 2d\Box\kappa = 2\Box\kappa \quad \overset{d\neq 1}{\implies} \quad \Box\kappa = 0 \\
\partial_{(a}\partial^n [\text{CKV}]_{b)n} \implies \quad & d\partial_a\partial_b\kappa + \underbrace{\Box\partial_{(a}\xi_{b)}}_{\propto \Box\kappa = 0} = 2\partial_a\partial_b\kappa \quad \overset{d\neq 2}{\implies} \quad \partial_a\partial_b\kappa = 0
\end{aligned}$$

Substituting this into ($\partial_{\bullet}$ of) Proposition 3.29 → p.59, we obtain

$$\partial_a \partial_b \xi_c = (\partial_a \kappa)\eta_{bc} + (\partial_b \kappa)\eta_{ca} - (\partial_c \kappa)\eta_{ab} \overset{\partial_{\bullet}}{\implies} \partial_a \partial_b \partial_c \xi_d = 0 ,$$

therefore $\xi^{\bullet}$ is at most quadratic in $x$. Expanding it as

$$\xi^a(x) = t^a + A^a{}_b x^b + S^a{}_{bc} x^b x^c , \quad \text{where } S^a{}_{bc} \equiv S^a{}_{(bc)} ,$$

substituting it back into [CKV], and comparing order by order we obtain:

**(1)** No restrictions on $t^{\bullet} \in \mathbb{R}^{p,q}$, so $t^a \boldsymbol{\partial}_a$ is CKV corresponding to *translations*.

**(2)** For linear part we get

$$2A_{(ab)} = \frac{2}{d} A^n{}_n \eta_{ab} .$$

Decomposing $A_{\bullet\bullet}$ into traceless symmetric, antisymmetric, and trace parts, we see that the condition implies traceless symmetric part is zero, so we are left with

$$A_{ab} = A_{[ab]} + \frac{1}{d} A^n{}_n \eta_{ab} \equiv \omega_{ab} + \sigma \eta_{ab} \implies A^a{}_b x^b \boldsymbol{\partial}_a = \omega^{ab} x_{[b} \boldsymbol{\partial}_{a]} + \sigma x^n \boldsymbol{\partial}_n \,.$$

The first part corresponds to *pseudorotations*, and the second to scaling transformations also called as *dilatations*.

**(3)** For the quadratic part we get

$$4 S_{(ab)c} x^c = \frac{2}{d} S^n{}_{nc} x^c \eta_{ab} \equiv 4 s_c x^c \eta_{ab} \iff S_{(ab)c} = \eta_{ab} s_c \,.$$

We can invert the relation (using $S_{abc} = S_{a(bc)}$) to get

$$S_{abc} = \eta_{ab} s_c + \eta_{ac} s_b - \eta_{bc} s_a \implies S^a{}_{bc} x^b x^c \boldsymbol{\partial}_a = s^a (2 x_a x^n \boldsymbol{\partial}_n - x^2 \boldsymbol{\partial}_a) \,.$$

This corresponds to *Special Conformal Transformations* (SCTs). □

**Remark 3.38.** As was mentioned in Remark 3.32→p.60, the cases with $d \leq 2$ are special. Any vector field in $d = 1$ is CKV, which is compatible with Example 3.14→p.55. In $d = 2$ we can have $\partial_\bullet \partial_\bullet \kappa \neq 0$, which allows infinite independent CKVs. For more detail see [4] and [39]. ⌟

Since hopefully everybody knows finite forms of translations and pseudo-rotations, and we encountered dilatations in Example 3.9→p.54, we will look more closely on SCTs. But first, it is convenient to discuss an important transformation not continuously connected to the identity component of the conformal group.

**Remark 3.39** (Inversion)**.** The (radial) *inversion* map is given by

$$\begin{aligned} \mathcal{I} \colon \overline{\mathbb{R}^d} &\longrightarrow \overline{\mathbb{R}^d} \\ x &\longmapsto \mathcal{I}(x) \equiv \frac{x}{|x|^2} \,. \end{aligned}$$

It maps 0 to $\infty$ and vice versa, since $\mathcal{I} = \mathcal{I}^{-1}$. Locally (infinitesimally) we have

$$\frac{\partial \mathcal{I}^a(x)}{\partial x^b} = \frac{\partial \mathcal{I}^a}{\partial x^b}\bigg|_x = \frac{\delta^a_b}{|x|^2} - \frac{2 x^a x_b}{|x|^4} = \frac{1}{|x|^2}\left(\delta^a_b - 2\frac{x^a x_b}{|x|^2}\right) \equiv \Omega_{\mathcal{I}}(x) \mathcal{I}^a_b(\hat{x}) \,,$$

where

$$\Omega_{\mathcal{I}}(x) \equiv \left|\frac{x}{|x|^2}\right|^2 = |\mathcal{I}(x)|^2 = \frac{1}{|x|^2} \,, \qquad \mathcal{I}^a_b(\hat{x}) \equiv (\delta^a_b - 2\hat{x}^a \hat{x}_b) \in \mathsf{O}(d) \,,$$

so it is rescaling and reflection along the radial direction $\hat{x} \equiv x/|x|$.

Induced action on a flat metric is

$$\begin{aligned} (\mathcal{I}^* \eta)_{ab}\Big|_x = \frac{\partial \mathcal{I}^c}{\partial x^a} \frac{\partial \mathcal{I}^d}{\partial x^b} \eta_{cd}\Big|_x &= \frac{1}{|x|^4} (\delta^c_a - 2\hat{x}^c \hat{x}_a)\left(\delta^d_b - 2\hat{x}^d \hat{x}_b\right) \eta_{cd} \\ &= \frac{1}{|x|^4} (\eta_{ab} - 4\hat{x}_a \hat{x}_b + 4\hat{x}_a \hat{x}_b) = \frac{1}{|x|^4} \eta_{ab} \equiv \Omega^2_{\mathcal{I}} \eta_{ab}\Big|_x \,, \end{aligned}$$

confirming that inversion is a conformal transformation.

Alternatively, after the inverse stereographic projection to the sphere $\mathbb{S}^d \subset \mathbb{R}^{d+1}$ Example 3.11 →p.54, it corresponds to the reflection $\mathsf{x}^0 \mapsto -\mathsf{x}^0$, which is clearly conformal. ⌟

**Remark 3.40** (Finite Special Conformal Transformations). Special Conformal Transformations (SCTs) can be understood as a composition of inversion, translation by $-b \in \mathbb{R}^d$, and inversion again — they are "translations around infinity". While SCTs are not defined everywhere on $\mathbb{R}^{p,q}$, but will be on appropriate conformal compactification.

Infinitesimally, generators of translations and SCTs are related by

$$(-\mathcal{I}\boldsymbol{p}_a\mathcal{I})\Big|_x = -\boldsymbol{\partial}_a\mathcal{I}\Big|_{\frac{x}{|x|^2}} = -\frac{\partial\mathcal{I}^b}{\partial x^a}\Big|_{\frac{x}{|x|^2}}\boldsymbol{\partial}_b = -|x|^2\Big(\delta_a^b - 2\hat{x}^b\hat{x}_a\Big)\boldsymbol{\partial}_b$$
$$= 2x_a\, x\cdot\boldsymbol{\partial} - |x|^2\boldsymbol{\partial}_a \equiv \boldsymbol{k}_a\Big|_x\,,$$

where the generators are understood as differential operators on functions, and $\mathcal{I}$ acts as $(\mathcal{I}f)(x) \equiv f(\mathcal{I}(x))$. Alternatively, we have

$$(-\mathrm{Ad}_{\mathcal{I}}\boldsymbol{p}_a)\Big|_x \equiv \frac{\mathbf{D}}{\mathrm{d}\varepsilon}\mathcal{I}\circ[\square\mapsto\square-\varepsilon\boldsymbol{e}_a]\circ\mathcal{I}(x)\Big|_{\varepsilon=0}$$
$$= \frac{\mathbf{D}}{\mathrm{d}\varepsilon}\frac{x-|x|^2\varepsilon\boldsymbol{e}_a}{1-2\varepsilon x_a+|\varepsilon|^2|x|^2}\Big|_{\varepsilon=0} = -|x|^2\boldsymbol{\partial}_a + 2x_a x\cdot\boldsymbol{\partial} \equiv \boldsymbol{k}_a\Big|_x\,.$$

So, for finite (exponentiated) action of SCTs has the form

$$\mathrm{e}^{b\cdot\boldsymbol{k}} = \mathrm{e}^{-\mathcal{I}b\cdot\boldsymbol{p}\mathcal{I}} = \mathcal{I}\,\mathrm{e}^{-b\cdot\boldsymbol{p}}\,\mathcal{I} = \mathcal{I}\circ[\square\mapsto\square-b]\circ\mathcal{I}\,,$$

that is

$$\mathrm{e}^{b\cdot\boldsymbol{k}}(x) = \mathcal{I}\circ[\square\mapsto\square-b]\left(\frac{x}{|x|^2}\right) = \mathcal{I}\left(\frac{x}{|x|^2}-b\right)$$
$$= \frac{\frac{x}{|x|^2}-b}{\left|\frac{x}{|x|^2}-b\right|^2} = \frac{x-|x|^2b}{1-2b\cdot x+|b|^2|x|^2}$$

Since translations completely preserve the flat metric, using the form of $\Omega_{\mathcal{I}}^2$ we obtain $\Omega$-factor of a SCT as

$$\Omega_b = \Omega_{\mathcal{I}}\left(\frac{x}{|x|^2}-b\right)\Omega_{\mathcal{I}}(x) = \frac{1}{\left|\frac{x}{|x|^2}-b\right|^2}\frac{1}{|x|^2} = \frac{1}{\left(1-2b\cdot x+|b|^2|x|^2\right)^2}\,.$$

⌟

## Conformal Algebra of Flat Space

Having the explicit form of CKVs, we can now compute their Lie brackets.

**Lemma 3.41** (Lie Brackets of CKVs in Flat Space)**.** Lie brackets of CKVs in flat space $\mathbb{R}^{p,q}$ ($d \equiv p+q \geq 3$) are given by

$$
\begin{gathered}
[\![\boldsymbol{m}_{ab}, \boldsymbol{p}_c]\!] = -\boldsymbol{p}_a \eta_{bc} + \boldsymbol{p}_b \eta_{ac}\,, \qquad [\![\boldsymbol{m}_{ab}, \boldsymbol{k}_c]\!] = -\boldsymbol{k}_a \eta_{bc} + \boldsymbol{k}_b \eta_{ac}\,, \\
[\![\boldsymbol{m}_{ab}, \boldsymbol{m}_{cd}]\!] = \boldsymbol{m}_{ac}\eta_{bd} - (a \leftrightarrow b) - (c \leftrightarrow d) + (a \leftrightarrow b, c \leftrightarrow d)\,, \\
[\![\boldsymbol{d}, \boldsymbol{d}]\!] = 0\,, \qquad [\![\boldsymbol{d}, \boldsymbol{m}_{ab}]\!] = 0\,, \qquad [\![\boldsymbol{p}_a, \boldsymbol{p}_b]\!] = 0\,, \qquad [\![\boldsymbol{k}_a, \boldsymbol{k}_b]\!] = 0\,, \\
[\![\boldsymbol{d}, \boldsymbol{p}_a]\!] = -\boldsymbol{p}_a\,, \qquad [\![\boldsymbol{d}, \boldsymbol{k}_a]\!] = \boldsymbol{k}_a\,, \qquad [\![\boldsymbol{k}_a, \boldsymbol{p}_b]\!] = -2\eta_{ab}\boldsymbol{d} + 2\boldsymbol{m}_{ab}\,.
\end{gathered}
$$

**Proof.** Can be verified by a straightforward computation. Most of the brackets with $\boldsymbol{k}_\bullet$ can be obtained from the ones with $\boldsymbol{p}_\bullet$ by utilizing (at least in $\mathbb{R}^d$)

$$
\begin{gathered}
\mathcal{I} \circ D_\lambda \circ \mathcal{I} = D_{1/\lambda} \implies \mathrm{Ad}_{\mathcal{I}}\boldsymbol{d} = -\boldsymbol{d}\,, \qquad \mathcal{I} \circ R \circ \mathcal{I} = R \implies \mathrm{Ad}_{\mathcal{I}}\boldsymbol{m}_{ab} = \boldsymbol{m}_{ab}\,, \\
\mathcal{I}^2 = \mathbf{1}\,, \qquad \mathrm{Ad}_{\mathcal{I}}\boldsymbol{p}_a = -\boldsymbol{k}_a\,, \qquad \mathrm{Ad}_\bullet [\![\circ, \circ]\!] = [\![\mathrm{Ad}_\bullet \circ, \mathrm{Ad}_\bullet \circ]\!]\,,
\end{gathered}
$$

where $R$ is any rotation. To derive these relations, the relevant places to look are Example 3.9→p.54, Remark 3.39→p.63, and Remark 3.40→p.64. □

It is convenient to introduce alternative basis for the conformal algebra.

**Definition 3.42** (Generators of Flat-Space Conformal Algebra)**.** An alternative basis of $\mathfrak{conf}(\mathbb{R}^{p,q})$ for $d \equiv p+q \geq 3$ is provided by the generators $\boldsymbol{l}_{AB} \equiv \boldsymbol{l}_{[AB]}$ with $A, B \in \{-1, 0, 1, \ldots, d\}$ (and $a, b \in \{1, \ldots, d\}$) defined as

$$
\begin{aligned}
\boldsymbol{l}_{-1,0} &= \boldsymbol{d}\,, \\
\boldsymbol{l}_{-1,b} &= \tfrac{1}{2}(\boldsymbol{p}_b - \boldsymbol{k}_b)\,, \\
\boldsymbol{l}_{0,b} &= \tfrac{1}{2}(\boldsymbol{p}_b + \boldsymbol{k}_b)\,, \\
\boldsymbol{l}_{ab} &= \boldsymbol{m}_{ab}\,,
\end{aligned}
\qquad
\boldsymbol{l}_{AB} = \begin{pmatrix} 0 & \boldsymbol{d} & \tfrac{1}{2}(\boldsymbol{p}_b - \boldsymbol{k}_b) \\ -\boldsymbol{d} & 0 & \tfrac{1}{2}(\boldsymbol{p}_b + \boldsymbol{k}_b) \\ \vdots & \vdots & \boldsymbol{m}_{ab} \end{pmatrix}.
$$

For capital letters we will use diagonal metric $\eta_{AB} \equiv \eta^{(p+1,q+1)}_{AB}$ with $\eta_{-1,-1} \equiv -1$, $\eta_{0,0} \equiv +1$, and $\eta_{ab} \equiv \eta^{(p,q)}_{ab}$, so we added one timelike and one spacelike dimension.

**Remark 3.43.** We usually work in the Lorentzian signature $(p,q) = (1, d-1)$ or in the Euclidean signature $(p,q) = (0, d)$ (often obtained by the Wick rotation). It is then common to use index "0" for the "time" direction, and $i \in \{1, \ldots, d-1\}$ for "space" directions. Notice that this is not the case in Definition 3.42→p.65. ⌟

**Theorem 3.44** (Conformal Algebra and Group of $\mathbb{R}^{p,q}$)**.** The generators $\boldsymbol{l}_{AB}$ of $\mathfrak{conf}(\mathbb{R}^{p,q})$ for $d \equiv p+q \geq 3$ have the following Lie brackets

$$
[\![\boldsymbol{l}_{AB}, \boldsymbol{l}_{CD}]\!] = \boldsymbol{l}_{AC}\eta_{BD} - (A \leftrightarrow B) - (C \leftrightarrow D) + (A \leftrightarrow B, C \leftrightarrow D)\,,
$$

so the conformal algebra of $\mathbb{R}^{p,q}$ is

$$
\mathfrak{conf}(\mathbb{R}^{p,q}) \cong \mathfrak{so}(p+1, q+1)\,.
$$

The corresponding conformal group of the conformal compactification $\overline{\mathbb{R}^{p,q}}$ is

$$\mathsf{Conf}(\overline{\mathbb{R}^{p,q}}) \cong \mathsf{O}(p+1, q+1)/\mathbb{Z}_2 \,.$$

**Idea of the Proof.** First part follows by expanding $\{\boldsymbol{l}_{AB}\}$ via Definition 3.42 → p.65 in terms of $\{\boldsymbol{d}, \boldsymbol{p}_a, \boldsymbol{k}_a, \boldsymbol{m}_{ab}\}$, and then using Lemma 3.41 → p.65. Second part can be understood by constructing the conformal compactification using the embedding/ambient space to be discussed shortly. For a more detailed Theorem statement and proof, see [39]. □

**Remark 3.45** (Conformal Group of Celestial Sphere)**.** Since sphere $\mathbb{S}^{d-2}$ is conformally equivalent to flat Euclidean space via stereographic projection, we have $\mathsf{Conf}(\mathbb{S}^{d-2}) = \mathsf{Conf}(\mathbb{R}^{d-2}) = \mathsf{SO}(1, d-1)$, so the conformal group of the celestial sphere is precisely the Lorentz group. Every light ray is transformed to another light ray by the Lorentz group, and looking at the corresponding points on the *celestial sphere*, they are transformed by the corresponding conformal transformation. Nice treatment of this can be found in [42]. ⌟

## Embedding Space

Both Theorem 3.44 → p.65 and Remark 3.45 → p.66 suggest that viewing given flat space as embedded in higher-dimensional flat space can lead to a simpler description of conformal group. This is the idea behind *embedding formalism* of conformal field theory. We will just briefly outline the construction, for more details see for example [11].

On $\mathbb{R}^{p+1,q+1}$ we have natural linear action of pseudo-orthogonal group $\mathsf{SO}(p+1, q+1)$ which commutes with dilatations, so it descends also to real projective space... Real projective spaces are compact, so also are their closed subspaces (example being the null cone), providing us with natural conformal compactifications of $\mathbb{R}^{p,q}$.

For $\mathbb{R}^{1,d-1}$, by this construction we obtain something like (assume $d > 2$)

$$\frac{\mathbb{S}^1 \times \mathbb{S}^{d-1}}{\mathbb{Z}_2} \simeq \mathbb{R}^{1,d-1} \cup \{\text{points at infinity}\}$$

with natural action of $\mathsf{SO}(2, d)$, which however contains closed time-like curves.

We can get rid of $\mathbb{Z}_2$ (antipodal) identification by projectivizing only with positive factor $\lambda > 0$ (oriented rays) — then it contains two copies of $\mathsf{Mink}_d$.

It is then common to unwrap the $\mathbb{S}^1$ piece (corresponding to light-like direction) and consider the universal covering having the form $\mathbb{R} \times \mathbb{S}^{d-1}$, on which acts the universal (infinitely-sheeted) cover $\widetilde{\mathsf{SO}}(2, d)$ which unwraps the $\mathsf{SO}(2)$ piece of $\mathsf{SO}(2, d) \supset \mathsf{SO}(2) \times \mathsf{SO}(d)$.

## Conformal Cross-Ratios

In the following Section 3.3→p.69 we will look at correlators in flat-space CFTs. It will be important to understand how the distances between points transform under conformal transformations, which is content of the following lemma.

**Lemma 3.46** (Transformation of Euclidean Distance under Conformal Maps). For any conformal transformation $g \in \mathsf{Conf}(\mathbb{R}^d) : x \mapsto x' \equiv g(x)$ we have

$$|x' - y'|^2 = \Omega_g(x)\Omega_g(y)|x-y|^2 \,,$$

where $\Omega_g$ is defined as in Remark 3.36→p.61.

**Proof.** For translations and rotations we have $\Omega = 1$, so it is clearly true. Also simple are dilatations, where we again have a constant scaling factor $\Omega_\lambda = \lambda$. It suffices to show it for the inversion $\mathcal{I}$, and then the relation for SCTs follows from Remark 3.40→p.64 and the chain rule for the Jacobian.

Direct computation gives (see Remark 3.39→p.63 for definitions)

$$\begin{aligned}|\mathcal{I}(x) - \mathcal{I}(y)|^2 &\equiv \left|\frac{x}{|x|^2} - \frac{y}{|y|^2}\right|^2 = \frac{1}{|x|^2|y|^2}\left(\hat{x}|y| - \hat{y}|x|\right)^2 \\ &\equiv \Omega_{\mathcal{I}}(x)\Omega_{\mathcal{I}}(y)\Big(|y|^2 - 2x\cdot y + |x|^2\Big) = \Omega_{\mathcal{I}}(x)\Omega_{\mathcal{I}}(y)|x-y|^2\end{aligned}$$

□

It is an interesting question how to construct conformal invariants from the positions of some number of points. The translation and rotation invariance restricts our attention to the distance between points, and the scaling invariance (together with special conformal transformations) suggests looking at their ratios. Starting with 4 points, we indeed find such conformally invariant quantities.

**Definition 3.47** (Conformal Cross-Ratios). Given 4 distinct points $x_1, x_2, x_3, x_4$ in $\mathbb{R}^d$, we define their *conformal cross-ratios* as

$$u \equiv \frac{x_{12}^2 x_{34}^2}{x_{13}^2 x_{24}^2} \,, \qquad v \equiv \frac{x_{23}^2 x_{14}^2}{x_{13}^2 x_{24}^2} = u|_{1\leftrightarrow 3} \,,$$

where $x_{ij} \equiv x_i - x_j$ and $x_{ij}^\alpha \equiv |x_{ij}|^\alpha$ for any $\alpha \in \mathbb{R}$.

**Corollary 3.48** (Conformal Cross-Ratios are Conformal Invariants). The conformal cross-ratios $u$ and $v$ are invariant under conformal transformations.

**Proof.** Follows directly from Lemma 3.46→p.67 and the fact that each point $x_i$ appears both in the numerator and denominator of the cross-ratios. □

**Remark 3.49.** We could choose different ways how to arrange the points in the definition of the conformal cross-ratios, but all of them turn out to be expressible in terms of $u$ and $v$. ⌟

## Conformal Frame

We can get better intuition about the cross-ratios by considering them in a specific *conformal frame*, where using the conformal transformations we place the points at special positions.

**Lemma 3.50** (3-transitivity of Conformal Group)**.** The action of $\mathsf{Conf}(\mathbb{R}^d)$ on $\overline{\mathbb{R}^d}$ is 3-transitive, that is, for any 3-tuple of distinct points $(x_1, x_2, x_3) \subset \overline{\mathbb{R}^d}$ there exists a conformal transformation $g \in \mathsf{Conf}(\mathbb{R}^d)$ such that $g(x_\bullet) = y_\bullet$ for any other 3-tuple of distinct points $(y_1, y_2, y_3) \subset \overline{\mathbb{R}^d}$.

**Proof.** We will show that any 3-tuple of distinct points can be mapped to the standard configuration $\{0, e, \infty\}$, where $e$ is a point corresponding to some unit vector. This is done by composing simple conformal transformations as

$$(x_1, x_2, x_3) \xmapsto{\text{SCT}} (x_1', x_2', \infty) \xmapsto{\text{translation}} (0, x_2'', \infty) \longmapsto \left(\widehat{x_2''} \equiv x_2''/|x_2''|\right)$$
$$\xmapsto{\text{dilatation}} (0, \widehat{x_2''}, \infty) \xmapsto{\text{rotation}} (0, e, \infty)\,.$$

In the first two steps we move two points to $\infty$ and 0 using the fact that SCTs are translations around $\infty$, and that $\infty$ is invariant under usual translations. Since both $\{0, \infty\}$ are invariant under dilatations and rotations, in the final steps we can move the third point to any position on the unit circle. □

**Remark 3.51.** For $d = 2$ such transformations between 3-tuples of points are unique, which can be seen from uniqueness of transformation to the specified standard configuration $(0, e, \infty)$. In higher dimensions, we still have freedom to make rotations that fix the axis pointing in the direction of $e$. ⌟

**Remark 3.52.** Conformal frame point of view clearly shows why we could not construct nontrivial conformal invariants from 3 or fewer points. ⌟

**Proposition 3.53** (Number of Independent Conformal Cross-Ratios)**.** Given a set of $n$ distinct points in $\mathbb{R}^d$, the number of independent conformal cross-ratios is given by (with understanding that for $n \leq 3$ it is zero)

$$\left|\begin{Bmatrix}\text{independent conformal invariants} \\ \text{build from } n \text{ distinct points in } \mathbb{R}^d\end{Bmatrix}\right| = \begin{cases} \frac{n(n-3)}{2} & \text{for } n \leq d+2\,, \\ nd - \frac{(d+2)(d+1)}{2} & \text{for } n \geq d+2\,. \end{cases}$$

**Proof.** For large enough set of points, the whole conformal group $\mathsf{Conf}(\mathbb{R}^d)$ acts effectively on the configuration space $C_n(\overline{\mathbb{R}^d})$ of $n$ distinct points in $\overline{\mathbb{R}^d}$. The number of independent conformal invariants is then given simply by subtracting the number of symmetries from the dimension of the configuration space, that is

$$\dim C_n(\overline{\mathbb{R}^d}) - \dim \mathsf{Conf}(\mathbb{R}^d) = nd - \frac{(d+2)(d+1)}{2}\,.$$

For small enough number of points, not all conformal transformations nontrivially act on the $n$-tuple of points, and thus do not reduce the number of independent conformal invariants.

We can transform any 3 points to the standard configuration $(0, e, \infty)$ using Lemma 3.50→p.68, leaving us with the subgroup $\mathsf{SO}(d-1) \subset \mathsf{Conf}(\mathbb{R}^d)$ acting on the remaining $n-3$ points. We find a (possibly nontrivial) stabilizer

$$\mathsf{Stab}\Big[\mathsf{SO}(d-1) \subsetneq C_{n-3}\Big(\overline{\mathbb{R}^d}\backslash\{0, e, \infty\}\Big)\Big] \cong \mathsf{SO}(d+2-n) \quad \text{for } n \leq d+2\,,$$

and adding its dimension to the previously oversubtracted expression we obtain

$$nd - \frac{(d+2)(d+1)}{2} + \frac{(d+2-n)(d+1-n)}{2} = \frac{n(n-3)}{2}\,. \qquad \square$$

**Remark 3.54.** Now we see that conformal cross-ratios $u$ and $v$ in Section 3.2→p.67 indeed capture all of the conformally-invariant information about the positions of 4 points in $\mathbb{R}^d$. After performing transformation $(x_1, x_2, x_3, x_4) \mapsto (0, x_2', e, \infty)$, where for simplicity we take $e \equiv (1, 0, \ldots, 0)$, we can utilize the remaining rotations to always bring $x_2'$ to the two-dimensional plane $x_2' \equiv (x, y, 0, \ldots, 0)$, so we are generally left with 2 degrees of freedom — in $d = 1$ it degenerates just to one.

Using such conformal frame, the cross-ratios are given by the squared distances of the transformed point $x_2'$ from $\{0, e\}$, that is (infinities cancel out)

$$u = |x_2'|^2 \equiv z\bar{z}\,, \quad v = |x_2' - e|^2 \equiv (1-z)(1-\bar{z})\,,$$

where $x_2'$ is often parametrized by a complex coordinate $z \equiv x + \mathring{\imath} y \in \mathbb{C}$.

In Euclidean signature the variables $z$ and $\bar{z}$ are complex conjugates, but when we analytically continue back to the Lorentzian signature using $\tau \mapsto \mathring{\imath} t$, they become two independent real variables. ⌟

## 3.3 Primary Operators and Descendants

Now that we understand the conformal structure of the flat space, we can start to study the implications of conformal symmetry on the properties of CFTs. By a CFT we understand either a Lagrangian QFT with traceless energy–momentum tensor, or even a "non-local" QFT with conformal symmetry as described below. In this section we will focus on constraining the form of the correlation functions by utilizing the methods of Section 1.3→p.22.

To each generator of the conformal algebra we can assign a corresponding conserved charge — a topological surface operator representing the symmetry on the quantum level. As usual, to simplify the analysis, we will decompose the set of all local operators into irreducible representations of the conformal group. This will enable us clearly see the consequences for the CFT correlators.

**Remark 3.55.** In the following we will exclusively consider CFTs Wick-rotated to the Euclidean flat space $\mathbb{R}^d$, unless stated otherwise. ⌟

## Quantum Conformal Algebra

When we are in local CFT, we have a local energy–momentum/stress tensor $\boldsymbol{T}^{\bullet}{}_{\bullet}$ which is conserved and traceless. As shown in Section 1.3 →p.22, we can obtain conserved charges corresponding to the conformal algebra by integrating $\boldsymbol{T}^{\bullet}{}_{\bullet}$ against a conformal Killing vector $\boldsymbol{\xi}$ over a $(d-1)$–dimensional surface $\Sigma$ as

$$Q_{\boldsymbol{\xi}} \equiv -\int_\Sigma \boldsymbol{dS}_a \, \boldsymbol{T}^a{}_b \boldsymbol{\xi}^b \,.$$

While we do not need to commit to any particular operator formalism yet, which will be the focus of the next Section 3.4 →p.78, it is natural to consider surfaces surrounding the origin.

Even if we do not have a local Lagrangian description in our CFT, and thus no energy–momentum tensor to explicitly write out the charges as surface integrals of local operators, the conformal symmetry — and in particular the conformal algebra — is still represented on the quantum level.

**Definition 3.56** (Quantum Representation of Conformal Algebra)**.** We denote the conserved charges corresponding to CKV $\boldsymbol{\xi}$ by $Q_{\boldsymbol{\xi}}$. Moreover, we require that they form a representation of the conformal algebra $\mathfrak{conf}(\mathbb{R}^d)$, that is

$$[Q_{\boldsymbol{\xi}}, Q_{\boldsymbol{\zeta}}] = Q_{-[\![\boldsymbol{\xi},\boldsymbol{\zeta}]\!]} \,.$$

For convenience we introduce the notation

$$\boldsymbol{L}_{AB} \equiv Q_{\boldsymbol{l}_{AB}} \,, \quad \boldsymbol{D} \equiv Q_{\boldsymbol{d}} \,, \quad \boldsymbol{P}_a \equiv Q_{\boldsymbol{p}_a} \,, \quad \boldsymbol{K}_a \equiv Q_{\boldsymbol{k}_a} \,, \quad \boldsymbol{M}_{ab} \equiv Q_{\boldsymbol{m}_{ab}} \,.$$

**Remark 3.57.** We included the minus sign such that it is a proper representation of $\mathsf{Lie}(\mathsf{Conf}(\mathbb{R}^d))$, since using Remark 3.22 →p.58 we have $[Q_{\boldsymbol{\xi}_X}, Q_{\boldsymbol{\xi}_Y}] = Q_{-[\![\boldsymbol{\xi}_X,\boldsymbol{\xi}_Y]\!]} = Q_{\boldsymbol{\xi}_{[X,Y]}}$ for any $X, Y \in \mathsf{Lie}(\mathsf{Conf}(\mathbb{R}^d))$. Alternatively, under the correlator we have

$$\Big[Q_1, [Q_2, \mathcal{O}]\Big] = \mathcal{D}_2[Q_1, \mathcal{O}] = \mathcal{D}_2\mathcal{D}_1\mathcal{O} \,,$$

where we used that commutator of a charge $Q_i$ with the operator $\mathcal{O}$ can be deformed to topological operator on a small sphere around the operator $\mathcal{O}$, which due to Ward–Takahashi identities Calculation 1.81 →p.27 acts as some differential operator on the correlator. We first need to shrink the inner commutator, and only then the outer, so the differential operators act in the reversed order. ⌟

**Remark 3.58** ($\boldsymbol{P}/\boldsymbol{K}$ as Raising/Lowering Operators)**.** From Lemma 3.41 →p.65 we immediately get commutation relations of these charges. For example, we see that $\boldsymbol{M}_{\bullet\bullet}$ generate the rotations $\mathsf{SO}(d)$, under which $\boldsymbol{P}_{\bullet}$ and $\boldsymbol{K}_{\bullet}$ transform as vectors,

↳ while $\boldsymbol{D}$ is a scalar/invariant. Of special interest are

$$[\boldsymbol{D}, \boldsymbol{P}_a] = \boldsymbol{P}_a\,, \quad [\boldsymbol{D}, \boldsymbol{K}_a] = -\boldsymbol{K}_a\,,$$

indicating that $\boldsymbol{P}_\bullet / \boldsymbol{K}_\bullet$ can be understood as raising/lowering operators for $\boldsymbol{D}$. ⌟

**Remark 3.59.** In $d = 2$ often all local CKVs are included — this in particular includes CKVs not associated with global conformal transformations, and even those with singularity at the origin. Corresponding conformal algebra on the quantum level is generally a *central extension* of such CKV algebra. ⌟

**Definition 3.60** (Quadratic Conformal Casimir)**.** The quadratic *Casimir* of the conformal algebra $\mathfrak{conf}(\mathbb{R}^d) \cong \mathfrak{so}(1, d+1)$ is given by

$$\mathfrak{Cas}_2 \equiv -\frac{1}{2}\boldsymbol{L}^{AB}\boldsymbol{L}_{AB} = -\frac{1}{2}\boldsymbol{M}^{ab}\boldsymbol{M}_{ab} - \frac{1}{2}\boldsymbol{P}\cdot\boldsymbol{K} - \frac{1}{2}\boldsymbol{K}\cdot\boldsymbol{P} + \boldsymbol{D}^2\,,$$

where we write it already represented by quantum charges.

**Remark 3.61.** Conformal Casimir is a distinguished element of the center of the universal enveloping algebra of $\mathfrak{conf}(\mathbb{R}^d)$, that is it commutes with all generators of the conformal algebra. Thus, by Schur's lemma, in any irreducible representation of $\mathfrak{conf}(\mathbb{R}^d)$ it is represented as a multiple of the identity operator, and the eigenvalues can be used to classify the representations. ⌟

## Conformal Representations of Local Operators

The classification of conformal representations in a certain sense parallels the classification of one-particle states in Poincaré-invariant QFTs — namely both utilize the notion of *induced representations* — so let us first briefly recall the familiar story from standard QFTs in $\mathsf{Mink}_\bullet$.

**Remark 3.62** (Classification of One-Particle States in Poincaré Invariant QFTs)**.** Since the translation subgroup is abelian, we can diagonalize it first, obtaining the momentum eigenstates. This is naturally connected with the choice of inertial foliation discussed in Example 1.88→p.31. The only thing left it to obtain the transformation properties under the action of the homogeneous Lorentz subgroup.

After restricting our attention to a fixed value of the translation Casimir $\boldsymbol{P}^2 = M^2$ — which partly classifies different irreps — we can always use boosts to transform the state to have some chosen *canonical* or *standard momentum*. It is therefore enough to understand the action of rotations for states with the given canonical momentum, that is specify a representation of the *little group* (stabilizer of the canonical momentum).

Transformation of a general state is then uniquely *induced* from this little group action, essentially just by using the group algebra. For more details see [25]. ⌟

**Derivation 3.63** (Conformal Algebra Representations of Local Operators)**.** We want to perform a similar analysis in the context of local operators in CFTs. Here it is natural to proceed directly in the position-space [43]:

**(1)** Since action of translations is transitive, we can fix any point to be the origin. Then we would like to specify the representation of its stabilizer, which is generated by $\{\boldsymbol{D}, \boldsymbol{M}_{\bullet\bullet}, \boldsymbol{K}_{\bullet}\}$.

**(2)** We choose to diagonalize the dilatations generated by $\boldsymbol{D}$, and since the rotations generated by $\boldsymbol{M}_{\bullet\bullet}$ commute with $\boldsymbol{D}$, we can specify a representation of the rotation subgroup $\mathsf{SO}(d)$ as well — or its double cover $\mathsf{Spin}(d)$ for spinor representations. So we have

$$[\boldsymbol{D}, \mathcal{O}(0)] = \Delta\, \mathcal{O}(0)\,,$$
$$[\boldsymbol{M}_{ab}, \mathcal{O}(0)] = -\mathcal{S}_{ab}\mathcal{O}(0)\,,$$

where $\Delta$ is the *scaling dimension* of the operator, $\mathcal{S}_{\bullet\bullet}$ are generators of the $\mathsf{SO}(d)$ representation, and we suppress the corresponding indices carried both by $\mathcal{S}$ and $\mathcal{O}$ (and their appropriate contraction). Since we chose $\mathcal{S}_{\bullet\bullet}$ to act from the left, and we want them to have the same commutation relations as $\boldsymbol{M}_{\bullet\bullet}$, we have to include a minus sign.

**(3)** We already know from Remark 3.58 →p.70 that all of $\boldsymbol{K}_{\bullet}$ are lowering operators for $\boldsymbol{D}$, which in this case means (using the Jacobi identity)

$$\Big[\boldsymbol{D}, [\boldsymbol{K}_{\bullet}, \mathcal{O}(0)]\Big] = \Big[[\boldsymbol{D}, \boldsymbol{K}_{\bullet}], \mathcal{O}(0)\Big] + \Big[\boldsymbol{K}_{\bullet}, [\boldsymbol{D}, \mathcal{O}(0)]\Big] = (\Delta - 1)[\boldsymbol{K}_{\bullet}, \mathcal{O}(0)]\,.$$

As we will see in Remark 3.82 →p.78, physically-admissible operators must have $\Delta \geq 0$, so there must be a lowest-weight state where this process eventually terminates with

$$[\boldsymbol{K}_a, \mathcal{O}(0)] = 0 \quad \forall a\,.$$

Such operators which are annihilated by all $\boldsymbol{K}_{\bullet}$ at the origin are called *primary operators*. The rest of operators in a given conformal representation — also called *conformal multiplet* — are obtained by acting with momentum generators $\boldsymbol{P}_a$ (raising operators of $\boldsymbol{D}$) as

$$\mathcal{O}(0) \longmapsto \Big[\boldsymbol{P}_{a_N}, \Big[\cdots, [\boldsymbol{P}_{a_1}, \mathcal{O}(0)]\cdots\Big]\Big] \simeq \boldsymbol{\partial}_{a_1}\cdots\boldsymbol{\partial}_{a_N}\mathcal{O}(0)\,,$$
$$\Delta \longmapsto \Delta + N\,,$$

which are called *descendant operators* of level $N \geq 1$.

**(4)** Action at a general point is then obtained via

$$\mathcal{O}(x) \equiv U_x \mathcal{O}(0) U_x^{-1} \equiv \mathrm{e}^{x\cdot\boldsymbol{P}}\, \mathcal{O}(0)\, \mathrm{e}^{-x\cdot\boldsymbol{P}} = \mathrm{e}^{[x\cdot\boldsymbol{P}, \Box]}\, \mathcal{O}(0)\,,$$

and the conformal/operator algebra. In particular, we see that a shifted primary operator is an infinite linear combination of the primary and its descendants at the origin.

⌟

**Proposition 3.64** (Action of Conformal Algebra on Primary Operators)**.** Let $\mathcal{O}$ be a primary operator with scaling dimension $\Delta$ and rotation generators $\mathcal{S}_{\bullet\bullet}$. Then the action of the conformal algebra is then given by

$$\begin{aligned}[\boldsymbol{P}_a, \mathcal{O}(x)] &= \boldsymbol{p}_a \mathcal{O}(x) \equiv \partial_a \mathcal{O}(x)\,, \\ [\boldsymbol{D}, \mathcal{O}(x)] &= (\boldsymbol{d} + \Delta)\mathcal{O}(x)\,, \\ [\boldsymbol{M}_{ab}, \mathcal{O}(x)] &= (\boldsymbol{m}_{ab} - \mathcal{S}_{ab})\mathcal{O}(x)\,, \\ [\boldsymbol{K}_a, \mathcal{O}(x)] &= (\boldsymbol{k}_a + 2\Delta x_a + 2x^b \mathcal{S}_{ab})\mathcal{O}(x)\,,\end{aligned}$$

which can be summarized for any CKV $\boldsymbol{\xi}$ as

$$[Q_{\boldsymbol{\xi}}, \mathcal{O}(x)] = \Big(\boldsymbol{\xi} + \Delta\,\tfrac{1}{d}(\partial\cdot\xi) + \tfrac{1}{2}(\partial^a\xi^b)\mathcal{S}_{ab}\Big)\mathcal{O}(x)\,.$$

**Proof.** Using a shorthand notation $[Q, \mathcal{O}] \to Q\mathcal{O}$, we have

$$\begin{aligned}\boldsymbol{D}\mathcal{O}(x) &= \boldsymbol{D}\,\mathrm{e}^{x\cdot\boldsymbol{P}}\mathcal{O}(0) = \mathrm{e}^{x\cdot\boldsymbol{P}}\Big(\mathrm{e}^{-x\cdot\boldsymbol{P}}\,\boldsymbol{D}\,\mathrm{e}^{x\cdot\boldsymbol{P}}\Big)\mathcal{O}(0) = \mathrm{e}^{x\cdot\boldsymbol{P}}\Big(\mathrm{e}^{[-x\cdot\boldsymbol{P},\,\square]}\,\boldsymbol{D}\Big)\mathcal{O}(0) \\ &= \mathrm{e}^{x\cdot\boldsymbol{P}}\Big(\boldsymbol{D} + \underbrace{[-x\cdot\boldsymbol{P}, \boldsymbol{D}]}_{x\cdot\boldsymbol{P}} + \underbrace{\cdots}_{\sim[\boldsymbol{P},\boldsymbol{P}]=0}\Big)\mathcal{O}(0) = (x\cdot\boldsymbol{\partial} + \Delta)\,\mathrm{e}^{x\cdot\boldsymbol{P}}\mathcal{O}(0) \\ &\equiv (\boldsymbol{d} + \Delta)\mathcal{O}(x)\,.\end{aligned}$$

We proceed similarly also for the other generators. The final formula just needs to be checked individually for each CKV defined in Proposition 3.37 →p.62. □

Alternatively, we could define primary/descendant operators by starting with the transformation properties under finite conformal maps.

**Definition 3.65** (Primary Operator and its Finite Transformation, Descendants)**.** An operator $\mathcal{O}$ is called a *primary operator* if its transformation under any conformal map $g \in \mathsf{Conf}(\mathbb{R}^d)$, $x \mapsto x' \equiv g(x)$, is given by its *scaling dimension* $\Delta$ and its representation $\rho$ under rotations $\mathsf{SO}(d)$ as

$$U_g\,\mathcal{O}(x)\,U_g^{-1} = \Omega_g(x)^{\Delta}\rho\Big(R_g(x)^{-1}\Big)\cdot\mathcal{O}(x')\,,$$

where we performed the decomposition Remark 3.36 →p.61 into local scaling and rotation. Often we will keep the $g$ subscript for $\Omega$ and $R$ implicit.

The corresponding *descendants* are operators of the form $\boldsymbol{P}_\bullet\cdots\boldsymbol{P}_\bullet\mathcal{O}$, whose transformation can be obtained from the transformation of the primary $\mathcal{O}$.

**Remark 3.66.** It is easy to confirm that transformations of the primary operators indeed form a representation, that is $U_g U_h = U_{gh}$ for each $g, h \in \mathsf{Conf}(\mathbb{R}^d)$.

We can think of the conformal representation corresponding to the primary $\mathcal{O}$ either as $\{\mathbb{C}\mathcal{O}(x)\,|\,x\in\mathbb{R}^d\}$, or alternatively as $\mathrm{Span}_{\mathbb{C}}\Big(\{\boldsymbol{P}^n_\bullet\mathcal{O}(0)\,|\,n\in\mathbb{N}_0\}\Big)$. ⌟

**Remark 3.67.** The finite transformation Definition 3.65 →p.73 and the infinitesimal action Proposition 3.64 →p.73 are related by $U_{g_t} \equiv \mathrm{e}^{tQ_{\boldsymbol{\xi}}}$, where the finite conformal transformation $g_t \equiv e^{t\xi} \in \mathsf{Conf}(\mathbb{R}^d)$ is obtained from the flow $\phi^{\xi}_{\circ}$ along CKV $\boldsymbol{\xi}$, that is $x'_t \equiv g_t(x) = e^{t\xi}(x) \equiv \phi^{\xi}_t(x)$.

To give an example, for translations, dilatations, and rotations we have

$$\begin{aligned} T_x &: \circ \mapsto e^{x\cdot \boldsymbol{p}}(\circ) = \circ + x \\ D_\lambda &: \circ \mapsto e^{\sigma \boldsymbol{d}}(\circ) = e^{\sigma}\circ \equiv \lambda \circ \\ R_\theta &: \circ \mapsto e^{\frac{1}{2}\theta^{ab}\boldsymbol{m}_{ab}}(\circ) = R_\theta(\circ) \end{aligned} \quad \rightsquigarrow \quad \begin{aligned} U_x &\equiv U_{T_x} \equiv e^{x\cdot \boldsymbol{P}}\,, \\ U_\lambda &\equiv U_{D_\lambda} \equiv e^{\sigma \boldsymbol{D}} \equiv \lambda^{\boldsymbol{D}}\,, \\ U_\theta &\equiv U_{R_\theta} \equiv e^{\frac{1}{2}\theta^{ab}\boldsymbol{M}_{ab}}\,. \end{aligned}$$

**Remark 3.68** (Scalars, Symmetric Traceless Tensors, Spin)**.** We will primarily consider scalar operators, or more generally the *symmetric traceless tensor* irreducible representations of $\mathsf{SO}(d)$. A primary operator $\mathcal{O}_{\Delta,J}$ with scaling dimension $\Delta$ and *spin* $J$ is understood to be a traceless symmetric tensor of rank $J$ (with $J$ indices). Scalars have $J = 0$, vectors have $J = 1$, and so on. We refer to operators with nonzero spin $J > 0$ as "spinning" operators.

**Example 3.69** (Identity Operator)**.** In every CFT there is a primary operator with $\Delta = J = 0$, so it is invariant under all conformal transformations. This is simply the identity operator $\mathbf{1}$, which we already discussed in Example 1.64→p.20. As we will see in Section 3.4→p.78, its uniqueness corresponds to the uniqueness of the vacuum state.

**Example 3.70** (Fundamental Fields)**.** The fundamental fields $\boldsymbol{\phi}$ over which we integrate in the path integral are always primary operators, and together with their descendants/derivatives they form the corresponding conformal multiplet.

**Example 3.71** (Double-Twist Operators)**.** Quite generally, but in particular for MFTs (see Example 2.26→p.48), given two primaries $\mathcal{O}_1$ and $\mathcal{O}_2$ — for example fundamental fields $\phi$ — there are primaries which are composites of $\mathcal{O}_1, \mathcal{O}_2$ in the schematic form

$$[\mathcal{O}_1\mathcal{O}_2]_{n,J} \simeq \mathcal{O}_1 \Box^n \partial^J \mathcal{O}_2 - (\text{traces})\,,$$

where the derivatives act appropriately on both operators to ensure $[\mathcal{O}_1\mathcal{O}_2]_{n,J}$ is a primary, and the traces are subtracted such that the final operator transforms in a spin $J$ symmetric traceless representation.

Their appearance from the thermal partition function point of view can be seen for example in [44]. See Appendix F [45] and Appendix C [46] for derivations of explicit formulas.

## Anomalous Dimensions

Scaling dimensions of primary operators are fundamental data of the theory which figure in basically all CFT calculations, in particular the correlation functions. Also other important quantities such as critical exponents are derived from them. Therefore, finding their values is one of the major tasks in the study of CFTs.

This is a highly nontrivial task for interacting CFTs. As we already discussed in Section 1.1→p.10, we can utilize perturbation theory around a certain "free" theory, where we have a complete knowledge of the scaling dimensions. In such approaches we calculate quantum corrections due to interactions as a series expansion in some appropriate parameter.

**Definition 3.72** (Anomalous Dimensions)**.** Suppose we are solving the theory in some kind of perturbative expansion with parameter $g$. The expansion of the scaling dimension of some primary operator has the form

$$\Delta = \Delta^{(0)} + \sum_{k=0}^{\infty} g^k \gamma^{(k)} \,,$$

where $\Delta^{(0)}$ is a known scaling dimension not containing quantum corrections, and the deviations $\gamma^{(k)}$ (of $k$-th order) are the so-called *anomalous dimensions*.

**Remark 3.73.** This is a slight generalization of the anomalous dimensions discussed in Remark 2.22→p.46. There we compared $\Delta$ to the classical mass dimensions corresponding to the free Gaussian fixed point. In this case, we consider $\Delta^{(0)}$ to be the scaling dimension in *some* CFT from which we are perturbing away. ⌟

**Remark 3.74** (Energy–Dimension Correspondence in AdS/CFT)**.** In the context of QFTs in AdS to be discussed in Chapter 5→p.100, the scaling dimensions in the CFT on the boundary correspond to energies in the bulk AdS.

In particular, the scaling dimensions of boundary operators dual to fundamental fields in AdS correspond to their bulk masses. In our treatment these will not obtain any anomalous dimensions, as our choice of renormalization scheme will fix them to be the same as in the free theory.

However, the double-twist operators discussed in Example 3.71→p.74 obtain non-trivial anomalous dimensions due to interactions, which can be interpreted as the binding energies of the corresponding two-particle states in the bulk. Calculation of these anomalous dimensions will be the main focus of the final Chapter 5→p.100. ⌟

## Correlation Functions of Primary Operators

Now that we know transformation properties of local operators in CFTs — in particular those of the primary operators — we can use them to constrain the form of their correlation functions. By taking appropriate derivatives we can obtain also correlators of their descendants.

For simplicity, we will mainly consider only scalar operators.

**Lemma 3.75** (Conformal Symmetry of Primary Scalar Correlators)**.** Consider a set $\{\mathcal{O}_1, \ldots, \mathcal{O}_n\}$ of primary scalar operators with scaling dimensions

$\{\Delta_1, \ldots, \Delta_n\}$. The conformal symmetry of the correlator reads

$$\begin{aligned}\Big\langle \mathcal{O}_1(x_1)\cdots\mathcal{O}_n(x_n)\Big\rangle &= \Big\langle (U_g\mathcal{O}_1(x_1)U_g^{-1})\cdots(U_g\mathcal{O}_n(x_n)U_g^{-1})\Big\rangle \\ &= \Omega(x_1)^{\Delta_1}\cdots\Omega(x_n)^{\Delta_n}\Big\langle \mathcal{O}_1(x'_1)\cdots\mathcal{O}_n(x'_n)\Big\rangle\end{aligned}$$

for any $g \in \mathsf{Conf}(\mathbb{R}^d)$, $x \mapsto x' \equiv g(x)$.

**Proof.** Follows directly from Definition 3.65 → p.73 and $\rho \equiv \mathbf{1}$ for scalars. □

**Proposition 3.76** (Constraints on Correlators of Primary Scalar Operators)**.** Correlators of scalar primary operators are functions only of the distances $|x_{ij}| \equiv |x_i - x_j|$ between the insertion points in the general form

$$\Big\langle \mathcal{O}_1(x_1)\cdots\mathcal{O}_n(x_n)\Big\rangle = \left[\prod_{i<j}\frac{1}{|x_{ij}|^{c_{ij}}}\right] f\Big(\{u, v, \ldots\}\Big)\,,$$

where $f(\cdot)$ is a function of the conformal cross-ratios $\{u, v, \ldots\}$ build from $\{x_1, \ldots, x_n\}$, and the powers $c_{ij} \equiv c_{(ij)}$ must satisfy

$$\forall i \in \{1, \ldots, n\} \qquad \Delta_i = \frac{1}{2}\sum_{j\neq i}^{n} c_{ij}\,.$$

**Proof.** Translation and rotation invariance implies that scalar correlators can depend only on the distances between the points — there is no preferred position or direction. Therefore, they assume a from as stated in the Proposition, or possibly a sum over such terms with different $c_{ij}$ and $f$.

Since the cross-ratios are conformal invariants, the conformal symmetry of the correlator Lemma 3.75 → p.75 requires (terms with different $c_{ij}$ can not cancel out)

$$\begin{aligned}\prod_{i<j}\frac{1}{|x_{ij}|^{c_{ij}}} &\overset{!}{=} \Omega(x_1)^{\Delta_1}\cdots\Omega(x_n)^{\Delta_n}\prod_{i<j}\frac{1}{|x'_{ij}|^{c_{ij}}} \\ &= \Omega(x_1)^{\Delta_1}\cdots\Omega(x_n)^{\Delta_n}\prod_{i<j}\frac{1}{\Omega^{c_{ij}/2}(x_i)\Omega^{c_{ij}/2}(x_j)}\,\frac{1}{|x_{ij}|^{c_{ij}}}\,,\end{aligned}$$

where we utilized Lemma 3.46 → p.67. Since $\Omega$ for SCTs depends nontrivially on the position — see Remark 3.40 → p.64 — the only way to satisfy the conformal symmetry condition is to have the $\Omega$-factors cancel out for each $x_i$ individually. This is indeed the stated condition for the powers $c_{ij}$.

Picking any allowed $\{c_{ij}\}$ and factoring $\prod |x_{ij}|^{-c_{ij}}$ out, the rest of the correlator must be a conformal invariant, and thus just a function of the cross-ratios. □

**Remark 3.77.** Here we ignore possible contact terms in the correlators, which can modify the correlator at coinciding points. ⌟

Conformal symmetry completely fixes (up to overall normalization) the form of scalar correlators with $n \leq 3$, since there are no conformal cross-ratios build from 3 or fewer points as shown in Proposition 3.53 →p.68 and before. Let us look at them in more detail.

**Corollary 3.78** (One-Point Functions). Let $\mathcal{O}$ be any primary operator not proportional to the identity operator $\mathbf{1}$. Then its one-point function vanishes

$$\left\langle \mathcal{O}(x) \right\rangle = 0\,,$$

as do one-point functions of all its descendants.

Only operators proportional to the identity operator have a nontrivial one-point function. Due to the normalization of the correlator we have

$$\left\langle \mathbf{1}(x) \right\rangle \equiv 1\,.$$

**Proof.** With one point insertion there are no distances between the points, and therefore it is generally not possible to satisfy condition in Proposition 3.76 →p.76 unless the correlator is trivially zero, or $\Delta = 0$.

The proof is concluded by noticing that the only operator with $\Delta = 0$ is the identity $\mathbf{1}$, and that correlators of descendants are given by derivatives of the primary one-point functions, which in all cases turn out to be zero. □

**Remark 3.79** (One-Point Functions in Different Backgrounds). Remember that we consider CFT in flat space $\mathbb{R}^d$. Things change for example in BCFTs (CFTs with a boundary) or when we consider the theory on space with compact dimensions (like torus). Then also other operators can have nontrivial one-point functions, depending on the distance from the boundary or the sizes of compact dimensions. ⌟

**Corollary 3.80** (Two-Point Functions of Scalar Primaries). The two-point function of primary scalar operators is fixed to be of the form

$$\left\langle \mathcal{O}_1(x)\mathcal{O}_2(y) \right\rangle = \frac{C_{\mathcal{O}_1\mathcal{O}_2}\,\delta_{\Delta_1,\Delta_2}}{|x-y|^{2\Delta_1}}\,,$$

where $C_{\mathcal{O}_1\mathcal{O}_2}$ is a constant. In particular, the correlator can be nonzero only if the scaling dimensions of the two operators are equal, that is $\Delta_1 = \Delta_2$.

**Proof.** The condition in Proposition 3.76 →p.76 gives us $\Delta_1 = \frac{1}{2}c_{12} = \Delta_2$. □

**Remark 3.81** (Normalization of Operators). Considering a basis of all (real) scalar primaries $\{\mathcal{O}_i\}$, the constants $\{C_{\mathcal{O}_i\mathcal{O}_j}\,\delta_{\Delta_i,\Delta_j}\}$ form a symmetric matrix. Since it can be diagonalized, we will usually consider a basis with $C_{\mathcal{O}_i\mathcal{O}_j}\,\delta_{\Delta_i,\Delta_j} = \delta_{ij}$, where we appropriately normalized the operators. From now on, any scalar primary operator $\mathcal{O}_\Delta$ with scaling dimension $\Delta$ has the two-point function

$$\left\langle \mathcal{O}_\Delta(x)\mathcal{O}_\Delta(y)\right\rangle = \frac{1}{|x-y|^{2\Delta}} \,.$$

**Remark 3.82.** Ignoring the identity operator, we must have $\Delta > 0$, otherwise the correlator would grow with increasing distance, which would violate the cluster decomposition principle. In the following Section 3.4→p.78 we will derive even stronger bounds on $\Delta$ in unitary theories.

**Corollary 3.83** (Three-Point Functions of Scalar Primaries)**.** The three-point function of primary scalar operators is fixed to be of the form

$$\left\langle \mathcal{O}_1(x_1)\mathcal{O}_2(x_2)\mathcal{O}_3(x_3)\right\rangle = \frac{f_{123}}{x_{12}^{\Delta_1+\Delta_2-\Delta_3}x_{13}^{\Delta_1+\Delta_3-\Delta_2}x_{23}^{\Delta_2+\Delta_3-\Delta_1}} \,.$$

where $f_{123}$ is a constant depending on the operators $\mathcal{O}_1, \mathcal{O}_2, \mathcal{O}_3$.

**Proof.** The condition in Proposition 3.76→p.76 gives us 3 equations for 3 unknowns $\{c_{12}, c_{13}, c_{23}\}$, which has a unique solution

$$\begin{aligned} \Delta_1 &= \tfrac{1}{2}(c_{12}+c_{13}) \\ \Delta_2 &= \tfrac{1}{2}(c_{12}+c_{23}) \\ \Delta_3 &= \tfrac{1}{2}(c_{13}+c_{23}) \end{aligned} \quad\Longrightarrow\quad \begin{aligned} c_{12} &= \Delta_1+\Delta_2-\Delta_3 \\ c_{13} &= \Delta_1+\Delta_3-\Delta_2 \\ c_{23} &= \Delta_2+\Delta_3-\Delta_1 \,. \end{aligned}$$

□

**Remark 3.84.** Since we already fixed the normalization of the operators by requiring the specific form of the two-point function in Remark 3.81→p.77, the coefficients $f_{ijk}$ are important physical data of the theory. We will discuss them in more detail in Section 3.5→p.86.

**Remark 3.85.** The proofs in this section utilized the finite form of conformal symmetry for the correlators, but we could equivalently use Ward–Takahashi identities for conformal transformations to obtain the same results.

The higher-point ($n \geq 4$) correlators in general have nontrivial dependence on cross-ratios, and their functional form is therefore not fixed by conformal symmetry. Still, as we will see in the following sections, they can be expressed in terms of data present in the two-point and three-point functions.

## 3.4 Radial Quantization

Up to now we studied the conformal symmetry/algebra and the correlation functions of local operators in CFTs on general grounds without choosing any particular operator formalism in the sense of Section 1.4→p.29. But without choosing some foliation of the spacetime manifold, and considering the corresponding Hilbert spaces, it is not clear how to investigate important questions pertaining to the unitarity/reflection positivity of the theory.

## Radial Foliation and Cylinder Picture

As already alluded to in Section 3.3→p.69, in CFTs it is natural to foliate the spacetime by concentric spheres around a chosen origin. Such *radial foliation* is adapted to the action of dilatations and rotations generated by $\boldsymbol{D}$ and $\boldsymbol{M}_{\bullet\bullet}$, which nicely matches with the way we classified the local operator conformal representations.

As the Hilbert spaces $\mathcal{H}_\Sigma$ on all different spheres $\Sigma \simeq \mathbb{S}^{d-1}$ around the origin are actually isomorphic — they are related by the scaling symmetry of the theory generated by the "radial Hamiltonian" $\boldsymbol{D}$ — we are able to formulate the theory on a single Hilbert space $\mathcal{H}$. Recall Remark 1.67→p.21 and Section 1.4→p.29 for the general story.

But how do we obtain the underlying Hilbert space, together with the inner product and corresponding conjugation? The idea is to start with the cylinder.

**Idea 3.86** (Cylinder Picture and Radial Quantization)**.** The *radial quantization* comes from canonically quantizing the theory on a cylinder $\mathbb{R} \times \mathbb{S}^{d-1}$, and then mapping the cylinder to the flat space $\mathbb{R}^d$ via a Weyl/conformal transformation.

In more detail, by expressing $x \in \mathbb{R}^d \cong \mathbb{R}_{\geq} \times \mathbb{S}^{d-1}$ as $x \equiv |x|\hat{x} \equiv r\mathbf{n}$, and then rescaling the radial coordinate using $\phi\colon \mathbb{R} \to \mathbb{R}_{>}, \tau \mapsto r = \phi(\tau) \equiv \mathrm{e}^\tau$, we obtain

$$\boldsymbol{g}_{\mathbb{R}^d} \equiv \mathbf{d}r^2 + r^2 \boldsymbol{g}_{\mathbb{S}^{d-1}} = r^2\left(\frac{\mathbf{d}r^2}{r^2} + \boldsymbol{g}_{\mathbb{S}^{d-1}}\right) \overset{\phi}{\simeq} \mathrm{e}^{2\tau}\left(\mathbf{d}\tau^2 + \boldsymbol{g}_{\mathbb{S}^{d-1}}\right) \equiv \mathrm{e}^{2\tau}\, \boldsymbol{g}_{\mathbb{R}\times\mathbb{S}^{d-1}}\,,$$

so after a Weyl transformation we obtain the flat metric on the cylinder $\mathbb{R} \times \mathbb{S}^{d-1}$. Note that only for CFTs we are able to perform such Weyl rescaling to relate the theory on flat space to the theory on the cylinder.

In the Figure 3.87→p.80 we clearly see that spheres of the same radius $r \equiv |x|$ become constant $\tau$ slices, and dilatations $x \mapsto \lambda x$ become shifts of cylinder time $\tau \mapsto \tau + \ln\lambda$. Therefore, $\boldsymbol{D}$ now generates isometries via $\mathrm{e}^{-\tau\boldsymbol{D}}$, and can indeed be interpreted as the Hamiltonian of the theory on the cylinder. Another point worth stressing is that operator insertions at the origin $x = 0$ correspond to the preparation of state at $\tau = -\infty$. ⌟

Using the Weyl transformation of a correlator Example 1.73→p.25 with $\Omega(x) = \mathrm{e}^\tau$, we obtain transformation of correlator between cylinder and flat space in the form (here $\{y_i\}$ are points on the cylinder $\mathbb{R} \times \mathbb{S}^{d-1}$)

$$\left\langle \mathcal{O}_1(y_1)\cdots\mathcal{O}_n(y_n)\right\rangle_{\boldsymbol{g}_{\mathbb{R}\times\mathbb{S}^{d-1}}} = \left(\prod_{i=1}^n \mathrm{e}^{\tau_i\Delta_i}\right)\left\langle \mathcal{O}_1(y_1)\cdots\mathcal{O}_n(y_n)\right\rangle_{\mathrm{e}^{2\tau}\boldsymbol{g}_{\mathbb{R}\times\mathbb{S}^{d-1}}}$$

where the right-side (after change of coordinates with $\phi$) is the correlator in flat space $\mathbb{R}^d$.

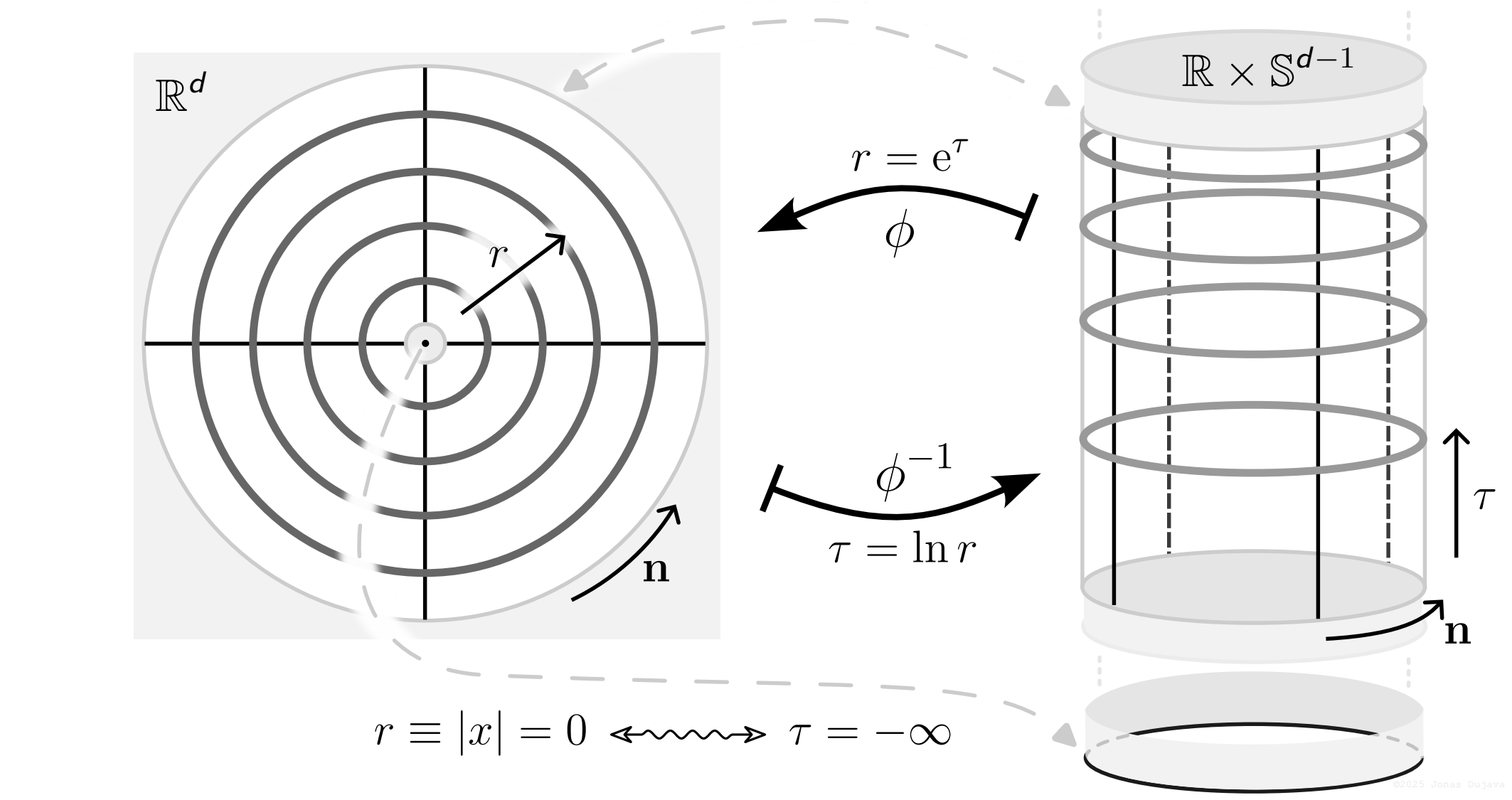

**Figure 3.87 /** Conformal map between (flat space $\mathbb{R}^d\backslash\{0\} \leftrightsquigarrow \mathbb{R}\times\mathbb{S}^{d-1}$ cylinder). By adding two additional points at the infinities of the cylinder we actually obtain a conformal compactification of the flat space $\overline{\mathbb{R}^d} \simeq (\mathbb{R}\times\mathbb{S}^{d-1}) \cup \{\pm\infty\}$.

**Definition 3.88** (Cylinder Operator, Flat Operator). The *cylinder* operator $\mathcal{O}_{\text{cyl.}}$ inserted at $(\tau,\mathbf{n}) \in \mathbb{R}\times\mathbb{S}^{d-1}$ corresponds to the *flat* operator $\mathcal{O}_{\text{flat}}$ inserted at $x \equiv \phi(\tau,\mathbf{n}) \equiv e^\tau\mathbf{n} \in \mathbb{R}^d$ by

$$\mathcal{O}_{\text{cyl.}}(\tau,\mathbf{n}) \leftrightsquigarrow e^{\tau\Delta}\,\mathcal{O}_{\text{flat}}(x \equiv e^\tau\mathbf{n})\,,$$

meaning that correlators of $\mathcal{O}_{\text{cyl.}}$ on the cylinder $\mathbb{R}\times\mathbb{S}^{d-1}$ can be calculated in the flat space $\mathbb{R}^d$ by considering their flat-space equivalents $e^{\tau\Delta}\,\mathcal{O}_{\text{flat}}$.

**Remark 3.89.** We often omit subscripts "cyl."/"flat" and rely on the coordinates to distinguish which operator or space we have in mind. ⌟

The canonical quantization on the (Lorentzian) cylinder $\mathbb{R}\times\mathbb{S}^{d-1}$ gives us a unitary CFT. In particular, the complex vector space of states $\mathcal{H}$ contains a distinguished vector $|\text{vac}\rangle \in \mathcal{H}$ invariant under all conformal transformations, and $\mathcal{H}$ carries a positive-definite inner product $\langle\cdot,\cdot\rangle$ (as usual $\langle\text{vac}| \equiv \langle|\text{vac}\rangle,\cdot\rangle \in \mathcal{H}^*$).

After performing the Wick rotation to the Euclidean cylinder we obtain a reflection-positive Euclidean CFT, where (recall Section 1.4→p.32) the conjugation acts as

$$\mathcal{O}^{a_1\ldots a_l}_{\text{cyl.}}(\tau,\mathbf{n})^\dagger = \Theta^{a_1}_{b_1}\cdots\Theta^{a_l}_{b_l}\,\mathcal{O}^{b_1\ldots b_l*}_{\text{cyl.}}(-\tau,\mathbf{n}) \equiv \Theta\Big[\mathcal{O}^{a_1\ldots a_l*}_{\text{cyl.}}(\tau,\mathbf{n})\Big]\,.$$

The radial quantization of CFT in $\mathbb{R}^d$ borrows all of this structure and thus becomes a reflection-positive CFT as well (with an appropriate conjugation). Using Definition 3.88→p.80, we see that the cylinder time-reflection $(\tau \mapsto -\tau)$ becomes the radial inversion $\mathcal{I}$ defined in Remark 3.39→p.63.

**Definition 3.90** (Conjugation in Radial Quantization)**.** A local operator with spin $l$ behaves under *radial conjugation* as

$$\mathcal{O}_{\text{flat}}^{a_1\ldots a_l}(x)^{\dagger_{\text{rad}}} = |x|^{-2\Delta}\,\mathcal{I}^{a_1}_{b_1}\cdots\mathcal{I}^{a_l}_{b_l}\,\mathcal{O}_{\text{flat}}^{b_1\ldots b_l *}\Big(\tfrac{x}{|x|^2}\Big) \equiv \mathcal{I}\Big[\mathcal{O}_{\text{flat}}^{a_1\ldots a_l *}(x)\Big]\,,$$

where $\mathcal{I}^a_b \equiv \delta^a_b - 2\hat{x}^a\hat{x}_b$, so the *radial inversion* operation $\mathcal{I}$ in addition to inverting the position argument also includes the Weyl factor $\Omega_{\mathcal{I}}(x)^{\Delta} \equiv |x|^{-2\Delta}$ and a reflection of the indices along the radial direction.

**Remark 3.91.** We have written above $\dagger_{\text{rad}}$ to distinguish the Hermitian conjugation in the radial quantization (corresponding to $\boldsymbol{D}$) from the conjugation in the usual inertial quantization (corresponding to $\boldsymbol{P}_0$).

In the following we will often drop the subscript "rad", as it should be clear when we are working in the radial quantization. ⌟

**Corollary 3.92** (Radial Conjugation of Conformal Generators)**.** The radial conjugation of a conserved conformal charge $Q_{\boldsymbol{\xi}}$, $\boldsymbol{\xi} \in \mathfrak{conf}(\mathbb{R}^d)$, is given by

$$Q_{\boldsymbol{\xi}}^{\dagger} = Q_{-\mathrm{Ad}_{\mathcal{I}}\boldsymbol{\xi}} \equiv Q_{-\mathcal{I}\boldsymbol{\xi}\mathcal{I}}\,,$$

so in particular

$$\boldsymbol{D}^{\dagger} = \boldsymbol{D}\,, \quad \boldsymbol{M}^{\dagger}_{\bullet\bullet} = -\boldsymbol{M}_{\bullet\bullet}\,, \quad \boldsymbol{P}^{\dagger}_{\bullet} = \boldsymbol{K}_{\bullet}\,.$$

**Idea of the Proof.** Obtained from the definition of radial conjugation and relations in the proof of Lemma 3.41 → p. 65, similarly as in Remark 1.100 → p. 34. □

**Definition 3.93** (Correlators in Radial Quantization Picture, Radial Ordering)**.** In the *radial quantization*, a correlation function of local operators is interpreted as a radially-ordered *vacuum expectation value* (VEV)

$$\Big\langle \mathcal{O}_1(x_1)\cdots\mathcal{O}_n(x_n)\Big\rangle = \langle \text{vac}|\,\mathcal{R}\Big(\mathcal{O}_1(x_1)\cdots\mathcal{O}_n(x_n)\Big)\,|\text{vac}\rangle\,,$$

with the *radial ordering* of operators inserted at $n$ distinct points defined as

$$\mathcal{R}\Big(\mathcal{O}_1(x_1)\cdots\mathcal{O}_n(x_n)\Big) \equiv \mathcal{O}_{\sigma(1)}(x_{\sigma(1)})\cdots\mathcal{O}_{\sigma(n)}(x_{\sigma(n)})\,,$$

where $\sigma \in S_n$ is a permutation such that $|x_{\sigma(1)}| \geq \cdots \geq |x_{\sigma(n)}|$.

**Remark 3.94.** We assume $\mathcal{O}_{\bullet}(x)$ and $\mathcal{O}_{\bullet}(y)$ commute if $x \neq y$ and $|x| = |y|$. ⌟

**Remark 3.95.** We do not explicitly distinguish between the operator insertion $\mathcal{O}(x)$ and its particular representation by an operator $\widehat{\mathcal{O}}(x) \in \mathrm{End}(\mathcal{H})$. It should be clear from the context which one we mean. ⌟

**Remark 3.96** (Change of Origin)**.** Choosing a different origin would amount to working with a different Hilbert space and associated operator representation. Nonetheless, it would be isomorphic to the original one, and calculation of appropriately radially-ordered correlators would yield the same results. ⌟

## State–Operator Correspondence

Consider the following assignments:

- **Operator $\mapsto$ State** — An insertion of a local operator $\mathcal{O}(0)$ at the origin $x = 0$ prepares a state $|\psi\rangle \in \mathcal{H}_\Sigma \simeq \mathcal{H}$ by performing the path integral over the interior of some sphere $\Sigma$ centered at the origin (essentially with arbitrary radius $r$ assuming no other operator insertions).
- **State $\mapsto$ Operator** — A state $|\psi\rangle \in \mathcal{H}_\Sigma \simeq \mathcal{H}$ on sphere $\Sigma$ of some radius can be "back-evolved" by dilatations — which are symmetry transformations of the theory — to an arbitrarily small sphere $\mathbb{S}^{d-1}_\varepsilon$ of radius $\varepsilon$ around the origin, so it defines a local operator $\mathcal{O}(0)$ inserted at the origin.

We have thus come to one of the central points of the CFT formalism, which as we will see has various far-reaching consequences.

**Definition 3.97** (State–Operator Correspondence)**.** A Euclidean CFT enjoys the *state–operator correspondence*, which provides a linear isomorphism $\mathsf{s}\colon V \simeq \mathcal{O}_0\mathbb{R}^d \xrightarrow{\sim} \mathcal{H}$ between local operators at the origin and quantum states in the radial quantization. This gives us the identification

$$\mathcal{O}(0) \quad \leftrightsquigarrow \quad |\mathcal{O}\rangle \equiv \mathcal{O}(0)|\text{vac}\rangle\,,$$

where $|\mathcal{O}\rangle \equiv \mathsf{s}(\mathcal{O})$ is the state obtained by inserting $\mathcal{O}$ at the origin.

**Remark 3.98** (FQFT Perspective)**.** Point operators in FQFT context were defined in Definition 1.61 →p.20 and Remark 1.63 →p.20 as a family of states living on infinitesimally small spheres around a point. Since CFTs have the scaling symmetry, spaces $\mathcal{H}_{\mathbb{S}^{d-1}_r}$ with different radii $r$ are isomorphic, implying a state on any sphere has a corresponding point operator at the center. ⌟

**Remark 3.99** (Hilbert Space Decomposition into Irreducible Representations)**.** Since vacuum is invariant under conformal transformations, that is $Q_{\boldsymbol{\xi}}|\text{vac}\rangle = 0$, the action of conformal algebra is given by

$$[Q_{\boldsymbol{\xi}}, \mathcal{O}(0)] \quad \leftrightsquigarrow \quad [Q_{\boldsymbol{\xi}}, \mathcal{O}(0)]|\text{vac}\rangle = Q_{\boldsymbol{\xi}}\mathcal{O}(0)|\text{vac}\rangle \equiv Q_{\boldsymbol{\xi}}|\mathcal{O}\rangle\,.$$

Therefore, studying quantum states in CFTs (in radial quantization) is same as studying the algebra of point operators. The whole analysis performed in Derivation 3.63 →p.72 automatically carries over also to states in $\mathcal{H}$, so we can schematically write the decomposition

$$\mathcal{H} \simeq \bigoplus_{\substack{\text{all local}\\ \text{operators } \mathcal{O}}} \mathbb{C}\mathcal{O}(x) \simeq \bigoplus_{\substack{\text{primary}\\ \text{operators } \mathcal{O}}} \left[\bigoplus_{n=0}^{\infty} \mathbb{C}\boldsymbol{P}_{\bullet}^{n}\mathcal{O}(0)\right].$$

**Remark 3.100.** It would be more precise to distinguish following spaces

$$\mathcal{H}^{\text{small}} \subset \mathcal{H}^{\text{Hilb}} \subset \mathcal{H}^{\text{big}},$$

by which we mean:

- First, $\mathcal{H}^{\text{small}}$ is to be identified (by the state–operator correspondence) with the complex vector space $\mathcal{O}_0\mathbb{R}^d \simeq V$ of all local observables at the origin. In particular, we consider here only *finite* linear combinations of point operators inserted at the origin. Even though it carries a Hermitian inner product, it is not a Hilbert space, since it is not complete with respect to it.
- Its completion $\mathcal{H}^{\text{big}}$ on which all operators $\mathcal{O}(x)$ act (at arbitrary points), so it contains all vectors of the form (with $\{x_i\}$ radially-ordered)

  $$\mathcal{O}_1(x_1)\cdots\mathcal{O}_n(x_n)|\text{vac}\rangle\,,$$

  which includes certain infinite linear combinations of operators at the origin. It too does not have a well-defined Hermitian structure, since it contains vectors with infinite norm — if $|x| \geq 1$, the conjugation maps it to $|\mathcal{I}(x)| \leq 1$, and the norm would be given by a VEV of non-radially-ordered operators, which is generally infinite in Euclidean signature.
- Finally, the $L^2$ completion $\mathcal{H}^{\text{Hilb}}$ of $\mathcal{H}^{\text{small}}$ with respect to its inner product gives us a Hilbert space. It contains vectors of the form as above, but all insertion points $\{x_i\}$ must be distinct and have $|x_i| < 1$.

This is similar to the *rigged Hilbert space* construction in ordinary QM.

**Example 3.101** (Identity Operator ⟷ Vacuum State)**.** We clearly have

$$\mathbf{1}(0) \;\longleftrightarrow\; |\mathbf{1}\rangle \equiv \mathbf{1}(0)|\text{vac}\rangle \equiv |\text{vac}\rangle\,,$$

which we already discussed in Example 1.64 →p.20. This works even when inserting $\mathbf{1}(x)$ at a general point $x$, since $[\boldsymbol{P}_{\bullet}, \mathbf{1}(0)] \equiv 0$ — we can shrink any topological surface operator to zero, which is equivalent to the invariance of $|\text{vac}\rangle$ under conformal transformations.

**Example 3.102** (Primary State, Descendant States)**.** From Remark 3.99 →p.82 we see that a primary operator $\mathcal{O}(0)$ corresponds to a *primary state* $|\mathcal{O}\rangle \equiv \mathcal{O}(0)|\text{vac}\rangle \in \mathcal{H}$. The whole conformal multiplet is then given by adding an infinite tower of corresponding *descendant states* above the primary

$$\left\{|\mathcal{O}\rangle, \boldsymbol{P}_{\bullet}|\mathcal{O}\rangle, \boldsymbol{P}_{\bullet}\boldsymbol{P}_{\bullet}|\mathcal{O}\rangle, \dots\right\} \equiv \left\{|\mathcal{O}\rangle, |\boldsymbol{P}_{\bullet}\mathcal{O}\rangle, |\boldsymbol{P}_{\bullet}\boldsymbol{P}_{\bullet}\mathcal{O}\rangle, \dots\right\}.$$

Their scaling dimensions (eigenvalues of $\boldsymbol{D}$) are given as $\Delta + N$, where $N$ is the *level* of the descendant (or $N = 0$ for the primary). This in particular means that states at different levels are orthogonal to each other. ⌟

**Example 3.103** (Dual State)**.** Considering (for simplicity) a scalar state $|\mathcal{O}\rangle$, the dual state is given by Definition 3.90 →p.81 as

$$\langle\mathcal{O}| \equiv \big(|\mathcal{O}\rangle\big)^\dagger \equiv \langle\text{vac}|\mathcal{O}(0)^\dagger = \lim_{x\to 0} |x|^{-2\Delta}\langle\text{vac}|\mathcal{O}^*\Big(\tfrac{x}{|x|^2}\Big) = \lim_{y\to\infty} |y|^{2\Delta}\langle\text{vac}|\mathcal{O}^*(y)\,,$$

where in the last step we exchanged $x$ with $y \equiv \mathcal{I}(x)$. ⌟

**Remark 3.104** (Casimir Eigenvalues for Symmetric Traceless Representations)**.** The eigenvalue of the Casimir for a symmetric traceless tensor $\mathcal{O}_{\Delta,J}$ is given by

$$\begin{aligned} \mathfrak{Cas}_2|\mathcal{O}_{\Delta,J}\rangle &= \lambda_{\Delta,J}|\mathcal{O}_{\Delta,J}\rangle\,, \\ \Big[\mathfrak{Cas}_2, \mathcal{O}_{\Delta,J}(x)\Big] &= \lambda_{\Delta,J}\mathcal{O}_{\Delta,J}(x)\,, \end{aligned} \qquad \lambda_{\Delta,J} \equiv \Delta(\Delta - d) + J(J + d - 2)\,,$$

where we could insert operator at arbitrary point since $\mathfrak{Cas}_2$ commutes with all generators of the conformal algebra, which also means that whole conformal multiplet shares the same eigenvalue.

It is thus simplest to calculate the eigenvalue for the primary state. The first term of Definition 3.60 →p.71 is the Casimir of the $\mathsf{SO}(d)$ subalgebra with eigenvalue $J(J + d - 2)$, and the rest acts on the primary as $\boldsymbol{D}(\boldsymbol{D} - d)$. ⌟

## Implications of Unitarity

Having developed a radial quantization formalism, we can now study the implications of the CFT unitarity — or maybe better "inversion positivity". We will just briefly explore some necessary conditions stemming from the non-negativity of the inner product on $\mathcal{H}$. Following will also illustrate how CFT computations can be reduced (in radial quantization) just to algebraic manipulations.

**Remark 3.105** (Norms of Scalar Primary Operators)**.** In Remark 3.81 →p.77 we considered a basis of all real scalar primaries $\{\mathcal{O}_i\}$, and their two-point functions. Now we know that in radial quantization they correspond to primary states $\{|\mathcal{O}_i\rangle\}$, and unitarity requires positive-definiteness of the inner product matrix

$$\begin{aligned} \langle\mathcal{O}_i|\mathcal{O}_j\rangle &\equiv \lim_{y\to\infty} |y|^{2\Delta_i}\langle\text{vac}|\mathcal{O}_i(y)\mathcal{O}_j(0)|\text{vac}\rangle \\ &= \lim_{y\to\infty} |y|^{2\Delta_i}\Big\langle\mathcal{O}_i(y)\mathcal{O}_j(0)\Big\rangle \equiv \lim_{y\to\infty} |y|^{2\Delta_i}\,\frac{C_{\mathcal{O}_i\mathcal{O}_j}\delta_{\Delta_i,\Delta_j}}{|y|^{2\Delta_i}} = C_{\mathcal{O}_i\mathcal{O}_j}\delta_{\Delta_i,\Delta_j}\,. \end{aligned}$$

So $C_{\mathcal{O}_i\mathcal{O}_j}\delta_{\Delta_i,\Delta_j}$ is not only symmetric, but also positive-definite, which allows us to diagonalize it to $\delta_{ij}$ without changing realness of $\{\mathcal{O}_i\}$. ⌟

What does the positivity imply when we consider the descendant states?

**Calculation 3.106** (Positivity of $N = 1$ Descendants of Scalar Operators)**.** Considering a level $N = 1$ descendant state $\boldsymbol{P}_a|\mathcal{O}\rangle$, its squared norm is

$$\left\|\boldsymbol{P}_a|\mathcal{O}\rangle\right\|^2 = \langle\mathcal{O}|\boldsymbol{P}_a^\dagger \boldsymbol{P}_a|\mathcal{O}\rangle = \langle\mathcal{O}|\boldsymbol{K}_a \boldsymbol{P}_a|\mathcal{O}\rangle$$
$$= \langle\mathcal{O}|[\boldsymbol{K}_a, \boldsymbol{P}_a]|\mathcal{O}\rangle = \langle\mathcal{O}|2\boldsymbol{D}|\mathcal{O}\rangle = 2\Delta\langle\mathcal{O}|\mathcal{O}\rangle \overset{!}{\geq} 0\,,$$

where we used Corollary 3.92→p.81, Definition 3.56→p.70 with Lemma 3.41→p.65, and that primary state satisfies $\boldsymbol{K}_\bullet|\mathcal{O}\rangle \equiv 0 = \langle\mathcal{O}|\boldsymbol{P}_\bullet$ and $\boldsymbol{D}|\mathcal{O}\rangle \equiv \Delta|\mathcal{O}\rangle$.

Assuming nontrivial $|\mathcal{O}\rangle$, unitarity implies $\Delta \geq 0$, as was already noted in Remark 3.82→p.78. For $\Delta = 0$, that is identity operator $\mathbf{1}$ and $|\mathbf{1}\rangle \equiv |\text{vac}\rangle$, the descendant becomes null, and unitarity consistently requires $\boldsymbol{P}_\bullet|\mathbf{1}\rangle \equiv 0$. ⌟

Similar calculation can be performed for spinning operators/states, see for example [11] [36] [34], and references therein. In the case of scalars, at level $N = 2$ one obtains a stronger condition than $\Delta \geq 0$, as mentioned below.

**Remark 3.107** (Unitarity Bounds for Traceless Symmetric Tensors)**.** For traceless symmetric tensors with spin $J$, the unitarity requires

$$\Delta = 0 \qquad \text{(unit operator), or}$$
$$\Delta \geq \begin{cases} \frac{d-2}{2} & J = 0\,, \\ d - 2 + J & J > 0\,. \end{cases}$$

Alternatively, these *unitarity bounds* — and that they cannot be strengthened — can be derived directly in the Lorentzian signature [47]. ⌟

Special things happen when the primary saturates the unitarity bound. In such a case, some descendant state becomes null, and needs to be removed (together with all of its descendants) from the original conformal multiplet — obtaining thus a so-called *short multiplet*.

**Example 3.108** (Free Scalar)**.** Example of such *short* multiplet is a scalar primary with $\Delta = \frac{d-2}{2}$, where the null state is (nice exercise to check explicitly)

$$\boldsymbol{P}^2|\mathcal{O}\rangle \equiv \boldsymbol{P}_a \boldsymbol{P}^a|\mathcal{O}\rangle = 0 \quad \Longleftrightarrow \quad \partial^2\mathcal{O}(x) = 0\,.$$

This means $\mathcal{O}$ satisfies the massless Klein–Gordon equation, and is therefore a *free massless scalar* — we indeed recognize the usual value $\Delta$, see for example Example 2.26→p.48. ⌟

**Example 3.109** (Conserved Currents)**.** Consider a spin $J$ primary $|\mathcal{O}^{a_1 \ldots a_J}\rangle$ with a null state

$$\boldsymbol{P}_n|\mathcal{O}^{n a_2 \ldots a_J}\rangle \overset{!}{=} 0 \quad \Longleftrightarrow \quad \partial_n \mathcal{O}^{n a_2 \ldots a_J}(x) = 0\,.$$

↳ Then we can calculate

$$
\begin{aligned}
0 &\overset{!}{=} \boldsymbol{K}_a \boldsymbol{P}_n |\mathcal{O}^{na_2\ldots a_J}\rangle = [\boldsymbol{K}_a, \boldsymbol{P}_n] |\mathcal{O}^{na_2\ldots a_J}\rangle = (2\delta_{an}\boldsymbol{D} - 2\boldsymbol{M}_{an}) |\mathcal{O}^{na_2\ldots a_J}\rangle \\
&= 2\Delta |\mathcal{O}^{aa_2\ldots a_J}\rangle - 2(\delta_{ab}\delta^n_n - \delta_{nb}\delta^n_a) |\mathcal{O}^{ba_2\ldots a_J}\rangle - 2\sum_{i=2}^{J} (\delta_{ab}\delta^{a_i}_n - \delta_{nb}\delta^{a_i}_a) |\mathcal{O}^{na_2\ldots b\ldots a_J}\rangle \\
&= 2\big(\Delta - (d-1) - (J-1)\big) |\mathcal{O}^{aa_2\ldots a_J}\rangle \quad \Longleftrightarrow \quad \Delta = d - 2 + J\,.
\end{aligned}
$$

In the first line we used that $|\mathcal{O}^{\bullet\cdots\bullet}\rangle$ is a primary and a little bit of conformal algebra. In the second line we employed the standard form of rotation generators for each vector index. Finally, we used symmetry and tracelessness of $|\mathcal{O}^{\bullet\cdots\bullet}\rangle$.

We thus verified that conserved currents correspond to a short multiplet saturating the unitarity bound in Remark 3.107→p.85. From the calculation above we clearly see also the opposite implication — a primary with $\Delta = d - 2 + J$ has necessarily a null state, making it a conserved current.

Important examples are global symmetry currents with $(J = 1, \Delta = d - 1)$, and the energy–momentum tensor with $(J = 2, \Delta = d)$. ⌟

In the following section we will mention one additional implication of unitarity — the reality of the 3-point function coefficients.

## 3.5 Operator Product Expansion

The state–operator correspondence has another essential consequence.

**Idea 3.110** (Operator Product Expansion)**.** Consider an insertion of two local operators close enough to each other, that we can draw a sphere separating them from all other operator insertions. They prepare some state on this sphere (in the corresponding radial quantization), which in turn corresponds to a local operator at the center — we can express product of two operators as a linear combination of primaries and descendants at the origin. This is called the *Operator Product Expansion* (OPE). ⌟

**Remark 3.111.** The notion of OPE can be introduced in a more general context of QFTs, but even when two point operators are brought really close together, it usually provides just an asymptotic series. Only in CFTs it is elevated to its powerful status, since it can be shown that conformal symmetry guarantees nonzero radius of convergence of OPE (for well separated operators) [48] [49]. ⌟

### Operator Algebra

As described in Idea 3.110→p.86, we can express the product of two operators as

$$
\mathcal{O}_i(x_1)\mathcal{O}_j(x_2) \sim \sum_{\mathcal{O}_k \in \mathcal{O}_i \times \mathcal{O}_j} C_{\mathcal{O}_i\mathcal{O}_j\mathcal{O}_k}(x_1, x_2, y, \boldsymbol{\partial}_y)\, \mathcal{O}_k(y)\,,
$$

where $\mathcal{O}_k$ runs through all primary operators included in the OPE, which we denote symbolically by $\mathcal{O}_i \times \mathcal{O}_j$. The function $C_{\mathcal{O}_1\mathcal{O}_2\mathcal{O}'}$ also includes contributions of the corresponding descendants through the action of the momentum/derivative operator $\boldsymbol{P}_y \simeq \boldsymbol{\partial}_y$ acting on $y$.

**Remark 3.112.** We keep all of the indices present for spinning operators implicit. ⌟

The OPE is understood as an operator equation, which holds inside the correlation functions independent of other insertions (assuming they are outside the sphere surrounding the operators, where the OPE converges). So we have

$$\left\langle \mathcal{O}_i(x_1)\mathcal{O}_j(x_2)\,\square\ldots\square \right\rangle = \sum_{\mathcal{O}_k\in\mathcal{O}_i\times\mathcal{O}_j} C_{\mathcal{O}_i\mathcal{O}_j\mathcal{O}_k}(x_1,x_2,y,\boldsymbol{\partial}_y)\left\langle \mathcal{O}_k(y)\,\square\ldots\square \right\rangle,$$

independent of the precise choice of radial foliation/quantization and the corresponding state–operator correspondence we used to derive/motivate it.

**Remark 3.113.** When possible, it is convenient for calculations to choose for example $y = x_2$, which we assume in the following. ⌟

## OPE Coefficients

The conformal symmetry greatly restricts the form of the OPE. For simplicity, we will take all of the operators $\mathcal{O}_i, \mathcal{O}_j, \mathcal{O}_k$ to be scalars.

The translation invariance gives us that $C_{\mathcal{O}_i\mathcal{O}_j\mathcal{O}_k}(x_1,x_2,\boldsymbol{\partial})$ can depend on positions only through the combination $x_{12} \equiv x_1 - x_2$. Utilizing also scaling and rotation invariance — or alternatively consistency of OPE with the action of $\boldsymbol{D}$ and $\boldsymbol{M}_{\bullet\bullet}$ — we obtain expansion for $|x_{12}| \ll 1$ of the form

$$C_{\mathcal{O}_i\mathcal{O}_j\mathcal{O}_k}(x_{12},\boldsymbol{\partial}) \propto \frac{1}{x_{12}^{\Delta_i+\Delta_j-\Delta_k}}\Big(1 + (\#)x_{12}\cdot\boldsymbol{\partial} + \ldots\Big)$$

Finally, consistency with the action of $\boldsymbol{K}_\bullet$ completely fixes $C_{\mathcal{O}_i\mathcal{O}_j\mathcal{O}_k}$ up to an overall constant.

**Remark 3.114** (Fixed Form of OPE, OPE Coefficients)**.** More elegant way to see that OPE is fixed up to a constant is to reduce the computation of 3-point correlator to 2-point functions as

$$\left\langle \mathcal{O}_i(x_1)\mathcal{O}_j(x_2)\mathcal{O}_k(x_3)\right\rangle = \sum_{k'} C_{\mathcal{O}_i\mathcal{O}_j\mathcal{O}_{k'}}(x_{12},\partial_2)\left\langle \mathcal{O}_{k'}(x_2)\mathcal{O}_k(x_3)\right\rangle.$$

Now we can utilize the fixed forms of 2-point and 3-point functions as shown in Corollary 3.80 →p.77 (in diagonal form Remark 3.81 →p.77) and Corollary 3.83 →p.78, giving us

$$\frac{f_{ijk}}{x_{12}^{\Delta_i+\Delta_j-\Delta_k}x_{13}^{\Delta_i+\Delta_k-\Delta_j}x_{23}^{\Delta_j+\Delta_k-\Delta_i}} \overset{!}{=} C_{\mathcal{O}_i\mathcal{O}_j\mathcal{O}_k}(x_{12},\partial_2)\,\frac{1}{x_{23}^{2\Delta_k}}\,.$$

We see that $C_{\mathcal{O}_i\mathcal{O}_j\mathcal{O}_k} \equiv f_{ijk}C_{ijk}$ is proportional to the 3-point function coefficient $f_{ijk}$, which from now on we call *OPE coefficient*. Matching the small $|x_{12}|/|x_{23}|$ expansion of both sides completely fixes the "reduced" function $C_{ijk}$ in terms of the operator dimensions $\Delta_i, \Delta_j, \Delta_k$ (if the operators were spinning, then also their spins), and the spacetime dimension $d$. ⌟

**Remark 3.115** (Recursive Calculation of Correlators Through OPE). More generally, using the Operator Algebra/OPE we can recursively reduce calculation of any correlator eventually down to a sum of two-point functions, or even one-point functions (then use Corollary 3.78→p.77). ⌟

From previous remarks we see that to calculate any correlator of local operators, it is enough to know all primary operators and their OPE coefficients. The contribution of descendants is fixed by the conformal symmetry.

**Definition 3.116** (CFT Data). Essential information about a given CFT is encoded in the corresponding

$$\mathsf{CFT}\ \text{data} \equiv \Big\{(\Delta_i, \rho_i), f_{ijk}\Big\}\,,$$

containing the conformal representations of all primary operators $\{\mathcal{O}_i\}$ — scaling dimensions $\Delta_i$ and their $\mathsf{SO}(d)$ representations $\rho_i$ — together with all of the OPE coefficients $f_{ijk} \equiv \mathsf{ope}[\mathcal{O}_i\mathcal{O}_j\mathcal{O}_k]$.

**Remark 3.117** (Associativity of OPE, Overcompleteness of CFT Data). We can reduce down the correlator by performing the OPE in different order, but we should always obtain the same result — the OPE is necessarily "associative". This induces relation in the CFT data, which tells us two things — an arbitrary collection of "CFT data" does not necessarily define a consistent CFT, and when it does, the CFT data are overcomplete. ⌟

**Remark 3.118.** The same OPE is even valid on any (locally) conformally flat background — for example we can consider boundaries or compact dimensions. Only thing that changes in the calculation of correlators is the form of one-point functions, as mentioned in Remark 3.79→p.77 ⌟

**Remark 3.119** (Unitarity, Reality of OPE Coefficients). Apart from the unitarity bounds on the scaling dimensions discussed in Remark 3.107→p.85, the unitarity of the CFT also implies that the OPE coefficients are real [11] [34]. ⌟

## 3.6 Conformal Blocks

As we already know from Section 3.3→p.75, the form of higher-point correlators is not completely fixed by conformal symmetry. In this section we will focus on 4-point functions of identical real scalars $\mathcal{O}_\phi$ with scaling dimension $\Delta_\phi$.

### Conformal Block Decomposition

Using Proposition 3.76→p.76 we can write

$$\Big\langle \mathcal{O}_1\mathcal{O}_2\mathcal{O}_3\mathcal{O}_4 \Big\rangle \equiv \Big\langle \mathcal{O}_\phi(x_1)\mathcal{O}_\phi(x_2)\mathcal{O}_\phi(x_3)\mathcal{O}_\phi(x_4) \Big\rangle \equiv \frac{g(u,v)}{x_{12}^{2\Delta_\phi} x_{34}^{2\Delta_\phi}}\,,$$

where the function $g(u,v)$ of the conformal cross-ratios depends on the dynamics of the theory. Alternatively, by pairing up the operators and performing the OPE (assuming points are configured appropriately), we can express it in terms of CFT data schematically as

$$\begin{aligned}\Big\langle(\mathcal{O}_1\mathcal{O}_2)(\mathcal{O}_3\mathcal{O}_4)\Big\rangle &= \sum_{\mathcal{O},\mathcal{O}'} f_{\phi\phi\mathcal{O}} f_{\phi\phi\mathcal{O}'} C_{\bullet}(x_{12},\partial_2) C_{\bullet}(x_{34},\partial_4)\Big\langle\mathcal{O}^{\bullet}(x_2)\mathcal{O}'^{\bullet}(x_4)\Big\rangle \\ &= \sum_{\mathcal{O}} f^2_{\phi\phi\mathcal{O}} C_{\bullet}(x_{12},\partial_2) C_{\bullet}(x_{34},\partial_4)\,\frac{I^{\bullet\bullet}(x_{24})}{x_{24}^{2\Delta}} \\ &\equiv \sum_{\substack{\text{primary } \mathcal{O} \\ \text{with } \Delta,J}} \mathsf{ope}^2\Big[\mathcal{O}_\phi\mathcal{O}_\phi\mathcal{O}\Big]\, G^{(s)}_{\Delta,J}\Big(\{x_i\}\Big)\,.\end{aligned}$$

In the last line we defined the *conformal block* $G^{(s)}_{\Delta,J}$, which encodes the contribution of exchanging a single conformal multiplet — a primary $\mathcal{O}$ with all of its descendants — in the $s$-channel $(\mathcal{O}_1\mathcal{O}_2)(\mathcal{O}_3\mathcal{O}_4)$.

**Remark 3.120.** Only spin-$J$ symmetric traceless tensors appear in OPE of two scalar fields. We denoted by $I^{\bullet\bullet}$ the corresponding (unique) 2-point tensor structure. ⌟

So knowing the CFT data, we can compute 4-point functions. We can also turn this around — having some alternative way of computing 4-point functions (perhaps in some perturbative expansion), we can try to extract the CFT data, which is particularly convenient after performing the conformal block decomposition. This is what we shall do in the following Chapter 5→p.100.

For more details, such as interpretation of the conformal block decomposition in the radial quantization, and an elegant way of computing conformal blocks through a differential equation coming from action of the conformal Casimir Definition 3.60→p.71, see [11] [34] and references therein.

Let us make a final remark about yet another 4-point function decomposition, which we shall discuss more in Section 5.3→p.114.

**Remark 3.121** (Conformal Block versus Conformal Partial Wave Decomposition)**.** The decomposition of 4-point function into conformal blocks can be understood as decomposition into "physical" irreducible representations of the conformal group $\mathsf{Conf}(\mathbb{R}^{1,d-1}) \simeq \widetilde{\mathsf{SO}}(2,d)$ unitarily represented on the Hilbert space $\mathcal{H}$, which we often consider Wick-rotated to the Euclidean signature.

Alternatively, when we have a Euclidean 4-point correlator, it is also natural to decompose it into irreducible representations of the Euclidean conformal group $\mathsf{SO}(1,d+1)$. This so-called *conformal partial wave* decomposition corresponds to Unitary Irreducible Representations (UIRs) of $\mathsf{SO}(1,d+1)$, but this time "unitary" corresponds to the Euclidean conformal group, which is different from the positivity conditions coming from the "physical" Lorentzian signature. ⌟

## Crossing Symmetry

We can express 4-point function in various channels, so for example we could perform OPE in the $t$-channel $(\mathcal{O}_1\mathcal{O}_3)(\mathcal{O}_2\mathcal{O}_4)$ or $u$-channel $(\mathcal{O}_1\mathcal{O}_4)(\mathcal{O}_2\mathcal{O}_3)$, and obtain the corresponding conformal block decomposition. All of these expressions should in the end agree, which leads to non-trivial constraints on the CFT data.

**Remark 3.122.** It can be shown that crossing symmetry of all 4-point functions is equivalent to the associativity of the OPE (see Remark 3.117→p.88). ⌟

**Remark 3.123.** One conclusion of the crossing symmetry is that any CFT must have an infinite number of primary operators. In the case of MFT, the exchange of identity operator in the $t$-channel corresponds to exchanges of the double-twist operators Example 3.71→p.74 in the $s$-channel. ⌟

## Conformal Bootstrap

We have various consistency constraints/conditions on a CFT data defining a consistent (unitary) CFT, which include:

- *Crossing Symmetry*, which is essentially the OPE associativity,
- *Unitarity/Reflection Positivity* — implies lower unitarity bounds on scaling dimensions, and a possibility to choose a real orthonormal operator basis where all OPE coefficients are real,
- (optionally) *Existence of Energy-Momentum Tensor* — conserved rank-2 symmetric traceless primary operator (with $\Delta = d$) with correlators obeying Ward–Takahashi identities.

One can then try to study these conditions — in particular the crossing equation — not necessarily by finding exact solutions, but by excluding regions of the CFT data which are inconsistent. In this way it is possible to obtain non-perturbative bounds on the scaling dimensions and OPE coefficients, which in some cases can be surprisingly strong. For a review of the *Conformal Bootstrap* program, see for example [34].

**Remark 3.124.** In some lucky cases, the theory which we are interested in lies at (or very close to) the boundary of the allowed region of CFT data. Then we are able to obtain precise data of the given theory — non-trivial example being the *Ising model* in $d = 3$.

If our physical theory of interest is not at the boundary, we can try to input some model-dependent information to the bootstrap procedure. ⌟

# 4 QFT in Anti–de Sitter Spacetime

In this chapter we briefly introduce the QFT in AdS spacetime, and mainly focus on the relation to corresponding CFT on the boundary of AdS.

More details can be found for example in [50] [33] [51] [52].

## 4.1 Anti–de Sitter Spacetime

We will introduce AdS spacetime in several ways, each offering various insights into its structure and properties. Often we will work in the Euclidean signature, that is with EAdS, and sometimes we refer to both as AdS.

**Remark 4.1** (Dimensions). We will work with $\mathsf{AdS}_{D\equiv d+1}$ spacetime, where $D$ is the number of bulk dimensions, and $d$ is the number of spatial dimensions, or alternatively the dimension of the asymptotic boundary. ⌟

### Maximally Symmetric Space and Solution of Einstein's Equations

Both Lorentzian and Euclidean AdS are maximally symmetric manifolds of constant negative curvature — they are homogeneous, isotropic, and admit a maximal number of Killing vectors. This will be easily seen in the "embedding formalism".

It is also a vacuum solution (zero energy–momentum tensor $\boldsymbol{T_{ab}} = 0$) of Einstein Field Equations $\mathbf{Ric}_{ab} - \frac{1}{2}\mathcal{R}\boldsymbol{g_{ab}} + \Lambda \boldsymbol{g_{ab}} = 0$ with negative cosmological constant $\Lambda \equiv -\frac{1}{\ell^2}\frac{d(d-1)}{2} < 0$, where $\ell$ is the *radius of curvature* of $\mathsf{AdS}_{d+1}$.

**Remark 4.2** (Curvature of Maximally Symmetric Spaces). To see this, we can utilize that a maximally symmetric space(time) $\mathcal{M}$ ($\dim \mathcal{M} \equiv D \equiv d+1$) — both *homogeneous* and *isotropic* — has no special point, so the Ricci scalar curvature $\mathcal{R}$ must be constant. Furthermore, we necessarily have [53]

$$\mathbf{Ric}_{ab} = \frac{\mathcal{R}}{d+1}\,\boldsymbol{g_{ab}}\,,$$
$$\boldsymbol{R_{abcd}} = \frac{\mathcal{R}}{d(d+1)}\,\boldsymbol{g}_{a[c}\,\boldsymbol{g}_{b]d} \equiv -\frac{1}{\ell^2}\,\boldsymbol{g}_{a[c}\,\boldsymbol{g}_{b]d}\,,$$

otherwise either the homogeneity or isotropy would be violated — there is no other invariant tensor (with respect to isometries) apart from $\boldsymbol{g}$ which could be used to construct the curvature.

Einstein's equations then reduce to relation between $\mathcal{R}$ and $\Lambda$

$$\mathcal{R} = 2\Lambda\,\frac{d+1}{d-1} \implies \mathbf{Ric} = \frac{2\Lambda}{d-1}\,\boldsymbol{g} \equiv -\frac{d}{\ell^2}\,\boldsymbol{g}\,,$$

so Einstein's equations are automatically satisfied by choosing the scale $\ell$ or $\Lambda$ appropriately. ⌟

**Remark 4.3** (de Sitter Spacetime). More interesting from the cosmological point of view is the case of positive cosmological constant $\Lambda > 0$ (which seems to be the case in our Universe), and corresponds to de Sitter spacetime dS. It is used to model the period of inflation, and also the late-time acceleration of the Universe (where effect of the matter content is negligible). Lately, there has been progress in understanding QFT in dS through techniques coming from AdS via analytic continuation, see for example [54]. ⌟

## Hyperboloid/Quadric/Quasi-Sphere

Perhaps the most convenient definition of $\mathsf{AdS}_{d+1}$ is as a (universal cover) of hyperboloid/quadric/quasi-sphere in the embedding/ambient space $\mathbb{R}^{2,d}$ with metric

$$\eta^{(2,d)} \equiv \operatorname{diag}(-,-,\underbrace{+,\ldots,+}_{d}).$$

The Anti–de Sitter spacetime (with compactified time coordinate) is defined as

$$\mathsf{AdS}_{d+1}/\mathbb{Z} \equiv \left\{ X \in \mathbb{R}^{2,d} \;\middle|\; X^2 \equiv \eta^{(2,d)}(X,X) = -\ell^2 \right\},$$

where $\ell$ is the *radius* (of curvature) of AdS. We often choose units such that $\ell = 1$. In the Euclidean signature we just work with $\eta^{(1,d+1)}$.

**Remark 4.4** (Isometries of AdS, AdS as a Homogeneous Space). From such description we immediately see that the isometry group of $\mathsf{AdS}_{d+1}$ is $\mathsf{Iso}(\mathsf{AdS}_{d+1}) = \mathsf{SO}(2,d)$, since both the metric and the hyperboloid are invariant under its action. Clearly, action of AdS isometry group is transitive — for example, each point can be brought to some canonical point, such as $X_* = (-1, 0, 0, \ldots, 0)$. Thus, by orbit-stabilizer theorem, we have (up to identifications)

$$\mathsf{AdS}_{d+1} \equiv \mathsf{Orbit}_{X_*} = \frac{\mathsf{SO}(2,d)}{\mathsf{Stab}_{X_*}} = \frac{\mathsf{SO}(2,d)}{\mathsf{SO}(1,d)}.$$

More generally, every so-called *homogeneous* manifold is a coset space $G/H$, where $G$ is the isometry group, and $H$ is isotropy group. ⌟

**Remark 4.5** (Closed Timelike Curves). Note that the hyperboloid defined above has closed non-contractible timelike curves associated with the action of $\mathsf{SO}(2) \subset \mathsf{SO}(2,d)$ in the (-1–0)-plane, for example we have $(X^{-1})^2 + (X^0)^2 = \ell^2$ for $X^i = 0$. Only after decompactifying (unrolling) along these cycles do we obtain the AdS as we understand it. ⌟

Asymptotically in the limit of large components of $X$ — after dividing the equation defining the hyperboloid by a large constant factor — the hyperboloid approaches the light-cone

$$C = \left\{ P \in \mathbb{R}^{2,d} \;\middle|\; P^2 \equiv \eta^{(2,d)}(P,P) = 0 \right\}.$$

We can thus define the asymptotic/conformal *boundary* of AdS as the set of light (half)rays going through the origin, that is

$$\partial\mathsf{AdS} = (C \backslash \{0\})/\sim,$$

where we identify $P \sim P' \iff P = \lambda P'$ with $\lambda > 0$. This is precisely the conformal compactification of flat space. For the illustration of this in Euclidean signature, see Figure 4.6→p.93.

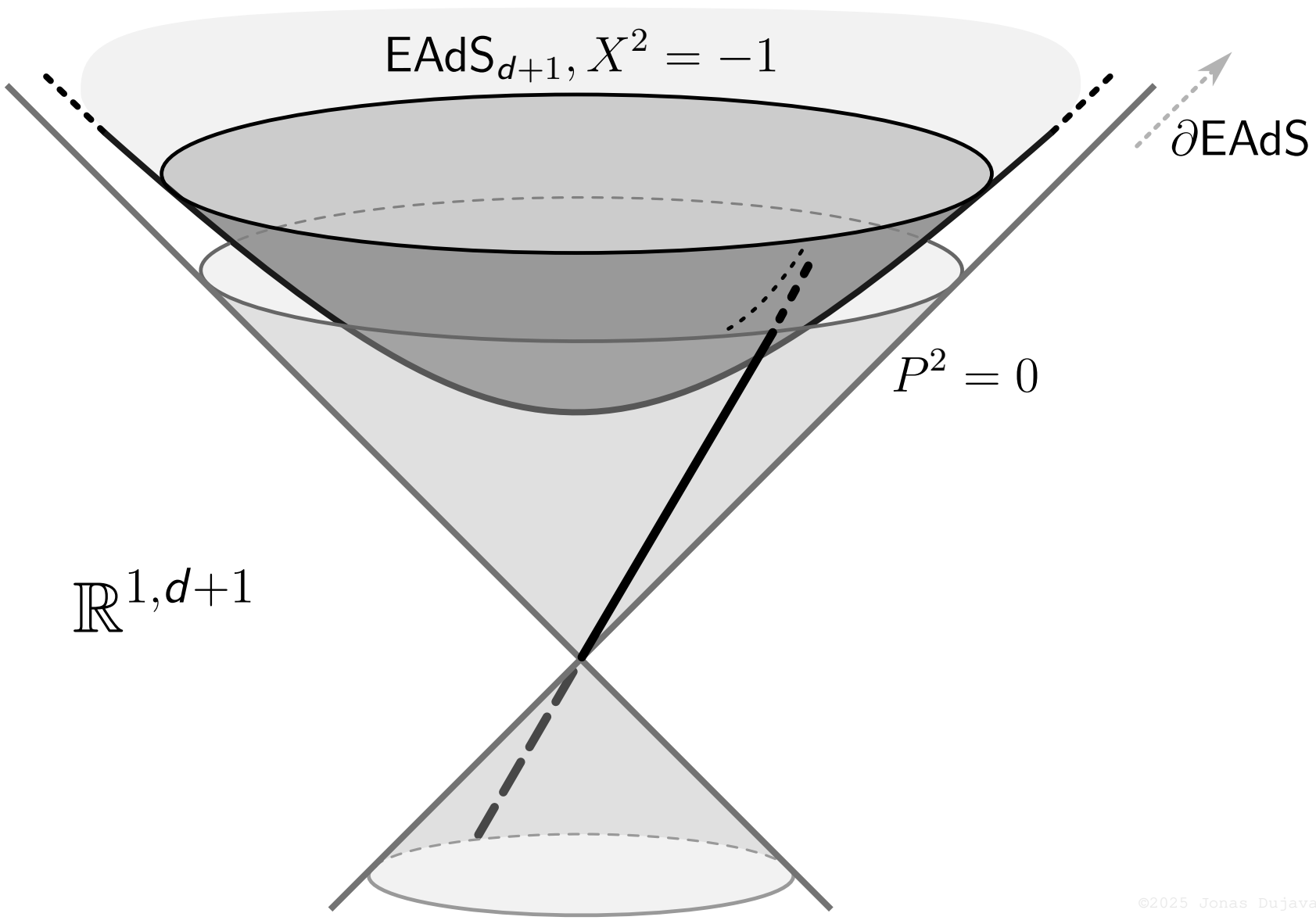

**Figure 4.6 /** $\mathsf{EAdS}_{d+1}$ embedded in $\mathbb{R}^{1,d+1}$ by the equation $X^2 = -1$. Its boundary is identified with the null rays $P^2 = 0$, $P \sim \lambda P$, which is precisely the conformal compactification of $\mathbb{R}^d$.

We usually work with certain representatives of the rays (equivalence classes) defining $\partial\mathsf{AdS}$ by choice of particular section of $C$. One convenient choice includes the *Poincaré section* intersecting the cone diagonally, that is fixing $P^+ \equiv P^{-1}+P^0 \overset{!}{=} 1$, which leads to interpretation of the boundary as a flat Minkowski. A horizontal section fixed by $P^{-1} = 1$ is adapted for global coordinates.

**Remark 4.7** (Change of Boundary Defining Section)**.** Changing the section by choosing different representatives of boundary points induces a conformal transformation. Indeed, having some parametrization of the boundary point representative $P$ (lying on a chosen section $\Sigma$ of the null cone), the induced metric is

$$\boldsymbol{g}_{\partial\mathsf{AdS},\Sigma} \equiv \boldsymbol{g}\Big|_{\mathsf{AdS}} \equiv \boldsymbol{\eta}\Big|_{\Sigma} = (\mathbf{d}X)^2\Big|_{\Sigma} = \boldsymbol{\eta}_{ab}\,\mathbf{d}X^a\,\mathbf{d}X^b\Big|_{\Sigma} = \boldsymbol{\eta}_{ab}\,\mathbf{d}P^a\,\mathbf{d}P^b \equiv (\mathbf{d}P)^2\,.$$

After a change of section $\Sigma \ni P \longmapsto P' \in \Sigma'$ with $P \sim P' \equiv \lambda(P)P$, using

$$P^2 \equiv 0 \implies \mathbf{d}(P^2) = 0 \implies P \cdot \mathbf{d}P = 0$$

we obtain

$$\begin{aligned}\boldsymbol{g}_{\partial\mathsf{AdS},\Sigma'} \equiv (\mathbf{d}P')^2 &= (P\,\mathbf{d}\lambda + \lambda\,\mathbf{d}P)^2\\ &= P^2\,\mathbf{d}\lambda^2 + 2P\cdot\mathbf{d}P\,\lambda\,\mathbf{d}\lambda + \lambda^2\,\mathbf{d}P^2 = \lambda^2(\mathbf{d}P)^2 \equiv \lambda^2\boldsymbol{g}_{\partial\mathsf{AdS},\Sigma}\,,\end{aligned}$$

which is precisely a Weyl transformation of the original boundary metric.

Considering that boundary of $\mathsf{AdS}$ is in infinity, and we can "really" obtain it only after conformal compactification of $\mathsf{AdS}$, it is natural that we can determine only

↳ the conformal class of $\partial$AdS, which has no canonical choice of metric structure. If we would like to put a QFT theory on the boundary, it thus seems natural that it will be a CFT. ⌟

We will further discuss the *embedding formalism* in Section 4.2 →p.95.

## Coordinate Systems

Now we will briefly discuss some coordinate systems on AdS, even though we will mainly work directly in the embedding space.

**Global coordinates.** We can parametrize the whole hyperboloid by $(t, r, \omega^\bullet)$ as (except for the usual polar-type coordinate singularities)

$$\begin{aligned} \iota\colon (\mathbb{S}^1, \mathbb{R}_{\geq 0}, \mathbb{S}^{d-1}) &\longrightarrow \mathsf{AdS}_{d+1}/\mathbb{Z} \\ (\, t \,,\; r \,,\; \omega^\bullet \,) &\longmapsto X = \iota(t,r,\omega^\bullet) \equiv \begin{cases} X^{-1} = \sqrt{\ell^2 + r^2}\cos(t/\ell), \\ X^0 = \sqrt{\ell^2 + r^2}\sin(t/\ell), \\ X^i = r\omega^i \quad \text{for } i \in \{1, \dots, d\}\,, \end{cases} \end{aligned}$$

where $t \in [0, 2\pi)$ is angle in (-1–0)-plane, $r \in \mathbb{R}_{\geq 0}$ is radial parameter, and $\omega$ is a unit vector parametrizing $\mathbb{S}^{d-1}$, so $\sum_i (\omega^i)^2 \equiv 1$.

The induced metric is then

$$\boldsymbol{g}_{\mathsf{AdS}_{d+1}} = \iota^*(\boldsymbol{\eta}^{(2,d)}) = \dots = -\left(1 + \frac{r^2}{\ell^2}\right)\mathbf{d}t^2 + \left(1 + \frac{r^2}{\ell^2}\right)^{-1}\mathbf{d}r^2 + r^2\boldsymbol{g}_{\mathbb{S}^{d-1}}\,.$$

Going to the universal cover means we just unwrap the circle $\mathbb{S}^1 \ni t$ and take $t \in \mathbb{R}$ from the whole real line.

Clearly, AdS is static with $\boldsymbol{\xi}_t \equiv \frac{\boldsymbol{\partial}}{\boldsymbol{\partial} t}$ being the timelike Killing vector, which corresponds to $\boldsymbol{J}_{-1,0} = \dots = \ell\frac{\boldsymbol{\partial}}{\boldsymbol{\partial} t}$.

**Poincaré/upper-half plane.** The idea is to find slicing of AdS which is conformally flat. We define $X^\pm \equiv X^d \pm X^{-1}$, and write $X^+ \equiv \frac{1}{z}$, such that $z \to 0_+$ we go to the asymptotic boundary. For $z \to \infty$ we approach the light-like surface $X^+ = 0$, which is called the Poincaré horizon. If we want to be conformally flat we choose $X^a = \frac{x^a}{z}$, and finally by $X^2 = -1 \equiv -\ell^2$ we must choose $X^- = \frac{x^2+z^2}{z}$. We can thus cover half of $\mathsf{AdS}/\mathbb{Z}$ by (now reinstating the radius $\ell$)

$$\begin{aligned} \iota\colon (\mathbb{R}_{\geq 0}, \mathbb{R}^{1,d-1}) &\longrightarrow \mathsf{AdS}_{d+1}/\mathbb{Z} \\ (\, z \,,\; x \,) &\longmapsto X = \iota(z,x) \equiv \begin{cases} X^{-1} = \ell\,\dfrac{1 + x^2 + z^2}{2z}\,, \\ X^a = \ell\,\dfrac{x^a}{z} \quad \text{for } a \in \{0, \dots, d-1\}\,, \\ X^d = \ell\,\dfrac{1 - x^2 - z^2}{2z}\,. \end{cases} \end{aligned}$$

The metric is then given by

$$\boldsymbol{g} = \frac{\ell^2}{z^2}\left(\mathbf{d}z^2 + \boldsymbol{\eta}\right) \equiv \frac{\ell^2}{z^2}\left(\mathbf{d}z^2 + \boldsymbol{\eta}_{ab}\,\mathbf{d}x^a\,\mathbf{d}x^b\right).$$

Similarly, we can cover the *whole* EAdS by

$$\begin{aligned} \iota\colon (\mathbb{R}_{\geq 0}, \mathbb{R}^{d}) &\longrightarrow \mathsf{EAdS}_{d+1} \\ (\ z\ ,\ x\ ) &\longmapsto X = \iota(z,x) \equiv \begin{cases} X^{-1} = \ell\,\dfrac{1+x^2+z^2}{2z}\,, \\ X^{0} = \ell\,\dfrac{1-x^2-z^2}{2z}\,, \\ X^{a} = \ell\,\dfrac{x^a}{z} \quad \text{for } a \in \{1,\dots,d\}\,. \end{cases} \end{aligned}$$

The metric has the same form as for Lorentzian Poincaré coordinates, just in the Euclidean signature, so it can be seen as a warped product of $\mathbb{R}_{\geq}$ and $\mathbb{R}^{d}$. In particular, this shows $\mathsf{EAdS}_{d+1}$ is conformal to $\mathbb{R}_{\geq} \times \mathbb{R}^{d}$, with the boundary at $z = 0$ being $\mathbb{R}^{d}$ (if we also consider $z = \infty$, we have the whole $\overline{\mathbb{R}_{\geq} \times \mathbb{R}^{d}}$).

These coordinates make explicit the subgroup

$$\mathsf{SO}(1,1) \times \mathsf{ISO}(d) \equiv \mathsf{SO}(1,1) \times \mathsf{Iso}(\mathbb{R}^{d}) \subset \mathsf{Iso}(\mathsf{EAdS})$$

corresponding to the dilatations and Poincaré transformations in $\mathsf{Conf}(\mathbb{R}^{d})$.

## 4.2 QFT in AdS

We will not extensively cover the QFT in AdS, so let us quickly mention some important points:

- Due to the maximal symmetry, similarly as for Minkowski [25], the natural choice of vacuum is invariant under isometry group Iso(AdS).
- The presence of timelike boundary — the light rays reach it in finite global coordinate time — means the Cauchy problem is ill-defined, and we need to additionally fix the boundary conditions [55].
- Particles can be categorized by irreducible representations of $\mathsf{Iso}(\mathsf{AdS}_{d+1})$, and since it is exactly the conformal group $\mathsf{Conf}(\mathsf{Mink}_{d})$, it is not surprising that there is a large overlap between the techniques used in CFT and QFT in AdS. In particular, a lot can be said just using the group/representation theory.
- The AdS background provides a natural IR regulator — it can be interpreted as a "gravitational box", which is however *really* symmetric. This also means the states have a discrete spectrum — similarly to how in CFTs the scaling dimensions of descendants increase discretely — and there is a possibility of slightly negative mass (BF bound) [56]
- Flat space limit of AdS can be interesting, since it would allow us to port CFT methods such as conformal bootstrap to study flat-space $S$-matrix [57].

### Embedding Formalism

It is very convenient to work directly in the embedding space, since the action of isometry group is very simple. Let us mention some important points:

- It is simple to obtain geodesics in AdS. In particular, the timelike geodesics turn out to be periodic with a specific period (depends on the chosen length-scale). This can be understood as a residual effect of obtaining AdS as a universal cover of the space with closed timelike loops (the geodesics are essentially the same for both spaces, we just unwrap them).
- It leads to a simple form of bulk-to-boundary propagator, as we will see in a moment. Moreover, the embedding formalism is also very efficient for calculating bulk-to-bulk propagators of arbitrary spin.

**Remark 4.8** (AdS Isometries, Quadratic Casimir)**.** Generators of AdS isometries are (their action on AdS functions does not depend on their extension to the embedding space, since they are "internal", as they fix the condition $X^2 = -R^2$)

$$\boldsymbol{J}_{AB} \equiv X_A \boldsymbol{\partial}_B - X_B \boldsymbol{\partial}_A \,,$$

giving us the quadratic Casimir operator

$$\begin{aligned} \mathfrak{Cas}_2^{\mathsf{Iso}(\mathsf{EAdS}_{d+1})} &\equiv -\frac{1}{2}\boldsymbol{J}_{AB}\boldsymbol{J}^{AB} = -(X^A\boldsymbol{\partial}^B - X^B\boldsymbol{\partial}^A)X_A\boldsymbol{\partial}_B \\ &= -X^2\boldsymbol{\partial}^2 + (d+1)X\cdot\boldsymbol{\partial} + X^AX^B\boldsymbol{\partial}_A\boldsymbol{\partial}_B \\ &= -X^2\boldsymbol{\partial}^2 + X\cdot\boldsymbol{\partial}(d + X\cdot\boldsymbol{\partial})\,, \end{aligned}$$

where $\boldsymbol{\partial}$ is the derivative acting in the embedding space $\mathbb{R}^{1,d+1}$.

Since action of Iso(EAdS) on the boundary is the same as $\mathsf{Conf}(\mathbb{R}^d)$, we obtain

$$\mathfrak{Cas}_2^{\mathsf{Conf}(\mathbb{R}^d)} = \mathfrak{Cas}_2^{\mathsf{Iso}(\mathsf{EAdS}_{d+1})}\Big|_{\partial\mathsf{EAdS}_{d+1}\simeq\mathbb{R}^d}\,,$$

where the left-hand side is Definition 3.60→p.71. ⌟

**Calculation 4.9** (d'Alembertian in Embedding Coordinates)**.** By foliating the flat space $\mathbb{R}^{2,d}$ with $\mathsf{AdS}_{d+1}$ hyperboloids of radii $R$, we have

$$\boldsymbol{g}_{\mathbb{R}^{2,d}} = -\mathbf{d}R^2 + \boldsymbol{g}_{\mathsf{AdS}_{d+1}} \quad\Longrightarrow\quad \mathfrak{g}^{1/2}_{\mathbb{R}^{2,d}} = \mathfrak{g}^{1/2}_{\mathsf{AdS}_{d+1}} \propto R^{d+1}\,.$$

This gives us the d'Alembertian operator on the flat space as

$$\begin{aligned} \Box_{\mathbb{R}^{2,d}} \equiv \boldsymbol{\partial}^2 &= \frac{1}{\mathfrak{g}^{1/2}}\,\boldsymbol{\partial}_A\Big(\mathfrak{g}^{1/2}\boldsymbol{g}^{AB}_{\mathbb{R}^{2,d}}\boldsymbol{\partial}_B\Big) \\ &= -\frac{1}{R^{d+1}}\,\boldsymbol{\partial}_R\Big(R^{d+1}\boldsymbol{\partial}_R\Big) + \Box_{\mathsf{AdS}_{d+1}} \\ &= -\frac{1}{R^2}\,X\cdot\boldsymbol{\partial}(d + X\cdot\boldsymbol{\partial}) + \Box_{\mathsf{AdS}_{d+1}}\,, \end{aligned}$$

where we used $X\cdot\boldsymbol{\partial} = R\boldsymbol{\partial}_R$.

Comparing with Remark 4.8→p.96, we see that (using $X^2 = -R^2 \equiv -\ell^2$)

$$R^2\Box_{\mathsf{AdS}} = \mathfrak{Cas}_2^{\mathsf{Iso}(\mathsf{AdS})} \simeq \mathfrak{Cas}_2^{\mathsf{Conf}(\partial\mathsf{AdS})}\,.$$

This is a particular case of a more general relation — the d'Alembertian operator on a homogeneous spacetime $\mathcal{M}$ is essentially the Casimir operator of $\mathsf{Iso}(\mathcal{M})$. ⌟

**Remark 4.10** (Mass–Scaling Dimension Relation)**.** Using the previous calculation and considering a free Klein–Gordon equation

$$\Big(-\Box_{\mathsf{AdS}} + m^2\Big)\phi(x) = 0 \quad \Longrightarrow \quad m^2 R^2 = \mathfrak{Cas}_2^{\mathsf{Conf}(\partial\mathsf{AdS})} = \Delta(\Delta - d)\,,$$

we identify the mass-squared $m^2$ (in the units of $1/R \equiv 1/\ell$) with the quadratic Casimir eigenvalue of the corresponding CFT representation on the boundary. Here we utilized Remark 3.104→p.84 for a scalar representation ($J = 0$). ⌟

## Bulk-to-Bulk Two-Point Function/Propagator

Consider a free scalar field on the EAdS background with Euclidean action

$$\begin{aligned}\mathcal{S}[\phi] &= \int_{\mathsf{EAdS}_{d+1}} \mathrm{d}X \left[\tfrac{1}{2}(\boldsymbol{\partial}\phi)^2 + \tfrac{1}{2}m^2\phi^2\right] \\ &\equiv \int_{\mathbb{R}^{1,d+1}} \mathrm{d}^{d+2}X\, \theta(X^{-1}) 2\delta(X^2 + \ell^2)\Big[\tfrac{1}{2}(\boldsymbol{\partial}\phi)^2 + \tfrac{1}{2}m^2\phi^2\Big]\,.\end{aligned}$$

For homogeneous space we can always absorb the possible coupling to the curvature into the mass term.

The free *bulk-to-bulk* two-point function $\langle\phi(X)\phi(Y)\rangle_{\text{free}} \equiv G^{bb}(X,Y)$ is given by the propagator solving the equation

$$\Big(-\Box_X + m^2\Big)G^{bb}(X,Y) = \delta(X,Y)\,.$$

Here, $\Box_X$ is the EAdS d'Alembertian operator acting on the first argument, where both $X$ and $Y$ are understood as points of EAdS.

The symmetry of the problem implies that $G^{bb}$ is a function of only invariant quantities such as $X \cdot Y$ or $\zeta \equiv (X - Y)^2/R^2$, which we call *chordal distance*. From this point on we set $R \equiv \ell \equiv 1$, and all lengths are measured in units of the AdS radius.

**Calculation 4.11** (Free Bulk-to-Bulk Propagator)**.** Using ansatz $G^{bb}(X,Y) \equiv G(\zeta)$, expressing d'Alembertian in embedding coordinates using Calculation 4.9→p.96 and Remark 4.8→p.96 we obtain (it does not make difference with which form of $\zeta \equiv (X-Y)^2 = -2(1 + X\cdot Y)$ we work, we just need to be consistent, and set $X^2 = Y^2 \equiv -1$ only at the end)

$$\begin{aligned}\Box_X G^{bb}(X,Y) &= \Big(\boldsymbol{\partial}_X^2 + X\cdot\boldsymbol{\partial}_X(d + X\cdot\boldsymbol{\partial}_X)\Big)G(\zeta) \\ &= \boldsymbol{\partial}_X\cdot[(-2Y)G'(\zeta)] + (d + X\cdot\boldsymbol{\partial}_X)\Big[(-2X\cdot Y)G'(\zeta)\Big] \\ &= (-2Y)^2 G''(\zeta) + (d+1)(-2X\cdot Y)G'(\zeta) + (-2X\cdot Y)^2 G''(\zeta) \\ &= \Big(-4 + (2+\zeta)^2\Big)G''(\zeta) + (d+1)(2+\zeta)G'(\zeta) \\ &= z(z-1)G''(z) + (d+1)(z - \tfrac{1}{2})G'(z)\,,\end{aligned}$$

where in the last step we performed substitution $z \equiv -\frac{\zeta}{4}$. Thus, we have obtained

↳ an equation in the form of the hypergeometric equation (for nonzero $z$)

$$\begin{aligned}
&\Big[-\Box_X + \Delta(\Delta - d)\Big] G(z) = 0 \\
&\implies \quad z(1-z)G''(z) + \left(\tfrac{d+1}{2} - (d+1)z\right)G'(z) - \Delta(d-\Delta)G(z) = 0 \\
&\qquad\quad\; z(1-z)G''(z) + \Big(c - (a+b+1)z\Big)G'(z) - abG(z) = 0\,,
\end{aligned}$$

where $a = \Delta$, $b = d - \Delta$, $c = \frac{d+1}{2}$, or same with $\Delta \longleftrightarrow d - \Delta$, which leaves the mass/conformal Casimir eigenvalue unchanged. By requiring the singularity at $X \to Y$ ($\zeta \to 0$) to match with the flat space propagator singularity, we obtain the solution

$$G^{bb}_{\Delta}(X,Y) \equiv G_{\Delta}(\zeta) \equiv \frac{\mathfrak{C}_{\Delta}}{\zeta^{\Delta}}\, {}_2\mathsf{F}_1\left[\begin{matrix}\Delta,\ \Delta - \frac{d}{2} + \frac{1}{2} \\ 2\Delta - d + 1\end{matrix}\;\middle|\; -\frac{4}{\zeta}\right]$$

where the normalization is

$$\mathfrak{C}_{\Delta} = \frac{\Gamma(\Delta)}{2\pi^{\frac{d}{2}}\,\Gamma\!\left(\Delta - \frac{d}{2} + 1\right)}\,.$$

⌟

**Remark 4.12** (Free Theory, Perturbation Theory)**.** Calculation of correlators in perturbation theory follows the general story as discussed in Section 1.1 →p.10 — we can use same Feynman diagram techniques, just considering EAdS specific differential operators (and corresponding propagators) and integrals. ⌟

## Boundary Limit

Since the correlators are invariant under $\mathsf{Iso}(\mathsf{EAdS}_{d+1})$ act on the boundary as $\mathsf{Conf}(\mathbb{R}^d)$, by taking their appropriate asymptotic/boundary limit, we obtain *boundary correlators* satisfying the $\mathsf{CFT}_d$ axioms (apart from the presence of the energy–momentum tensor).

As shown in Remark 4.10 →p.97, the (physical) mass of the bulk field $m_\phi$ is related to the scaling dimension of the corresponding operator in the boundary theory. To avoid certain subtleties, in the following we consider the positive "Dirichlet" branch

$$\Delta_\phi = \Delta_+ \equiv \frac{d}{2} + \sqrt{\frac{d^2}{4} + m_\phi^2}\,,$$

which in Poincaré coordinates $(z, y) \in \mathbb{R}_{\geq} \times \mathbb{R}^d$ corresponds to the boundary condition $\phi \sim z^{\Delta_+}$ as $z \to 0$ (such choice is consistent with the EAdS isometries).

To have well-defined boundary correlators, we define the dual operator by taking an appropriately scaled boundary limit of the bulk field.

**Definition 4.13** (Dual Boundary Operator)**.** The boundary operator $\mathcal{O}_\phi$ dual to the bulk field $\phi$ is defined as the limit

$$\mathcal{O}_\phi(P) \equiv \frac{1}{\sqrt{\mathfrak{C}_{\Delta_\phi}}} \lim_{s\to\infty} s^{\Delta_\phi} \phi\Big(X \equiv sP + O(1/s)\Big) \,,$$

where the bulk point $X \in \mathsf{EAdS}$ approaches the boundary point $P \in \partial\mathsf{EAdS}$.

**Remark 4.14** (Bulk-to-Boundary Propagator)**.** Taking the appropriate boundary limit of one operator in the bulk-to-bulk propagator Calculation 4.11 →p.97 we obtain the *bulk-to-boundary propagator* $G^{b\partial}_{\Delta_\phi}$, which has a much simpler form

$$\Big\langle \phi(X)\mathcal{O}_\phi(P) \Big\rangle_{\text{free}} \equiv G^{b\partial}_{\Delta_\phi}(X,P) = \frac{\sqrt{\mathfrak{C}_{\Delta_\phi}}}{(-2X\cdot P)^{\Delta_\phi}} \,,$$

since taking $Y \simeq sP$ to the boundary by $s\to\infty$, we have $\zeta \simeq X\cdot(sP)\to\infty$, so the hypergeometric function becomes just 1. Note how the boundary limit scaling in the definition of $\mathcal{O}_\phi$ precisely cancels the scaling of $\zeta^{-\Delta_\phi}$ in $G^{bb}_{\Delta_\phi}$. ⌟

**Remark 4.15** (Boundary-to-Boundary Propagator)**.** Sending also the second operator to the boundary, one finds the *boundary-to-boundary propagator* $G^{\partial\partial}_{\Delta_\phi}$ with the usual form of the 2-point function in flat-space CFT of scalar operator with scaling dimension $\Delta_\phi$, that is

$$\Big\langle \mathcal{O}_\phi(P_1)\mathcal{O}_\phi(P_2) \Big\rangle_{\text{free}} \equiv G^{\partial\partial}_{\Delta_\phi}(P_1,P_2) = \frac{1}{(-2P_1\cdot P_2)^{\Delta_\phi}} = \frac{1}{|y_1-y_2|^{2\Delta_\phi}} \,,$$

where $y_1, y_2 \in \mathbb{R}^d$ are flat-space coordinates corresponding to the points $P_1, P_2 \in \partial\mathsf{EAdS}_{d+1}$. We now see that the normalization in the definition of the boundary operator was chosen such that the boundary 2-point function is conventionally normalized. ⌟

# 5 O($N$) Model in Anti–de Sitter

**Abstract.** We determine the scaling dimensions in the boundary $\mathsf{CFT}_d$ corresponding to the $\mathsf{O}(N)$ model in $\mathsf{EAdS}_{d+1}$. The CFT data accessible to the 4-point boundary correlator of fundamental fields are extracted in $d = 2$ and $d = 4$, at a finite coupling, and to the leading nontrivial order in the $1/N$ expansion. We focus on the non-singlet sectors, namely the anti-symmetric and symmetric traceless irreducible representations of the $\mathsf{O}(N)$ group, extending the previous results that considered only the singlet sector. Studying the non-singlet sector requires an understanding of the crossed-channel diagram contributions to the $s$-channel conformal block decomposition. Building upon an existing computation, we present general formulas in $d = 2$ and $d = 4$ for the contribution of a $t$-channel conformal block to the anomalous dimensions of $s$-channel double-twist operators, derived for external scalar operators with equal scaling dimensions. Up to some technical details, this eventually leads to the complete picture of $1/N$ corrections to the CFT data in the interacting theory.

## 5.1 Introduction

Recently, various QFT models have been studied in AdS (or dS) spacetime at finite coupling. Rather than the usual weak coupling expansion, the large $N$ expansion together with AdS/CFT intuition was employed as an alternative handle to perform explicit calculations. This movement started with $\mathsf{O}(N)$ and Gross–Neveu Model [59], and was continued also for scalar QED [60].

While these studies have addressed key questions such as the phase structure, exact propagators, and the boundary correlators (4-point functions) in the large $N$ limit, they have primarily focused on the *singlet spectrum* of the CFT on the boundary.

Largely unexplored is the *non-singlet spectrum* corresponding to operators transforming in nontrivial representations of the global internal (large $N$) symmetry group. We will thus primarily focus on extending the analysis of the $\mathsf{O}(N)$ model in AdS at finite coupling also to the non-singlet sector — namely the *rank-2 anti-symmetric* and *symmetric traceless* representations of $\mathsf{O}(N)$ group.

Building on the already established results for the singlet sector, we were able to extract anomalous dimensions of operators in the non-singlet sector by utilizing the 6j–symbols (also known as the *crossing kernel*) [61]. It appears that these methods (in particular the Lorentzian inversion formula) missed some contributions to anomalous dimensions of scalar (spin zero) operators. Further discussion of this subtlety can be found in Section 5.5→p.141.

More broadly, there are essentially three main pieces of motivation in this line of research. They were nicely summarized in the Introduction sections of [59] [60], which the reader is invited to check out for more details. Even after receiving some attention, numerous issues require deeper investigation:

- Free theory in AdS bulk corresponds to GFF/MFT at the asymptotic boundary, that is a CFT with all correlation functions simply given just by products of 2-point functions. While a great deal has been said about small deformations in a bulk coupling, less explored are large deformations of the CFT data away from MFT. Large $N$ techniques are helpful in this respect, since they retain more of the nonlinear structure of the exact theory compared to the ordinary lowest-order perturbation theory. As such, they can shed some light on interesting phenomena like appearance of new operators in the spectrum (corresponding to the existence of bound states), AdS analogs of resonances, or level crossing.

- For a QFT in $\mathsf{AdS}_{d+1}$ with a nontrivial phase structure, it is interesting to determine how $\mathsf{CFT}_d$ data of the dual theory at the boundary change, in particular as borderlines separating different phases of the bulk theory are crossed (for instance by deforming the critical bulk theory at a given fixed point of the renormalization group by a relevant operator, $S_{\text{CFT}} \mapsto S_{\text{CFT}} + g \int \mathcal{O}_\Delta^{(\text{bulk})}$, triggering an RG-flow to the new desired phase). Bulk couplings $g$ are generically dimensionful, as exactly marginal couplings corresponding to families of bulk CFTs are very rare. Since AdS provides a natural length scale, its radius $\ell$, it is convenient to form dimensionless combinations $\widehat{g} = g\ell^{\Delta-(d+1)}$. Thus the bulk RG-flow of $g$ can be thought of as a flow in the AdS radius $\ell$.

  The framework for studying it, with a particular emphasis on the flat space limit $\ell \to \infty$, was laid out in [57]. More detailed studies about constraining the bulk RG-flow by boundary CFT axioms followed. By now there is already a significant amount of literature, for an invitation to this topic with further references included, see [62] [63] [64]. The role of certain 3-point correlation functions (AdS form factors) of two boundary operators with one bulk field in the change of the AdS length scale $\ell$ was investigated in [65]. The important special case of the bulk operator being the stress tensor was done in [66].

  A distinct instance of the above discussion occurs when the flow is initiated in a free conformal bulk theory (for example a conformally coupled scalar) corresponding to a MFT at the boundary, which ubiquitously possesses higher spin currents (all the double-twist families saturate unitarity bound and form thus short conformal multiplets). Then one can naturally wonder about their fate once interactions are turned on. General theorems constraining their existence in arbitrary interacting CFTs were given in [67] [68]. On the other hand, one can ask a similar question, just not at the boundary but instead in the bulk. A complete breaking of bulk higher spin currents by interactions in $\mathsf{AdS}_2$ was analyzed in [69].

- In the special case when the bulk theory in $\mathsf{AdS}_{d+1}$ is critical (becoming a $\mathsf{CFT}_{d+1}$), one can perform (in the Euclidean signature) a Weyl symmetry transformation from EAdS to flat half-space $\mathbb{R}^d \times \mathbb{R}_{\geq}$. This is easily done using the Poincaré coordinates, which cover EAdS globally. The CFT with a boundary — $\mathsf{BCFT}_{d+1}$ — on this flat half-space can thus be studied using AdS methods, whose recent advances have made such approach rather efficient.

  Moreover, one can generalize it from boundaries (BCFT) to defects (DCFT). A Weyl equivalence between an $(n-1)$-dimensional defect in $\mathbb{R}^{n+m}$ and $\mathsf{AdS}_n \times \mathbb{S}^m$ allowed [70] to use results of [59] for the singlet spectrum to extract interesting DCFT data. More concretely, they studied the critical O($N$) model in the presence of a localized magnetic external field understood as an 1-dimensional defect, thus $\mathsf{AdS}_2$ was relevant. Such data is important to understand phase transitions of real-world systems.

- Finally, there are attempts to replace the LSZ axioms for the flat-space $S$-matrix by the flat-space limit of asymptotic boundary observables in AdS [71] [72] [73] [74]. Asymptotic boundary correlators for massive QFTs in $\mathsf{AdS}_{d+1}$ obey $\mathsf{CFT}_d$ axioms — except for the existence of a stress tensor — that are mathematically more rigorous than the current ones for the $S$-matrix in flat space. These asymptotic correlators are by definition holographic as is the flat-space $S$-matrix, and reduce to it in the flat-space limit (in a sense that still requires a more rigorous definition).

  Since the curvature of AdS also acts as an IR regulator, it could potentially cure the possible IR divergences, which make the flat $S$-matrix an ill-defined object [75] [76]. In practice, one often has to recourse to IR safe inclusive observables, which however require taking the experimental setup into account, and are thus not clean theoretical observables. More discussion about the IR issues of the flat $S$-matrix and another attempt how to solve it can be found in [77].

Our work focuses on a specific topic in the above vast landscape. We examined large deformations of the free MFT spectrum for the O($N$) model in the unbroken phase. While primary outcome is the computation and analysis of the non-singlet spectrum, along the way we also obtained some additional results.

First, we completed the singlet CFT data by extracting the OPE coefficients.

Furthermore, we discussed extension of the spectrum (even the singlet one) to $d = 4$, where the necessary regularization complicates the picture. To partly deal with the associated renormalization scheme ambiguity — finite part of a counterterm — we resorted to the critical bulk theory describing an ordinary phase transition.

Formally extending the critical point beyond its upper critical dimension — also considered in Section 3.1 and Figure 6 [78] — we discovered an intriguing pattern in the singlet spectrum. It is related with the appearance of emergent operators at a strong enough coupling, which are not connected with the usual MFT-type spectrum found at weak coupling. Based on this analysis, a formula for the critical singlet spectral function in all even (boundary) dimensions reproducing the desired critical spectrum was proposed.

The next natural step would be to deal with the broken phase and especially with the critical theory in the bulk, whose analysis could result in extraction of non-singlet DCFT data along the lines of [70]. Hopefully, our results can add a concrete small piece to the above complex mosaic.

**Outline of the Paper.** The subsequent sections are organized as follows.

In Section 5.2 →p.104 we define the O($N$) model, mainly focusing on its formulation suitable for a systematic large $N$ expansion. Then we introduce the main observable — the 4-point boundary correlator of fundamental fields $\phi^{\circ}$ in AdS. The rest of the section is devoted to fixing our conventions, particularly those for OPE channels, and some comments about renormalization scheme are made as well.

In Section 5.3 →p.114 we will review relevant generalities of CFTs, focusing on the extraction of the CFT data from the 4-point correlator. In particular, we will present general formulas (in $d = 2$ and $d = 4$) for the contribution of a single $t$-channel conformal block to anomalous dimensions of $s$-channel double-twist operators, applicable for arbitrary twists and spins.

These formulas will be utilized in Section 5.4 →p.137 to calculate the leading $1/N$ contributions to the non-singlet scaling dimensions of the O($N$) model in AdS at finite coupling. This computation requires the singlet spectrum as an input, whose properties are summarized in the previous parts of Section 5.4 →p.123, together with the decomposition of the 4-point correlator into O($N$) irreducible representations. Criticality in the bulk is also briefly discussed in Section 5.4 →p.134, which suggests a possible extension beyond its upper critical dimension.

In Section 5.5 →p.138 we present the results for the non-singlet spectrum, mostly in the form of various plots. The main one is the twist–spin plot showing a characteristic organization of the spectrum into Regge trajectories. Various limits are studied, specifically a cross-check regarding the large spin asymptotics.

An executive summary and synthesis of results is given in Section 5.6 →p.151, together with possible future directions.

Various detailed computations and the implementation of main formulas can be found in the accompanying NOTEBOOK — links to a GitHub repository [jdujava/ONinAdS].

## 5.2 Review of the O($N$) Model in AdS

We start by introducing the O($N$) model in Section 5.2 →p.104. Some general features are reviewed, in particular how its large $N$ expansion enables calculations at finite coupling. In Section 5.2 →p.108 we define our main observable — the boundary 4-point correlator in AdS — which can be viewed as an observable in the CFT living on the boundary of AdS. In the following Section 5.2 →p.112 we briefly describe how the spectral representation can be utilized to express the boundary correlator in a form particularly suitable for the study of the CFT spectrum.

While most of the material in this section is well-known for experts in the field, with some more specific details thoroughly discussed already in [59], we believe that an alternative account (stressing some additional points) can be beneficial. In the meanwhile, we also fix the notation and conventions that will be used throughout the paper.

### Generalities of the O($N$) Model

The Euclidean action of the O($N$) model on a $(d+1)$–dimensional Riemannian manifold $\mathcal{M}$ takes the form (we implicitly contract both spacetime and internal indices)

$$\mathcal{S}[\phi^\bullet] = \int_{\mathcal{M}} \mathrm{d}^{d+1}x \, \sqrt{g} \left[ \frac{1}{2}(\partial\phi^\bullet)^2 + \frac{1}{2}m^2(\phi^\bullet)^2 + \frac{\lambda}{2N}\left((\phi^\bullet)^2\right)^2 \right], \tag{5.1}$$

where $\phi^\bullet = (\phi^1, \ldots, \phi^N)$ is an $N$-tuple of real scalar fields transforming in the vector representation of the global internal O($N$) symmetry group. In the case of maximally symmetric spacetimes — in our case EAdS — the possible coupling to the curvature can be absorbed into the mass term $m^2$. Results for Lorentzian AdS can be obtained by analytical continuation, so we will use EAdS/AdS interchangeably.

The interaction term was introduced in such a way that the model admits a large $N$ expansion with the $\lambda$ coupling fixed and finite. We will review details relevant to us. It was also outlined in Section 2 [59], and more thorough accounts in flat space can be found in a specialized review [79] or in the original paper [80].

Feynman rules for (5.1) assign a propagator to each internal line, and the interaction vertex can be diagrammatically represented as (up to numerical factors)

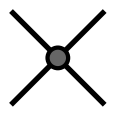 $\sim \frac{\lambda}{N}$ 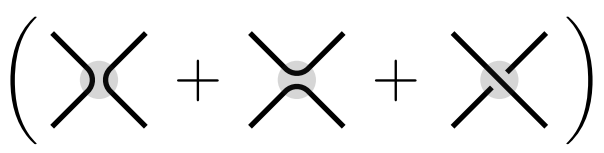 , (5.2)

where the connected lines on the right-hand side indicate Kronecker deltas in the indices of the corresponding $\phi$ fields. As usual, all internal vertices throughout the paper are to be integrated over $\mathcal{M}$ with the geometric measure including the $\sqrt{g}$ density.

Order of $1/N$ in a given diagram is not simply given just by the number of interaction vertices, since after their expansion (5.2) a number of closed index loops can form. Each such index loop carries an additional factor of $N$, since all $\{\phi^i\}_{i=1}^N$ circulating in the loop contribute equally.

**Hubbard–Stratonovich Transformation.** For the purposes of large $N$ expansion, it turns out more efficient to proceed via a reparametrization of the interaction term under the path integral by introducing an auxiliary Hubbard–Stratonovich field $\sigma$ — equivalent on-shell (up to normalization) to the composite operator $(\phi^{\bullet})^2$ — thus obtaining the action

$$\mathcal{S}_{\mathrm{HS}}[\phi^{\bullet},\sigma] = \int_{\mathcal{M}} \mathrm{d}^{d+1}x\,\sqrt{g}\left[\frac{1}{2}(\partial\phi^{\bullet})^2 + \frac{1}{2}m^2(\phi^{\bullet})^2 - \frac{1}{2\lambda}\sigma^2 + \frac{1}{\sqrt{N}}\sigma(\phi^{\bullet})^2\right]. \quad (5.3)$$

The interaction vertex (5.2) is now substituted by the following rules

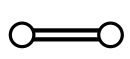 $\equiv -\lambda\mathbf{1}\,,$ 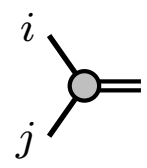 $\equiv \frac{2}{\sqrt{N}}\,\delta^{ij}\,,$ (5.4)

first representing the free $\sigma$-propagator, and second the $\sigma\phi^2$ vertex. We suppressed the position dependence of 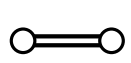 $\equiv \langle\sigma\sigma\rangle_{\mathrm{free}}$ by writing it — up to factor $(-\lambda)$ — as an operator $\mathbf{1}$ acting on functions through the convolution with the kernel $\mathbf{1}(x,y) \equiv \delta(x,y)$. Here we use the $\delta$-distribution normalized with respect to the geometric measure, so the $\mathbf{1}$ operator is really acting on functions as an identity.

**Effective Action.** Now, we want to "integrate out the loops" of the $\phi$ fields, which contribute at the leading order in the large $N$ expansion. Since our new Lagrangian $\mathcal{L}_{\mathrm{HS}}$ is only quadratic in $\phi$, the $\phi$-loop subdiagrams have the form of a loop with arbitrary number of $\sigma$ lines attached to it, thus inducing non-local contributions to the 1PI effective interaction vertices between $\sigma$ fields of the form

$\boldsymbol{\Gamma}^{\text{1-loop}}[\sigma] = \sum_{n=0}^{\infty}$ 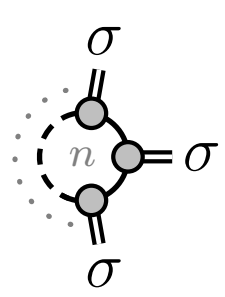

$$= \mathrm{const.} - N\sum_{n=1}^{\infty}\frac{(-1)^n}{2n}\,\mathrm{Tr}\left[\left(\frac{1}{(-\Box+m^2)\mathbf{1}}\circ\frac{2}{\sqrt{N}}\,\sigma\right)^n\right]$$

$$= \frac{N}{2}\,\mathrm{Tr}\ln\!\left((-\Box+m^2)\mathbf{1}+\frac{2}{\sqrt{N}}\,\sigma\right) = -\ln\mathrm{Det}^{-\frac{N}{2}}\!\left((-\Box+m^2)\mathbf{1}+\frac{2}{\sqrt{N}}\,\sigma\right). \quad (5.5)$$

The prefactor of $N$ comes from the already performed trace over the indices in the closed loop. Also other combinatorial/symmetry factors are properly accounted for. This can be alternatively seen as a functional determinant coming from Gaussian integration over $\phi$ rewritten as contribution to the Euclidean effective action.

Note, that in (5.5) we understand the argument of $\operatorname{Tr}\ln(\cdot)$ as an operator acting on functions, which we can represent as a convolution with an associated kernel. In particular, the kernel corresponding to $(-\Box + m^2)\mathbf{1}$ is $[(-\Box + m^2)\mathbf{1}](x,y) \equiv (-\Box_x + m^2)\delta(x,y)$, and the action of pointwise multiplication with $\sigma$ has the kernel $\sigma(x,y) \equiv \sigma(x)\delta(x,y)$.

We can now introduce a (Euclidean) effective action

$$\begin{aligned}\Gamma[\phi^{\circ},\sigma] &\equiv \Gamma^{\text{0-loop}}[\phi^{\circ},\sigma] + \Gamma^{\text{1-loop}}[\sigma] \equiv \mathcal{S}_{\text{HS}}[\phi^{\circ},\sigma] + (5.5)\\ &= \int_{\mathcal{M}} \mathrm{d}^{d+1}x\,\sqrt{g}\left[\frac{1}{2}(\partial\phi^{\circ})^2 + \frac{1}{2}m^2(\phi^{\circ})^2 - \frac{1}{2\lambda}\sigma^2 + \frac{1}{\sqrt{N}}\sigma(\phi^{\circ})^2\right]\\ &\qquad + \frac{N}{2}\operatorname{Tr}\ln\left((-\Box + m^2)\mathbf{1} + \frac{2}{\sqrt{N}}\sigma\right),\end{aligned} \tag{5.6}$$

which is indeed the effective 1PI action including all of the leading $1/N$ contributions. Proceeding further, we should first determine the ground state of the theory by extremizing (preferably minimizing) the effective potential, given by the effective action $\Gamma$ evaluated for constant classical fields and divided by the volume of the spacetime which factorizes. Afterward, one expands the effective action in terms of the shifted fields $\delta\phi$ and $\delta\sigma$ which vanish at the found extremum.

The coefficients of this expansion are the exact propagators (quadratic terms) and 1PI vertices (higher order terms) in the leading $1/N$ approximation. The linear “tadpole” terms in $\delta\phi$ and $\delta\sigma$ are absent by virtue of expanding around the extremum of the effective potential. Observables can be then calculated using these leading forms of exact propagators and 1PI vertices, but all diagrams which contain $\phi$-loops should be omitted, since those are already included in the effective $\sigma$ self-interactions.

Alternatively, this can be viewed as a large $N$ saddle point analysis, since all terms in the effective action are of the same order of magnitude in the neighborhood of the effective potential extremum — they are all of order $O(N)$.

**Phase Structure.** Phases of the O($N$) model have been thoroughly investigated on flat space of dimension $D \equiv d+1$. The dynamics of the theory significantly differs for the ranges $2 < D < 4$ and $D \geq 4$. Since we will do computations for $d = 2 \Leftrightarrow D = 3$ ($\mathsf{AdS}_3$) and $d = 4 \Leftrightarrow D = 5$ ($\mathsf{AdS}_5$), we need to briefly summarize both cases, in order to clearly specify which phase we treat in this work.

In the first case $2 < d+1 < 4$, the flat-space theory is asymptotically free. The free unstable UV fixed point $\mathsf{CFT}_{\mathrm{UV}}$ has two relevant operators (the mass term and the interaction term) that trigger an RG flow. When appropriately tuned, it ends in a semi-stable interacting (the strength increases as $D$ decreases from 4) IR fixed point — the Wilson–Fisher $\mathsf{CFT}_{\mathrm{IR}}$ [81] — with one relevant operator corresponding to the mass term. The RG flow triggered by it, depending on the sign of the deformation, leads either to a trivially gapped phase with unbroken global $\mathsf{O}(N)$ symmetry or to spontaneous symmetry breaking $\mathsf{O}(N) \to \mathsf{O}(N-1)$, whose low energy dynamics is governed by a non-linear sigma model of $(N-1)$ Goldstone bosons (with a free fixed point $\mathsf{CFT}_{\mathrm{IR}}$ in the deep IR).

In the second case $d+1 \geq 4$, the flat-space theory is IR free. For large enough $N$, a perturbatively unitary UV fixed point was found in [82]. The authors proposed a UV completion by a cubic theory in $D = 6$ with an IR fixed point, whose equivalence to the UV fixed point of the $\mathsf{O}(N)$ model was checked using $\varepsilon$-expansions. Later, it was shown [83] that this fixed point is non-unitary beyond perturbation theory. In particular, scaling dimensions receive exponentially suppressed (in large $N$) imaginary parts caused by instantons existing in either formulation of the fixed point. This phenomenon is known as complex CFT [84]. The gapped phase as well as the spontaneously broken phase exist as in the previous case.

The phase structure of $\mathsf{O}(N)$ model in $\mathsf{AdS}_{d+1}$ for dimensions ranging in $2 < d+1 < 4$ was analyzed in Section 3 [59], and differences from flat space were summarized there. For example, in $d = 2$ ($\mathsf{AdS}_3$), there exists a region of the parameter space where both the unbroken and broken phase seem to coexist. In this work, we treat almost exclusively the simplest of the phases — the unbroken phase with the $\mathsf{O}(N)$ symmetry preserved. We leave the broken phase and deeper analysis of the critical point for future work.

Moreover, the final results will be mainly presented for $d = 2$ ($\mathsf{AdS}_3$), where formulas simplify enough to actually perform the calculations. Nevertheless, at a certain point a numerical evaluation is necessary. In addition $d = 4$ ($\mathsf{AdS}_5$) will be also included, which however still requires an independent phase structure analysis.

Choice of the unbroken $\mathsf{O}(N)$–symmetric phase specializes to the vicinity of the saddle point $\left(\phi^{\bullet}(x) = 0,\, \sigma(x) = \sqrt{N}\sigma_\star\right)$, where $\sigma_\star$ is a constant parametrizing the vacuum expectation value (VEV) of the $\sigma$ field as $\langle\sigma\rangle \equiv \sqrt{N}\sigma_\star$. The expansion of the effective action $\Gamma$ (5.6) around this saddle point just gets rid of the $\sigma$-tadpole term, and the $\phi$-field mass-squared is shifted to $m_\phi^2 \equiv m^2 + 2\sigma_\star$. Since everything else stays the same, from now on we take all "free" $\phi$-field propagators with the effective mass-squared $m_\phi^2$, and use $\sigma$ to refer mostly just to the deviation $\delta\sigma \equiv \sigma - \langle\sigma\rangle$.

Finally, let us remark that the unbroken phase is present when the (effective)

mass is above the Breitenlohner–Freedman (BF) bound $m_\phi^2 > -\frac{d^2}{4}$ [56] [85], a point which will be also discussed in Section 5.2→p.108.

**Exact $\sigma$-Propagator.** Since the interaction terms in the expansion of $\boldsymbol{\Gamma}$ are of the order $O(1/\sqrt{N})$ or higher, the leading $O(1)$ form of the exact $\sigma$-propagator can be obtained by inverting the kernel of the quadratic $\sigma$-part of $\boldsymbol{\Gamma}$ expanded around the saddle point, or equivalently by summing up the geometric series

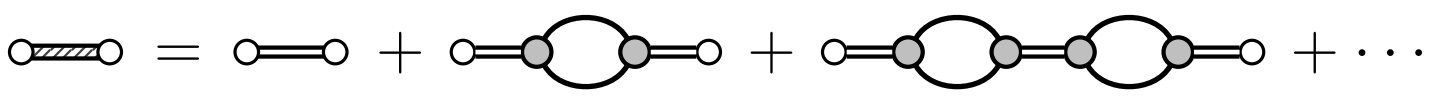

$$= (-\lambda\mathbf{1}) + (-\lambda\mathbf{1}) \circ 2B \circ (-\lambda\mathbf{1}) + \cdots = -\lambda \sum_{n=0}^{\infty} (-2\lambda B)^n = -\left[\frac{\mathbf{1}}{\lambda} + 2B\right]^{-1}, \tag{5.7}$$

where $B$ is the kernel corresponding to the (half of) "bubble" diagram given by

$$B(x,y) \equiv \frac{1}{2}\; x \quad y \equiv \left[\frac{1}{(-\Box + m_\phi^2)\mathbf{1}}(x,y)\right]^2 . \tag{5.8}$$

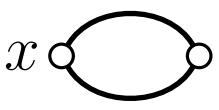

Note that for each bubble in (5.7) the $(1/\sqrt{N})^2$ from the vertices canceled out with the $N$ coming from the index loop. The diagram in (5.8) is really a free correlator of a scalar composite field $\phi^2 \equiv :(\phi^i)^2:$ being the normal-ordered square of $\phi^i$ for any fixed index $i$, that is $\langle \phi^2(x)\phi^2(y)\rangle_{\text{free}} = 2G_\phi(x,y)^2 \equiv 2B(x,y)$. The factor of 2 comes from two different ways of contracting legs at both vertices.

From the point of view of the initial action (5.1), the $\sigma$-propagator resums an infinite class of Feynman diagrams contributing in the leading order of $1/N$ expansion. For example, the leading $s$-channel connected contribution to the 4-point function of $\phi$ fields is given by

$$\sum_{n=0}^{\infty} \quad = \quad \sim \frac{1}{N} . \tag{5.9}$$

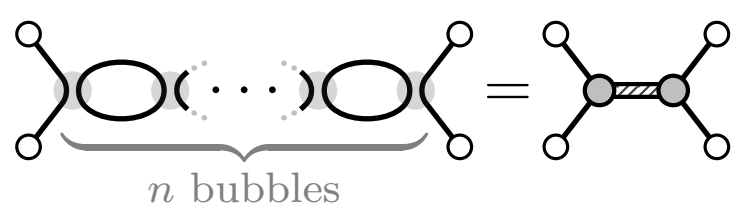

To increase clarity, from now on, we will explicitly write out the $N$-dependence in front of the diagrams instead of including it in the diagrams themselves.

## CFT on the Boundary of AdS

Now we will specialize to the case of $\mathsf{AdS}_{d+1}$ spacetime. The isometries of AdS act on its asymptotic/conformal boundary as conformal transformations, and by performing an appropriate boundary limit of the correlators one obtains *boundary correlators* satisfying the $\mathsf{CFT}_d$ axioms (apart from the presence of the stress tensor operator) [86].

The boundary operator $\mathcal{O}_\phi^i \equiv \mathcal{O}_{\phi^i}$ corresponding (or *dual*) to the scalar field $\phi^i$ has a scaling dimension $\Delta_\phi$ satisfying the equation $m_\phi^2 = \Delta_\phi(\Delta_\phi - d)$ [87]. We measure all dimensionful quantities in units of the AdS radius $\ell$, which we throughout set to $\ell \equiv 1$.

To avoid certain subtleties, in the following we consider the positive "Dirichlet" branch

$$\Delta_\phi = \Delta_+ \equiv \frac{d}{2} + \sqrt{\frac{d^2}{4} + m_\phi^2}\,, \tag{5.10}$$

which in Poincaré coordinates $(z, y) \in \mathbb{R}_\geq \times \mathbb{R}^d$ corresponds to the boundary condition $\phi \sim z^{\Delta_+}$ as $z \to 0$. The dual operator is then defined by the boundary limit

$$\mathcal{O}^i_\phi(P) \equiv \frac{1}{\sqrt{\mathfrak{C}_{\Delta_\phi}}} \lim_{s\to\infty} s^{\Delta_\phi} \phi^i\Big(X \equiv sP + O(1/s)\Big)\,, \tag{5.11}$$

where $P$ is a point on the boundary of $\mathsf{EAdS}_{d+1}$ (a future directed null vector in $\mathbb{R}^{1,d+1}$) and $X$ is a point of EAdS in the embedding formalism [51] approaching $P$ in the limit $s \to \infty$, where the $O(1/s)$ term enables the EAdS condition $X^2 = -\ell^2 \equiv -1$ to be satisfied while $P^2 = 0$. The normalization constant $\mathfrak{C}_{\Delta_\phi}$ is given by

$$\mathfrak{C}_{\Delta_\phi} = \frac{\Gamma(\Delta_\phi)}{2\pi^{\frac{d}{2}}\,\Gamma\Big(\Delta_\phi - \frac{d}{2} + 1\Big)}\,. \tag{5.12}$$

Free propagator of $\phi^\bullet$ fields in EAdS is given by the *bulk-to-bulk propagator* $G^{bb}_{\Delta_\phi}$ expressible in terms of the chordal distance $\zeta(X, Y) \equiv (X - Y)^2 = -2 - 2X \cdot Y$ between two points in AdS

$$\begin{aligned}\Big\langle \phi^i(X)\phi^j(Y)\Big\rangle_{\text{free}} &\equiv \delta^{ij} G^{bb}_{\Delta_\phi}(X, Y) \\ &= \delta^{ij}\, \frac{\mathfrak{C}_{\Delta_\phi}}{\zeta(X, Y)^{\Delta_\phi}}\, {}_2\mathsf{F}_1\left[\begin{matrix}\Delta_\phi,\ \Delta_\phi - \frac{d}{2} + \frac{1}{2} \\ 2\Delta_\phi - d + 1\end{matrix}\ \middle|\ -\frac{4}{\zeta(X, Y)}\right].\end{aligned} \tag{5.13}$$

Taking the appropriate boundary limit of one operator — in the sense of (5.11) — one obtains the *bulk-to-boundary propagator* $G^{b\partial}_{\Delta_\phi}$, which has a much simpler form

$$\Big\langle \phi^i(X)\mathcal{O}^j_\phi(P)\Big\rangle_{\text{free}} \equiv \delta^{ij} G^{b\partial}_{\Delta_\phi}(X, P) = \frac{\sqrt{\mathfrak{C}_{\Delta_\phi}}\,\delta^{ij}}{(-2X \cdot P)^{\Delta_\phi}}\,. \tag{5.14}$$

Sending also the second operator to the boundary, one finds the *boundary-to-boundary propagator* $G^{\partial\partial}_{\Delta_\phi}$ with the usual form of the 2-point function in flat-space CFT of scalar operator with scaling dimension $\Delta_\phi$, that is

$$\Big\langle \mathcal{O}^i_\phi(P_1)\mathcal{O}^j_\phi(P_2)\Big\rangle_{\text{free}} \equiv \delta^{ij} G^{\partial\partial}_{\Delta_\phi}(P_1, P_2) = \frac{\delta^{ij}}{(-2P_1 \cdot P_2)^{\Delta_\phi}} = \frac{\delta^{ij}}{|y_1 - y_2|^{2\Delta_\phi}}\,, \tag{5.15}$$

where $y_1, y_2 \in \mathbb{R}^d$ are flat-space coordinates corresponding to the points $P_1, P_2 \in \partial\mathsf{EAdS}_{d+1}$. We now see that the normalization in the definition of the boundary operator (5.11) was chosen such that the boundary 2-point function is conventionally normalized.

**Main Observable — Boundary 4-Point Correlator.** Let us define the main observable that we will use to probe the spectrum of the $\mathsf{CFT}_d$ emerging at the asymptotic/conformal boundary of $\mathsf{AdS}_{d+1}$. It is the boundary limit of the 4-point correlator $\langle\phi\phi\phi\phi\rangle$ of fundamental fields $\phi^{\bullet}$ in the vector representation of the global $\mathsf{O}(N)$ symmetry, that is

$$\left\langle \mathcal{O}_1^i \mathcal{O}_2^j \mathcal{O}_3^k \mathcal{O}_4^l \right\rangle \equiv \left\langle \mathcal{O}_\phi^i(P_1)\, \mathcal{O}_\phi^j(P_2)\, \mathcal{O}_\phi^k(P_3)\, \mathcal{O}_\phi^l(P_4) \right\rangle \equiv \tag{5.16}$$

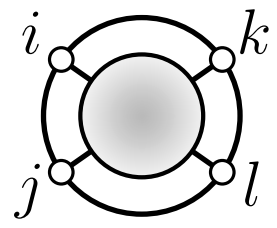

where $P_\bullet$ are points (suppressed in diagrams) lying on the boundary of $\mathsf{AdS}_{d+1}$, which is represented diagrammatically by a circle. Considering it up to the order $1/N$ in the large $N$ expansion, but to all orders in the coupling $\lambda$, its Witten diagram representation is given by

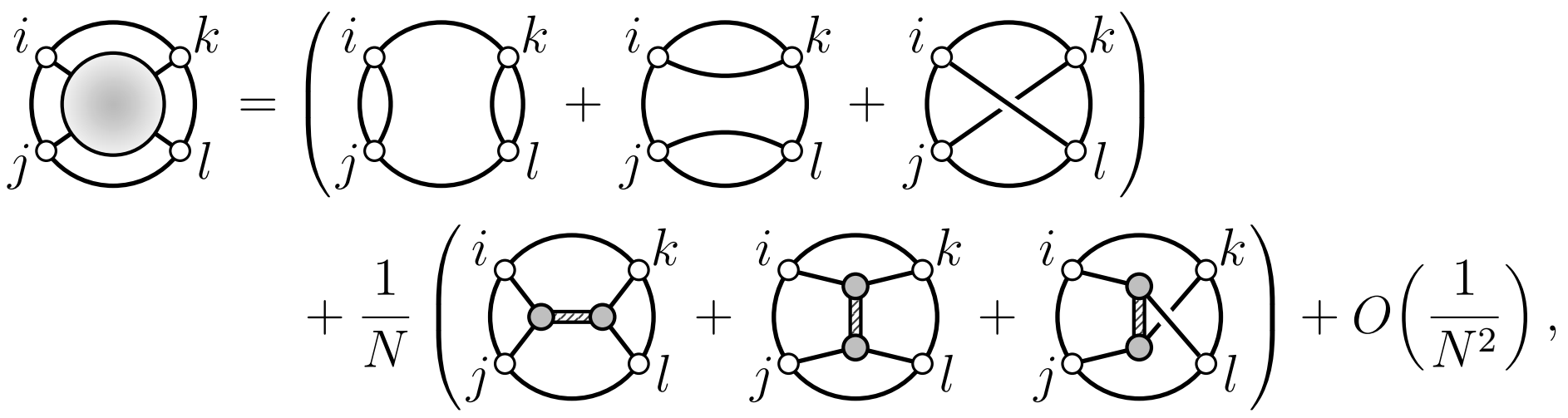

(5.17)

where we used the leading forms of the 1PI $\sigma\phi^2$ vertex (5.4) and exact $\sigma$-propagator (5.7). The $N$-dependence is now explicitly written in front of the diagrams. Lines with one or both ends on the boundary are bulk-to-boundary (5.14) or boundary-to-boundary (5.15) propagators, respectively. Using these ingredients, one can easily compose explicit expressions for diagrams figuring in (5.17), where as usual we integrate over the bulk points.

**Convention for Channels.** Our naming convention for both the OPE and also Witten diagrams of the 4-point correlator is

$$\begin{array}{ccc} s\text{-channel} & t\text{-channel} & u\text{-channel} \\ \left(\mathcal{O}_1^i\mathcal{O}_2^j\right)\left(\mathcal{O}_3^k\mathcal{O}_4^l\right), & \left(\mathcal{O}_1^i\mathcal{O}_3^k\right)\left(\mathcal{O}_2^j\mathcal{O}_4^l\right), & \left(\mathcal{O}_1^i\mathcal{O}_4^l\right)\left(\mathcal{O}_2^j\mathcal{O}_3^k\right). \end{array} \tag{5.18}$$

They correspond to the diagrams in (5.17), in respective order at each order of $1/N$. This is in agreement with the conventions used in [59] and (2.53) [88], but $t$-channel and $u$-channel are swapped in (3.5)/(3.8) [61]. As we will comment at appropriate places in Section 5.3 →p.114, one just needs to be careful with including an additional factor of $(-1)^J$ for 6j–symbol compared to their final formulas.

**Renormalization Scheme.** In (5.17) we drew only nontrivial Witten diagrams contributing to the 4-point correlator. Those just modifying the normalization of the $\phi$-field and shifting its mass — connected with the normalization of the dual boundary operator $\mathcal{O}_\phi$ and shifting the scaling dimension via the relation $m_\phi^2 = \Delta_\phi(\Delta_\phi - d)$ — were omitted, so let us comment on them here. Intuition about UV renormalization can be borrowed from flat space, since when distances among points are infinitesimally small, any spacetime looks approximately flat. Explicit renormalization in AdS in weak coupling perturbation theory, especially for a $\phi^4$-theory relevant for us, was done in [89] [90] [91] [92].

We work with the 4-point correlator computed up to (and including) the order $1/N$. We will describe the renormalization of the classical action (5.3), but sometimes it is useful to reiterate what various steps mean from the perspective of the original action (5.1).

First, let us once again qualify the term "free" in connection with the $\langle\phi\phi\rangle$ propagator. The discussion is intimately related with quantization around the nontrivial saddle point $\sigma(x) = \sqrt{N}\sigma_\star$ of the effective action (5.6). This step produces an $O(1)$ exact $\phi$-propagator (in the large $N$ expansion), which is from the point of view of (5.1) given by a resummation of all "cactus" diagrams. Those renormalize just the mass $m$ of the free $\phi$-field and shift it to the value $m_\phi^2 = m^2 + 2\sigma_\star$. Thus, the "free" $\phi$-propagator is associated with a free equation of motion, but with a shifted mass.

Such $\phi$-propagator is exact at $O(1)$ as we stated, therefore it is sufficient for all diagrams that are already of order $O(1/N)$, in particular for the exchanges of the $\sigma$-field in the second line of (5.17).

However, it is not sufficient for the disconnected diagrams of order $O(1)$ in the first line of (5.17). In those diagrams, the renormalization of the $\phi$-propagator needs to be carried out to order $O(1/N)$. Such a renormalization can affect both the normalization of the $\phi$-field and a shift in its mass. Luckily, the type of the propagator subjected to it is the boundary-boundary one, which is completely fixed by conformal symmetry on the boundary. We choose the two counterterms in such a way that it remains in the "free" form (5.15). This finishes renormalization of the $\phi$-propagator.

Now we start discussing renormalization of the $\langle\sigma\sigma\rangle$ bulk-bulk propagator. Its form (5.7) is $O(1)$ exact and from the perspective of (5.3) renormalizes the "mass-squared" $\lambda^{-1}$ of the free $\sigma$-field. In particular, the spectral representation — see next Section 5.2 →p.112 — of the $\sigma$-propagator $\left(\lambda^{-1} + 2\widetilde{B}\right)^{-1}$ requires a nontrivial renormalization in $d \geq 3$, where the bubble diagram acquires a UV divergence. The bubble function must be then somehow regularized, and with the UV divergence absorbed by a counterterm we are left with the finite combination $\lambda^{-1} + 2\widetilde{B}$. The poles of such renormalized spectral function determine the physical scaling dimensions of boundary operators associated with the bulk field $\sigma$. We will comment on fixing the subtraction scheme employed in defining the regularized bubble function $\widetilde{B}$ in Section 5.4 →p.134.

Let us also note that the renormalization of the $\sigma$-propagator described above,

when reinterpreted from the point of view of (5.1), would correspond to renormalization of the 1PI vertex $\phi^4$ associated with the coupling $\lambda$. Classically it starts at order $O(1/N)$, and since all "bubble-chain" diagrams contribute at the same order, they need to be resummed. This is done in (5.9), which precisely leads to the $O(1)$ exact form (5.7) of the $\sigma$-propagator.

Finally, let us comment about the 1PI vertex $\sigma\phi^2$, which is of order $O(1/\sqrt{N})$ classically. It appears twice in the $\sigma$-exchange diagrams, thus making them of order $O(1/N)$. Consequently, to the maximal order we consider, it does not require any renormalization, as all corrections to its classical part would be subleading in the large $N$ expansion.

This finishes the description of the renormalization. The renormalization scheme defined above might be called *on-shell*, since it is fully determined in terms of physical scaling dimensions $\Delta_\phi$, $\Delta_{\widehat{\sigma}_0}$ — $\widehat{\sigma}_0$ being the lowest-dimensional boundary operator associated to the $\sigma$-field — and the standard normalization of the boundary operator $\mathcal{O}_\phi$, leading to the canonical form of the CFT 2-point function (5.15).

## Utilizing the Spectral Representation

To proceed further with the calculation of the 4-point correlator (5.16), mainly the evaluation of bulk integrals in the $\sigma$-exchange diagrams, it is convenient to employ the so-called *spectral representation* of the $\sigma$-propagator. It is an integral expansion into *harmonic functions* — eigenfunctions of the AdS Laplacian — which form a continuous basis for integrable functions depending only on the geodesic distance between two points. This is in precise analogy to the (radial) Fourier representation of the propagator in flat space, which is also adapted to the isometries (translations and rotations) of the theory. In the limit of large AdS radius, it actually reduces to the flat-space (radial) Fourier transform.

While we will not explicitly use this technology — the relevant results from [59] will be cited in Section 5.4 →p.126 — it is rather crucial in calculation and deserves a brief mention. More details can be found in Appendix 4.C [50] Appendix B [45] and Appendix B [59].

Suppose we know the spectral representation $\widetilde{B}(\Delta)$ of the bubble function $B(x,y)$ defined in (5.8), that is we have a decomposition into harmonic functions $\Omega_\Delta$

$$B(x,y) = \int_{\mathbb{R}} \mathrm{d}\nu\, \widetilde{B}(\Delta)\Omega_\Delta(x,y) \equiv \int_{\frac{d}{2}+\mathring{\imath}\mathbb{R}} \frac{\mathrm{d}\Delta}{\mathring{\imath}}\, \widetilde{B}(\Delta)\Omega_\Delta(x,y)\,, \tag{5.19}$$

where we use $\Delta \equiv \frac{d}{2} + \mathring{\imath}\nu \Leftrightarrow \nu \equiv -\mathring{\imath}(\Delta - \frac{d}{2})$ interchangeably. While it is common to use the spectral parameter $\nu$ to index the harmonic functions and also to denote the functional dependence of spectral representations, we will mostly prefer using the dimension $\Delta$ directly for later notational convenience.

Ignoring the index structure, the calculation can be schematically represented as

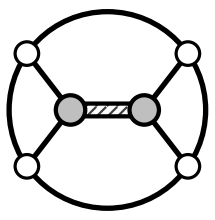

$$= 4\int_{\mathbb{R}} \mathrm{d}\nu \left(\frac{-1}{\lambda^{-1}+2\widetilde{B}(\Delta)}\right)$$

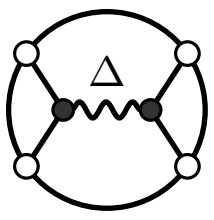

$$= 4\int_{\mathbb{R}} \mathrm{d}\nu \left(-\frac{1}{\lambda^{-1}+2\widetilde{B}(\Delta)}\right)\sqrt{\mathfrak{C}_{\Delta}\mathfrak{C}_{\widetilde{\Delta}}}\,\frac{\nu^2}{\pi}$$

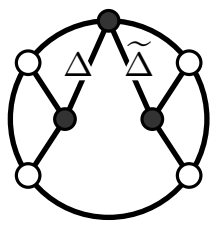

$$= \int_{\frac{d}{2}+i\mathbb{R}_{\geq 0}} \frac{\mathrm{d}\Delta}{2\pi i}\left(-\frac{1}{\lambda^{-1}+2\widetilde{B}(\Delta)}\right)\left(\frac{\Gamma^2_{\cdots}\cdots\Gamma_{\cdots}}{\cdots\Gamma^2_{\cdots}\cdots\Gamma_{\cdots}}\right) \tag{5.20}$$

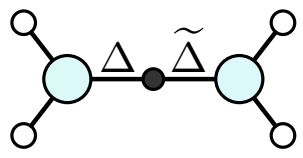

$$\equiv \int_{\frac{d}{2}+i\mathbb{R}_{\geq 0}} \frac{\mathrm{d}\Delta}{2\pi i}\left(-\frac{1}{\lambda^{-1}+2\widetilde{B}(\Delta)}\right)\left(\frac{\Gamma^2_{\cdots}\cdots\Gamma_{\cdots}}{\cdots\Gamma^2_{\cdots}\cdots\Gamma_{\cdots}}\right)$$

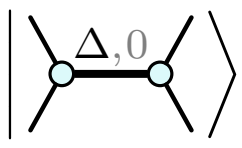

Line by line, we utilized the following properties and relations — for more details see Appendix C [59]:

**(1)** Similarly to the Fourier transform, the spectral representation transforms a convolution of functions into a product of their spectral functions. Thus, the "operator" inversion figuring in the $\sigma$-propagator (5.7) is represented just as a numeric inversion of the corresponding spectral function. The overall factor of 4 comes from the two $\sigma\phi^2$ vertices (5.4), we implicitly integrate over the black bulk points, and the wavy line represents the harmonic function.

**(2)** The harmonic function can be further rewritten by means of the *split representation* [51] [93], which up to a certain prefactor is given by two bulk-to-boundary propagators — one with dimension $\Delta$ and the other with the shadow dimension $\widetilde{\Delta} \equiv d - \Delta$ — with the black boundary point implicitly integrated over $\partial$AdS. Remember that the rest of bulk-to-boundary propagators have the dimension $\Delta_\phi$.

**(3)** Now we can perform the bulk integrations. For fixed boundary points, both of the bulk integrals individually must result in a multiple of the unique CFT 3-point structure of the corresponding scalar operators — these are diagrammatically represented by the blue blobs. We are thus left with a convolution of two such 3-point structures, since we have yet to integrate over the boundary point.

**(4)** Such a convolution is actually the *shadow representation* of the *Conformal Partial Wave* (CPW) [94] [95] [96] [97] [98] [61], which we denote by 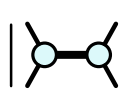.

In the end, the computation results in the *Conformal Partial Wave decomposition* of the $s$-channel $\sigma$-exchange diagram. The $t$-channel/$u$-channel diagrams are given by the same expression, just with the CPWs in the corresponding channels.

We will discuss such decompositions in more detail in the next section. The associated coefficient/weight function (also called the *spectral function*) is the main object of interest. It encodes the CFT data accessible to the 4-point correlator $\left\langle \mathcal{O}_\phi \mathcal{O}_\phi \mathcal{O}_\phi \mathcal{O}_\phi \right\rangle$.

## 5.3 CFT Generalities — 4-Point Correlators

As the boundary limits of correlators in AdS enjoy a conformal symmetry, it is only natural to study them using the language and methods of CFTs. Calculation of the exchange diagram contribution to the 4-point correlator — directly relevant for the O($N$) singlet CFT spectrum — naturally resulted in its *conformal partial wave* decomposition, which we will review in Section 5.3 →p.115.

Things get more involved when one wants to extract the non-singlet spectrum. We will see in Section 5.4 →p.124 that crossed-channel diagrams — take for example the $t$-channel — turn out to be essential for this task. Even though it is not viable to calculate directly the $s$-channel decomposition of the $t$-channel interaction diagram, its $t$-channel decomposition is basically known after resolving the singlet spectrum. In Section 5.3 →p.117 we will discuss how to translate the $t$-channel decomposition into the $s$-channel one via the 6j–symbols.

To fully prepare for the extraction of the leading $1/N$ corrections to the non-singlet spectrum, in Section 5.3 →p.119 we will present the general formulas (in $d = 2$ and $d = 4$) for the contribution of a $t$-channel conformal block to the anomalous dimensions of $s$-channel double-twist operators, applicable for external scalar operators with equal scaling dimensions.

For simplicity — and because it is the relevant case for us — in the following we assume all external operators to be scalars $\mathcal{O}_\phi$ with equal scaling dimensions $\Delta_\phi$. To avoid clutter, we will suppress throughout the possible global symmetry index structure of the operators, and the dependence of correlators/conformal blocks/conformal partial waves on the scaling dimensions and positions of external operators.

It is enough to concentrate only on the nontrivial crossing between $s$-channel and $t$-channel, since the $u$-channel is then given by including an additional factor of $(-1)^J$. Here $J$ is the spin of the exchanged operator belonging to the symmetric traceless representation of the rotation group SO($d$). This follows from a simple relation between $t$-channel and $u$-channel conformal blocks for equal external scaling dimensions (59) [34], or can be deduced directly from the property of the OPE coefficients for two scalar operators $\mathcal{O}_1, \mathcal{O}_2$ and a spin $J$ operator $\mathcal{O}_J$ — $\mathsf{ope}[\mathcal{O}_J \mathcal{O}_1 \mathcal{O}_2] = (-1)^J \, \mathsf{ope}[\mathcal{O}_J \mathcal{O}_2 \mathcal{O}_1]$ — see (25) [34].

To summarize, the starting point is MFT, where it is known that the identity operator in the $t$-channel induces a double-twist family of operators $\mathcal{O}_{n,J}$ in the $s$-channel decomposition, with scaling dimensions $\Delta^{(\mathrm{MFT})}_{n,J} \equiv 2\Delta_\phi + 2n + J$. The question that we want to answer in the following is: Given a $t$-channel contribution of an operator $\mathcal{O}'$ with scaling dimension $\Delta'$ and spin $J'$, what anomalous dimensions (and corrections to OPE coefficients) does it induce for the double-twist family in the $s$-channel decomposition? Supposing that the exchange contribution is of order $O(1/N)$, we will compute the leading form of the corrected scaling dimensions $\Delta_{n,J} \equiv \Delta_{n,J}^{(\mathrm{MFT})} + \gamma_{n,J} \equiv 2\Delta_\phi + 2n + J + \gamma_{n,J}$, where the anomalous dimensions $\gamma_{n,J}$ are also of order $O(1/N)$. Thus, as expected, this family of operators reduces to the MFT ones in the limit $N \to \infty$. The dimension $\Delta'$ of the $t$-channel exchanged primary does not need to be parametrically close to an MFT value in the large $N$ expansion. In fact, it actually turns out that to contribute nontrivially at leading order, it should not be.

## Conformal Block and Conformal Partial Wave Decompositions

By grouping the operators in pairs and using the *Operator Product Expansion* (OPE) twice, we can write the 4-point correlator as a discrete sum over contributions of exchanged conformal families, weighted by the (squares of) OPE coefficients. Such contributions are called *Conformal Blocks* (CBs), and they correspond to an exchange of a physical primary operator together with all of its descendants. Since only symmetric traceless tensors appear in the OPE of two scalars, the exchanged families are labeled by their scaling dimension $\Delta_\star$ and spin $J_\star$, giving us the *Conformal Block Decomposition*

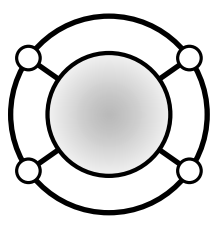

$$= \sum_{\substack{\text{primary } \mathcal{O}_\star \\ \text{with } \Delta_\star, J_\star}} \mathsf{ope}^2\Big[\mathcal{O}_\phi \mathcal{O}_\phi \mathcal{O}_\star\Big] \Big| G^{(s)}_{\Delta_\star, J_\star} \Big\rangle = \left(\begin{smallmatrix}\text{analogously in} \\ t\text{-channel}\end{smallmatrix}\right), \tag{5.21}$$

where $\mathsf{ope}\Big[\mathcal{O}_\phi \mathcal{O}_\phi \mathcal{O}_\star\Big]$ is the OPE coefficient for operator $\mathcal{O}_\star$ appearing in the $\mathcal{O}_\phi \times \mathcal{O}_\phi$ OPE, and the corresponding $s$-channel conformal block is denoted by $\Big| G^{(s)}_{\Delta_\star, J_\star} \Big\rangle$. We suppress the dependence of CBs on conformal representations and positions of the external operators.

Alternatively, there is an expansion into Unitary Irreducible Representations (UIRs) of the (Euclidean) conformal group $\mathsf{SO}(d+1,1)$, including mainly the *Principal Series* (traceless symmetric) representations with integer spin $J \in \mathbb{N}_0$ but "unphysical" complex dimensions $\Delta \in \frac{d}{2} + i\mathbb{R}_{\geq}$. The associated eigenfunctions of conformal Casimir are called *Conformal Partial Waves* (CPWs), which furthermore form a complete basis of "normalizable" functions [95]. They come from the *harmonic analysis* of the Euclidean conformal group $\mathsf{SO}(1, d+1)$, and are analogous to plane waves, which are eigenfunctions of translations in the flat space.

Ignoring non-normalizable contributions such as the exchange of the identity operator (see [98] for some discussion), we have the *Conformal Partial Wave*

*Decomposition*

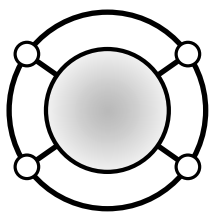
$$= \sum_{J=0}^{\infty} \int_{\frac{d}{2}+i\mathbb{R}_{\geq}} \frac{\mathrm{d}\Delta}{2\pi i}\, \mathsf{Spec}_s\left[\begin{matrix}\Delta \\ J\end{matrix}\,\middle|\, \right.$$
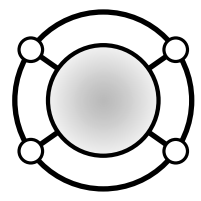
$$\left.\right]\Big|$$
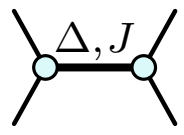
$$\Big\rangle = \left(\begin{matrix}\text{analogously in} \\ t\text{-channel}\end{matrix}\right), \qquad (5.22)$$

where the $s$-channel conformal partial wave is denoted by $|$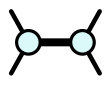$\rangle$, and the spectral function $\mathsf{Spec}_s[\cdots]$ represents the coefficients of the corresponding CPWs. Again, we suppress the dependence of CPWs on conformal representations and positions of the external operators.

If we were to perform decomposition in the $t$-channel, we would use $t$-channel spectral function $\mathsf{Spec}_t[\cdots]$ and $t$-channel CPWs $|$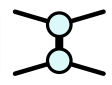$\rangle$. In [61] [98] they use either $\rho$ or $I/n$ for our $\mathsf{Spec}_s$, where the normalization $n$ is defined in (5.28). Notice that the integration in (5.22) is over the half-line $\Delta \in \frac{d}{2} + i\mathbb{R}_{\geq}$, since the *shadow* scaling dimensions $\widetilde{\Delta} \equiv d - \Delta \in \frac{d}{2} + i\mathbb{R}_{\leq}$ correspond to equivalent representations of the conformal group.

These two decompositions are closely related, as CPWs are (up to normalization) shadow-symmetric combinations of CBs

$$\Big|$$
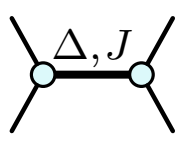
$$\Big\rangle = K_{\widetilde{\Delta},J}\Big|G^{(s)}_{\Delta,J}\Big\rangle + K_{\Delta,J}\Big|G^{(s)}_{\widetilde{\Delta},J}\Big\rangle,$$
$$\Big|$$
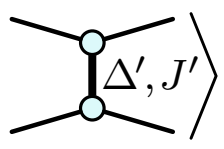
$$\Big\rangle = K_{\widetilde{\Delta}',J'}\Big|G^{(t)}_{\Delta',J'}\Big\rangle + K_{\Delta',J'}\Big|G^{(t)}_{\widetilde{\Delta}',J'}\Big\rangle, \qquad (5.23)$$

where the normalization coefficients are given by (2.16) [61] (A.5), (A.6) [98]

$$K_{\Delta,J} = \frac{\pi^{\frac{d}{2}}}{(-2)^J}\, \frac{\Gamma_{\Delta-\frac{d}{2}}\, \Gamma_{\Delta+J-1}}{\Gamma_{\Delta-1}\, \Gamma_{d-\Delta+J}} \left(\frac{\Gamma_{\frac{\widetilde{\Delta}+J}{2}}}{\Gamma_{\frac{\Delta+J}{2}}}\right)^2 . \qquad (5.24)$$

In the $t$-channel we will usually use primed quantities for improved distinction. Also, to make the expressions compact, we will often utilize the abbreviation $\Gamma_{\circ} \equiv \Gamma(\circ)$.

Since CPWs are shadow-symmetric, it is natural to choose the spectral function to be shadow-symmetric as well. Substituting (5.23) into (5.22), using the aforementioned shadow-symmetry of $\mathsf{Spec}$ to extend the integration from half-line $\frac{d}{2} + i\mathbb{R}_{\geq}$ to $\frac{d}{2} + i\mathbb{R}$ together with taking only the first conformal block term in CPW, we obtain an integral decomposition in terms of conformal blocks as

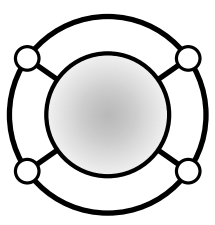
$$= \sum_J \int_{\frac{d}{2}+i\mathbb{R}} \frac{\mathrm{d}\Delta}{2\pi i}\, \mathsf{Spec}_s\left[\begin{matrix}\Delta \\ J\end{matrix}\,\middle|\, \right.$$
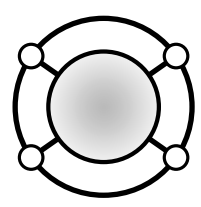
$$\left.\right] K_{\widetilde{\Delta},J}\Big|G^{(s)}_{\Delta,J}\Big\rangle . \qquad (5.25)$$

After enclosing the integration in the right-half plane, the residue theorem enables us to write it as a discrete sum over physical poles of $\mathsf{Spec}_s$, thus obtaining the usual conformal block decomposition (5.21). Certain subtleties of this procedure, in particular appearance and cancellation of additional spurious poles, are discussed in Appendix B [98].

We see that (at least the accessible part of) the CFT data are encoded in the spectral function $\mathsf{Spec}_s$, namely the scaling dimensions of primary operators are given by the positions of $\mathsf{Spec}_s$ poles, and the corresponding squared OPE coefficients in the $s$-channel are given by the (minus, since the contour is clockwise) residues as

$$\mathsf{ope}^2\left[\mathcal{O}_\phi \mathcal{O}_\phi \mathcal{O}_\star\right] = -\operatorname{Res}_{\Delta=\Delta_\star}\left( K_{\widetilde{\Delta},J_\star}\, \mathsf{Spec}_s\left[\begin{matrix}\Delta \\ J_\star\end{matrix} \,\middle|\, \right.\right.$$ 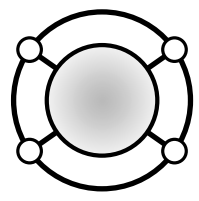 $$\left.\left.\right]\right) . \tag{5.26}$$

## CPW Orthogonality and Completeness, 6j–Symbol

As already mentioned, CPWs (along the principal series) form a complete basis of functions, which furthermore are orthogonal with respect to an appropriate conformally-invariant pairing involving integration over all four external points (1.3) [98], or alternatively a closely related inner product (A.27) [98]. We will use the bra-ket/inner-product notation in the following, that is

$$\Big\langle$$ 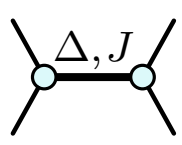 $$\Big|$$ 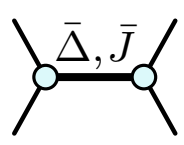 $$\Big\rangle = n_{\Delta,J}\, 2\pi\delta(\nu-\bar{\nu})\,\delta_{J\bar{J}} \,, \tag{5.27}$$

where $\Delta \equiv \frac{d}{2} + \mathring{\imath}\nu$ and $\bar{\Delta} \equiv \frac{d}{2} + \mathring{\imath}\bar{\nu}$ with $\nu, \bar{\nu} \geq 0$ are scaling dimensions in the principal series, and $n_{\Delta,J}$ is the normalization (2.35) [61]

$$n_{\Delta,J} \equiv \frac{K_{\widetilde{\Delta},J}\, K_{\Delta,J}\, \mathrm{vol}(\mathbb{S}^{d-2})}{2^d\, \mathrm{vol}(\mathsf{SO}(d-1))}\, \frac{(2J+d-2)\pi\, \Gamma_{J+1}\, \Gamma_{J+d-2}}{2^{d-2}\, \Gamma^2_{J+\frac{d}{2}}} \,. \tag{5.28}$$

Note that it includes an extra $2^{-d}$ compared to (A.14), (A.15) [98]. We can thus express the completeness relation (for normalizable functions) in terms of CPWs as

$$\mathbf{1} = \sum_J \int_{\frac{d}{2}+\mathring{\imath}\mathbb{R}_\geq} \frac{\mathrm{d}\Delta}{2\pi\mathring{\imath}} \Big|$$ 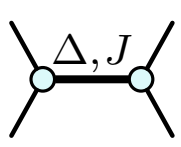 $$\Big\rangle\, \frac{1}{n_{\Delta,J}}\, \Big\langle$$ 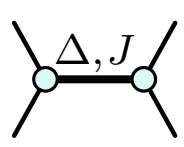 $$\Big| \,. \tag{5.29}$$

Consider now a certain contribution to 4-point correlator, for which we are able to calculate its $t$-channel spectral function $\mathsf{Spec}_t$, that is we know the decomposition

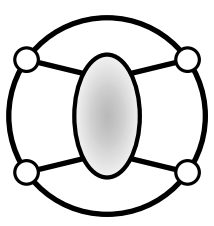 $$= \sum_{J'} \int_{\frac{d}{2}+\mathring{\imath}\mathbb{R}_\geq} \frac{\mathrm{d}\Delta'}{2\pi\mathring{\imath}}\, \mathsf{Spec}_t\left[\begin{matrix}\Delta' \\ J'\end{matrix} \,\middle|\, \right.$$ 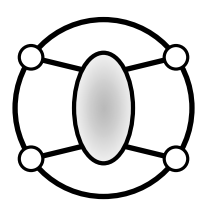 $$\left.\right] \Big|$$ 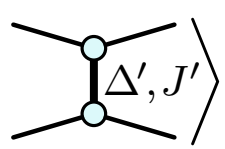 $$\Big\rangle \tag{5.30}$$

into $t$-channel CPWs. To extract the associated contribution to the CFT data, we first need to translate this decomposition into the $s$-channel. By inserting the completeness relation (5.29) into the $t$-channel decomposition, or by directly utilizing the orthogonality (5.27), we obtain the $s$-channel spectral function as

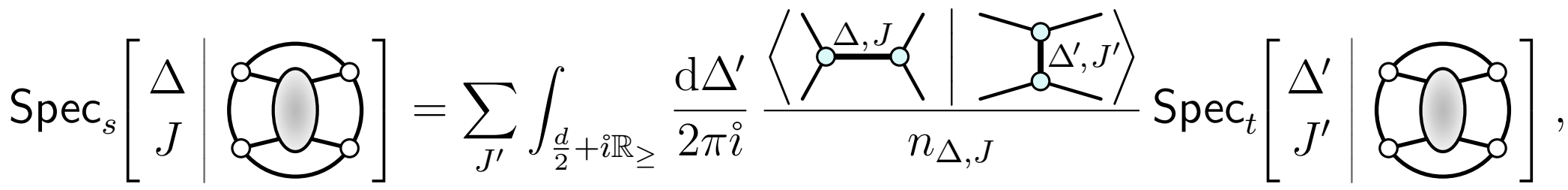

(5.31)

where the $(s \leftarrow t)$–channel translation is being performed (up to the normalization $n_{\Delta,J}$) by the "Clebsch-Gordan coefficient" for the conformal group called 6j–symbol (3.5) [61]

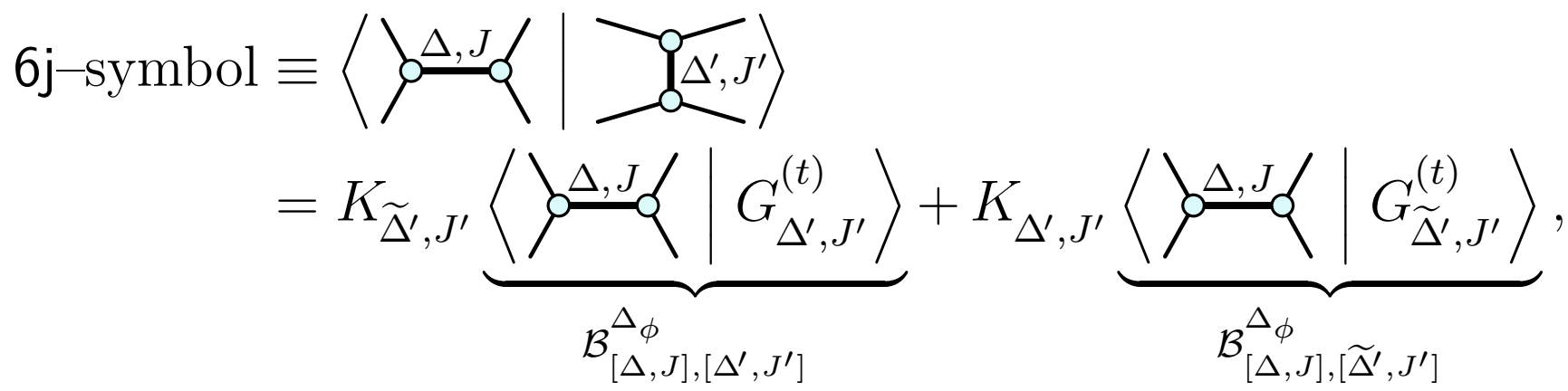

(5.32)

where $\mathcal{B}$ is the notation used in (3.35), (3.41) [61], and corresponds to the $s$-channel spectral function of a single $t$-channel conformal block.

Alternatively, we could start with the $t$-channel conformal block decomposition of the contribution (5.30), given by sum over $t$-channel conformal blocks with some coefficients

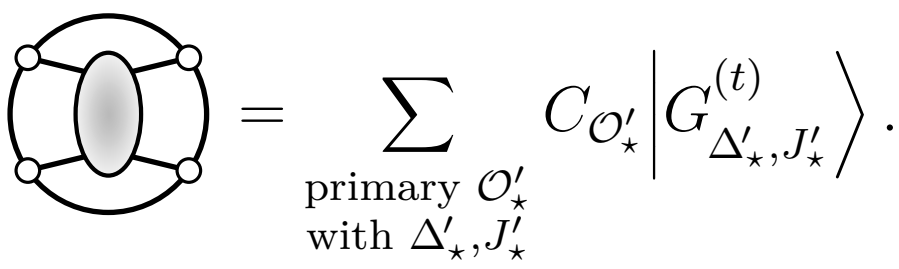

(5.33)

Again, utilizing the orthogonality (5.27), the corresponding $s$-channel spectral function is given by

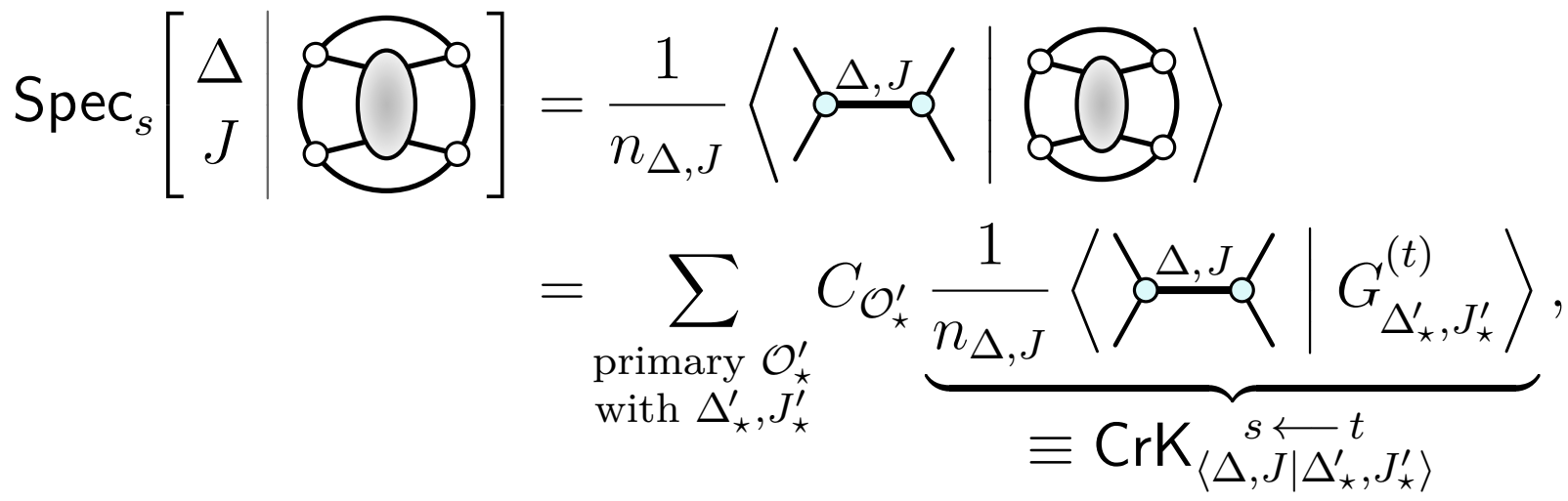

(5.34)

where we introduced the "crossing kernel" $\mathsf{CrK}^{s\leftarrow t}$ notation for the $s$-channel spectral function of a single $t$-channel conformal block (including the proper normalization), which in the following will be slightly more notationally convenient than $\mathcal{B}$.

Strictly speaking, due to the limited validity range of the Lorentzian inversion formula utilized to calculate $\mathsf{CrK}^{s\leftarrow t}$ in [98], the $t$-channel conformal block inversion in (5.34) seems to work only for $J > J'$. We will touch on this issue a little bit more in Section 5.5 →p.141, also see remarks in Section 4 [61].

From extensive studies of the Lorentzian inversion formula [99] [98] [100], it is well known that $\mathsf{CrK}^{s\leftarrow t}$ has double zeros in $\Delta'$ at the locations of MFT double-twist dimensions, therefore only non-MFT operators contribute to the $s$-channel spectral function $\mathsf{Spec}_s$. Direct consequence of this fact can be explicitly seen in the formulas for the anomalous dimensions presented in the following Section 5.3→p.119.

## Contribution of $t$-Channel CBs to Anomalous Dimensions

Suppose we solve the theory in some kind of perturbative expansion — we will be formulating everything in the context of large $N$ expansion, but it is applicable more generally. We therefore organize correlators and CFT data into expansion in powers of $1/N$, for example consider that certain part of 4-point correlator is given by

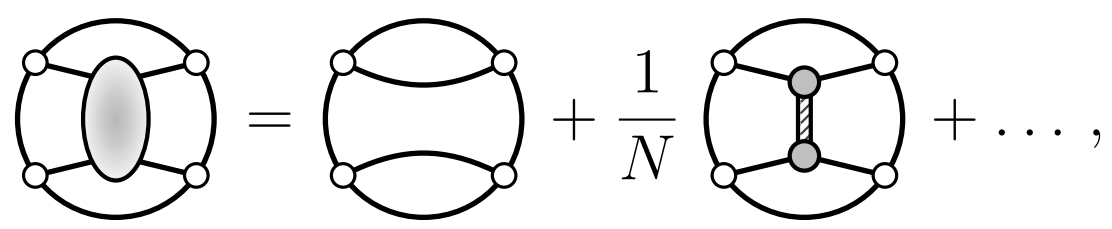

(5.35)

where the first disconnected term is the $t$-channel identity contribution, and the second term is some interaction contributing at $O(1/N)$ order, for which we are able to calculate its $t$-channel conformal block decomposition. Our goal is to extract the respective correction to the CFT spectrum.

**Disconnected $t$-Channel Spectral Function.** Spectral function of the $t$-channel identity ($t$-channel disconnected 4-point MFT correlator) in the $s$-channel was computed in its general form in [46] and later reproduced within an elegant harmonic analysis formalism [88]

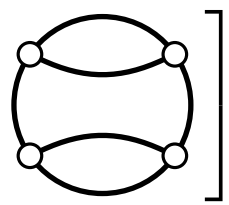

$$\mathsf{Spec}_s\left[\begin{matrix}\Delta\\ J\end{matrix}\,\middle|\,\cdots\right] = \frac{2^{J-1}}{S_{\widetilde{\Delta},J}} \frac{\Gamma_{\Delta-1}\,\Gamma^2_{\frac{d}{2}-\Delta_\phi}\,\Gamma_{\frac{d}{2}+J}\,\Gamma_{\widetilde{\Delta}+J}}{\Gamma^2_{\Delta_\phi}\,\Gamma_{J+1}\,\Gamma_{\Delta-\frac{d}{2}}\,\Gamma_{\Delta+J-1}} \frac{\Gamma^2_{\frac{J+\Delta}{2}}\,\Gamma_{\frac{J-\Delta}{2}+\Delta_\phi}\,\Gamma_{\frac{\Delta+J-d}{2}+\Delta_\phi}}{\Gamma^2_{\frac{J+\widetilde{\Delta}}{2}}\,\Gamma_{\frac{2d+J-\Delta}{2}-\Delta_\phi}\,\Gamma_{\frac{\Delta+J+d}{2}-\Delta_\phi}}\,, \tag{5.36}$$

where $S_{\widetilde{\Delta},J} \equiv (-2)^J K_{\widetilde{\Delta},J}$.

As expected, it contains poles — which come from $\Gamma\left(\frac{J-\Delta}{2}+\Delta_\phi\right)$ — located at the dimensions $\Delta^{(\mathrm{MFT})}_{n,J} = 2\Delta_\phi + 2n + J$ corresponding to the family of MFT double-twist operators schematically given by $\mathcal{O}^{(\mathrm{MFT})}_{n,J} = [\mathcal{O}_\phi \Box^n \partial^J \mathcal{O}_\phi - \text{traces}]$. Corresponding squared OPE coefficients can be easily calculated using (5.26).

There are also some spurious poles coming from $\Gamma(\widetilde{\Delta}+J) \equiv \Gamma(d+J-\Delta)$, but these are resolved as discussed in Appendix B [98].

Thus, from the $s$-channel conformal block decomposition point of view, at the leading $O(1)$ order the $t$-channel identity gives rise to the aforementioned double-twist operators. At the following $O(1/N)$ order, the connected interaction term modifies the CFT data, in particular it induces anomalous dimensions $\gamma_{n,J}$ of these operators. This is what we will focus on in the following.

**Contribution of $t$-Channel Exchange.** Recalling (5.26), the appearance of an operator $\mathcal{O}_\star$ in the $s$-channel conformal block decomposition of the 4-point correlator is reflected in the spectral function as a simple pole of the form (now given as series in $1/N$)

$$\frac{-C_\star\left(\frac{1}{N}\right)}{\Delta-\Delta_\star\left(\frac{1}{N}\right)} \in K_{\widetilde{\Delta},J}\,\mathsf{Spec}_s\left[\begin{array}{c}\Delta\\ J\end{array}\right|$$
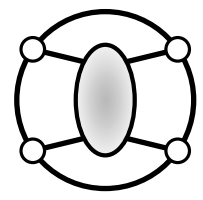
$$\Bigg] . \tag{5.37}$$

Defining the leading corrections to the squared OPE coefficients and scaling dimensions as

$$\mathsf{ope}^2\left[\mathcal{O}_\phi\mathcal{O}_\phi\mathcal{O}_\star\right] \equiv C_\star\left(\frac{1}{N}\right) = C_\star^{(\mathrm{MFT})} + \frac{1}{N}\,C_\star^{(1)} + O\left(\frac{1}{N^2}\right), \tag{5.38}$$

$$\Delta_\star\left(\frac{1}{N}\right) = \Delta_\star^{(\mathrm{MFT})} + \frac{1}{N}\,\gamma_\star^{(1)} + O\left(\frac{1}{N^2}\right), \tag{5.39}$$

the expansion of (5.37) to the first order in $1/N$ reads

$$\frac{-C_\star\left(\frac{1}{N}\right)}{\Delta-\Delta_\star\left(\frac{1}{N}\right)} = \frac{-C_\star^{(\mathrm{MFT})}}{\Delta-\Delta_\star^{(\mathrm{MFT})}} + \frac{1}{N}\left[\frac{-C_\star^{(1)}}{\Delta-\Delta_\star^{(\mathrm{MFT})}} + \frac{-C_\star^{(\mathrm{MFT})}\gamma_\star^{(1)}}{\left(\Delta-\Delta_\star^{(\mathrm{MFT})}\right)^2}\right] + \ldots . \tag{5.40}$$

Anomalous dimensions are thus encoded in the coefficients of double poles in the spectral function, divided by the corresponding MFT squared OPE coefficients. Since the MFT squared OPE coefficients are given by residues of the $t$-channel identity, we can express the ($O(1/N)$ part of) anomalous dimensions as

$$\gamma_{n,J}^{(1)} = \mathrm{Res}_{\Delta=2\Delta_\phi+2n+J}\left(\frac{\mathsf{Spec}_s\left[\begin{array}{c}\Delta\\ J\end{array}\right|\;\;\Bigg]}{\mathsf{Spec}_s\left[\begin{array}{c}\Delta\\ J\end{array}\right|\;\;\Bigg]}\right). \tag{5.41}$$
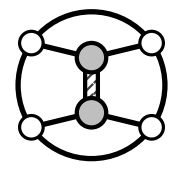

If we wanted to extract the $O(1/N)$ contributions to the OPE coefficients, we would just calculate the (minus) residue without dividing by the spectral function of $t$-identity.

For simplicity, consider that the interaction term in (5.41) is composed of a single $t$-channel conformal block with scaling dimension $\Delta'$ and spin $J'$. Utilizing (5.34), the corresponding contribution to the anomalous dimensions is given by

$$\gamma_{n,J}^{(1)}\Bigg|_{\substack{t\text{-channel}\\ \text{exchange}\\ \Delta',J'}} = \mathrm{Res}_{\Delta=2\Delta_\phi+2n+J}\left(\frac{\mathsf{CrK}^{s\leftarrow t}_{\langle\Delta,J|\Delta',J'\rangle}}{\mathsf{Spec}_s\left[\begin{array}{c}\Delta\\ J\end{array}\right|\;\;\Bigg]}\right). \tag{5.42}$$
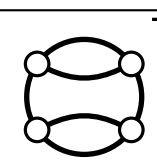

Simple poles of the expression inside residue (5.42) — or equivalently, double poles of 6j–symbol or $\mathsf{CrK}^{s\leftarrow t}$ — are not present for generic non-equal external scaling dimensions. Nonetheless, for pairwise-equal external dimensions the crossing kernel develops double poles (3.48) and below [61], and hence the anomalous dimensions obtain nontrivial contributions.

**General Formulas for $d = 2$ and $d = 4$.** The 6j–symbol and thus $\mathsf{CrK}^{s\leftarrow t}$ was explicitly computed in (3.36), (3.42) [61] for $d = 2$ and $d = 4$ by methods relying on the Lorentzian inversion formula [99]. Compared to our conventions specified in (5.18), their $t$-channel and $u$-channel are swapped, so we include a factor of $(-1)^J$ when taking over their results for 6j–symbol. Moreover, they calculated also the $t$-channel conformal block contributions to the leading-twist $(n = 0)$ anomalous dimensions (3.55), (3.56) [61].

Here we present the general formulas applicable for arbitrary (assuming $J > J'$) double-twist operators $\mathcal{O}_{n,J}$ in $d = 2$ and $d = 4$. For the detailed calculation see the accompanying NOTEBOOK. The final formulas read

$$\gamma^{(1)}_{n,J}\Big|_{\substack{t\text{-channel}\\ \text{exchange}\\ \Delta',J'}} \overset{d=2}{=\!=} -\frac{2\,\Gamma^2_{\Delta_\phi}\,\Gamma^2_{\Delta_\phi+J+n}}{\Gamma_{2(\Delta_\phi+J+n)}}\,\frac{\sin^2\!\Big(\pi\Big[\Delta_\phi-\frac{\Delta'-J'}{2}\Big]\Big)}{\pi^2}\,\frac{\Gamma_{J+n+1}}{\Gamma_{2\Delta_\phi+J+n-1}}\times$$
$$\times\frac{1}{1+\delta_{0,J'}}\left[\frac{\Gamma_{\Delta'+\widetilde{J}'}}{\Gamma^2_{\frac{\Delta'+\widetilde{J}'}{2}}}\,\mathsf{F}^{(d=2)}_{n,\frac{\Delta'+\widetilde{J}'}{2}}\,\Omega_{\Delta_\phi+J+n,\frac{\Delta'+J'}{2},\Delta_\phi}+\Big(J'\longleftrightarrow\widetilde{J}'\equiv -J'\Big)\right], \tag{5.43}$$

$$\gamma^{(1)}_{n,J}\Big|_{\substack{t\text{-channel}\\ \text{exchange}\\ \Delta',J'}} \overset{d=4}{=\!=} \frac{2\,\Gamma^2_{\Delta_\phi}\,\Gamma^2_{\Delta_\phi+J+n}}{\Gamma_{2(\Delta_\phi+J+n)}}\,\frac{\sin^2\!\Big(\pi\Big[\Delta_\phi-\frac{\Delta'-J'}{2}\Big]\Big)}{\pi^2}\,\frac{\Gamma_{J+n+1}}{\Gamma_{2\Delta_\phi+J+n-1}}\times$$
$$\times\frac{J+n+1}{J+1}\,\frac{2\Delta_\phi+J+n-2}{2\Delta_\phi+J+2n-2}\times$$
$$\times\left[\frac{\Gamma_{\Delta'+\widetilde{J}'}}{\Gamma^2_{\frac{\Delta'+\widetilde{J}'}{2}}}\,\mathsf{F}^{(d=4)}_{n,\frac{\Delta'+\widetilde{J}'}{2}}\,\Omega_{\Delta_\phi+J+n,\frac{\Delta'+J'}{2},\Delta_\phi-1}-\Big(J'\longleftrightarrow\widetilde{J}'\equiv -2-J'\Big)\right]. \tag{5.44}$$

The function $\Omega$ appearing in the above formulas is given in (3.38) [61], which after simplification for equal external dimensions takes the form

$$\Omega_{h,h',p} = \frac{\Gamma_{2h'}\,\Gamma^2_{h+p-1}\,\Gamma_{h'-h-p+1}}{\Gamma^2_{h'}\,\Gamma_{h'+h+p-1}}\;{}_4\mathsf{F}_3\left[\begin{matrix}h,\ h,\ \ h+p-1,\ \ h+p-1\\ 2h,\ \ h'+h+p-1,\ h-h'+p\end{matrix}\,\middle|\,1\right]$$
$$+\Big(h\longleftrightarrow h',\ p\longleftrightarrow 2-p\Big). \tag{5.45}$$

To further simplify the expressions, we defined (for $d = 2$ and $d = 4$)

$$\mathsf{F}^{(d)}_{n,\alpha} \overset{d\in\{2,4\}}{=\!=} \frac{(-1)^n}{n!}\,\frac{\Gamma_{d-2\Delta_\phi-n}}{\Gamma_{d-2\Delta_\phi-2n}}\,\frac{\Gamma^2_{\frac{d}{2}-\Delta_\phi-n}}{\Gamma^2_{\frac{d}{2}-\Delta_\phi}}\,\frac{\Gamma_{\alpha+n}}{\Gamma_{\alpha-n}}\times$$
$$\times\;{}_4\mathsf{F}_3\left[\begin{matrix}-n,\ -n,\ \frac{d}{2}-\Delta_\phi-n,\ \frac{d}{2}-\Delta_\phi-n\\ d-2\Delta_\phi-2n,\ 1-\alpha-n,\ \alpha-n\end{matrix}\,\middle|\,1\right]. \tag{5.46}$$

Using $\mathsf{F}^{(d)}_{0,\alpha} = 1$, we can quickly check that the original leading-twist $(n = 0)$ results are indeed reproduced.

Note that $d = 2$ formula (3.55) [61] considers exchange of a $t$-channel operator with general quantum numbers $(h', \overline{h}')$, while we directly wrote the exchange of symmetric traceless tensor $\mathsf{STT}_{J'}$ with spin $J' = |h' - \overline{h}'|$, which for nonzero $J'$ reduces as $\mathsf{STT}_{J'} = (h', \overline{h}') \oplus (\overline{h}', h')$ Section 2 [98].

Interestingly, both formulas (5.43)/(5.44) resemble each other quite well with minor modifications. Note the appearance of terms related by the "spin shadow" affine Weyl reflection given by $\widetilde{J}' \equiv 2 - d - J'$ [100], which is already present at the level of conformal blocks. It would be interesting to see if such similar structure persists for higher dimensions as well, or perhaps whether a simple master formula for general (even) dimensions can be derived.

**Summary.** Now we come back to the case of the interaction term having a nontrivial $t$-channel conformal block decomposition. The overall contribution to the anomalous dimensions of double-twist operators is then simply given by a sum of contributions for each appearing conformal block — in $d = 2/d = 4$ given by (5.43)/(5.44) — weighted by the corresponding (squared OPE) coefficients, that is

$$\gamma_{n,J} = \sum_{\text{primary } \mathcal{O}'_\star} C^{\text{int}}_{\mathcal{O}'_\star} \, \gamma^{(1)}_{n,J}\Big|_{\substack{t\text{-channel} \\ \text{exchange of } \mathcal{O}'_\star}} . \tag{5.47}$$

The $1/N$ factor associated with interaction/exchange diagram — see (5.35) — is included in the weighting coefficients $C^{\text{int}}$, so the anomalous dimensions $\gamma_{n,J}$ are of order $O(1/N)$.

Note the appearance of the $\sin(\pi[\cdots])$ factor in both of the formulas (5.43)/(5.44). They have therefore zeros at the MFT dimensions $\Delta' = 2\Delta_\phi + 2n' + J', n' \in \mathbb{N}_0$, so only non-MFT operators in the crossed channels contribute to the anomalous dimensions. This directly reflects the property of $\mathsf{CrK}^{s\leftarrow t}$ mentioned at the end of Section 5.3 →p.117.

The only *theory-specific* information is thus dimensions and spins of non-MFT operators together with corresponding squared OPE coefficients, or equivalently the poles and residues of $t$-channel spectral function of $t$-channel exchange. Conformal symmetry then dictates the form of the anomalous dimensions through the structure of 6j–symbol/$\mathsf{CrK}^{s\leftarrow t}$.

**Large Spin Asymptotics.** Finally, we will briefly discuss the large spin asymptotics of the anomalous dimensions, and in particular their $n$-dependence. More details (with references) can be found in Section 5.5 →p.148.

One can check that large $J$ asymptotics of the contributions (5.43)/(5.44) to the anomalous dimensions are given by

$$\gamma^{(1)}_{n,J}\Big|_{\substack{t\text{-channel} \\ \text{exchange} \\ \Delta', J'}} \sim J^{-\tau'} \equiv J^{-(\Delta' - J')} \quad \Longrightarrow \quad \gamma_{n,J} \simeq -\frac{c_n}{J^{\tau_{\min}}} + \ldots , \tag{5.48}$$

so the leading asymptotics of anomalous dimensions (5.47) are governed by the exchanged operator with the lowest twist $\tau_{\text{min}} = \min \tau' \equiv \min(\Delta' - J')$.

The point we want to make here is that whatever the coefficient $c_0$ for the leading-twist operators is — for explicit formula see (5.90), which is actually independent of the dimension — inspection of (5.43)/(5.44) leads to

$$c_n \overset{d\in\{2,4\}}{=\!=\!=} \mathsf{F}^{(d)}_{n,\frac{\Delta'+J'}{2}}\, c_0\,, \tag{5.49}$$

where we take the operator with $(\Delta', J')$ to be the one with the lowest twist, and $\mathsf{F}^{(d)}_{n,\alpha}$ was defined in (5.46).

This follows from the fact that the “spin-shadow” terms — with $\Omega_{\Delta_\phi+J+n,\frac{\Delta'-J'}{2},\Delta_\phi}$ in $d = 2$ and $\Omega_{\Delta_\phi+J+n,\frac{\Delta'-J'}{2}-1,\Delta_\phi-1}$ in $d = 4$ — are dominant for large $J$.

Since rest of their prefactors are the same for $n = 0$, the dimension-independence of $c_0$ results from their identical asymptotics. Going now to nonzero $n$, the modifications are:

**(1)** in $d = 4$ we have extra rational factors which go to 1 for $J \to \infty$,

**(2)** there is extra factor of $\mathsf{F}^{(d)}_{n,\frac{\Delta'+J'}{2}}$ before the “spin-shadow” dominant term,

**(3)** in the rest of formula we replace $J \mapsto J + n$, which does not change the asymptotics of the form $J^{-\tau'}$.

The conclusion is that only **(2)** changes the asymptotics, leading to (5.49).

Just to be clear, we have defined $\mathsf{F}^{(d)}_{n,\alpha}$ in (5.46) only for $d \in \{2, 4\}$ by bundling similar factors appearing in both dimensions. Whether such elegant structure for general-twist anomalous dimensions persists also for higher dimensions is an open question.

## 5.4 Spectrum of the O($N$) Model in AdS

Naturally, as the first step, in Section 5.4 →p.124 we decompose the 4-point boundary correlator into irreducible representations of the global symmetry group O($N$). This effectively solves the O($N$) index structure, and enables us to focus on each irrep separately.

The singlet spectrum was investigated in [59]. In Section 5.4 →p.126 we summarize their main results and also extend them in some ways — in addition to $d = 2$ we consider also $d = 4$, where a new operator can appear at strong enough coupling. We showcase the coupling dependence of both the anomalous dimensions and OPE coefficients, and also discuss the striking pattern of the spectrum when the bulk is tuned to the criticality. The role of singlet spectrum does not end here, since it provides crucial input for the non-singlet spectrum via methods presented in Section 5.3→p.117 and Section 5.3→p.119. We treat their specific implementation for non-singlet spectrum of O($N$) model in Section 5.4→p.137.

## Decomposition into O($N$) Irreducible Representations

We want to study various operators appearing in the $\mathcal{O}_\phi^\bullet \times \mathcal{O}_\phi^\bullet$ OPE, where $\mathcal{O}_\phi^\bullet$ is the $\mathsf{CFT}_d$ operator dual to the elementary field $\phi^\bullet$ in the $\mathsf{EAdS}_{d+1}$ bulk. It is well known that primaries appearing in the MFT limit (in our case $N \to \infty$) — apart from the identity operator — are the double-twist operators schematically given by $\mathcal{O}_{n,J}^{\bullet\bullet} = [\mathcal{O}_\phi^\bullet \mathcal{O}_\phi^\bullet]_{n,J} \equiv [\mathcal{O}_\phi^\bullet \Box^n \partial^J \mathcal{O}_\phi^\bullet - \text{traces}]$. They come from the crossed-channel identities, as was already mentioned in Section 5.3→p.119.

In order to alleviate the struggle of carrying the O($N$) indices around, we will organize the CFT operators by their O($N$) irreps, and correspondingly decompose the 4-point correlator. Without loss of generality, we choose the $s$-channel as the one in which we perform both the OPE and the O($N$)-irrep decomposition.

**Double-Twist Operators.** Since each $\mathcal{O}_\phi^\bullet$ transforms in the vector representation $V$ of O($N$), the $\mathcal{O}_\phi^\bullet \times \mathcal{O}_\phi^\bullet$ OPE decomposition under O($N$) follows from the standard

$$V \otimes V = \underbrace{\mathbf{1}}_{(\mathtt{S})} \oplus \underbrace{\textstyle\bigwedge^2 V}_{(\mathtt{AS})} \oplus \underbrace{\odot^2 V}_{(\mathtt{ST})} , \tag{5.50}$$

where the O($N$) irreps appearing are the singlet (S), the anti-symmetric (AS), and the symmetric traceless (ST) representation. The corresponding projectors are given by

$$\left(\mathcal{P}_{(\mathtt{S})}\right)^{ij}_{kl} = \frac{1}{N}\,\delta^{ij}\delta_{kl}\,, \qquad \left(\mathcal{P}_{(\mathtt{AS})}\right)^{ij}_{kl} = \delta^{[i}_{[k}\delta^{j]}_{l]}\,, \qquad \left(\mathcal{P}_{(\mathtt{ST})}\right)^{ij}_{kl} = \delta^{(i}_{(k}\delta^{j)}_{l)} - \frac{1}{N}\,\delta^{ij}\delta_{kl}\,. \tag{5.51}$$

Thus, the double-twist operators can be organized into these three O($N$) irreps, and are schematically given as

$$\mathcal{O}_{n,J}^{\bullet\bullet} \cdots \begin{cases} (\mathtt{S}) & \mathcal{O}_{n,J}^{(\mathtt{S})} \equiv \mathcal{P}_{(\mathtt{S})}\Big([\mathcal{O}_\phi^\bullet \mathcal{O}_\phi^\bullet]_{n,J}\Big) = \dfrac{1}{N}\displaystyle\sum_i [\mathcal{O}_\phi^i \mathcal{O}_\phi^i]_{n,J}\,, \\ (\mathtt{AS}) & \mathcal{O}_{n,J}^{[ij]} \equiv \mathcal{P}_{(\mathtt{AS})}\Big([\mathcal{O}_\phi^i \mathcal{O}_\phi^j]_{n,J}\Big) = [\mathcal{O}_\phi^{[i} \mathcal{O}_\phi^{j]}]_{n,J}\,, \\ (\mathtt{ST}) & \mathcal{O}_{n,J}^{\{ij\}} \equiv \mathcal{P}_{(\mathtt{ST})}\Big([\mathcal{O}_\phi^i \mathcal{O}_\phi^j]_{n,J}\Big) = [\mathcal{O}_\phi^{(i} \mathcal{O}_\phi^{j)}]_{n,J} - \dfrac{1}{N}\,\delta^{ij}\displaystyle\sum_k [\mathcal{O}_\phi^k \mathcal{O}_\phi^k]_{n,J}\,. \end{cases} \tag{5.52}$$

**Decomposition of the 4-Point Correlator.** From the form of the $\sigma\phi^2$ interaction vertex (5.4) that couples only two identical $\phi^\bullet$ fields, the correlator (5.17) clearly

takes the form

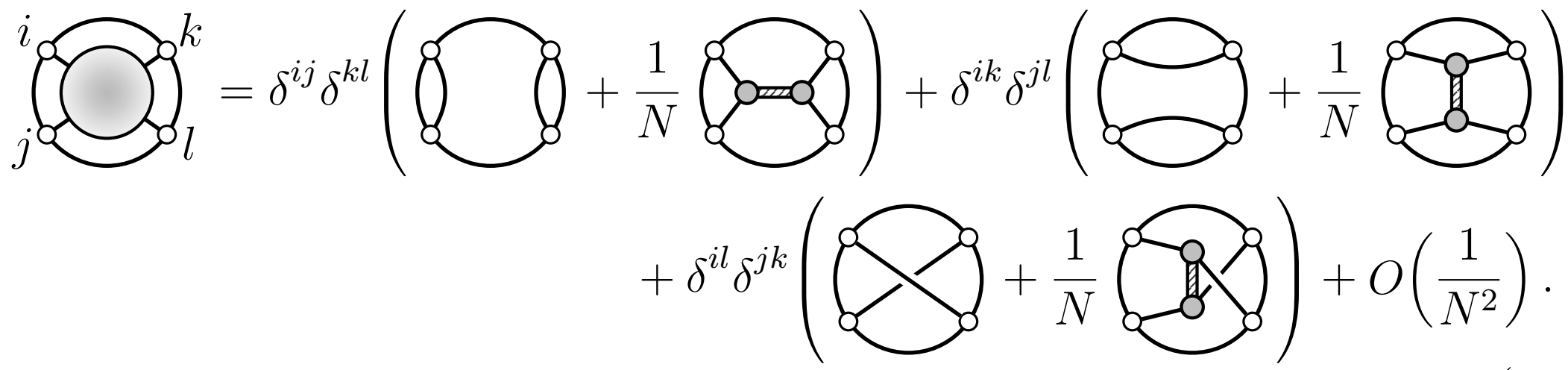

(5.53)

Diagrams on the right-hand side without indices are meant to be evaluated by taking any (but fixed) field $\phi^i$ on all external legs. We have thus decoupled the global group-theoretic index structure of the correlator from the dynamics carried by exchange of the $\sigma$-field.

By irreducibility (Schur's lemma or equivalently the fact that a product of non-equal projectors vanishes), only matching O($N$) irreps on both sides of the $s$-channel decomposition can combine to contribute nontrivially to the 4-point correlator. Therefore, its O($N$)-irrep decomposition reads

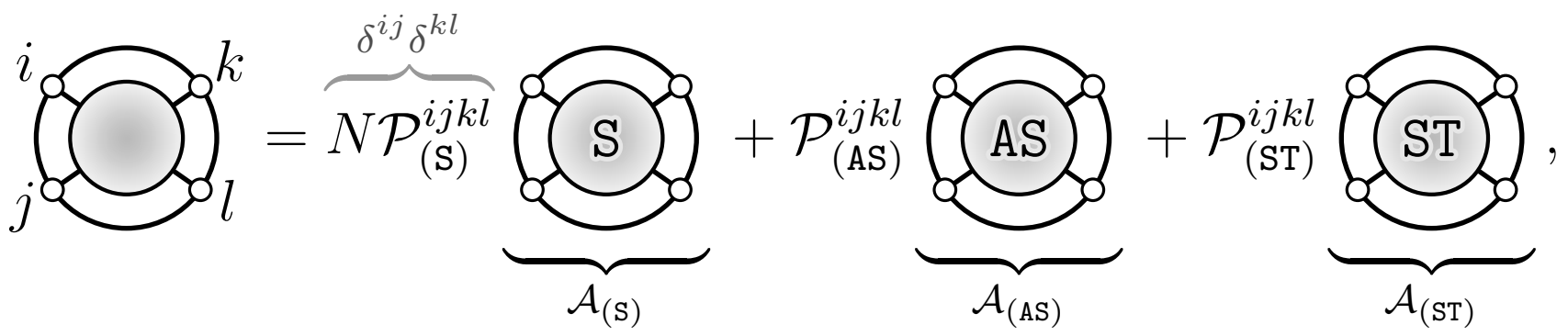

(5.54)

where indices of projectors (5.51) were raised using the Kronecker delta $\delta^{\bullet\bullet}$. Together with the singlet projector $\mathcal{P}_{(\mathtt{S})}$ we introduced an explicit factor of $N$, such that together they are of order $O(1)$, same as non-singlet projectors.

Solving (5.53) and (5.54) for $\mathcal{A}_{(\mathtt{S})}$, $\mathcal{A}_{(\mathtt{AS})}$ and $\mathcal{A}_{(\mathtt{ST})}$ yields

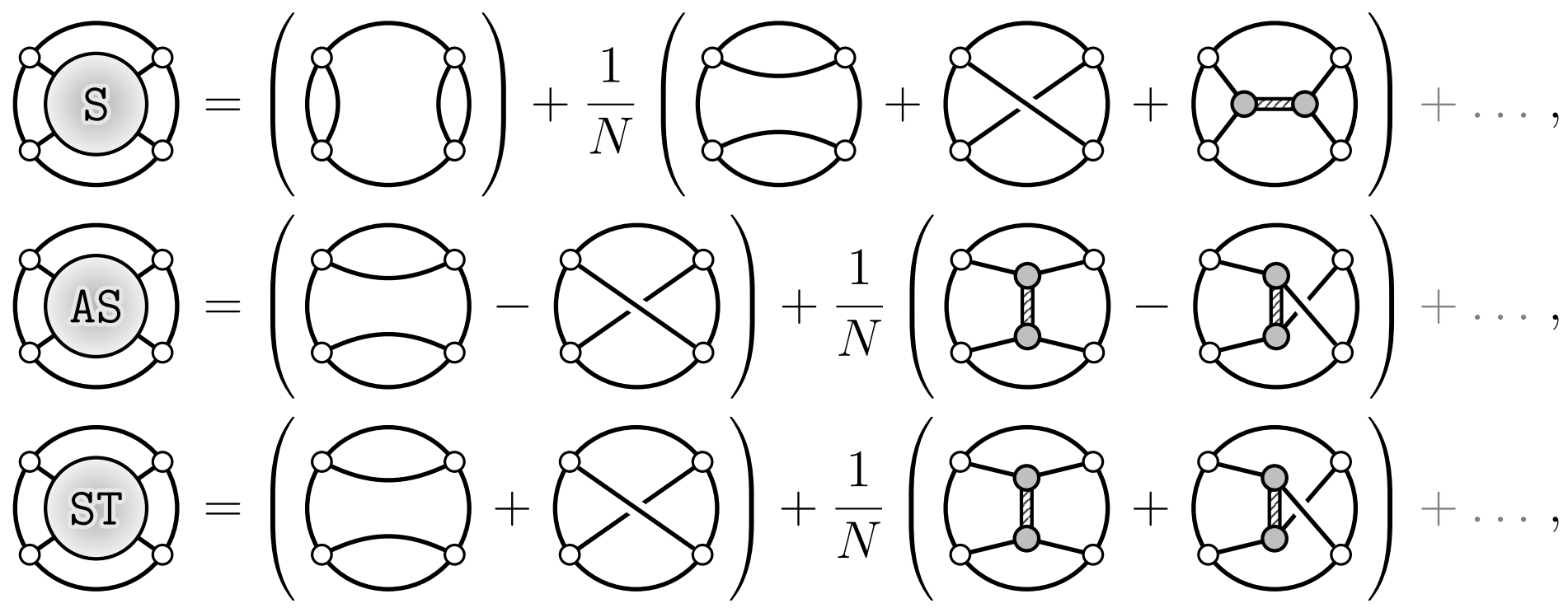

(5.55)

(5.56)

(5.57)

which represent the three projections of the correlator onto O($N$) irreps. Since the whole correlator is itself of order $O(1)$, their leading order is also $O(1)$. Next are then $O(1/N)$ corrections — mainly coming from the interactions — which we are about to study. As we know, the boundary CFT spectrum decomposes into the same O($N$) irreps, and each sector can be analyzed by its associated projection of the correlator.

Authors of [59] paid thorough attention to the singlet spectrum encoded in the singlet projection $\mathcal{A}_{(\mathtt{S})} = (5.55)$. The goal of this paper is to supplement their efforts by the analysis of the remaining rank-2 non-singlet sectors governed by $\mathcal{A}_{(\mathtt{AS})} = (5.56)$ and $\mathcal{A}_{(\mathtt{ST})} = (5.57)$.

However, before embarking on this journey, we need to recall the results of singlet sector, as it will serve as an input for computing the anomalous dimensions for the remaining two O($N$) irreps via crossing relations discussed in Section 5.3 →p.119.

## Singlet Spectrum

The leading $O(1)$ term in singlet sector (5.55) is just the $s$-channel identity contribution — two identical scalar operators $\mathcal{O}^i_\phi \mathcal{O}^i_\phi$ always contain the identity operator $\mathbf{1}$ in their OPE.

Substantially more interesting is the subleading $O(1/N)$ term, where the $t$-channel and $u$-channel disconnected diagrams meet together with the $s$-channel exchange diagram. To understand this correction to the MFT picture, we need to analyze the spectral function

$\mathsf{Spec}_s$ [ 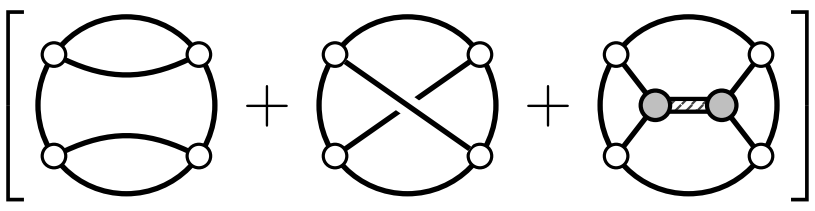 ] . (5.58)

**Disconnected Contributions.** The first two disconnected Witten diagrams correspond in the boundary CFT to GFF/MFT contributions, which were already showcased in Section 5.3 →p.119. Slightly more generally (for even/odd combination of $t$-channel and $u$-channel) we have

$\mathsf{Spec}_s \left[ \begin{matrix} \Delta \\ J \end{matrix} \middle| \right.$ 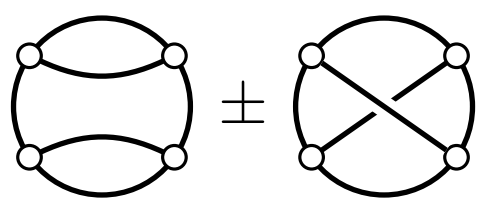 $\left. \right] = \left(1 \pm (-1)^J\right)\Big((5.36)\Big)\,,$ (5.59)

with the two terms differing just by the overall sign $(-1)^J$, see the beginning of Section 5.3 →p.114.

In the $O(1/N)$ singlet sector spectral function (5.58) we encounter the even combination of (5.59), so only even $J$ survive. Thus, the $O(1/N)$ disconnected part of the singlet sector generates poles at the MFT dimensions $\Delta^{(\mathrm{MFT})}_{n,J} = 2\Delta_\phi + 2n + J$ for even spins $J$, and the associated squared OPE coefficients are $2/N$ times the expression one would get just from the $t$-channel identity (5.36). Since the squared OPE coefficients have a factor of $1/N$, the OPE coefficients themselves are of order $O(1/\sqrt{N})$.

The conformal block decomposition of the disconnected contributions to (5.55) is therefore

$\frac{1}{N}\Big($ 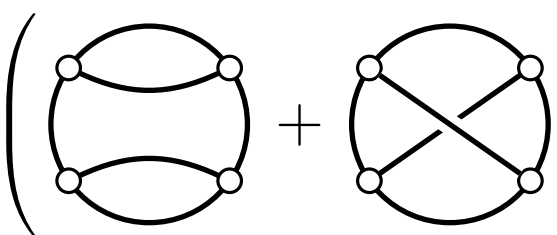 $\Big) = \sum_{n=0}^{\infty} \sum_{\substack{J=0 \\ J \text{ even}}}^{\infty} \frac{2}{N}\, C^{(\mathrm{MFT})}_{n,J} \Big| G^{(s)}_{2\Delta_\phi+2n+J,J} \Big\rangle\,,$ (5.60)

where $C_{n,J}^{(\mathrm{MFT})}$ are MFT squared OPE coefficients coming from the $t$-channel identity.

**Exchange of $\sigma$-Field.** The next contribution in (5.58) comes from the $s$-channel exchange diagram, whose calculation was schematically outlined in (5.20). The corresponding spectral function — including all numeric and $\Gamma$-factors — is obtained by comparing the result of (4.18) [59] with our convention for the integral CB decomposition (5.25), yielding

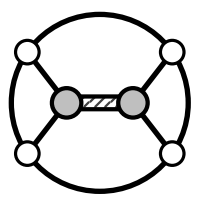

$$\mathsf{Spec}_s\left[\begin{array}{c}\Delta\\ J\end{array}\middle|\quad\right] = -\delta_{J,0}\,\frac{1}{\lambda^{-1}+2\widetilde{B}(\Delta)}\,\frac{\Gamma^2_{\Delta_\phi-\frac{\Delta}{2}}\,\Gamma^2_{\Delta_\phi-\frac{\widetilde{\Delta}}{2}}\,\Gamma^2_{\frac{\Delta}{2}}\,\Gamma^2_{\frac{\widetilde{\Delta}}{2}}}{4\pi^d\,\Gamma^2_{\Delta_\phi}\,\Gamma^2_{1-\frac{d}{2}+\Delta_\phi}\,\Gamma_{\Delta-\frac{d}{2}}\,\Gamma_{\widetilde{\Delta}-\frac{d}{2}}}\,. \tag{5.61}$$

Compared to [59], we use $\Delta_\phi$ for external scaling dimension instead of theirs $\Delta$, and we usually prefer writing formulas directly in terms of $\Delta \equiv \frac{d}{2} + \mathring{\imath}\nu$ instead of $\nu$. We easily see that (5.61) is shadow-symmetric — $\Delta \longleftrightarrow \widetilde{\Delta}$ — provided that the bubble function $\widetilde{B}$ is shadow-symmetric as well, which holds generally for spectral representations of functions.

The first thing to notice is that (5.61) has support only on spin $J = 0$ operators. Already at this point we can see that $J > 0$ operators in the singlet sector have zero anomalous dimensions — a statement valid to the $O(1)$ order actually considered, since the squared OPE coefficients already have a factor of $1/N$, and we do not consider $O(1/N^2)$ corrections. Concluding this observation, to the order $O(1/N)$, the singlet $J > 0$ operators appearing in the $\mathcal{O}_\phi^\bullet \times \mathcal{O}_\phi^\bullet$ OPE are $\mathcal{O}_{n,J}^{(\mathsf{S})}$ with even $J$, MFT dimensions $\Delta_{n,J>0}^{(\mathsf{S})} = \Delta_{n,J}^{(\mathrm{MFT})} \equiv 2\Delta_\phi + 2n + J$, and squared OPE coefficients as in (5.60).

**Bootstrap Idea — Consistency of the Spectrum.** The case of $J = 0$ is more intricate. The factor of $\Gamma^2\big(\Delta_\phi - \frac{\Delta}{2}\big)$ generates a set of double-poles at the MFT dimensions $\Delta = 2\Delta_\phi + 2n$, which is something one would expect from perturbative corrections to the scaling dimensions of $\mathcal{O}_{n,0}^{(\mathsf{S})}$ from the interaction. Indeed, expanding (5.61) in $\lambda$, and comparing it to the perturbative expansion of poles in $\mathsf{Spec}_s$ — as in (5.40), where instead of $1/N$ we now consider series in $\lambda$ — one sees that $O(\lambda)$ term coming from single insertion of $\lambda\phi^4$ vertex has double poles and generates $O(\lambda)$ anomalous dimensions for $\mathcal{O}_{n,0}^{(\mathsf{S})}$.

Furthermore, at $O(\lambda^k)$ we would expect appearance of $(k+1)$–degree poles at MFT dimensions. Such a behavior can come only from higher and higher powers of bubble function $\widetilde{B}$ in the $\lambda$-expansion of (5.61). It is thus plausible to anticipate that $\widetilde{B}$ has poles at MFT dimensions, such that diagrams (5.9) with more bubbles contribute to anomalous dimensions at corresponding orders of $\lambda$. At the same time, if $\widetilde{B}$ happened to have additional poles at non-MFT dimensions, it would indicate appearance of new operators already at the $O(\lambda^2)$ order. Such operator spectrum changes are generally not expected to happen perturbatively for weak coupling, so we can conclude that the $\widetilde{B}$ function should have poles only at MFT dimensions.

However, in the resummed situation — that is finite $\lambda$ — in addition to MFT double-poles still coming from $\Gamma^2\left(\Delta_\phi - \frac{\Delta}{2}\right)$ there appear poles coming from zeros of the $\lambda^{-1} + 2\widetilde{B}$ denominator. These poles depend continuously on the coupling $\lambda$, so in order for this to be consistent — without sudden appearance of a whole bunch of new operators once we turn on the coupling — the remnants of the MFT poles must be canceled between the disconnected and exchange diagrams. For such cancellation, the $\Gamma^2$ double-poles must be accompanied by simple zeros of the $(\lambda^{-1} + 2\widetilde{B})^{-1}$, that is simple poles of $\widetilde{B}$ at MFT dimensions, as was already anticipated. Such arguments allowed the authors of [59] to "bootstrap" the bubble function $\widetilde{B}$ (or more precisely its spectral representation).

**Bubble Function.** This requirement of precise cancellation of the MFT poles — together with the nice behavior at infinity for low enough dimension $d$ — leads to the following sum over the $\Delta$ poles (4.26), (4.27) [59]

$$\begin{aligned}\widetilde{B}(\Delta) &= \frac{1}{4(4\pi)^{\frac{d}{2}}} \sum_{n=0}^{\infty} \frac{1}{\Delta_\phi - \frac{\Delta}{2} + n} \frac{\Gamma_{\frac{d}{2}+n}\,\Gamma_{\Delta_\phi+n}\,\Gamma_{\Delta_\phi-\frac{d}{2}+\frac{1}{2}+n}\,\Gamma_{2\Delta_\phi-\frac{d}{2}+n}}{\Gamma_{\frac{d}{2}}\,\Gamma_{\Delta_\phi+\frac{1}{2}+n}\,\Gamma_{\Delta_\phi-\frac{d}{2}+1+n}\,\Gamma_{2\Delta_\phi-d+1+n}} \frac{1}{n!} + \left(\Delta \leftrightarrow \widetilde{\Delta}\right) \\ &= \frac{\Gamma_{\Delta_\phi}\,\Gamma_{\Delta_\phi-\frac{d}{2}+\frac{1}{2}}\,\Gamma_{2\Delta_\phi-\frac{d}{2}}}{4(4\pi)^{\frac{d}{2}}} \times \\ &\times \left( \begin{array}{c} \Gamma_{\Delta_\phi-\frac{\Delta}{2}}\, {}_5\widetilde{\mathbf{F}}_4\left[ \begin{array}{c} \Delta_\phi - \frac{\Delta}{2},\ \frac{d}{2},\ \Delta_\phi,\ \Delta_\phi - \frac{d}{2} + \frac{1}{2},\ 2\Delta_\phi - \frac{d}{2} \\ \Delta_\phi - \frac{\Delta}{2} + 1,\ \Delta_\phi + \frac{1}{2},\ \Delta_\phi - \frac{d}{2} + 1,\ 2\Delta_\phi - d + 1 \end{array} \middle| 1 \right] \\ + \left(\Delta \xleftrightarrow{\text{exchange}} \widetilde{\Delta}\right) \end{array} \right),\end{aligned} \tag{5.62}$$

where after extracting some $\Gamma$-factors in front we recognized the *regularized* generalized hypergeometric function ${}_5\widetilde{\mathbf{F}}_4$. The bubble function $\widetilde{B}(\Delta)$ — even for unequal masses — has been previously computed also by different methods in [101] (see also [102]).

The summands in (5.62) behave asymptotically as $n^{d-4}$ for $n \to \infty$, so the sum becomes divergent for $d + 1 \geq 4$. Same as in flat space, this corresponds to the UV divergence of the bubble diagram, and one must regularize it somehow. By subtracting sufficient number of terms (up to and including degree $d - 3$) in the Taylor expansion of the summands around $\nu = 0 \Leftrightarrow \Delta = \frac{d}{2}$, the series becomes convergent and can be summed up in principle. Note that only even powers of $\nu \equiv -\mathring{\imath}(\Delta - \frac{d}{2})$ are present, since (5.62) is shadow-symmetric. At the end, to account for the subtraction, an even polynomial in $\nu$ of the corresponding degree with arbitrary coefficients (in principle depending on $\Delta_\phi$) must be added back to the regularized sum. This was also discussed for the spin 1 bubble function in Section 3.3 [60].

The sum simplifies in even dimensions (see NOTEBOOK for details). For example, taking $d = 2$, the bubble function does not need any regularization, and it evaluates to

$$\widetilde{B}(\Delta) \overset{d=2}{=\!=} \frac{\mathring{\imath}}{4\pi} \frac{\psi\left(\Delta_\phi - \frac{1+\mathring{\imath}\nu}{2}\right) - \psi\left(\Delta_\phi - \frac{1-\mathring{\imath}\nu}{2}\right)}{2\nu} \equiv \frac{1}{4\pi} \frac{\psi_{\Delta_\phi - \frac{\Delta}{2}} - \psi_{\Delta_\phi - \frac{\widetilde{\Delta}}{2}}}{\widetilde{\Delta} - \Delta}\,, \quad (5.63)$$

where $\psi_z \equiv \psi(z) \equiv \frac{\mathrm{d}}{\mathrm{d}z} \ln \Gamma(z)$ is the digamma function.

In the case of $d = 4$, the bubble function is UV divergent and requires a regularization as we described above. Due to the shadow-symmetry, it is enough to subtract the constant term in the expansion of the summands around $\Delta = \frac{d}{2}$, and the result is

$$\widetilde{B}(\Delta) \overset{d=4}{=\!=} \frac{\nu\left[2(2\Delta_\phi - 5)\nu - \mathring{\imath}\left(4(\Delta_\phi - 2)^2 + \nu^2\right)\left(\psi_{\Delta_\phi - \frac{\Delta}{2}} - \psi_{\Delta_\phi - \frac{\widetilde{\Delta}}{2}}\right)\right]}{128\pi^2(1+\nu^2)} + a_0(\Delta_\phi)\,. \quad (5.64)$$

There are no clear physical constraints which would enable us to fix the subtraction ambiguity in the calculation of the regularized bubble function, in the case of $d = 4$ being just the undetermined constant $a_0$. This can be seen from the fact that $\lambda$ itself is not renormalization scheme independent, and only the combination $\lambda^{-1} + 2\widetilde{B}$ in (5.61) has invariant meaning directly connected with the shape of the physical spectrum.

In the following we will set $a_0 \equiv -\frac{1}{16\pi^2}$. In Section 5.4 →p.134 we will discuss the rationale behind this choice. Choosing different $a_0$ can always be reabsorbed in the coupling $\lambda$.

One question still remains — what is the range of physically admissible values of $\lambda$? Without a more detailed analysis of the phase structure in $d = 4$ we are unable to provide a definitive answer, and will assume that any non-negative value of $\lambda$ is allowed.

## Analysis of the Singlet Sector

Now that we presented all relevant formulas for the singlet sector, we can analyze it in more detail. Since the $J > 0$ operators are unaffected by the $s$-channel exchange diagram, in the following we will focus solely on the $J = 0$ operators.

In particular, we can extract the scaling dimensions of the singlet operators contributing in the subleading $O(1/N)$ order from the poles of the spectral function (5.58), and also the corresponding squared OPE coefficients from the residues of these poles.

**Scaling Dimensions in the Singlet Sector.** As we already discussed, due to the interplay of the disconnected diagrams and the $s$-channel exchange, the $J = 0$

MFT poles $\Delta^{(\text{MFT})}_{n,0} \equiv 2\Delta_\phi + 2n$ are canceled in the complete spectral function (5.58). Instead, they are replaced by their finite-shifted counterparts $\Delta^{(\text{S})}_{\bullet,0}$, which are roots of

$$\lambda^{-1} + 2\widetilde{B}\left(\Delta^{(\text{S})}_{\bullet,0}\right) = 0\,, \tag{5.65}$$

such that (5.61) has a pole at the corresponding location. We use $\bullet$ to indicate the place for indexing this new (infinite) family of singlet $J = 0$ non-MFT primary operators denoted suggestively as $\widehat{\sigma}_\bullet$, since they are induced by the exchange of the $\sigma$-field. As we will see soon, not all of them need to be continuously connected to the MFT operators.

For generic $\lambda$ and $\Delta_\phi$, the key equation (5.65) is transcendental, and its roots have to be found numerically. Nevertheless, this can be easily done to a very high precision using standard numerical methods, for example with the help of Wolfram Mathematica.

Examples of the bubble functions $\widetilde{B}$ in $d = 2$ and $d = 4$, together with some particular choices of $\lambda$, are displayed in Figure 5.66 →p.131. Intersection points in the plots correspond to singlet scalar operators $\widehat{\sigma}_\bullet$. At small $\lambda$, we can clearly identify them as finite deformations of the MFT operators $\mathcal{O}^{(\text{S})}_{n,0} \simeq [\mathcal{O}_\phi^{\bullet}\Box^n\mathcal{O}_\phi^{\bullet}]^{(\text{S})}$. With increasing $\lambda$, the anomalous dimensions grow, and eventually become of order $O(1)$ in both $\lambda$ and $1/N$.

There is a subtlety in $d = 4$, where for a strong enough coupling $\lambda$ a new operator possibly appears that is not continuously connected with the MFT spectrum. Such an operator would be then associated to a bound state in AdS. As we will see, its emergence is crucial for the bulk theory to be critical, which will be discussed in Section 5.4 →p.134.

At large conformal dimensions $\Delta \gg 1$, which concerns the operators $\mathcal{O}^{(\text{S})}_{n,0}$ with $n \gg 1$, the bubble functions have asymptotics (dots include subleading $1/\Delta$ terms)

$$\widetilde{B}(\Delta) \overset{d=2}{=\!=} \frac{\cot\left(\pi(\Delta_\phi - \frac{\Delta}{2})\right)}{8\Delta} + \ldots\,, \qquad \widetilde{B}(\Delta) \overset{d=4}{=\!=} \frac{\Delta\cot\left(\pi(\Delta_\phi - \frac{\Delta}{2})\right)}{128\pi} + \ldots\,. \tag{5.67}$$

For finite $\lambda$, the inspection of (5.65) gives us following asymptotic values of the anomalous dimensions (governed by the infinities or zeros of $\cot(\cdots)$ in $d = 2$ or $d = 4$, respectively)

$$(0 < \lambda < \infty) \qquad \lim_{n\to\infty} \gamma^{(\text{S})}_{n,0} = \begin{cases} 0 & \text{for } d = 2\,, \\ 1 & \text{for } d = 4\,. \end{cases} \tag{5.68}$$

Of course, for $\lambda = 0$ all anomalous dimensions vanish, and for $\lambda = \infty$ the different $\Delta$ scaling in (5.67) does not play a role, and we have $\lim_{n\to\infty} \gamma^{(\text{S})}_{n,0} = 1$ for both $d = 2$ and $d = 4$.

In summary, the complete singlet spectrum given by the poles of the spectral function (5.58) consists of non-MFT scalar ($J = 0$) operators supplemented by MFT operators supported at even spins $J \geq 2$. The corresponding singlet twist–spin plot is displayed in Figure 5.69→p.132, with the twist being defined as $\tau^{(\mathrm{S})}_{n,J} \equiv \Delta^{(\mathrm{S})}_{n,J} - J \equiv 2\Delta_\phi + 2n + \gamma^{(\mathrm{S})}_{n,J}$.

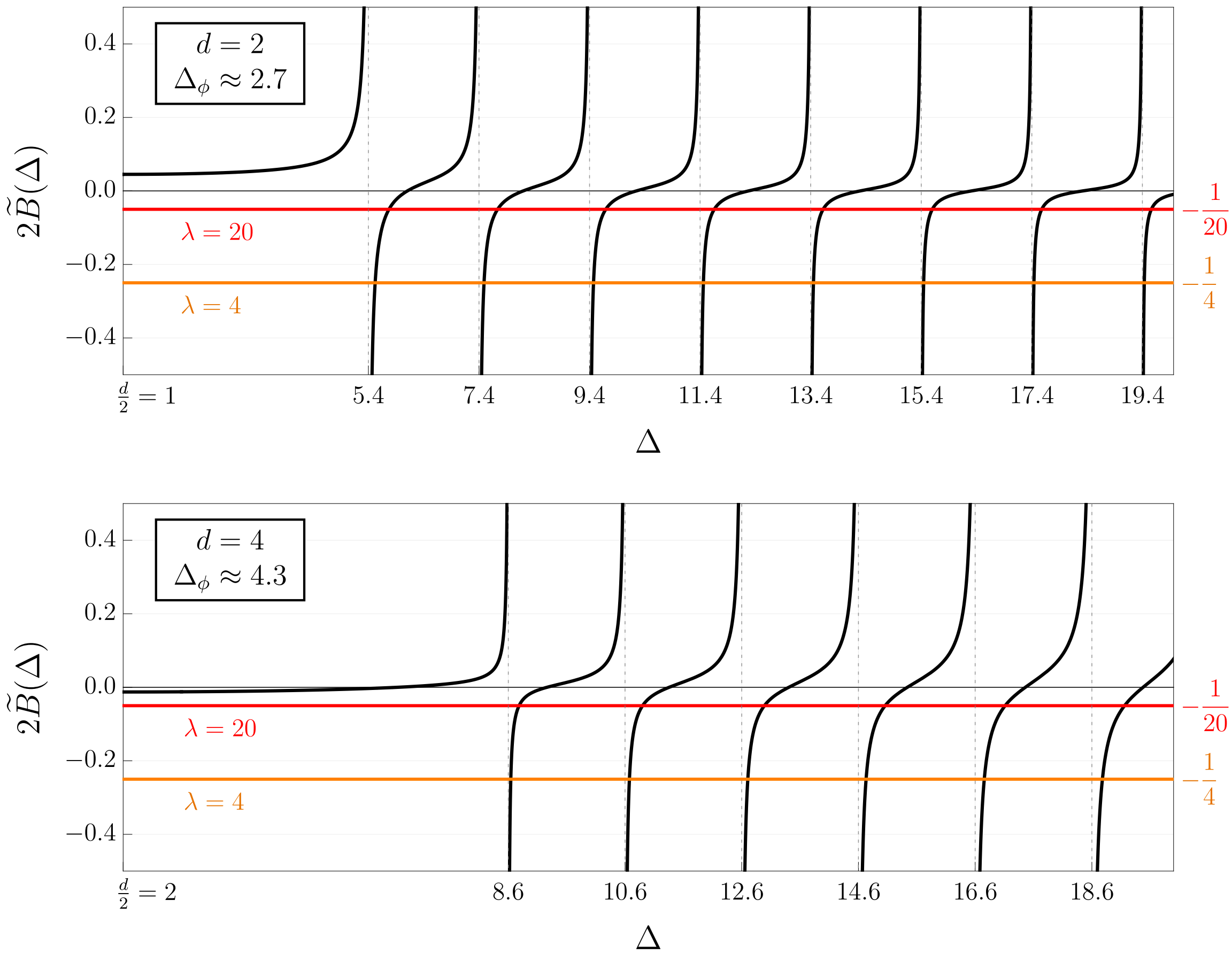

**Figure 5.66 /** Graphical representation of scalar singlet spectrum equation (5.65) in $d = 2$ and $d = 4$. The graph of (twice) the bubble function $\widetilde{B}$ (5.63)/(5.64) is drawn by solid black lines. The term $-\lambda^{-1}$ for $\lambda = 4$ and $\lambda = 20$ is represented by the orange and red lines, respectively. Their intersection points with black lines correspond to the $J = 0$ operators $\widehat{\sigma}_\bullet$ in the singlet sector. Most of these operators (all in $d = 2$) are associated with the MFT spectrum as can be seen by their asymptotic convergence to MFT values (gray dashed lines) for $\lambda \to 0$. In the case of $d = 4$, there is however a potential emergent operator at strong enough coupling in the region $\Delta \lesssim 2\Delta_\phi$, where the graph of $\widetilde{B}$ is reaching slightly below the horizontal axis. The plot was evaluated at indicated external scaling dimensions, corresponding to a massive $\phi$-field with Dirichlet boundary conditions.

The precise dependence of the singlet scalar anomalous dimensions on the coupling is plotted in Figure 5.70→p.133. The plot mainly focuses on the strong coupling region, and shows in what fashion the finite constant values of the anomalous dimensions are approached at infinite coupling. The weak coupling regime is not entirely captured in these plots, but the asymptotics in that region are simple. Anomalous dimensions are approximately linear in the coupling there, which follows from standard perturbation theory, namely the $\phi^4$ contact Witten diagram contributes to the $J = 0$ anomalous dimensions at order $O(\lambda)$.

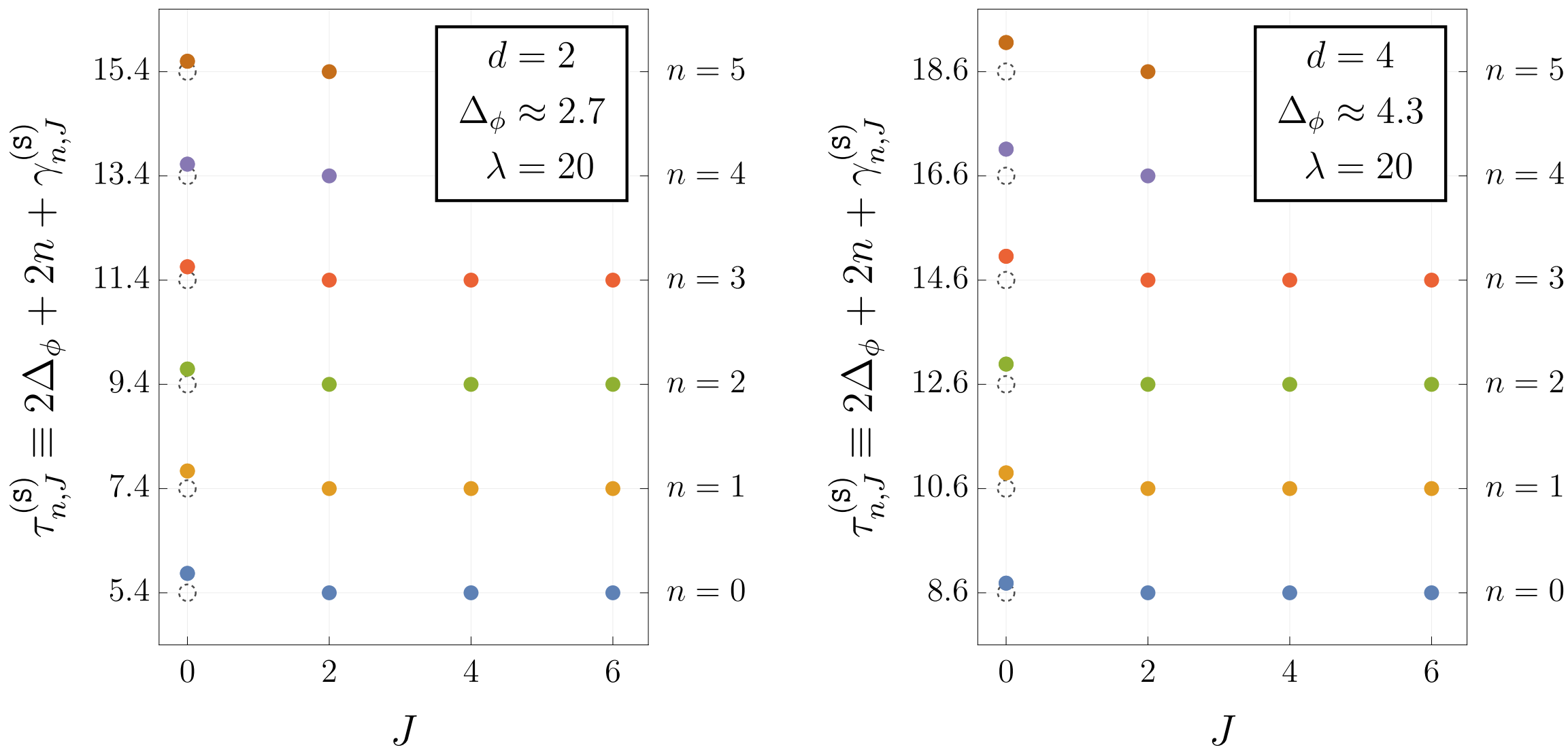

**Figure 5.69 /** Twist–spin plots of the singlet spectrum for $d = 2$ and $d = 4$ given by the poles of the complete spectral function (5.58). Only spin $J = 0$ operators get $O(1)$ anomalous dimensions in large $N$ expansion. The plots correspond to Figure 5.66 →p.131, so the non-MFT scaling dimensions of scalar operators are precisely given by the intersection points of black and red lines in that figure.

**Squared OPE Coefficients in the Singlet Sector.** Having identified the locations of physical poles $\Delta^{(s)}_{\bullet,0}$ in the spectral function (5.61), we can now compute the squared OPE coefficients of the corresponding singlet operators $\mathcal{O}^{(s)}_{\bullet,0}$.

Excluding the isolated case where $\lambda$ is tuned precisely to the critical value when a new operator emerges at $\Delta = \frac{d}{2}$, all physical poles at solutions of (5.65) are simple. Since the rest of the remaining factors in (5.61) together with $K_{\widetilde{\Delta},0}$ are holomorphic at these poles, the corresponding squared OPE coefficients are given by (5.26) as

$$\mathsf{ope}^2\left[\mathcal{P}_{(s)}\left(\mathcal{O}_\phi^{\bullet}\mathcal{O}_\phi^{\bullet}\right)\mathcal{O}^{(s)}_{\bullet,0}\right] = -\operatorname{Res}_{\Delta=\Delta^{(s)}_{\bullet,0}}\left(K_{\widetilde{\Delta},0}\,\frac{1}{N}\,\mathsf{Spec}_s\left[\begin{array}{c|c}\Delta & \\ 0 & \end{array}\right.\right.$$
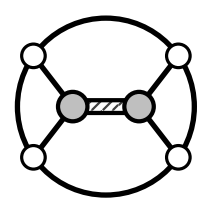
$$\left.\left.\right]\right) + O\left(\frac{1}{N^2}\right)$$
$$= \frac{1}{N}\,\frac{1}{2\widetilde{B}'(\Delta)}\,\frac{\Gamma^2_{\Delta_\phi-\frac{\Delta}{2}}\,\Gamma^2_{\Delta_\phi-\frac{\widetilde{\Delta}}{2}}\,\Gamma^4_{\frac{\Delta}{2}}}{4\pi^{\frac{d}{2}}\,\Gamma^2_{\Delta_\phi}\,\Gamma^2_{1-\frac{d}{2}+\Delta_\phi}\,\Gamma_{\Delta-\frac{d}{2}}\,\Gamma_\Delta}\Bigg|_{\Delta=\Delta^{(s)}_{\bullet,0}} + O\left(\frac{1}{N^2}\right), \tag{5.71}$$

where $\widetilde{B}' \equiv \frac{\mathrm{d}\widetilde{B}}{\mathrm{d}\Delta}$ denotes the derivative of the bubble function. Note that we included the factor of $1/N$ with which the exchange diagram enters into the correlator.

Plots for the coupling dependence of squared OPE coefficients (5.71) for some of the $J = 0$ singlet operators $\mathcal{O}^{(\mathsf{S})}_{n,0}$ associated to MFT ones in the $\lambda \to 0$ limit — we do not show possible new operator in $d = 4$ — is displayed in Figure 5.72→ p.134.

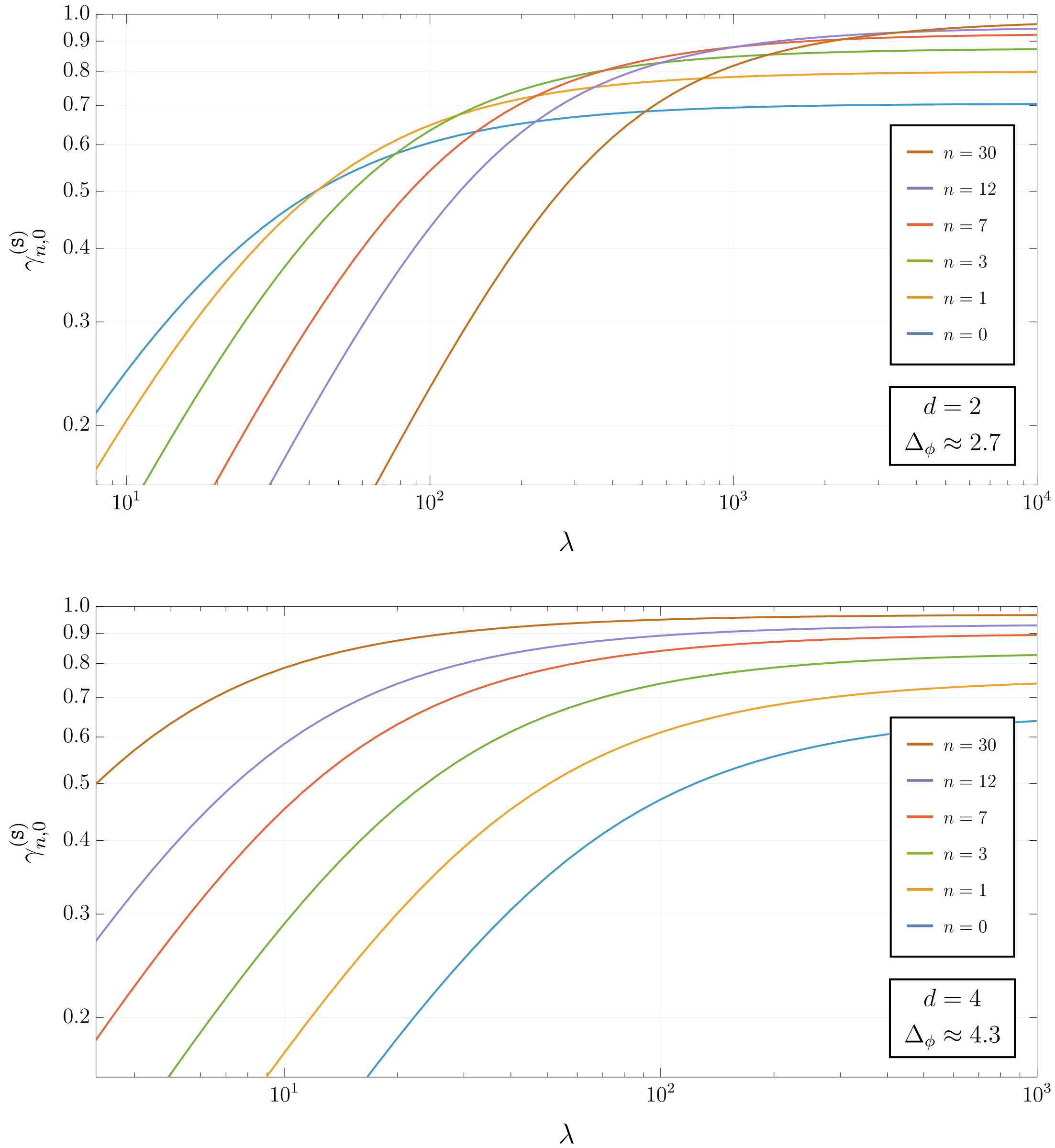

**Figure 5.70 /** Coupling dependence of the singlet scalar anomalous dimensions at indicated fixed external scaling dimensions coinciding with previous plots. In the free limit $\lambda = 0$, they correspond to primary MFT operators of the schematic form $[\mathcal{O}_\phi \Box^n \mathcal{O}_\phi]^{(\mathsf{S})}$ (we do not include here the possible new operator in $d = 4$). For any fixed $n$, the anomalous dimensions in the singlet sector are all positive and approach a constant value (bounded above by 1) at sufficiently strong coupling. Positiveness of anomalous dimensions indicates that the interaction in bulk AdS has repulsive character in the singlet sector. Different behavior when increasing $n$ for constant $\lambda$ in $d = 2$ and $d = 4$ follows from the large conformal dimension asymptotics discussed in (5.68).

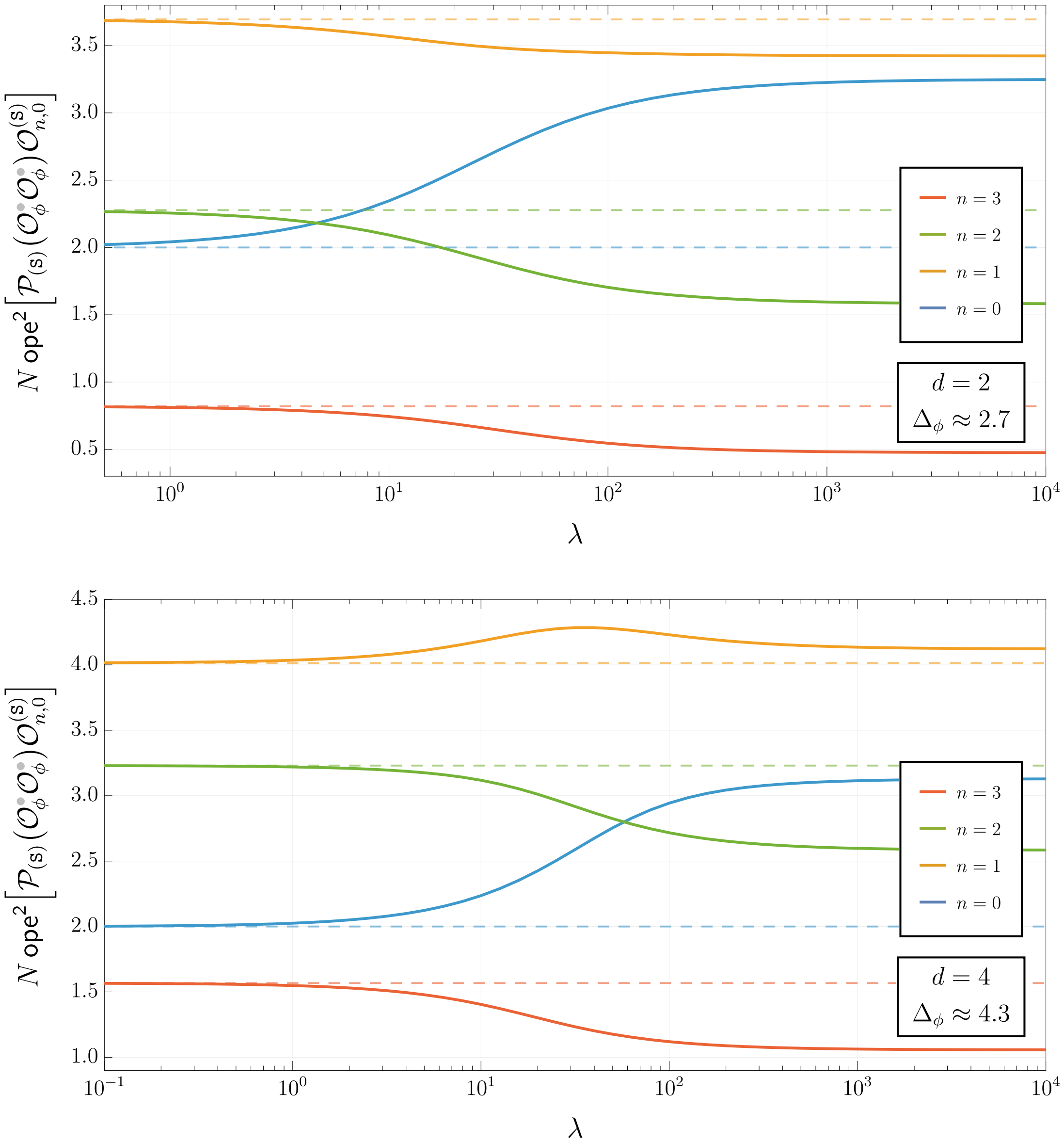

**Figure 5.72 /** Coupling dependence of squared OPE coefficients of singlet scalar operators appearing in the $\mathcal{O}_\phi^\circ \times \mathcal{O}_\phi^\circ$ OPE. They are of order $1/N$ as is clear from (5.71), so we plot the expression multiplied by this factor. In the free limit $\lambda \to 0$, they reduce to MFT values (dashed lines) given by minus residues of (5.36), as they should. All OPE coefficients approach constant values at strong coupling, with $O(1)$ corrections to the corresponding MFT values. No regularities can be inferred from these plots, neither in $d = 2$ nor in $d = 4$. Corrections of both signs occur, their strength is not ordered by scaling dimensions of the singlet scalar operators, moreover their magnitudes can cross.

## Criticality in the Bulk

By appropriately tuning the couplings of the theory, a *critical point* in the bulk of AdS can be reached. Its existence and the evidence for the bulk conformal symmetry was first discussed in Section 5 [59]. Since the critical theory in $\mathsf{EAdS}_{d+1}$ describes an interacting $\mathsf{BCFT}_{d+1}$ by performing a Weyl transformation to a flat half-space $\mathbb{R}^d \times \mathbb{R}_{\geq}$, we start by recalling the results for the large $N$ critical $\mathsf{O}(N)$ model obtained there [103].

As already mentioned in Section 5.2 →p.104, an appropriately tuned $\mathsf{O}(N)$ model in $\mathbb{R}^{d+1}$ with dimension ranging in $2 < d+1 < 4$ flows in the IR to an interacting $\mathsf{CFT}_{\mathrm{IR}}$ describing the second-order phase transition separating the broken and the unbroken phase. Considering now the theory on a flat half-space, we can obtain different BCFTs in the IR by imposing different conformal boundary conditions — for the definition of the *ordinary*, *special*, and *extraordinary* transitions of the $\mathsf{O}(N)$ model on half-space see [104] [105] [78]. Our choice of boundary conditions in AdS corresponds to the Dirichlet boundary conditions leading to the ordinary transition.

As explained for example in [78], the ordinary transition can be reached by setting $\Delta_\phi = d-1$ and sending the coupling $\lambda \to \infty$. In the critical point, the scaling dimensions of boundary operators $\widehat{\sigma}_n$ induced by the bulk $\sigma$-field are given by $\Delta_{\widehat{\sigma}_n} = d+1+2n$. The leading boundary operator $\widehat{\sigma}_0$ corresponds to the displacement operator generally present in any BCFT with a protected scaling dimension $\Delta_{\widehat{\sigma}_0} = d+1$ — see Section 3.2 [106].

As was shown already in [59] for $d = 2$ ($\mathsf{AdS}_3$), the singlet spectrum — or equivalently the spectral representation of the $\sigma$-propagator (5.7) or the bubble function (5.63) — simplifies greatly in the critical point ($\Delta_\phi = 1, \lambda \to \infty$). Recalling the equation governing the singlet spectrum (5.65), we just need to find the zeros of the simplified bubble function

$$\widetilde{B}(\Delta)\Big|_{\Delta_\phi=1} \overset{d=2}{=\!=} -\frac{\cot\left(\frac{\pi}{2}\Delta\right)}{8(\Delta-1)} \,. \tag{5.73}$$

Thus, the induced boundary singlet scalar operators $\widehat{\sigma}_n$ indeed turn out to have the expected scaling dimensions $\Delta_{\widehat{\sigma}_n} = 3+2n$.

Although the preceding discussion assumed $2 < d+1 < 4$, and in particular $d = 2$, we found a similar striking pattern exhibited also for other even $d$, suggesting the continuation of the ordinary transition above its upper critical dimension $d+1 = 4$. Here we however need to discuss the subtraction ambiguity in the definition of bubble function (5.62), which for example in $d = 4$ is just the constant $a_0$ in (5.64).

This turns out to be tied with the appearance of a new operator in $d = 4$. Taking $\Delta_\phi = d-1 = 3$, all singlet scalar operators continuously connected with the MFT spectrum have scaling dimensions $\Delta > 2\Delta_\phi = 6$, so the only candidate for the displacement operator with $\Delta_{\widehat{\sigma}_0} = 5$ is the one coming from the branch reaching slightly below the horizontal axis for small $\Delta$, see Figure 5.66 →p.131.

If we want to obtain the critical behavior at $\lambda \to \infty$, the requirement of $\Delta_{\widehat{\sigma}_0} = 5$ fixes the subtraction constant as $a_0 = -\frac{1}{16\pi^2}$ — we choose it independent of $\Delta_\phi$, such that the critical value of the coupling $\lambda_\star \equiv 8\pi^2$ when the new operator emerges is fixed. It just so happens that the rest of operators $\widehat{\sigma}_n$ with $n \geq 1$ have scaling dimensions $\Delta_{\widehat{\sigma}_n} = 5 + 2n$, so they complete the family of boundary operators induced by the bulk $\sigma$-field.

A similar story is true also for higher dimensions. Let us first present the observed formula for simplified bubble functions at critical point in even dimensions, and comment on the fixing of the subtraction ambiguity after. We found

$$\widetilde{B}(\Delta)\Big|_{\Delta_\phi = d-1} \overset{d \text{ even}}{=\!=} -\frac{\cot\left(\frac{\pi}{2}\Delta\right)}{2^{2d-1}\pi^{\frac{d}{2}-1}\Gamma\left(\frac{d}{2}\right)} \left( \prod_{\substack{a=4-d \\ a \text{ even}}}^{2(d-2)} (\Delta - a) \Big/ \prod_{\substack{b=1 \\ b \text{ odd}}}^{d-1} (\Delta - b) \right), \tag{5.74}$$

so for example we have explicitly

$$\widetilde{B}(\Delta)\Big|_{\Delta_\phi = d-1} = \begin{cases} \overset{d=2}{=\!=} -\dfrac{\cot\left(\frac{\pi}{2}\Delta\right)}{8} \dfrac{1}{(\Delta-1)}, \\ \overset{d=4}{=\!=} -\dfrac{\cot\left(\frac{\pi}{2}\Delta\right)}{128\pi} \dfrac{(\Delta-0)(\Delta-2)(\Delta-4)}{(\Delta-1)(\Delta-3)}, \\ \overset{d=6}{=\!=} -\dfrac{\cot\left(\frac{\pi}{2}\Delta\right)}{4096\pi^2} \dfrac{(\Delta+2)(\Delta-0)(\Delta-2)(\Delta-4)(\Delta-6)(\Delta-8)}{(\Delta-1)(\Delta-3)(\Delta-5)}. \end{cases} \tag{5.75}$$

Plots of the bubble functions at the critical point can be seen in Figure 5.76 → p.137.

Expressions for $d = 2$ and $d = 4$ in (5.75) can be directly obtained from general formulas (5.63) and (5.64) (with the appropriate choice of $a_0$ already discussed) by substituting $\Delta_\phi = d - 1$. In the case of $d = 6$, the subtraction ambiguity is a polynomial of the form $a_0 + a_1\nu^2$. There are now two emergent operators, and requiring their dimensions to be $\Delta_{\widehat{\sigma}_0} = 7$ and $\Delta_{\widehat{\sigma}_1} = 9$ fixes the subtraction constants $a_0$ and $a_1$. As before, continuous deformations of MFT operators then complete the family $\widehat{\sigma}_n$ with scaling dimensions $\Delta_{\widehat{\sigma}_n} = 7 + 2n$.

Similar procedure was applied to all even dimensions up to $d = 12$. Although we were not able to perform an explicit resummation of the critical bubble function for dimensions $d \geq 8$ at general $\Delta$ to explicitly compare with (5.74), direct evaluation at any chosen $\Delta$ confirms the expected structure.

The product in the numerator of (5.74) generates shadow-symmetric zeros at even integers below the MFT dimension $\Delta = 2\Delta_\phi = 2(d-1)$, canceling the "spurious" poles of $\cot\left(\frac{\pi}{2}\Delta\right)$. The product in the denominator generates shadow-symmetric poles at odd integers below the displacement operator dimension $\Delta = d + 1$, canceling the "spurious" zeros of $\cot\left(\frac{\pi}{2}\Delta\right)$. We are thus left with poles at MFT dimensions and zeros at $\Delta_{\widehat{\sigma}_n} = d + 1 + 2n$.

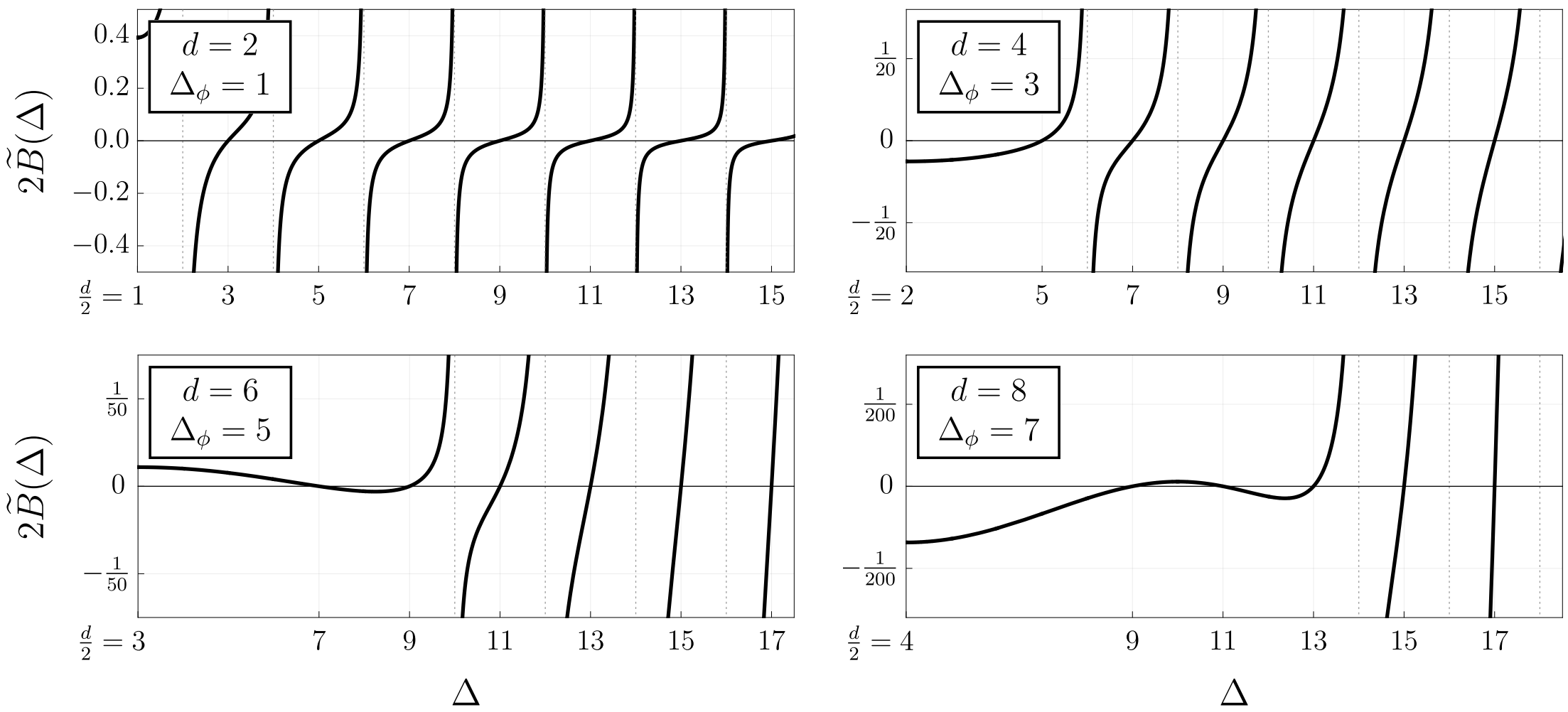

**Figure 5.76 /** Bubble functions for $\Delta_\phi = d-1$ in even dimensions up to $d=8$. The critical point is reached by sending the coupling $\lambda \to \infty$, that is by looking at the zeros of the bubble function. The scale of $y$-axis is adapted in each dimension to make the nuanced appearance of $\frac{d}{2}-1$ new operators more visible. All zeros are at dimensions $\Delta = d+1+2n$, the first corresponding to the displacement operator, and possibly some other emergent operators below the MFT dimension $2\Delta_\phi = 2(d-1)$. The rest are finite deformations of the MFT operators.

## Non-Singlet Spectrum

The non-singlet part of the spectrum is described by the spectral decomposition of (5.56) and (5.57). Both (ST) and (AS) cases can be treated simultaneously, as we have

$\mathsf{Spec}_s\Big[$ 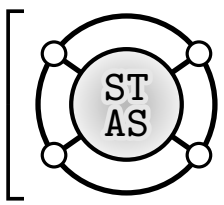 $\Big] = \big(1 \pm (-1)^J\big)\Big(\mathsf{Spec}_s\Big[$ 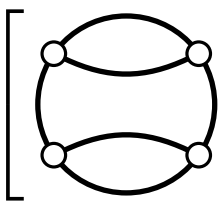 $\Big] + \frac{1}{N}\,\mathsf{Spec}_s\Big[$ 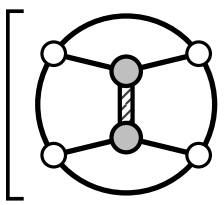 $\Big] + \dots\Big),$ (5.77)

where similarly to (5.59) we used that $u$-channel diagrams are related to the $t$-channel ones by a factor of $(-1)^J$. Now we will apply methods of Section 5.3 →p.117 and Section 5.3 →p.119 just focusing on $t$-channel diagrams of (5.77), and the subsequent addition of $u$-channel diagrams merely multiplies all of the squared OPE coefficients by $\big(1 \pm (-1)^J\big)$. We instantly see, that non-singlet sectors (ST)/(AS) contain only even/odd spins, respectively.

Decomposition of the $t$-channel exchange diagram into $t$-channel conformal blocks is clearly the same as the $s$-channel conformal block decomposition of the $s$-channel exchange diagram. Alternatively, the $t$-channel spectral function $\mathsf{Spec}_t$ of $t$-channel exchange diagram is given by the right-hand side of (5.61).

Even though (5.61) has two types of poles/operators — MFT ones originating from the $\Gamma^2$-factor and the non-MFT ones located at solutions of (5.65) — only the non-MFT poles/operators contribute to the $s$-channel spectral function $\mathsf{Spec}_s$ of $t$-channel exchange diagram. This was discussed after equation (5.34), which we now use to write the $s$-channel spectral function of the $t$-channel exchange diagram (including the $1/N$ prefactor) as

$$\frac{1}{N}\,\mathsf{Spec}_s\left[\begin{array}{c}\Delta\\ J\end{array}\right|$$
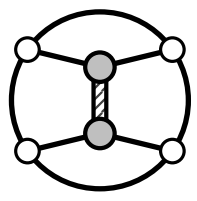
$$\Bigg] = \int_{\frac{d}{2}+i\mathbb{R}} \frac{\mathrm{d}\Delta'}{2\pi i}\,\mathsf{CrK}^{s\leftarrow t}_{\langle\Delta,J|\Delta',0\rangle}\, K_{\widetilde{\Delta}',0}\,\frac{1}{N}\,\mathsf{Spec}_t\left[\begin{array}{c}\Delta'\\ 0\end{array}\right|$$
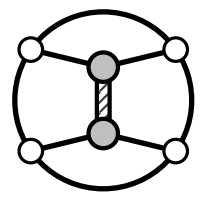
$$\Bigg] = \sum_{\mathcal{O}^{(\mathtt{S})}_{\bullet,0}} \mathsf{ope}^2\left[\mathcal{O}_\phi\mathcal{O}_\phi\mathcal{O}^{(\mathtt{S})}_{\bullet,0}\right]\mathsf{CrK}^{s\leftarrow t}_{\langle\Delta,J|\Delta^{(\mathtt{S})}_{\bullet,0},0\rangle}\,, \tag{5.78}$$

where the sum runs only over the non-MFT operators exchanged in the $t$-channel, all of which are scalar singlets $\mathcal{O}^{(\mathtt{S})}_{\bullet,0}$ with dimensions given by roots of (5.65) and corresponding squared OPE coefficients given by (5.71).

**Anomalous Dimensions in the Non-Singlet Sector.** We are precisely in the setting of Section 5.3 →p.119, and the formula for the anomalous dimensions of double-twist operators from the $t$-channel exchange (5.47) applied to the non-singlet sector of O($N$) model reads

$$\gamma^{(\mathtt{ST})/(\mathtt{AS})}_{n,J} = \sum_{\mathcal{O}^{(\mathtt{S})}_{\bullet,0}} \mathsf{ope}^2\left[\mathcal{O}_\phi\mathcal{O}_\phi\mathcal{O}^{(\mathtt{S})}_{\bullet,0}\right]\gamma^{(1)}_{n,J}\Big|_{\substack{t\text{-channel}\\ \text{exchange of } \mathcal{O}^{(\mathtt{S})}_{\bullet,0}}}\,, \tag{5.79}$$

where again only the non-MFT operators contribute. Definition of $\gamma^{(1)}_{n,J}$ was given in (5.42) and explicit expressions in $d = 2$ and $d = 4$ were presented in (5.43) and (5.44). Remember that the squared OPE coefficients (5.71) of the exchanged non-MFT operators are of order $O(1/N)$, and thus are also the anomalous dimensions $\gamma^{(\mathtt{ST})/(\mathtt{AS})}_{n,J}$. Even though the same formula (5.79) is applicable for both (ST)/(AS) sectors, only operators with even/odd spins $J$ actually appear in the (ST)/(AS) sector, respectively.

The whole next section is dedicated to the analysis of the $O(1/N)$ non-singlet spectrum corrections governed by (5.79).

## 5.5 Analysis of the Non-Singlet Sector

Methods established in previous sections allow us to compute the *complete non-singlet spectrum* occurring in the $\mathcal{O}_\phi^{\bullet} \times \mathcal{O}_\phi^{\bullet}$ OPE to the order $1/N$ considered. Almost all ingredients in the master formula (5.79) determining the non-singlet spectrum contributions are known analytically, exception being the precise values of singlet scalar scaling dimensions $\Delta^{(\mathtt{S})}_{\bullet,0}$, which need to be found numerically.

Therefore, in Section 5.5 →p.139 we first describe how we chose to truncate the sum initially going over infinitely many singlet operators. See the accompanying Notebook for the code implementing all of the calculations.

We then analyze the obtained non-singlet spectrum in a series of plots. The principal one — a twist–spin plot — is discussed in Section 5.5 →p.140 and verifies a known Regge trajectory structure of the spectrum. Asymptotic behaviors of two important sections of this plot are investigated further. First, Section 5.5 →p.145 studies the asymptotics at large twist for a fixed spin. Second, the large spin asymptotics for a fixed twist family is investigated in Section 5.5 →p.148, where a theorem determining the asymptotic behavior is verified as a consistency cross-check of the obtained spectral data.

Finally, dependence on the external scaling dimension and the coupling is examined in Section 5.5 →p.141. In particular, strong and weak coupling asymptotics are explored, from which it is recognized that scaling dimensions of (ST) scalar operators were not computed correctly due to a known limitation of the Lorentzian inversion formula entering the derivation of (5.79).

## Numerical Calculation of Anomalous Dimensions

As anticipated, the practical problem with the sum (5.79) is its infinite support on numerical solutions to (5.65) labeled by non-negative integers. The exchanged $t$-channel singlet $J' = 0$ operators will be labeled by $k$ as $\mathcal{O}^{(\mathtt{S})}_{k,0}$, and we reserve $n$ to label the non-singlet twist family whose anomalous dimensions we are computing. For simplicity, we exclude the possible contribution of the emergent operator in $d = 4$.

For numerical evaluation, we chose to simply truncate the sum over exchanged singlet operators at some maximal value $k_{\max}$. This corresponds to a pole of the $t$-channel spectral function of some maximal scaling dimension $\Delta^{(\mathtt{S})}_{k_{\max}}$. To be completely explicit, let us reproduce here the practical truncated formula for the non-singlet anomalous dimensions

$$\gamma^{(\mathtt{ST})/(\mathtt{AS})}_{n,J} = \sum_{k=0}^{k_{\max}} \mathsf{ope}^2\left[\mathcal{O}_\phi \mathcal{O}_\phi \mathcal{O}^{(\mathtt{S})}_{k,0}\right] \gamma^{(1)}_{n,J}\Big|_{\substack{t\text{-channel} \\ \text{exchange of } \mathcal{O}^{(\mathtt{S})}_{k,0}}} . \tag{5.80}$$

Before discussing its behavior on the choice of $k_{\max}$, or equivalently the convergence of original sum (5.79), let us make a remark about $\gamma^{(1)}_{n,J}$ provided in (5.43)/(5.44). Its evaluation sometimes involves performing a difference of two terms that are numerically huge but almost equal. This forced us to keep a high `WorkingPrecision` in the NOTEBOOK, namely we set it to a value of 100, sufficient for our range of computations.

We are still left with a legitimate question — how large should $k_{\max}$ be, such that the anomalous dimensions (5.80) of non-singlet operators are calculated with sufficient precision?

**Convergence.** Obtaining the asymptotic behavior of the summands in (5.79) is not straightforward — in particular for general $n$ — so we chose a more pedestrian approach. To assess the convergence, we plotted individual terms in (5.80) and analyzed their behavior for large $k$. See for example the $n = 0$ case shown in Figure 5.81→p.140.

By creating more examples of such log–log scale plots, where a power law falloff can be fitted conveniently, we observed an asymptotic behavior $\sim k^{-2(1+J)}$ for large $k$. Hence, the convergence improves drastically with increasing spin, and not many $t$-channel terms are required to obtain sufficiently precise results. Based on this analysis, we chose the value $k_{\text{max}} = 50$ for spinning ($J \geq 1$) non-singlet operators and $k_{\text{max}} = 200$ for scalar ($J = 0$) operators, where the convergence is much slower.

As will be explained later, (ST) operators with $J = 0$ are problematic for yet another reason — a limitation present already in the derivation of (5.79) — so we did not attempt to employ resummation techniques that would take care of the neglected tail of the sum.

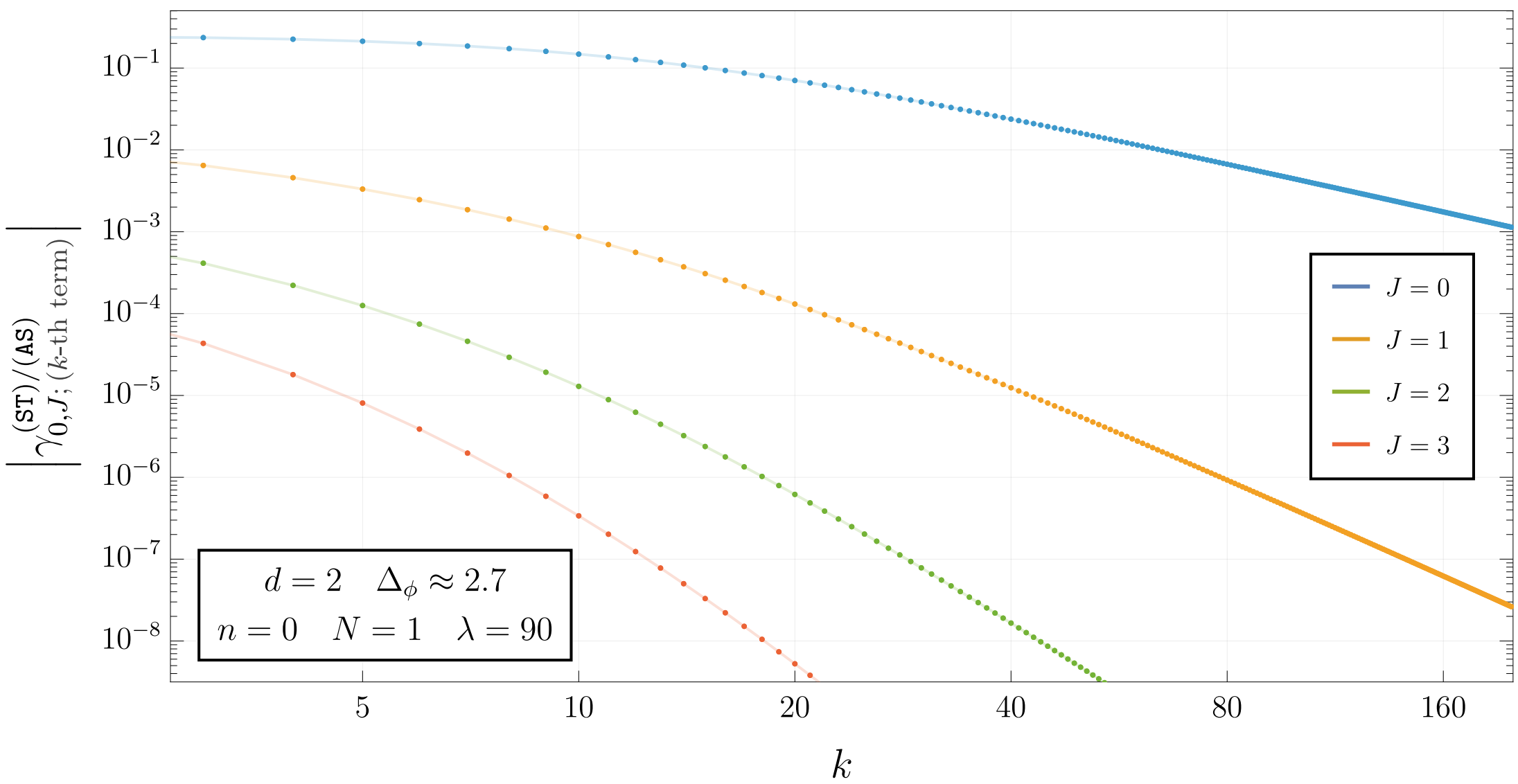

**Figure 5.81 /** Log–log plot of magnitudes of individual terms in (5.80) helping to assess the convergence of the sum calculating the non-singlet anomalous dimensions. Fixed external parameters are summarized in the boxes. For large $k$, the summands falloff as $\sim k^{-2(1+J)}$.

## Twists in the Non-Singlet Sector

Roughly since [107] [108] [109], it was known that the proper organizing principle for a CFT spectrum is in terms of twist/Regge trajectories labeled by $n \in \mathbb{N}_0$ that are analytic in spin. This idea was definitely confirmed by [99]. Therefore, we present the main plots of the non-singlet spectrum in the twist–spin plane, for a fixed value of the coupling and external scaling dimension. The plot for $d = 2$ ($\mathsf{AdS}_3$) is shown in Figure 5.83→p.142, while the one for $d = 4$ ($\mathsf{AdS}_5$) is

in Figure 5.84 →p.143.

To prepare the plots we had to choose also a particular value for $N$. It was fixed to a rather small unphysical value $N = \frac{1}{20}$, otherwise corrections of twists by anomalous dimensions (multiplied by a factor $1/N$) from their MFT values would be barely visible. So the functional dependence of the anomalous dimensions on $(n, J)$ is faithfully displayed, but their absolute value is exaggerated by extrapolating them far beyond the validity of the large $N$ expansion.

Both plots clearly exhibit the discussed organization of the spectrum into twist trajectories labeled by $n \in \mathbb{N}_0$ that are concave as a function of the spin $J$ and asymptote to the MFT spectrum for $J \to \infty$. This asymptotic shape is valid above some minimal spin $J_{\text{crit}}$ that depends on a concrete theory (in our case $J_{\text{crit}} = 1$).

These claims are by now a theorem valid in any CFT, which was established in a series of papers [108] [109] [110] [111] [112]. Applied to our setting, it states that if $\mathcal{O}^i_\phi$ is in the spectrum, then so must be (at least for $J \geq 1$) the double-twist families of schematic form $[\mathcal{O}^i_\phi \Box^n \partial^J \mathcal{O}^j_\phi]$.

For the MFT theory at $N \to \infty$, they all have twists $\tau_{n,J} \equiv \Delta_{n,J} - J = 2\Delta_\phi + 2n$. Deforming from the strict large $N$ limit, they receive anomalous dimensions and the large spin asymptotics of their twist trajectories was shown in (1.7) [108] and (1) [109] to be of the form

$$\tau_{n,J} \simeq 2\Delta_\phi + 2n - \frac{c_n}{J^{\tau_{\min}}} + \cdots . \tag{5.82}$$

This theorem serves us as a check of consistency done in Section 5.5 →p.148, where we also discuss which “minimal twist” operator governs the asymptotics in our case.

## Dependence of Anomalous Dimensions on Coupling

The main advantage of the large $N$ approach is that it outputs observables as exact functions of the coupling $\lambda$. We will exploit the known functional dependence in this section. However, we should emphasize that the coupling dependence is complicated — entering implicitly via numeric solutions to (5.65) — thus the best we can do is sample over a finite set of coupling values.

Of particular interest are the limiting cases. At which rate do the anomalous dimensions approach weak/strong coupling? Such question is best answered by a log–log plot, where a power law approach is captured by the slope of the graph.

We explore this question for $d = 2$ in Figure 5.85 →p.144. It shows the dependence of anomalous dimensions on the coupling for (ST)/(AS) operators with the lowest spins $J = 0/J = 1$ belonging to the first two leading Regge trajectories. As is clear from the plot, non-singlet anomalous dimensions approach a constant value at very strong coupling. On the other hand, the slope at very weak coupling is two, thus non-singlet anomalous dimensions start to decrease at a quadratic rate from their vanishing values in the free theory. This is expected for the $J = 1$ ($\mathtt{AS}$) operators, as the $O(\lambda)$ contact Witten diagram does not contribute to anomalous dimensions of $J \geq 1$ operators. However, we should see $O(\lambda)$ contribution to $J = 0$ operators, an issue to which we turn next.

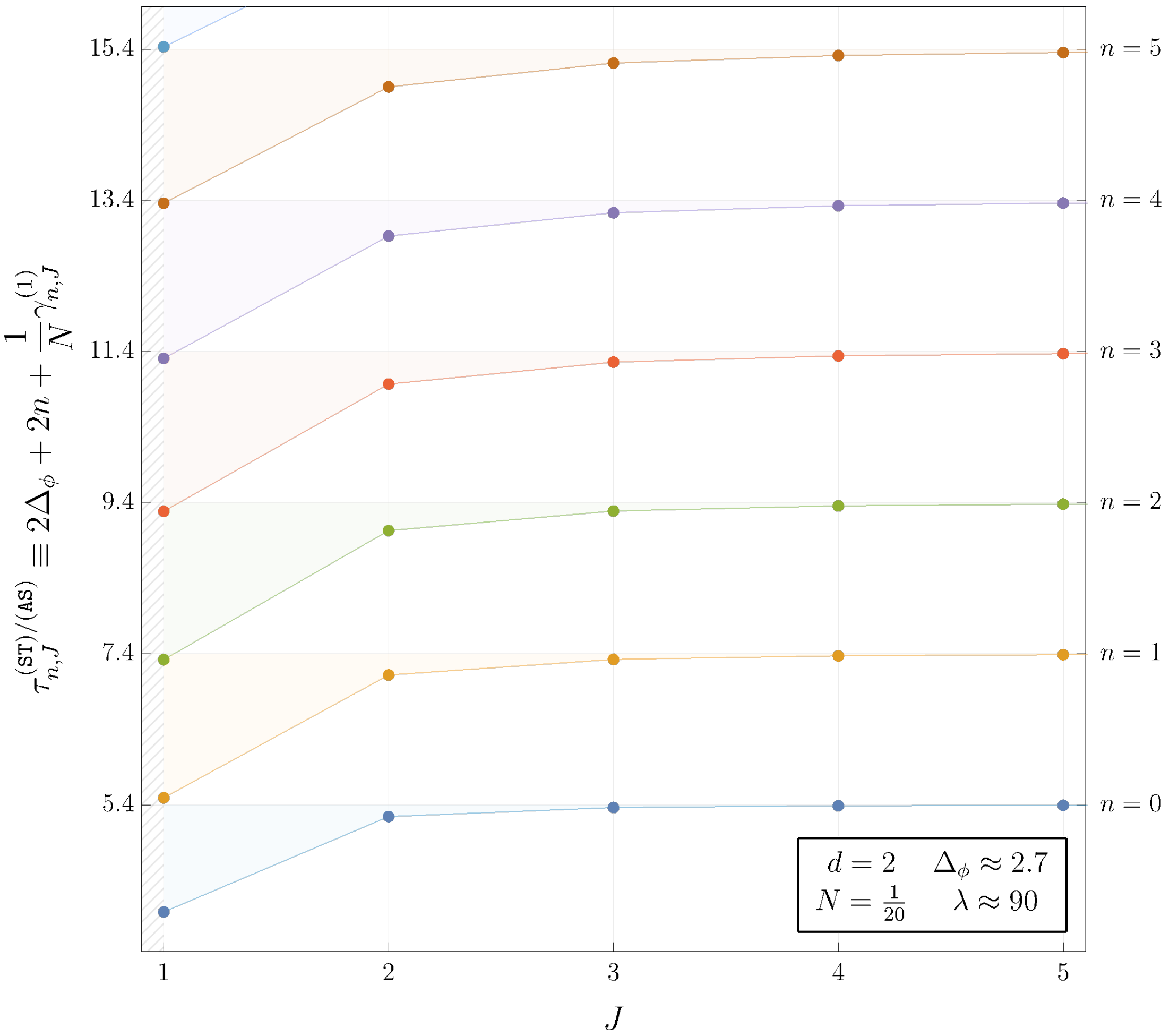

**Figure 5.83 /** The twist–spin plot of non-singlet spectrum in $d = 2$ at finite coupling $\lambda \approx 90$. The chosen dimension $\Delta_\phi$ corresponds to a massive $\phi$-field in $\mathsf{AdS}_3$ with Dirichlet boundary conditions. The plot shows operators in both non-singlet irreps of $\mathsf{O}(N)$ — ($\mathtt{ST}$) operators are supported on even spins, while ($\mathtt{AS}$) operators on odd spins. Anomalous dimensions for $J = 0$ are comparatively huge compared to those for $J \geq 1$, so we excluded the scalar operators from this plot (furthermore, they are not reliably computed by (5.79) due to a limitation in its derivation). The essential feature of the spectrum — its organization into Regge trajectories concave in spin — is discussed in the main text.

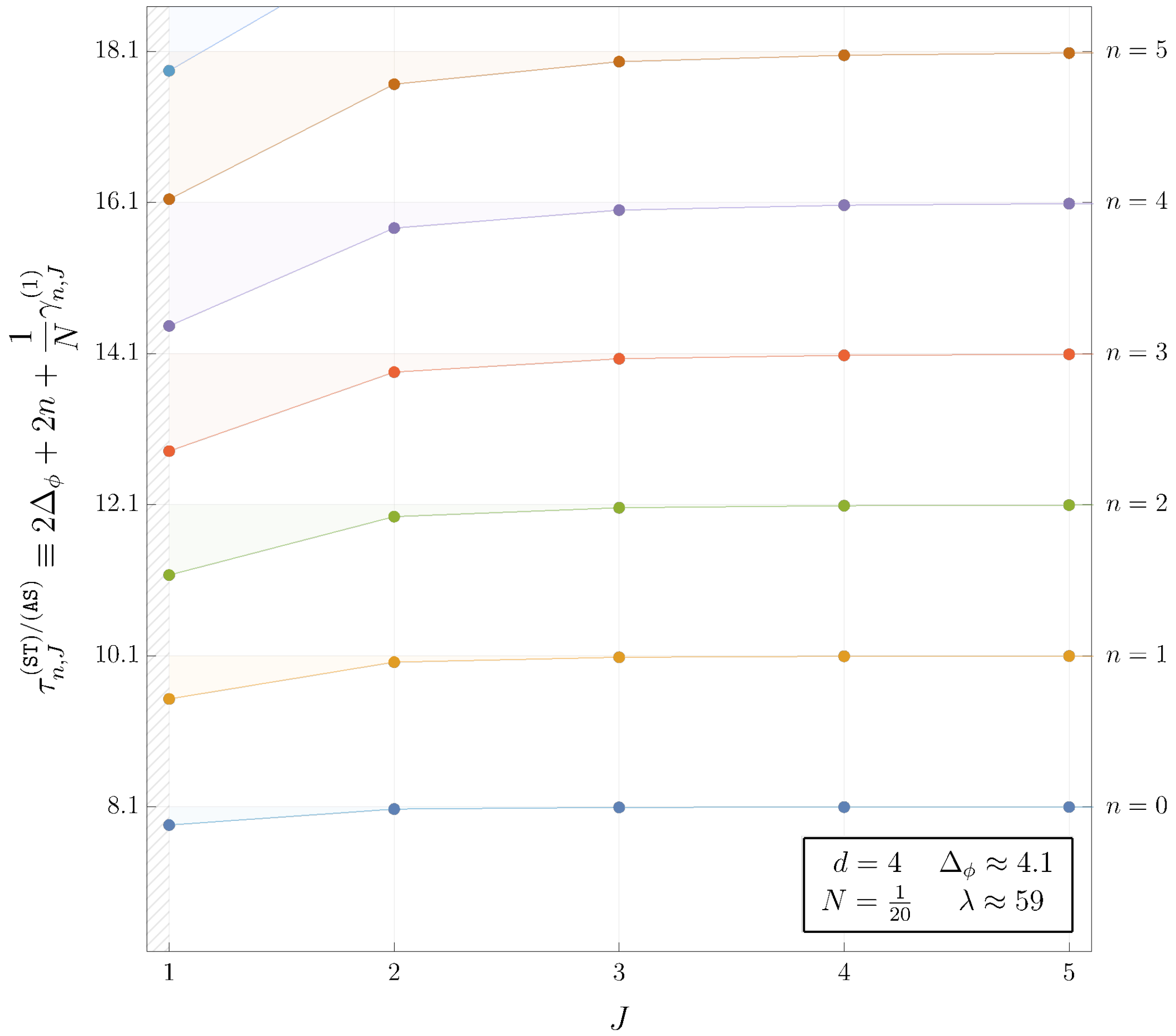

**Figure 5.84** / The twist–spin plot of non-singlet spectrum in $d = 4$ at finite coupling $\lambda \approx 59$. The chosen dimension $\Delta_\phi$ corresponds to a massive $\phi$-field in $\mathsf{AdS}_5$ with Dirichlet boundary conditions. The plot shows operators in both non-singlet irreps of $\mathsf{O}(N)$ — ($\mathtt{ST}$) operators are supported on even spins, while ($\mathtt{AS}$) operators on odd spins.

**Contribution of contact Witten diagram to ($\mathtt{ST}$) spin zero operators.** As we will review, the contact diagram contributes to anomalous dimensions of $J = 0$ ($\mathtt{ST}$) operators, thus their weak coupling asymptotics must be linear. It implies that the method on which we based our computations missed at least this contribution for this particular class of operators.

To locate the error, let us briefly comment on the Lorentzian inversion formula (LIF) that was used in [61] to derive the key formulas on which we build upon in Section 5.3 →p.114. Its purpose is to output the $s$-channel spectral density (encoding the anomalous dimensions we are after) from the $t$-channel scalar exchange Witten diagram.

The first step of LIF extracts from a given diagram (4-point correlator) its double-discontinuity, an operation that has a property of projecting out all contributions of MFT operators. However, since the conformal block decomposition of the contact diagram — the only $O(\lambda)$ contribution — contains solely MFT operators, it is completely annihilated by LIF. Below, in (5.87), we will provide its contribution to anomalous dimensions, which would otherwise be missed by LIF.

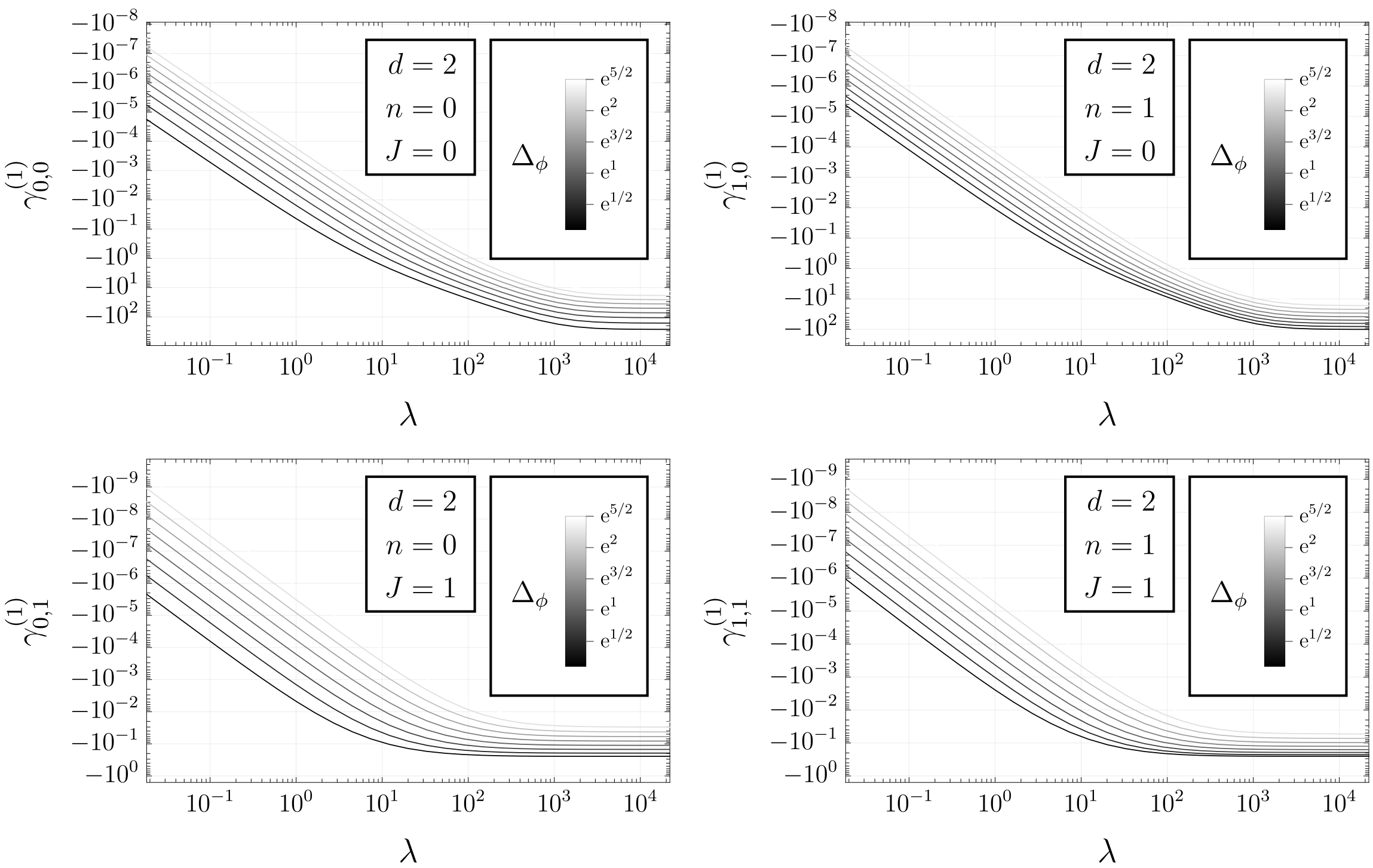

**Figure 5.85 /** Dependence of anomalous dimensions $\gamma^{(1)}_{n,J}$ on the coupling in $d = 2$. The top row shows the $J = 0$ operators in the (ST) irrep for the first two Regge trajectories — $n = 0$ and $n = 1$. The bottom row displays the same for the $J = 1$ operators in the (AS) irrep. Note, that the plots are in a log–log scale and cover a range of indicated external scaling dimensions starting at the Breitenlohner–Freedman bound, above which Dirichlet boundary conditions are applicable.

Higher order diagrams have a nonvanishing double-discontinuity, and are therefore taken into account by LIF. Still, there might be a subtlety connected with convergence. LIF integrates the double-discontinuity against a certain kernel and this integral might not converge. In fact, inverting $J' = 0$ operators in the $t$-channel, it is guaranteed to converge only for spin $J \geq 1$ operators in the $s$-channel. Yet, when extracting the anomalous dimensions from the spectral function computed by LIF, a brief comment was made in [61] that such combined operation should yield convergent integrals for all spins. If this is true, the contact diagram would be the only omission, and its inclusion in (5.88) would give correct anomalous dimensions also for $J = 0$ (ST) operators.

Having motivated the necessity of including the contact diagram contribution by hand, we proceed by recalling its contributions to anomalous dimensions. As is well known, it affects only scaling dimensions of spin zero operators, both (S) and (ST). Its conformal block decomposition determining also anomalous dimensions was studied in detail, and an explicit formula for (S) operators can be found for instance in (4.81), (4.85–4.87) [78]. Adapted to our conventions, the final expression reads (ignoring subleading $1/N$ corrections)

$$\gamma^{(\mathtt{S})\text{contact}}_{n,J=0} = \lambda \frac{1}{(4\pi)^{\frac{d}{2}} n!} \frac{\Gamma_{\frac{d}{2}+n}\,\Gamma_{\Delta_\phi+n}\,\Gamma_{\Delta_\phi-\frac{d}{2}+\frac{1}{2}+n}\,\Gamma_{2\Delta_\phi-\frac{d}{2}+n}}{\Gamma_{\frac{d}{2}}\,\Gamma_{\Delta_\phi+\frac{1}{2}+n}\,\Gamma_{\Delta_\phi-\frac{d}{2}+1+n}\,\Gamma_{2\Delta_\phi-d+1+n}} \tag{5.86}$$

where we used the dictionary $\lambda_{\text{[78]}} = \frac{2}{N}\lambda_{\text{us}}$ and $d_{\text{[78]}} = d_{\text{us}} + 1$.

Alternatively, we could obtain (5.86) without the explicit form of the contact diagram CB decomposition by looking at (5.65) in the regime of weak coupling, where to the first order only a single pole of bubble function (5.62) at $\Delta = 2\Delta_\phi + 2n$ plays a role. Ignoring other poles in $\widetilde{B}$ we immediately land at the same expression — easily checked by the identical appearance of Γ-factors — providing a successful cross-check.

By comparing the (ST)/(S) decompositions (5.57)/(5.55) and focusing only on terms up to $O(\lambda)$, we readily obtain the contact contribution for $J = 0$ (ST) operators as

$$\gamma^{(\mathtt{ST})\text{contact}}_{n,J=0} = \frac{2}{N}\gamma^{(\mathtt{S})\text{contact}}_{n,J=0}\,. \tag{5.87}$$

If the contact Witten diagram contribution is the only needed correction to LIF — which seems plausible based on comments in [61] — then the complete anomalous dimensions of $J = 0$ (ST) operators are given by

$$\gamma^{(\mathtt{ST})}_{n,J=0} = \gamma^{(\mathtt{ST})\text{LIF}}_{n,J=0} + \gamma^{(\mathtt{ST})\text{contact}}_{n,J=0} = \frac{1}{N}\left(\gamma^{(1)\text{LIF}}_{n,J=0} + 2\gamma^{(\mathtt{S})\text{contact}}_{n,J=0}\right) + O\left(\frac{1}{N^2}\right), \tag{5.88}$$

where the first piece is computed by LIF, see (5.79).

## Dependence of Anomalous Dimensions on Trajectory Label *n*

Inspecting a cut through Figure 5.83 →p.142 at fixed spin $J \geq 1$, it appears that the anomalous dimensions are growing (in absolute value) as a function of the twist trajectory label $n$. To establish their fate at large $n$, it is interesting to study the anomalous dimensions as a function of the twist trajectory label at fixed spin.

A representative plot of this dependence is shown in Figure 5.89 →p.146 for the case $d = 2$ and spin $J = 2$ corresponding to the Figure 5.83 →p.142 (for various values of coupling $\lambda$), but conclusions are generic. The anomalous dimensions (in the absolute value) increase at first until reaching a maximum for a critical twist family $n_*$ — the precise value depends on $\lambda$, for example in the case of Figure 5.83 →p.142 it is slightly higher than shown — and then start monotonically decreasing. This asymptotic fall off for very subleading Regge trajectories ($n \to \infty$) is in fact what one would expect for the asymptotically free O($N$) model in AdS$_3$.

Recalling (5.79), it is clear that dependence on $n$ resides only in the second piece inside the sum, given by (5.43) or (5.44). Those are known analytic functions whose large $n$ asymptotics can in principle be determined exactly.

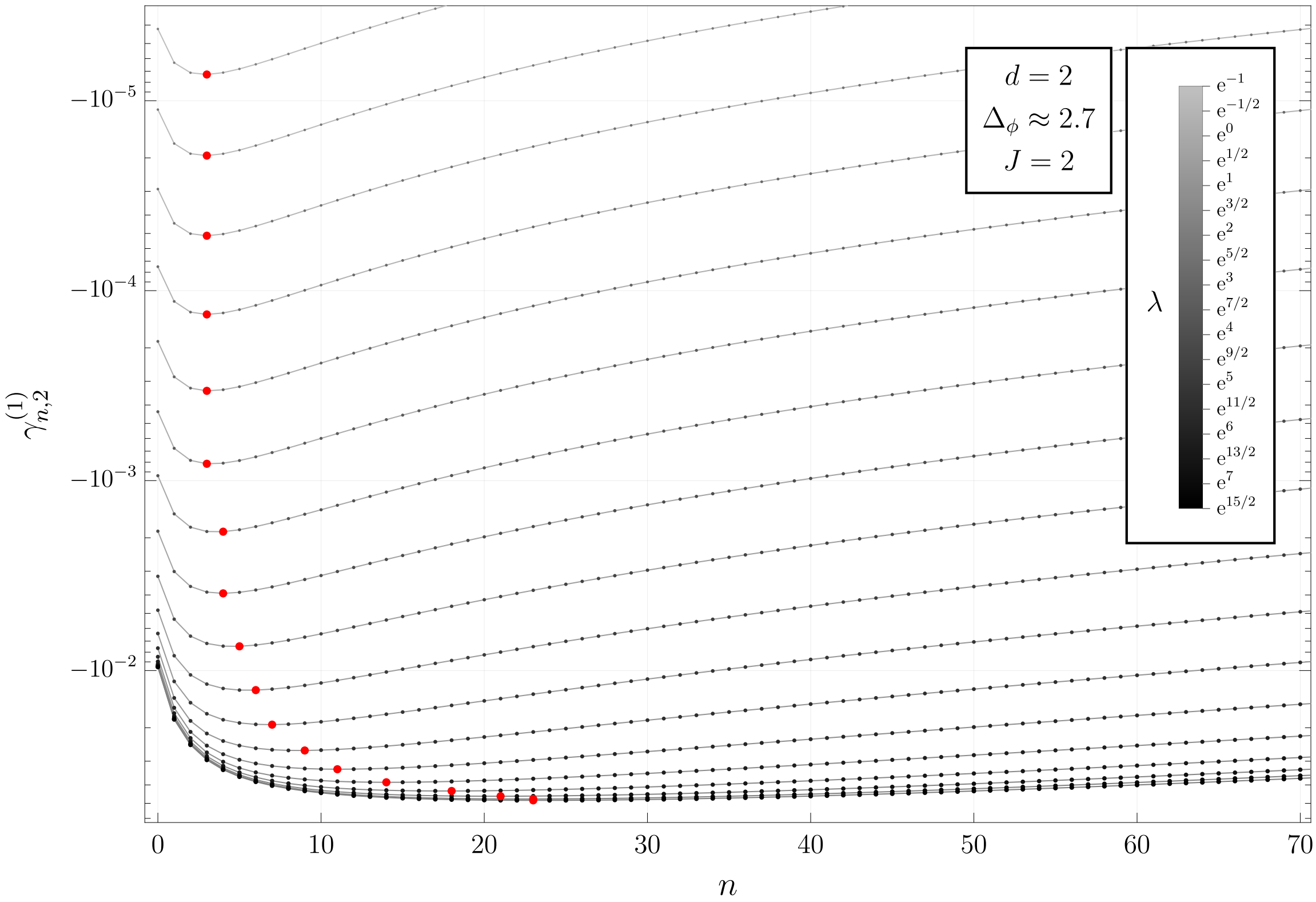

**Figure 5.89 /** Dependence of anomalous dimensions on the twist trajectory label $n$ in $d = 2$ for varying values of $\lambda$. A fixed spin $J = 2$ was chosen — operators are transforming in the (ST) irrep of O($N$) — and $\Delta_\phi$ corresponds to the value in Figure 5.83 →p.142. For the chosen parameters, the weak-coupling shape valid until $\lambda \approx e^2$ has the minimum at $n_* = 4$. For higher $\lambda$ we see a finite coupling deformation to take place, eventually (at $\lambda \approx e^7$) reaching a $\lambda \to \infty$ shape with the minimum at $n_* = 23$.

**Qualitative features of the *n*-dependence.** Dependence of anomalous dimensions on the Regge trajectory label $n$, in particular their aymptotic behavior for large $n$, was studied in detail in [113]. It should be emphasized that their analysis was at leading order at weak coupling, based on tree level unitarity in AdS. In fact, one of the most prominent suggestions for future directions, was extension to loops. In our case we work with a partially resummed perturbation series at large $N$, hence the details of their computations cannot be imported without modifications. Nevertheless, two main qualitative conclusions of their analysis should remain valid:

- The fall-off/growth of anomalous dimensions at large $n$ should be connected with relevance/irrelevance of interactions of the AdS theory understood as an EFT. They worked around free theory (Gaussian fixed point of the renormalization group) in AdS. The prime example in their paper was the 4-point function of light scalars $\phi$ with a tree level exchange of a heavy scalar $\chi$ governed by the interaction $g\phi^2\chi$. For instance in $\mathsf{AdS}_3$, canonical dimension of the relevant coupling $g$ is $[g] = \frac{3}{2}$ and the exchange diagram (only $s$-channel analyzed) has two powers of the coupling, thus the anomalous dimensions of the dual exchanged operators fall-off at large $n$ as $n^{-2[g]} = n^{-3}$. This simple dimensional analysis has to be modified in our case, as we do not work around a free AdS theory, but still the asymptotic large $n$ behavior of the anomalous dimensions should be connected with renormalized dimensions of operators/couplings in the AdS interaction Lagrangian.

- It was convincingly demonstrated on a number of examples that the large $n$ asymptotics (when the external dimensions are also taken large at the same time) of anomalous dimensions encodes the $S$-matrix of Minkowski spacetime when the flat space limit $\mathsf{AdS}_{d+1} \to \mathbb{R}^{1,d}$ is taken. To be precise, the asymptotic behavior of anomalous dimensions captures (up to precisely defined kinematic factors) partial waves associated with the flat space scattering amplitudes. Exchange of a particle — for example we can think about the heavy scalar $\chi$ — in AdS then manifests itself as a sharp peak in the $n$-dependence of anomalous dimensions.

At the moment, we can offer only a very preliminary study of the features mentioned above (without showing all the relevant plots):

**(1)** $\mathsf{AdS}_3$ ($d = 2$) — We observe peaks in the $n$-dependence of anomalous dimensions both for (ST)/(AS) (all couplings) and (S) (for sufficiently strong coupling), followed by a fall-off at large $n$. They are not as sharp as for the heavy scalar $\chi$ in [113], and based on the discussion above they should probably be interpreted as exchanges of 2-particle bound states associated with the large $N$ Hubbard-Stratonovich field $\sigma \sim \phi^2$.

**(2)** $\mathsf{AdS}_5$ ($d = 4$) — For (S)-sector anomalous dimensions we observe a monotonic growth that asymptotes to a constant value 1 at large $n$, no peaks seem to be present at any strength of the coupling. The case of (ST)/(AS) anomalous dimensions reaches a very shallow minimum/plateau followed by what seems to be a really slow fall-off of their absolute values at large $n$.

The above points should not be taken as conclusive results, but hopefully could help to guide potential future investigations. It would be interesting to make the connection between the large $n$ asymptotic behavior of the anomalous dimensions and (the partial wave expansion of) the flat space $S$-matrix at large $N$ more precise.

### Large Spin Asymptotics

Similarly to the previous subsection, also the large spin asymptotics — taking $n$ fixed — is fully specified by the second piece in (5.79), given by analytic formulas (5.43) and (5.44) for $d = 2$ and $d = 4$, respectively.

The large spin asymptotics were already derived in [108] [109], as presented in (5.82). The authors even computed the coefficient $c_0$ determining the leading-twist deviation from MFT at large spin (1.8) [108]. It is given in terms of OPE data associated with the operator of minimal twist $\tau_{\min} \equiv \Delta_{\min} - J_{\min}$ and spin $J_{\min}$. Since it turns out to be important for determining the correct "minimal twist" operator $\mathcal{O}^{(\mathtt{R})}_{\Delta_{\min},J_{\min}}$ governing the large spin asymptotics, we reproduce it here in the form

$$c_0 = \frac{2\,\Gamma_{\tau_{\min}+2J_{\min}}\,\Gamma^2_{\Delta_\phi}}{\Gamma^2_{\frac{\tau_{\min}}{2}+J_{\min}}\,\Gamma^2_{\Delta_\phi-\frac{\tau_{\min}}{2}}}\,\mathsf{ope}^2\Big[\mathcal{P}_{(\mathtt{R})}\Big(\mathcal{O}_\phi^{\bullet}\mathcal{O}_\phi^{\bullet}\Big)\,\mathcal{O}^{(\mathtt{R})}_{\Delta_{\min},J_{\min}}\Big]\,. \tag{5.90}$$

We take all operators involved to have 2-point functions normalized to unity, and compared to the cited formula we left out a factor of $2^{-J_{\min}}$ due to different normalization of conformal blocks.

Using (5.43)/(5.44) we were able to derive $c_n$ even for non-leading $(n > 0)$ twist families, see (5.49) for the expression in terms of $c_0$.

Prefactors in (5.90) can be traced back to (5.43)/(5.44), in particular $\Gamma^{-2}_{\Delta_\phi-\frac{\tau_{\min}}{2}}$ is related via Γ-function identities to $\sin^2\!\Big(\pi\Big[\Delta_\phi - \frac{\Delta'-J'}{2}\Big]\Big)$ after identification $\tau_{\min} \leftrightarrow \Delta' - J'$. These factors are responsible for producing double zeros at values of MFT operators, which will be important in the following.

Let us now recall possible candidates for the minimal twist operators, and discuss their respective orders at which they contribute in (5.90). The candidates are clearly the three minimal twist families $[\mathcal{O}^i_\phi \partial^J \mathcal{O}^j_\phi]^{(\mathtt{R})}$, $(\mathtt{R}) \in \{(\mathtt{S}), (\mathtt{AS}), (\mathtt{ST})\}$ (up to a possible emergent operator in $d = 4$ that would take over the role of a minimal twist operator at sufficiently strong coupling):

(S) The discussion in singlet sector splits into scalar and spinning operators (see Figure 5.69 →p.132).

- Scalar operators get $O(1)$ deformations from MFT scaling dimensions, thus their coefficients $c_n$ are $O(1/N)$ due to the squared OPE coefficients in the singlet sector (5.71).
- Spinning operators have MFT scaling dimensions (at $O(1)$ order), therefore their $c_n$ coefficients are strongly suppressed by the double zeros emphasized above. Given they receive some $O(1/N)$ anomalous dimensions by extending the computation of the correlator to order $O(1/N^2)$, their contributions to coefficient $c_n$ turn out to be $O(1/N^3)$ — a $O(1/N^2)$ suppression coming from the double-pole factor, and an additional $O(1/N)$ from their squared OPE coefficients.

(ST) (AS) Both types of non-singlet operators have $O(1/N)$ anomalous dimensions, therefore their contributions to $c_n$ coefficients are of order $O(1/N^2)$, since their leading OPE coefficients are $O(1)$.

The important corollary of this analysis is that although there are the (ST)/(AS) leading-twist families with $\tau^{(\mathtt{ST})/(\mathtt{AS})} < 2\Delta_\phi$ needed for unitarity of the boundary CFT, their contribution to large spin asymptotics (5.82) is strongly suppressed as $O(1/N^2)$. Hence, the large spin asymptotics are governed by the singlet scalar operator $\widehat{\sigma}_0 \sim [\mathcal{O}_\phi^\circ \mathcal{O}_\phi^\circ]^{(\mathtt{S})}$ of "minimal" non-MFT twist $\tau_{\min} = 2\Delta_\phi + \gamma_{0,0}^{(\mathtt{S})}$.

By evaluating (5.90, 5.49) for $\left(\tau_{\min} = 2\Delta_\phi + \gamma_{0,0}^{(\mathtt{S})}, J_{\min} = 0\right)$, the formula (5.82) provides a complete theoretical prediction for the large spin asymptotics of the (ST)/(AS) spectrum, that can be checked against data generated by our code.

We turn to describing the setup of these tests and presenting their results in the rest of this section. Starting from (5.82), taking its logarithm and plugging in the relation $\tau_{n,J} \simeq 2\Delta_\phi + 2n + \gamma_{n,J}$, we considered the following functional dependence for the asymptotics

$$\log_{10} |\gamma_{n,J}| \sim \log_{10} c_n - \tau_{\min} \log_{10} J\,. \tag{5.91}$$

The left-hand side was evaluated for $\log_{10} J \in [0, 4]$ with a step $\frac{1}{5}$, and last few data points were then fitted. The results for first two twist families are shown in Figure 5.92→p.150, but we performed the same analysis also for higher twist families with similar results.

The leading asymptotics $J^{-\tau_{\min}}$ given by the slope of the linear fit in these log–log plots is the same for all Regge trajectories. As explained, the value $\tau_{\min}^{(\text{fit})}$ determined by the fits should be compared against the scaling dimension $\Delta_{0,0}^{(\mathtt{S})}$ of the non-MFT "minimal" twist operator $\widehat{\sigma}_0$. We confirmed almost a perfect match, which only improves with higher $J$. Similarly, the coefficients $c_n$ show a good agreement with theoretical values (5.90), (5.49).

In spirit, this verifies that general formulas (5.43)/(5.44) have an appropriate form reproducing the large spin asymptotics theorem. The specific results for the O($N$) model then necessarily follow.

It is worthwhile to highlight another feature implemented in Figure 5.92→p.150. The black points represent "complete" anomalous dimensions (5.80) summed up to $k_{\max} = 20$, while the colored (blue, yellow, green) points correspond in order to contributions of the first few members $\{\widehat{\sigma}_0,\, \widehat{\sigma}_1,\, \widehat{\sigma}_2\}$ of the scalar singlet family. It is satisfying to visually see how the "minimal" twist operator $\widehat{\sigma}_0$ completely saturates the asymptotics, and how the subdominant contributions are suppressed at large $J$. Even if one resummed the whole infinite tower of remaining operators $\widehat{\sigma}_{n\geq 1}$, their asymptotics would still remain subleading compared to the one associated with $\widehat{\sigma}_0$.

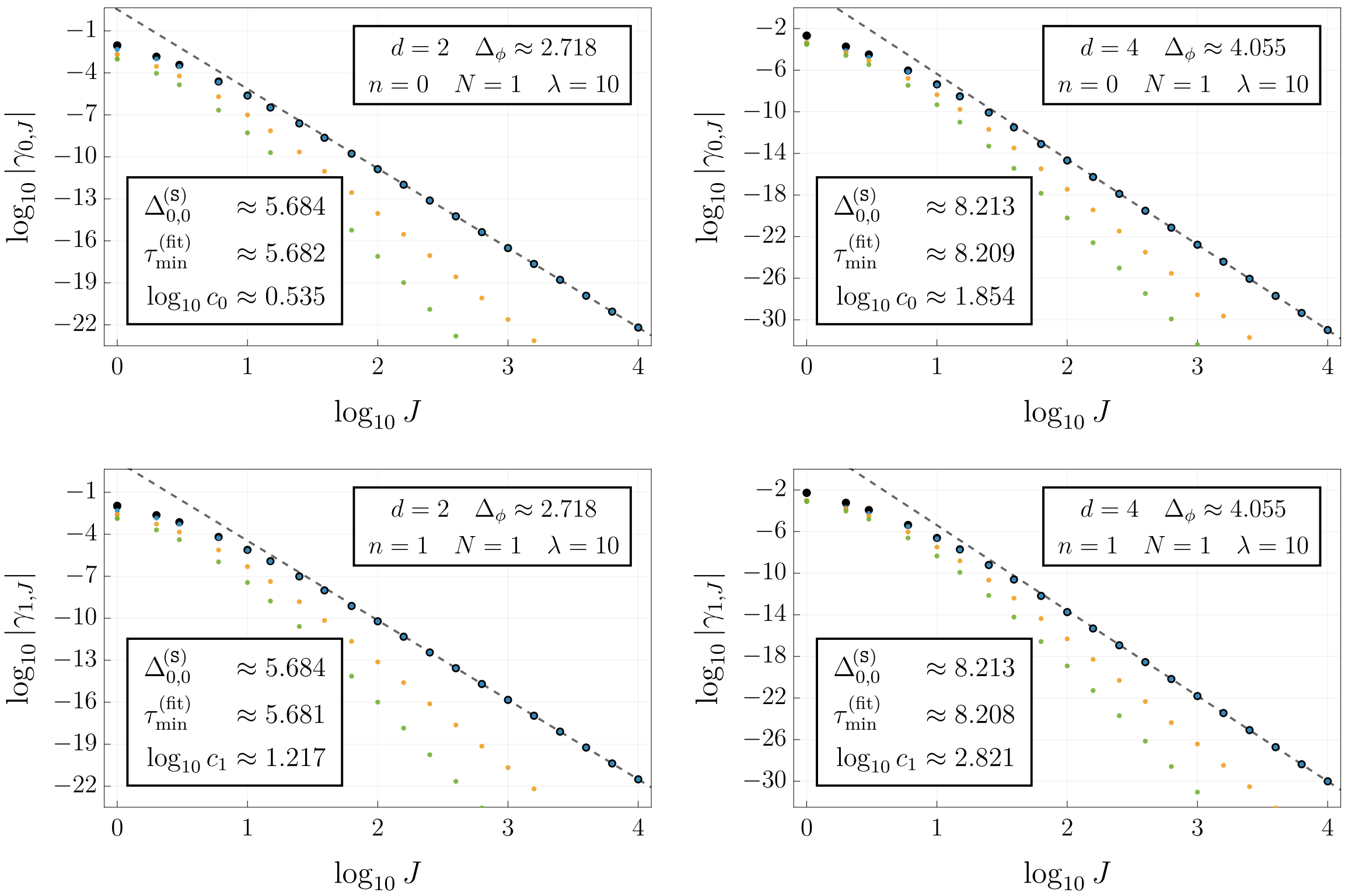

**Figure 5.92 /** Log–log plots displaying the large spin asymptotics of non-singlet anomalous dimensions for first two twist trajectories ($n = 0$ and $n = 1$) in $d = 2$ and $d = 4$. Choices of the parameters used in each plot are shown in the upper right boxes. The last few data (black) points were fitted by a linear function (5.91) (dashed line) with parameters given in the bottom left boxes. The slopes of dashed lines represent the calculated minimal twists, and are in good agreement with the theory value, thus verifying the large spin asymptotics theorem.

Finally, we should comment that the theorem is applicable for a generic $\mathsf{CFT}_d$ only in $d > 2$. The problem with $d = 2$ is that the stress tensor is in the same conformal Virasoro multiplet as the identity operator (it is its first nontrivial descendant at level two) and thus has the same twist $\tau = 0$. Without a twist gap between the identity and an operator of minimal twist, the theorem does not apply. Yet, we are dealing with a holographic theory without dynamical gravity in the bulk, and therefore with a non-local dual $\mathsf{CFT}_d$ on the boundary without a stress tensor. This implies that the whole Virasoro multiplet of the identity operator reduces just to the identity operator, and explains why we are able to use the asymptotic behavior (5.82) also in $d = 2$, which Figure 5.92 →p.150 confirms.

## 5.6 Summary and Outlook

This project arose from the desire to examine associativity of the OPE (crossing) in a CFT that is at finite distance (in the sense of its spectral and OPE data) from an MFT. Holographic theories are a fruitful playground, since especially in $d > 2$ there are not many other interacting examples that can be handled analytically, without resorting to the numerical conformal bootstrap.

We chose the O($N$) model at finite coupling in the bulk, as it is a reasonably simple theory and its input data for crossing, in the form of the singlet spectrum, were already available thanks to [59]. The set of techniques employed is applicable to any other theory as long as one has partial knowledge of its CFT data.

In the course of this research we obtained two classes of results which we summarize in the following. We remind the interested reader that various detailed computations and the implementation of main formulas can be found in the accompanying NOTEBOOK.

**Model-Independent — *t*-Channel CB Contribution to Anomalous Dimensions.** Imagine a situation when one has certain control of the CFT data in one channel, say the $t$-channel. To complete the analysis, one should try to deduce from it CFT data in the nontrivial crossed $s$-channel (as $u$-channel is related in a simple way). In particular, for analyzing the CFT spectrum it is important to know how a single $t$-channel conformal block contributes to $s$-channel anomalous dimensions. This contribution needs to be further weighted by $t$-channel (squared) OPE coefficients to obtain the final result for $s$-channel anomalous dimensions, hence both scaling dimensions and OPE coefficients are needed as input in one of the crossing channels.

This computation requires the knowledge of conformal blocks, whose analytic expressions are known in $d = 2$ and $d = 4$ dimensions. For these spacetime dimensions, [61] provided a closed-form formula for anomalous dimensions of leading-twist ($n = 0$) double-twist operators $[\mathcal{O}^{\bullet}_{\phi}\mathcal{O}^{\bullet}_{\phi}]_{n,J}$, once CFT data associated with the exchange of an operator in the crossed channel are specified. In this paper we present the generalization of these formulas to arbitrary twists $n \in \mathbb{N}_0$, namely see (5.43) and (5.44). Moreover, we found an elegant formula (5.49) for the coefficients of the corresponding large spin asymptotics.

Quite generally, once the direct channel spectrum is resolved, contributions from the crossed channel can be incorporated by the techniques presented in Section 5.3→p.114. We have demonstrated them in the context of the O($N$) model, but they are applicable also for other CFTs, perhaps after some generalization. The closest candidates are CFTs on the boundary of AdS in the large $N$ expansions of the Gross–Neveu model or the scalar QED.

**Model-Specific — Non-Singlet Spectrum of O(N) Model at Finite Coupling.** The main output of this paper is the structure of the OPE of two boundary operators $\mathcal{O}_\phi^\bullet$ associated with the fundamental fields $\phi^\bullet$ in the bulk, considering the phase with unbroken $\mathsf{O}(N)$ global symmetry. It was obtained at a finite coupling $\lambda$ and up to the first nontrivial order of $1/N$ in the large $N$ expansion.

It can be viewed as a deformation of the two limiting cases where the quartic interaction in the action (5.1) is turned off — either by setting $N=\infty$ or $\lambda=0$. In both cases, all correlators are just products of two-point functions, and the corresponding MFT OPE can be schematically decomposed into $\mathsf{O}(N)$ irreps (denoted by the superscripts) as

$$\begin{aligned}
\mathcal{O}_\phi^i \times \mathcal{O}_\phi^j &\overset{N=\infty}{\approx} \mathbf{1}^{(\mathtt{S})} \oplus \left[\mathcal{O}_\phi^{[i}\Box^n\partial^J_{\text{odd}}\mathcal{O}_\phi^{j]}\right]^{(\mathtt{AS})}_{\text{MFT}} \oplus \left[\mathcal{O}_\phi^{\{i}\Box^n\partial^J_{\text{even}}\mathcal{O}_\phi^{j\}}\right]^{(\mathtt{ST})}_{\text{MFT}}, \\
\mathcal{O}_\phi^i \times \mathcal{O}_\phi^j &\overset{\lambda=0}{\approx} \mathbf{1}^{(\mathtt{S})} \oplus \frac{1}{\sqrt{N}}\left[\mathcal{O}_\phi^\bullet\Box^n\partial^J_{\text{even}}\mathcal{O}_\phi^\bullet\right]^{(\mathtt{S})}_{\text{MFT}} \oplus \\
&\qquad \oplus \left[\mathcal{O}_\phi^{[i}\Box^n\partial^J_{\text{odd}}\mathcal{O}_\phi^{j]}\right]^{(\mathtt{AS})}_{\text{MFT}} \oplus \left[\mathcal{O}_\phi^{\{i}\Box^n\partial^J_{\text{even}}\mathcal{O}_\phi^{j\}}\right]^{(\mathtt{ST})}_{\text{MFT}}.
\end{aligned} \tag{5.93}$$

Both of them are obtained from the disconnected parts of the 4-point function decomposition in (5.55, 5.56, 5.57). As the singlet sector double-twist operators have a factor of $1/N$ in their squared OPE coefficients, OPE coefficients are of order $O(1/\sqrt{N})$. For $N=\infty$ they vanish, and the singlet double-twist operators are transferred fully to the $(\mathtt{ST})$ part, see also (5.52).

At finite $\lambda$ and large (but finite) $N$, this picture is modified — only leading corrections in large $N$ are shown — to the following schematic form

$$\begin{aligned}
\mathcal{O}_\phi^i \times \mathcal{O}_\phi^j \sim \mathbf{1}^{(\mathtt{S})} &\oplus \frac{1}{\sqrt{N}}\left[\widehat{\sigma}_\bullet \sim \text{“}\mathcal{O}_\phi^\bullet\Box^n\mathcal{O}_\phi^\bullet\text{”}\right]^{(\mathtt{S})}_{\text{fin}} \oplus \frac{1}{\sqrt{N}}\left[\mathcal{O}_\phi^\bullet\Box^n\partial^{J>0}_{\text{even}}\mathcal{O}_\phi^\bullet\right]^{(\mathtt{S})}_{\text{MFT}} \oplus \\
&\oplus \left[\mathcal{O}_\phi^{[i}\Box^n\partial^J_{\text{odd}}\mathcal{O}_\phi^{j]}\right]^{(\mathtt{AS})}_{1/N} \oplus \left[\mathcal{O}_\phi^{\{i}\Box^n\partial^J_{\text{even}}\mathcal{O}_\phi^{j\}}\right]^{(\mathtt{ST})}_{1/N}.
\end{aligned} \tag{5.94}$$

Discussion of singlet operators in the first line of (5.94) splits according to the spin. Scalar boundary operators $\widehat{\sigma}_\bullet$ are induced by the composite interacting $\sigma$-field in the bulk — they are given by the poles of the spectral function associated with the exact $\sigma$-propagator. As indicated, their scaling dimensions get finite shifts from MFT, and their OPE coefficients are $O(1/\sqrt{N})$ to the order considered. Spinning singlet operators do not receive such $O(1)$ corrections to their MFT scaling dimensions. The formula for scaling dimensions of the $\mathsf{O}(N)$ singlet sector was already given in [59].

The non-singlet operators in the second line of (5.94) get $1/N$ shifts to their scaling dimensions. Leading OPE coefficients in this sector are of order $O(1)$, and acquire $O(1/\sqrt{N})$ deformations. Both of these corrections correspond to a $1/N$ modification of the correlator. In this work we did not compute the corrections to the OPE coefficients, even though the setup is ready for it. It is just a matter of computing residues of fairly complicated functions at known poles, and we might come back to it in the future.

The main new contribution of this work are the scaling dimensions for the non-singlet double-twist families transforming in the symmetric traceless (ST) and anti-symmetric (AS) irreps of the O($N$) group. They were obtained for $d = 2$ ($AdS_3$) and $d = 4$ ($AdS_5$), as a function of the twist/Regge trajectory label $n$, the spin $J$, the external scaling dimension $\Delta_\phi$, and the coupling $\lambda$.

It is important to realize that they are given as a sum of partly factorized expressions — the model-dependent squared OPE coefficients, times a model-independent contribution of a crossed channel block, which however still needs to be evaluated at model-dependent values. The relevant formula is (5.79), which we reproduce here with a particular emphasis on the functional dependence

$$\gamma_{n,J}^{(\mathtt{ST})/(\mathtt{AS})}(\Delta_\phi,\lambda) = \sum_{\mathcal{O}^{(\mathtt{S})}_{\bullet,0}(\Delta_\phi,\lambda)} \mathsf{ope}^2\Big[\mathcal{O}_\phi\mathcal{O}_\phi\mathcal{O}^{(\mathtt{S})}_{\bullet,0}\Big](\Delta_\phi,\lambda)\,\gamma^{(1)}_{n,J}(\Delta_\phi)\Big|_{\substack{t\text{-channel}\\ \text{exchange of } \mathcal{O}^{(\mathtt{S})}_{\bullet,0}(\Delta_\phi,\lambda)}} . \tag{5.95}$$

The dependence on the external scaling dimension $\Delta_\phi$ and the coupling $\lambda$ is complicated, in particular due to an implicit dependence in $\gamma^{(1)}$ via solutions of a transcendental equation (5.65). However, the dependence on the Regge trajectory label $n$ and the spin $J$ is completely isolated in the model-independent piece. Especially, asymptotic behavior for large $J$ or $n$ can be in principle analytically determined from (5.43) or (5.44), and is displayed in Figure 5.92 →p.150 and Figure 5.89 →p.146. The large spin asymptotics are verified by a theorem independently predicting it.

The most significant projection of these data — after choosing some fixed coupling and desired external scaling dimension — is a plot in the twist–spin plane. A combined plot of the anti-symmetric and symmetric traceless double-twist families is shown in Figure 5.83 →p.142 for $d = 2$, and in Figure 5.84 →p.143 for $d = 4$. Both plots confirm organization of the non-singlet double-twist spectrum into Regge trajectories concave in spin.

Unlike in the singlet sector where anomalous dimensions are strictly positive for $J = 0$ operators — indicating a repulsive interaction in the bulk — and zero otherwise, anti-symmetric and symmetric traceless double-twist families have negative anomalous dimensions — indicating an attractive interaction in the bulk.

In the process of computing non-singlet anomalous dimensions, the (squared) OPE coefficients (5.71) in the scalar singlet sector were needed. We plotted them in Figure 5.72 →p.134 as a function of the coupling. In the limit $\lambda \to 0$ they approach MFT values as they must, which provides another successful test of implementation. Compared to anomalous dimensions displaying fixed sign and convexity properties, OPE coefficients show a rather unorganized behavior.

Different $\widehat{\sigma}_{\bullet}$ operators get corrected by different signs and their relative strength changes with the coupling.

Based on the self-consistency scrutiny of the spectrum summarized in the previous paragraphs and also additional checks done in [59] we believe that the results are reliable and ready to be used in applications (at least in $d = 2$ which has a well understood phase structure and no UV divergence).

**Future Directions.** There are still various unresolved questions and possible future directions. One concerns the anomalous dimensions of $J = 0$ (ST) operators, for which we have indications that they were not properly captured by our general method of computation. While it is possible to manually take into account the missed contribution of the contact Witten diagram, see (5.88), it is still unclear whether 6j–symbol (employing under the hood the Lorentzian inversion formula) is not missing some other contributions.

Another challenge (of a technical nature) is the computation of the OPE coefficients for the non-singlet operators. Since in principle we can calculate anomalous dimensions in the non-singlet sector for any $n$ and $J$, this opens a possibility to perform a more detailed analysis of the spectrum and its properties.

Following [114], it would be interesting to calculate the non-singlet spectrum of the $\mathsf{O}(N)$ model in *de Sitter* (dS) background.

A further valuable extension of our work concerns the critical theory in the bulk. Its careful investigation particularly attracts our attention for a future project — either in the $\mathsf{BCFT}_{d+1}$ setting [78] linked directly to $\mathsf{AdS}_{d+1}$ or in the $\mathsf{DCFT}_{n+m}$ setting [70] corresponding to a theory on $\mathsf{AdS}_n \times \mathbb{S}^m$ (and requiring thus the Kaluza–Klein reduction on the sphere). We hope to report on some additional observables in these theories that could be inferred from the results obtained in this work.

# Conclusion

Since we have covered a broad range of topics in this thesis, let us take a moment to summarize the main narrative.

In Chapter 1 → p.1, we began by reviewing the QFT formalism, taking the path integral as a powerful tool for both constructing and analyzing QFTs. We then formalized its key features within the FQFT framework, which gave us a firmer grip on the formulation of QFTs in general backgrounds. This allowed us to better appreciate the role of symmetries, and clarified the conceptual meaning of various operator formalisms associated with different spacetime foliations.

In Chapter 2 → p.37, we discussed various types of effective actions. In particular, we introduced the notion of the RG flow, which has profound implications for the structure of QFTs and physics in general. It provides a framework for understanding physical behavior at different energy scales, and explains the origin of universality observed at second-order phase transitions of statistical systems, which correspond to the fixed points of the RG flow described by CFTs.

In Chapter 3 → p.51, we built the foundations of $(d \geq 3)$ CFTs from the ground up. We first looked at the conformal structure of manifolds in general, and later we focused on flat space. Identifying the conformal group and algebra, we studied their representations and the initial implications of symmetry for correlation functions. Finally, a particularly important feature of CFT was discussed — the existence of state–operator correspondence, and the Operator Product Expansion.

In Chapter 4 → p.91, we turned our attention to the AdS space. Since the isometry group of AdS acts on its asymptotic boundary as a flat-space conformal group, QFTs in AdS admit a set of asymptotic observables that are conformally invariant, so much of the CFT methodology can be used to study them. This falls under the umbrella of the AdS/CFT correspondence, which more generally (after taking the bulk metric to be dynamic) attempts to define a Quantum Gravity in asymptotically AdS spacetime through a dual (holographic) CFT living on its boundary.

In Chapter 5 → p.100, we studied the $\mathsf{O}(N)$ model in AdS at finite coupling in the large $N$ expansion. Much of the formalism developed in the previous chapters was used, either directly or indirectly. The main focus was to extend the known results from the analysis of the singlet sector also to the non-singlet sector, thereby obtaining a more complete picture of the spectrum at order $1/N$. For a thorough summary of the results, we refer to Section 5.6 → p.151.

The ideas presented here reflect a deep interplay between various fields, such as *Mathematics*, *Quantum Field Theory*, *Statistical Physics*, and *Gravity*. I hope you found it as fascinating as I did.

# Notation & Conventions

## Miscellaneous

Imaginary unit is denoted as $\imath \equiv \sqrt{-1}$ (taking principal square root), and the Euler's number or natural logarithm base as "e".

Throughout we use units where the speed of light is unity. Sometimes, we also set reduced Planck's constant to be $\hbar \equiv 1$.

We will work in the "mostly plus" signature, that is in $\mathbb{R}^{1,d-1}$ we have

$$\eta^{(1,d-1)}_{\bullet\bullet} = \mathrm{diag}(-1, \overbrace{+1, +1, \ldots, +1}^{d-1}),$$

and in $\mathbb{R}^{d} \equiv \mathbb{R}^{0,d}$ we have

$$\eta^{(0,d)}_{\bullet\bullet} = \mathrm{diag}(\overbrace{+1, +1, \ldots, +1}^{d}).$$

Corresponding Lorentz groups are $\mathsf{SO}(1, d-1)$ and $\mathsf{SO}(d) \equiv \mathsf{SO}(0, d)$, respectively.

## Differential Geometry

While we assume familiarity with various concepts from differential geometry, including Lie groups, we briefly summarize the notation we will use in this thesis.

Tangent bundle is denoted as $\boldsymbol{T}\mathcal{M}$, the cotangent bundle as $\boldsymbol{T}^*\mathcal{M}$, and the bundle of $(p,q)$-tensors is denoted as $\boldsymbol{T}^p_q\mathcal{M}$. Corresponding vector, covector, and tensor fields are denoted as $\mathcal{T}\mathcal{M} \equiv \mathrm{Sect}\,\boldsymbol{T}\mathcal{M}$, $\mathcal{T}^*\mathcal{M} \equiv \mathrm{Sect}\,\boldsymbol{T}^*\mathcal{M}$, and $\mathcal{T}^p_q\mathcal{M} \equiv \mathrm{Sect}\,\boldsymbol{T}^p_q\mathcal{M}$, respectively.

Bold indices denote abstract indices [42] [115], for example $\boldsymbol{\xi}^{\boldsymbol{a}} \in \mathcal{T}\mathcal{M}$, $\boldsymbol{g}_{\boldsymbol{ab}} \in \mathcal{T}_2\mathcal{M}$. Repeated indices denote contraction. Indices are raised or lowered automatically using the metric tensor $\boldsymbol{g}_{\bullet\bullet}$ or its inverse $\boldsymbol{g}^{\bullet\bullet}$, for example $\boldsymbol{A} \cdot \boldsymbol{B} \equiv \boldsymbol{A}_{\boldsymbol{a}}\boldsymbol{B}^{\boldsymbol{a}} \equiv \boldsymbol{g}_{\boldsymbol{ab}}\boldsymbol{A}^{\boldsymbol{a}}\boldsymbol{B}^{\boldsymbol{b}} = \boldsymbol{g}^{\boldsymbol{ab}}\boldsymbol{A}_{\boldsymbol{a}}\boldsymbol{B}_{\boldsymbol{b}}$.

Indices which are not bold denote components of a tensor in a given coordinate system (choice of which is usually obvious from context), for example $g_{ab} \equiv \boldsymbol{g}_{ab}$. Repeated indices denote summation.

As already shown, we sometimes omit abstract/coordinate indices, and indicate the contraction/summation by a dot $\bullet/\cdot$. For example, we have $x^2 \equiv x \cdot x \equiv x_a x^a$, $\boldsymbol{g}_{\boldsymbol{ab}} \equiv g_{kl}\,\mathbf{d}_{\boldsymbol{a}}x^k\,\mathbf{d}_{\boldsymbol{b}}x^l \Leftrightarrow \boldsymbol{g} \equiv g_{kl}\,\mathbf{d}x^k\,\mathbf{d}x^l$ where $g_{kl} \equiv \boldsymbol{g}\Big(\frac{\boldsymbol{\partial}}{\partial x^k}, \frac{\boldsymbol{\partial}}{\partial x^l}\Big)$, or $\boldsymbol{\xi}^{\boldsymbol{a}} \equiv \xi^k \frac{\boldsymbol{\partial}^{\boldsymbol{a}}}{\partial x^k}$ with $\xi^k \equiv \boldsymbol{\xi}^k \equiv \boldsymbol{\xi} \cdot \mathbf{d}x^k \equiv \mathbf{d}x^k(\boldsymbol{\xi})$.

When we want to indicate the type of the object, or the place where some argument/object should be inserted, we use for example the following placeholders whose meaning should be clear from the context — $\boldsymbol{a}^{\bullet}$, $\boldsymbol{b}_{\bullet}$, $A_{\circ}$, $[\![\bullet, \bullet]\!]$.

The generator of the flow $\phi_{\circ *}$ (tangent vector field to the curves generated by the flow) is defined as $\boldsymbol{\xi} \equiv \frac{\mathbf{D}}{\mathrm{d}\varepsilon}\phi_\varepsilon|_{\varepsilon=0} \equiv \frac{\mathbf{D}\phi_\varepsilon}{\mathrm{d}\varepsilon}|_{\varepsilon=0} \in \mathcal{TM}$, where we distinguished $\mathbf{D}$ to be capital and bold — compared to the ordinary derivative — to indicate that the resulting object is a vector field.

Metric density (of weight 1) is denoted as $\mathfrak{g}^{1/2} \equiv |\mathrm{Det}\,\boldsymbol{g}|^{1/2} \in \mathfrak{Dens}^1\mathcal{M}$, which in given coordinate system can be expressed as $\mathfrak{g}^{1/2} = \sqrt{|\det \boldsymbol{g}_{\bullet\bullet}|}\,\mathrm{d}^d x$, where $\mathrm{d}^d x \equiv \left|\mathbf{d}x^0 \wedge \mathbf{d}x^1 \wedge \ldots \wedge \mathbf{d}x^{d-1}\right|$ is the coordinate density.

In the "tetrad" formalism, we will denote the vector field frame as $\{\boldsymbol{e}_k^{\boldsymbol{a}}\} \subset \mathcal{TM}$, where $\boldsymbol{a}$ is the abstract index and $k$ is the "enumerating" index. The dual (covector) tetrad frame will be denoted as $\{\boldsymbol{e}^k_{\boldsymbol{a}}\} \subset \mathcal{T}^*\mathcal{M}$. It is chosen such that $\boldsymbol{e}_k^{\boldsymbol{a}}\boldsymbol{e}^l_{\boldsymbol{a}} = \delta^l_k$ and $\boldsymbol{e}_k^{\boldsymbol{a}}\boldsymbol{e}^k_{\boldsymbol{b}} = \boldsymbol{\delta}^{\boldsymbol{a}}_{\boldsymbol{b}}$.

Frame is always taken to be orthonormal, that is $\boldsymbol{g}_{\boldsymbol{ab}} = \eta_{ij}\boldsymbol{e}^i_{\boldsymbol{a}}\boldsymbol{e}^j_{\boldsymbol{b}}$, where $\eta_{ij}$ are coordinates of the canonical metric in $\mathbb{R}^{p,q}$ (with corresponding signature). We can easily see (using $|\det \eta_{\bullet\bullet}| = 1$) that metric density is then $\mathfrak{e} = \mathfrak{g}^{1/2} = |\mathrm{Det}\,\boldsymbol{e}^{\bullet}_{\bullet}| \in \mathfrak{Dens}^1\mathcal{M}$.

We will always use Levi-Civita connection $\boldsymbol{\nabla}$, which is torsion-free ($\mathsf{Tor}[\boldsymbol{\nabla}] = 0$) and metric-compatible ($\boldsymbol{\nabla}\boldsymbol{g} = 0$). Alternatively, we use $\boldsymbol{\partial}$ when acting on scalars, since their covariant derivative does not depend on the connection.

# References

Back-references to the pages where the publication was cited are given by [•].